\documentclass[twoside,11pt]{Latex/Classes/PhDthesisPSnPDF}

\newcommand{\figuremacroW}[4]{
	\begin{figure}[!t]
		\centering
		\includegraphics[width=#4\textwidth]{#1}
		\caption[#2]{\textbf{#2} - #3}
		\label{#1}
	\end{figure}
}

\usepackage{rotating}
\usepackage{Latex/StyleFiles/ctable}
\usepackage[final]{pdfpages}

\ifpdf  
    \pdfcatalog { /PageMode (/UseOutlines)
                  /OpenAction (fitbh)  }
\fi

\title{Asteroseismology: Data Analysis Methods and Interpretation for Space and Ground-based Facilities}

\ifpdf
  \author{Tiago L.~Campante}
  \collegeordept{Faculdade de Ci\^encias e Centro de Astrof\'isica da Universidade do Porto}
  \collegeordepttwo{Institut for Fysik og Astronomi, Aarhus Universitet, Danmark}

  \crest{\includegraphics[width=5.5cm]{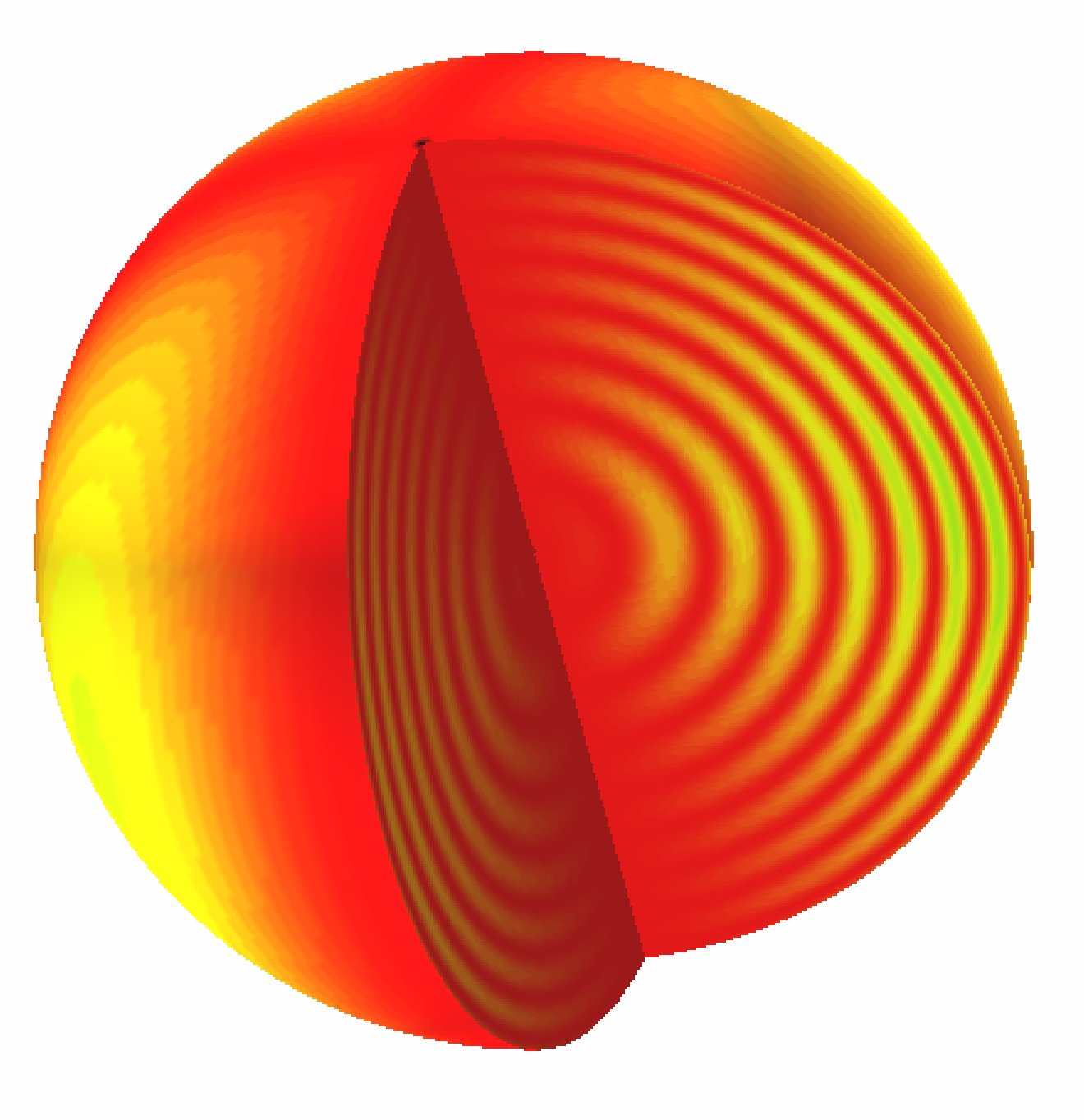}}
  
\else
  \author{}
  \collegeordept{}
  \university{}
  \crest{\includegraphics[width=4cm]{logo}}
\fi

\degree{Doutor em Astronomia (3\textordmasculine Ciclo da Universidade do Porto)}
\degreedate{November 2011}

\begin{document}


\renewcommand\baselinestretch{1.2}
\baselineskip=18pt plus1pt


\maketitle  


\newpage
\thispagestyle{empty}
\mbox{}

\vspace{195mm}
Cover figure taken from \url{http://science.au.dk/}.






\frontmatter
\begin{abstractslong}

\pagestyle{plain}

\noindent This dissertation has been submitted to the \emph{Faculdade de Ci\^encias da Universidade do Porto} in partial fulfillment of the requirements for the PhD degree in Astronomy. The scientific results presented herein follow from the research activity performed under the supervision of Dr.~M\'ario Jo\~ao Monteiro at the \emph{Centro de Astrof\'isica da Universidade do Porto} and Dr.~Hans Kjeldsen at the \emph{Institut for Fysik og Astronomi, Aarhus Universitet}.

The dissertation is mainly composed of three chapters and a list of appendices. Chapter \ref{Ch1} serves as an unpretentious and rather general introduction to the field of asteroseismology of solar-like stars. It starts with an historical account of the field of asteroseismology followed by a general review of the basic physics and properties of stellar pulsations. Emphasis is then naturally placed on the stochastic excitation of stellar oscillations and on the potential of asteroseismic inference. The chapter closes with a discussion about observational techniques and the observational status of the field. Given my exclusive role as a data analyst, I have devoted Chapter \ref{Ch2} to the subject of data analysis in asteroseismology. This is an extensive subject, therefore I have opted for presenting a compilation of relevant data analysis methods and techniques employed contemporarily in asteroseismology of solar-like stars, and of which I have made recurrent use. Special attention has been drawn to the subject of statistical inference both from the competing Bayesian and frequentist perspectives, a matter that I consider to be currently in vogue. The chapter ends with a description of the implementation of a pipeline for mode parameter analysis of \emph{Kepler} data. In the course of these two first chapters, reference is made to a series of published articles that have greatly benefited from my contribution and are, for that reason, collected in Appendices \ref{vichinature} to \ref{campanteautocov}. Chapter \ref{Ch3} then goes on to present a series of additional published results to which my contribution has been significant, although in a somewhat less determinant way. The compendium of scientific results presented in this dissertation is, to my mind, representative of my research activity and technical expertise.

The dawn of a new and prosperous era for the field of asteroseismology coincided with the development of space-based missions using the technique of ultra-high-precision photometry. The advent of the French-led \emph{CoRoT} and NASA \emph{Kepler} space missions had finally provided the possibility of carrying out long and almost uninterrupted observations of a multitude of targets, being at the same time capable of detecting faint solar-like oscillations in main-sequence stars. The main goal of my research activity has been the development of innovative data analysis methods and subsequent interpretation of the results in the context of space-based asteroseismology. That being said, the development of two pipelines for the analysis of \emph{Kepler} asteroseismic data, together with the development of a Bayesian peak-bagging tool based on Markov chain Monte Carlo techniques, constitute some of the main outcomes of my research work. 

An active membership of the \emph{Kepler} Asteroseismic Science Consortium (KASC) has allowed me not only to exchange technical skills and relevant knowledge with other members of the consortium, but also to take part in -- and even lead -- a series of workpackages covering a diversity of scientific aims, ranging from the comprehensive analysis of single objects to ensemble and differential asteroseismology. In this regard, I would highlight the analysis of multi-month time-series data on four evolved Sun-like stars, the very first results from ensemble asteroseismology based on a large cohort of solar-type field stars, and the observational confirmation of the presence of solar-like oscillations in a $\delta$ Sct star. 

Despite having focused my efforts on \emph{Kepler}-related investigations, I have still managed to sporadically contribute to the analysis of targets observed by \emph{CoRoT} (e.g., establishing a definite mode identification for the F-type star HD~49933, or characterizing the exoplanet-host solar-like star HD~52265 using both spectroscopic and seismic data) or during ground-based campaigns (e.g., a multi-site campaign dedicated to Procyon, or an asteroseismic and interferometric study of the solar twin 18 Scorpii), thus widening the scope of my research.

\end{abstractslong}

\begin{abstractslong2}

\pagestyle{plain}

\noindent Esta disserta\c{c}\~ao foi submetida \`a Faculdade de Ci\^encias da Universidade do Porto no cumprimento parcial dos requisitos necess\'arios \`a obten\c{c}\~ao do grau de Doutor. Os resultados cient\'ificos aqui apresentados decorrem da actividade de investiga\c{c}\~ao realizada sob a orienta\c{c}\~ao do Dr.~M\'ario Jo\~ao Monteiro do Centro de Astrof\'isica da Universidade do Porto e do Dr.~Hans Kjeldsen do \emph{Institut for Fysik og Astronomi, Aarhus Universitet}.

A disserta\c{c}\~ao \'e composta principalmente de tr\^es cap\'itulos e de uma lista de anexos. O primeiro cap\'itulo serve de introdu\c{c}\~ao despretensiosa e bastante geral ao campo da asterossismologia de estrelas do tipo solar. Come\c{c}a com um relato hist\'orico do campo da asterossismologia seguido de uma revis\~ao da f\'isica b\'asica e propriedades das pulsa\c{c}\~oes estelares. \'E ent\~ao dado natural \^enfase \`a excita\c{c}\~ao estoc\'astica de oscila\c{c}\~oes estelares e ao potencial da infer\^encia asteross\'ismica. O cap\'itulo termina com uma discuss\~ao sobre t\'ecnicas de observa\c{c}\~ao e sobre o estado actual do campo em termos observacionais. Dado o meu papel exclusivo como analista de dados, consagrei o segundo cap\'itulo ao tema da an\'alise de dados em asterossismologia. Este \'e um tema extenso e, por esse motivo, optei por apresentar uma compila\c{c}\~ao de m\'etodos e t\'ecnicas de an\'alise de dados relevantes, utilizados contemporaneamente na asterossismologia de estrelas do tipo solar, e dos quais fiz uso recorrente. Foi dada especial aten\c{c}\~ao ao tema da infer\^encia estat\'istica tanto de uma perspectiva Bayesiana como de uma perspectiva frequentista, um assunto que considero estar actualmente em voga. O cap\'itulo termina com a descri\c{c}\~ao da implementa\c{c}\~ao de um \emph{pipeline} usado na an\'alise dos par\^ametros de modos de oscila\c{c}\~ao presentes em dados do sat\'elite \emph{Kepler}. Ao longo destes dois primeiros cap\'itulos, \'e feita refer\^encia a uma s\'erie de artigos publicados que beneficiaram de modo determinante da minha contribui\c{c}\~ao e que, por essa raz\~ao, aparecem compilados na lista de anexos. O terceiro cap\'itulo passa ent\~ao a apresentar uma s\'erie adicional de resultados publicados, para a obten\c{c}\~ao dos quais a minha contribui\c{c}\~ao foi significativa, embora de forma n\~ao t\~ao determinante. O comp\^endio de resultados cient\'ificos apresentados nesta disserta\c{c}\~ao \'e, a meu ver, representativo da minha actividade de investiga\c{c}\~ao e conhecimento t\'ecnico.

O alvorecer de uma nova e pr\'ospera era para o campo da asterossismologia coincidiu com o desenvolvimento de miss\~oes espaciais empregando a t\'ecnica de fotometria de muito alta precis\~ao. O advento da miss\~ao espacial francesa \emph{CoRoT} e do sat\'elite \emph{Kepler} da NASA, tornou finalmente poss\'ivel a realiza\c{c}\~ao de observa\c{c}\~oes demoradas e quase ininterruptas de um sem-n\'umero de estrelas, sendo ao mesmo tempo essas miss\~oes capazes de detectar t\'enues oscila\c{c}\~oes do tipo solar em estrelas da sequ\^encia principal. O objectivo principal da minha actividade de investiga\c{c}\~ao consistiu no desenvolvimento de m\'etodos inovadores de an\'alise de dados e subsequente interpreta\c{c}\~ao dos resultados no \^ambito da asterossismologia espacial. Dito isto, o desenvolvimento de dois \emph{pipelines} usados na an\'alise de dados provenientes do sat\'elite \emph{Kepler}, juntamente com o desenvolvimento de uma ferramenta Bayesiana a usar na an\'alise de espectros de pot\^encia, constituem alguns dos principais resultados do meu trabalho de investiga\c{c}\~ao.

Uma participa\c{c}\~ao activa no \^ambito do KASC (\emph{Kepler} Asteroseismic Science Consortium) permitiu-me n\~ao s\'o a partilha de compet\^encias t\'ecnicas e conhecimentos relevantes com os demais membros do cons\'orcio, mas tamb\'em integrar, e at\'e mesmo liderar, uma s\'erie de grupos de trabalho abrangendo uma diversidade de fins cient\'ificos, desde a an\'alise detalhada de objectos individuais \`a pr\'atica estat\'istica da asterossismologia com base em grupos numerosos de estrelas. A este respeito, gostaria de destacar a an\'alise das s\'eries temporais, com a dura\c{c}\~ao de v\'arios meses, de quatro estrelas evolu\'idas do tipo solar, os primeiros resultados provenientes da pr\'atica estat\'istica da asterossismologia com base num grande n\'umero de estrelas de campo do tipo solar, e a confirma\c{c}\~ao observacional da presen\c{c}a de oscila\c{c}\~oes do tipo solar numa estrela $\delta$ Sct.

Apesar de ter centrado os meus esfor\c{c}os em investiga\c{c}\~oes relacionadas com o \emph{Kepler}, pude ainda contribuir esporadicamente para a an\'alise de estrelas observadas pela miss\~ao espacial \emph{CoRoT} (por exemplo, estabelecendo de modo definitivo a identifica\c{c}\~ao dos modos de oscila\c{c}\~ao da estrela HD~49933, de tipo espectral F, ou ao caracterizar a estrela de tipo solar HD~52265, que tamb\'em alberga um exoplaneta, atrav\'es do uso de dados s\'ismicos e espectrosc\'opicos) ou durante campanhas de observa\c{c}\~ao feitas a partir do solo (por exemplo, um campanha envolvendo v\'arios telesc\'opios dedicada a Procyon, ou um estudo s\'ismico e interferom\'etrico de 18 Scorpii, uma estrela muito semelhante ao nosso Sol), tendo assim alargado o \^ambito da minha actividade de investiga\c{c}\~ao.

\end{abstractslong2}
\begin{dedication}

\thispagestyle{plain}

To Isabel, my parents, and my sister.

\end{dedication}
\newpage
\thispagestyle{plain}
\mbox{}
\begin{acknowledgementslong}

\thispagestyle{plain}

\noindent First of all, I wish to thank both my supervisors, M\'ario Jo\~ao Monteiro and Hans Kjeldsen, for their invaluable advice and unconditional support throughout the past three years or so. 

I am grateful to my colleagues at CAUP, namely, Isa Brand\~ao, Margarida Cunha and Micha{\"e}l Bazot. I wish to thank all my former colleagues at Aarhus University, namely, G{\"u}lnur Do{\u{g}}an, J{\o}rgen Christensen-Dalsgaard, Christoffer Karoff, Torben Arentoft, Frank Grundahl, Rasmus Handberg and S{\o}ren Frandsen. A word of appreciation goes also to the people with whom I had the privilege of maintaining close collaborations, namely, Tim Bedding, Bill Chaplin, Thierry Appourchaux, Rafa Garc\'ia, Savita Mathur, Vichi Antoci and Othman Benomar.

I should also acknowledge the \emph{Funda\c{c}\~ao para a Ci\^encia e a Tecnologia} for the financial support provided in the course of my research work. 

\vspace{20mm}
\hspace{80mm}\emph{Porto, 30 October 2011}

\end{acknowledgementslong}

\newpage
\thispagestyle{plain}
\mbox{}


\setcounter{secnumdepth}{3} 
\setcounter{tocdepth}{3}    
\tableofcontents            


\chapter{List of Publications}
\markboth{LIST OF PUBLICATIONS}{LIST OF PUBLICATIONS}

\section*{Refereed papers}
\begin{itemize}

\item Benomar, O., Baudin, F., Campante, T.~L., et al.~2009, A\&A, 507, L13 \\
URL: \url{http://adsabs.harvard.edu/abs/2009A%26A...507L..13B}

\item Bedding, T.~R., Kjeldsen, H., Campante, T.~L., et al.~2010, ApJ, 713, 935 \\
URL: \url{http://adsabs.harvard.edu/abs/2010ApJ...713..935B}

\item Chaplin, W.~J., Appourchaux, T., Elsworth, Y., et al.~2010, ApJ, 713, L169 \\
URL: \url{http://adsabs.harvard.edu/abs/2010ApJ...713L.169C}

\item Bedding, T.~R., Huber, D., Stello, D., et al.~2010, ApJ, 713, L176 \\
URL: \url{http://adsabs.harvard.edu/abs/2010ApJ...713L.176B}

\item Campante, T.~L., Karoff, C., Chaplin, W.~J., et al.~2010, MNRAS, 408, 542 \\
URL: \url{http://adsabs.harvard.edu/abs/2010MNRAS.408..542C}

\item de Meulenaer, P., Carrier, F., Miglio, A., et al.~2010, A\&A, 523, A54 \\
URL: \url{http://adsabs.harvard.edu/abs/2010A%26A...523A..54D}

\item Metcalfe, T.~S., Monteiro, M.~J.~P.~F.~G., Thompson, M.~J., et al.~2010, ApJ, 723, 1583 \\
URL: \url{http://adsabs.harvard.edu/abs/2010ApJ...723.1583M}

\item Bazot, M., Ireland, M.~J., Huber, D., et al.~2011, A\&A, 526, L4 \\
URL: \url{http://adsabs.harvard.edu/abs/2011A%26A...526L...4B}

\item Handberg, R.~\& Campante, T.~L.~2011, A\&A, 527, A56 \\
URL: \url{http://adsabs.harvard.edu/abs/2011A%26A...527A..56H}

\item Chaplin, W.~J., Kjeldsen, H., Christensen-Dalsgaard, J., et al.~2011, Science, 332, 213 \\
URL: \url{http://adsabs.harvard.edu/abs/2011Sci...332..213C}

\item Chaplin, W.~J., Kjeldsen, H., Bedding, T.~R., et al.~2011, ApJ, 732, 54 \\
URL: \url{http://adsabs.harvard.edu/abs/2011ApJ...732...54C}

\item Chaplin, W.~J., Bedding, T.~R., Bonanno, A., et al.~2011, ApJ, 732, L5 \\
URL: \url{http://adsabs.harvard.edu/abs/2011ApJ...732L...5C}

\item Ballot, J., Gizon, L., Samadi, R., et al.~2011, A\&A, 530, A97 \\
URL: \url{http://adsabs.harvard.edu/abs/2011A%26A...530A..97B}

\item Mathur, S., Handberg, R., Campante, T.~L., et al.~2011, ApJ, 733, 95 \\
URL: \url{http://adsabs.harvard.edu/abs/2011ApJ...733...95M}

\item Silva Aguirre, V., Chaplin, W.~J., Ballot, J., et al.~2011, ApJ, 740, L2 \\
URL: \url{http://adsabs.harvard.edu/abs/2011ApJ...740L...2S}

\item Campante, T.~L., Handberg, R., Mathur, S., et al.~2011, A\&A, 534, A6 \\
URL: \url{http://adsabs.harvard.edu/abs/2011A%26A...534A...6C}

\item Verner, G.~A., Chaplin, W.~J., Basu, S., et al.~2011, ApJ, 738, L28 \\
URL: \url{http://adsabs.harvard.edu/abs/2011ApJ...738L..28V}

\item Verner, G.~A., Elsworth, Y., Chaplin, W.~J., et al.~2011, MNRAS, 415, 3539 \\
URL: \url{http://adsabs.harvard.edu/abs/2011MNRAS.415.3539V}

\item Antoci, V., Handler, G., Campante, T.~L., et al.~2011, Nature, 477, 570 \\
URL: \url{http://adsabs.harvard.edu/abs/2011Natur.477..570A}

\item White, T.~R., Bedding, T.~R., Stello, D., et al.~2011, ApJ, 742, L3 \\
URL: \url{http://adsabs.harvard.edu/abs/2011ApJ...742L...3W}

\item Huber, D., Bedding, T.~R., Stello, D., et al.~2011, ApJ, 743, 143 \\
URL: \url{http://adsabs.harvard.edu/abs/2011ApJ...743..143H}

\item Creevey, O.~L., Do{\u g}an, G., Frasca, A., et al.~2012, A\&A, 537, A111 \\
URL: \url{http://adsabs.harvard.edu/abs/2012A%26A...537A.111C}

\item Appourchaux, T., Benomar, O., Gruberbauer, M., et al.~2012, A\&A, 537, A134 \\
URL: \url{http://adsabs.harvard.edu/abs/2012A%26A...537A.134A}

\item Howell, S.~B., Rowe, J.~F., Bryson, S.~T., et al.~2011, ApJ, 746, 123 \\
URL: \url{http://adsabs.harvard.edu/abs/2012ApJ...746..123H}

\end{itemize}

\section*{Peer-reviewed conference proceedings}
\begin{itemize}

\item Do{\u g}an, G., Bonanno, A., Bedding, T.~R., et al.~2010, Astronomische Nachrichten, 331, 949 \\
URL: \url{http://adsabs.harvard.edu/abs/2010AN....331..949D}

\item Karoff, C., Chaplin, W.~J., Appourchaux, T., et al.~2010, Astronomische Nachrichten, 331, 972 \\
URL: \url{http://adsabs.harvard.edu/abs/2010AN....331..972K}

\end{itemize}

\section*{Preprints}
\begin{itemize}

\item Mathur, S., Metcalfe, T.~S., Woitaszek, M., et al.~2012, ApJ, in press [arXiv:1202.2844v1] \\
URL: \url{http://arxiv.org/abs/1202.2844}

\item Garc\'ia, R.~A., Ceillier, T., Campante, T.~L., et al.~2011, Astronomical Society of the Pacific, in press [arXiv:1109.6488v1] \\
URL: \url{http://arxiv.org/abs/1109.6488}

\item Mathur, S., Campante, T.~L., Handberg, R., et al.~2011, Astronomical Society of the Pacific, in press [arXiv:1110.0135v1] \\
URL: \url{http://arxiv.org/abs/1110.0135}

\end{itemize}

\section*{Conference proceedings without referee}
\begin{itemize}

\item Campante, T.~L., Grigahc\`ene, A., Su\'arez, J.~C., et al.~2010, Astronomische Nachrichten, in press [arXiv:1003.4427v1] \\
URL: \url{http://arxiv.org/abs/1003.4427}

\item Karoff, C., Campante, T.~L., \& Chaplin, W.~J.~2010, Astronomische Nachrichten, in press [arXiv:1003.4167v1] \\
URL: \url{http://arxiv.org/abs/1003.4167}

\end{itemize}

\listoffigures	

\listoftables  







\mainmatter




\newcommand\ion[2]{#1$\;${\scshape{#2}}}

\chapter{Asteroseismology of solar-like stars}\label{Ch1}

\ifpdf
    \graphicspath{{1/figures/PNG/}{1/figures/PDF/}{1/figures/}}
\else
    \graphicspath{{1/figures/EPS/}{1/figures/}}
\fi

This chapter introduces the field of asteroseismology of solar-like stars by presenting and discussing a series of key concepts that are essential for a complete understanding of the remaining of this dissertation. Following a brief historical account of the field of asteroseismology, an overview of the origin and nature of stellar pulsations is presented to the reader. The basic properties of oscillation modes are discussed next, before particular attention is paid to the process of stochastic excitation of oscillations and to the potential of asteroseismic inference. The chapter ends with summaries of the main observational techniques used in the field and of its observational status. 

The current chapter is by no means intended as a thorough review of the field. To serve that purpose I would strongly recommend the book by \citet{BookAstero}, J.~Christensen-Dalsgaard's \emph{Lecture Notes on Stellar Oscillations}\footnote{\url{http://users-phys.au.dk/jcd/oscilnotes/}}, and the reviews by \citet{JCDreview04}, \citet{Cunhareview07} and \citet{BeddingWS}, on which the following discussion is somewhat based. 

\section{A brief encounter with history}
The longest known case of a pulsating star is that of $o$ Ceti (Mira), the discovery of its variability being attributed to a Lutheran pastor and amateur astronomer named David Fabricius in 1596 \citep[e.g.,][]{Mira}. The star was then practically forgotten until Johann Fokkens Holwarda rediscovered it in 1638 and found that its magnitude varied periodically with a period of eleven months. By the time of the second centennial of its discovery, 1796, eleven variables had been discovered, four of them of the Mira type. However, firm establishment that such variability is in many cases due to intrinsic stellar pulsations came only in the twentieth century. In this regard \citet{Shapley1914} wrote: ``The main conclusion is that the Cepheid and cluster variables are not binary systems, and that the explanation of their light-changes can much more likely be found in a consideration of internal or surface pulsations of isolated stellar bodies.''

Early studies of pulsating stars were obviously restricted to large-amplitude pulsators such as the Cepheids and the long-period variables. The simple pulsatory behavior of these stars was interpreted in terms of pulsations in the fundamental radial mode, characterized by expansion of the star followed by its contraction, while preserving spherical symmetry. The discovery of the period-luminosity relation for the Cepheids by Henrietta Swan Leavitt \citep{Leavitt1908,LeavittPickering} supplied the foundation for the measurement of extragalactic distances. The decades that followed saw emphasis being placed on understanding the mechanism driving the pulsations, which would first be arrived at independently by \citet{CoxWhitney} and \citet{Zhevakin1963}. The latter reference provides a review of the early developments of such studies.  

The first detections of the oscillatory motion in the atmosphere of the Sun (as local modes), with periods of approximately five minutes, were made in the early 1960s by \citet{Leighton62}, paving the way for the development of helioseismology, by then an entire new field of research. The first detection and identification of these oscillations as global modes is attributed to \citet{Claverie}. Helioseismology has ever since proved to be extremely successful in probing the physics and dynamics of the solar interior. The vast amount of data on solar oscillations made available in the last two decades led to a considerably accurate determination of the solar sound speed, detailed testing of the equation of state and inference of the solar internal rotation \citep[e.g.,][and references therein]{JCDhelio,BasuAntia,ChaplinBasu,Howe09}.

The Sun is, however, a single star at a specific evolutionary stage, and it is further structurally simple if compared with certain other stars. A logical consequence was therefore the advent of asteroseismology, whereby one would expect to be able to probe the interiors of stars other than the Sun through the use of their intrinsic oscillations. The limited spatial resolution with which we can observe distant stars poses, however, a serious obstacle. Moreover, the very small number of oscillation frequencies observed for most of the pulsating stars renders them unsuitable for the pursuit of asteroseismic studies. Nonetheless, the field thrived and we are today able to proudly answer Sir Arthur Eddington's famous lament \citep{EddingtonBook}: ``What appliance can pierce through the outer layers of a star and test the conditions within?''

The definite detection of solar-like oscillations in stars other than the Sun had long been an illusory goal due to their very small amplitudes, particularly for main-sequence stars. However, the development of very stable techniques for radial-velocity observations, promoted by the hunt for extrasolar planets, produced a major breakthrough in the field by the turn of the millennium and led to the detection of solar-like oscillations in several stars \citep[e.g.,][]{BeddingKjeldsen03}. That was only the beginning of an exciting and successful journey.

\section{Overview of the origin and nature of stellar pulsations}
In order to conduct an asteroseismic study and to fully explore the diagnostic potential of the observed oscillations, one has to understand first their origin and physical nature. A relevant timescale in understanding the properties of oscillations is the dynamical timescale:
\begin{equation}
\label{t_dyn}
t_{\rm{dyn}}=\left(\frac{R^3}{GM}\right)^{1/2}\propto\left(G\,\overline{\rho}\right)^{-1/2} \, ,
\end{equation}
where $R$ and $M$ are the surface radius and mass of the star, respectively, $G$ is the gravitational constant, and $\overline{\rho}$ is the mean stellar density. The periods of the oscillations generally scale as $t_{\rm{dyn}}$. More specifically, $t_{\rm{dyn}}$ expresses the time the star needs to go back into hydrostatic equilibrium whenever some dynamical process disrupts the balance between pressure and gravitational force. Pressure modes (see below) may be the cause of such a disturbance and so their oscillation periods should not exceed $t_{\rm{dyn}}$. It is remarkable how the measurement of a period of oscillation immediately provides us with an estimate of an intrinsic property of the star, namely, its mean density.

Many stars, including the Sun, pulsate in more complex ways than the Cepheids do, being ubiquitous for more than one mode of oscillation to be excited simultaneously. These modes may include radial overtones, in addition to the fundamental radial mode, as well as non-radial modes, whose motion does not preserve spherical symmetry. 

The physical nature of the oscillations concerns the restoring force at play: modes are of the nature of either standing acoustic waves (p modes) with pressure acting as the restoring force or internal gravity waves (g modes; these are exclusively non-radial modes) with buoyancy acting as the restoring force. There is a clear separation between these two classes of modes in unevolved stars. This, however, may not be the case in evolved stars. Due to strong internal gradients in the chemical composition or large gravitational acceleration in a compact core, modes of mixed p- and g-mode character may occur in evolved stars. In addition, the Sun also displays surface gravity waves of large horizontal wave number (f modes).

When talking about their origin, one means the mechanism responsible for driving the oscillations. Oscillations can be either intrinsically unstable or intrinsically stable. In the former case, oscillations result from the amplification of small disturbances by means of a heat-engine mechanism converting thermal into mechanical energy in a specific region of the star, usually a radial layer. This region is heated up during the compressional phase of the pulsation cycle while being cooled off during expansion. An amplitude-limiting mechanism then sets in at some point, determining the final amplitude of the growing disturbance. Such a region inside the star is typically associated with opacity ($\kappa$) features and the resulting driving mechanism is thus known as the $\kappa$ mechanism. The $\kappa$ mechanism is responsible for the oscillatory behavior in Cepheids, RR Lyrae stars, $\delta$ Sct stars, $\beta$ Cep stars, and in most of the pulsating classes displayed in Fig.~\ref{HR}. A particularly important area depicted in that figure is the Cepheid (or classical) instability strip, where pulsating class members are believed to have their oscillations driven by an opacity mechanism associated with the second helium-ionization zone. In order to cause overall excitation of the oscillations, the region associated with the driving has to be placed at an appropriate depth inside the star, thus providing an explanation for the specific location of the resulting instability belt in the H-R diagram. This type of oscillations are generally known as classical oscillations. 

On the other hand, intrinsically stable oscillations, such as the solar five-minute oscillations, are stochastically excited by the vigorous near-surface convection. This type of oscillations, having first been detected in the Sun, are referred to as solar-like oscillations. Solar-like oscillations are predicted for all stars cool enough to harbor an outer convective envelope, and are thus found among main-sequence core, and post-main-sequence shell, hydrogen-burning stars, residing on the cool side of the Cepheid instability strip. Besides main-sequence stars with masses up to about $1.5\:{\rm{M}}_\odot$, solar-like oscillations are also expected to occur from the end of the main sequence up to the giant and asymptotic giant branches. The resulting mode amplitudes are considerably smaller than those generally found in classical pulsators. However, the stochastic process is characterized by varying slowly with frequency and hence modes tend to be excited to comparable amplitudes within a substantial frequency range. This happens in contrast to the distribution of mode amplitudes of classical pulsators which is highly irregular over the range of unstable modes.

\figuremacroW{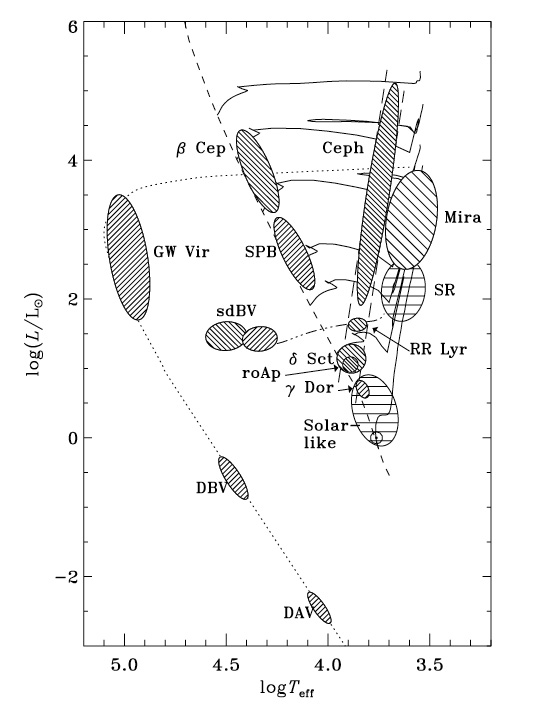}{Pulsating stars across the Hertzsprung-Russell diagram}{Several classes of pulsating stars, for which asteroseismology is possible, have been located. $T_{\rm{eff}}$ and $L$ are the effective temperature and stellar luminosity, respectively. The dashed line indicates the zero-age main sequence (ZAMS), the solid curves represent selected evolutionary tracks (for 1, 2, 3, 4, 7, 12, and $20\:{\rm{M}}_\odot$), the triple-dot-dashed line indicates the horizontal branch and the dotted curve follows the white-dwarf cooling track. The parallel long-dashed lines enclose the Cepheid instability strip. From \citet{Bedding07}.}{0.84}

\section{Basic properties of oscillation modes}
\subsection{Describing the oscillations}
Small-amplitude oscillations of a spherically symmetric star depend on co-latitude $\theta$ and longitude $\phi$ in terms of a spherical harmonic $Y_l^m(\theta,\phi)$. Use is made of spherical polar coordinates, $(r,\theta,\phi)$, where $r$ is the distance to the center of the star. The degree $l$ specifies the number of nodal surface lines, i.e., the complexity of the mode, better understood by defining the surface horizontal wave number, $k_{\rm{h}}^{(\rm{surf})}$:
\begin{equation}
k_{\rm{h}}^{(\rm{surf})}=\frac{\sqrt{l(l+1)}}{R} \, .
\end{equation}
Radial modes have $l\!=\!0$, whereas for non-radial modes $l\!>\!0$. The azimuthal order is represented by $m$, with $|m|$ specifying how many of the nodal surface lines are lines of longitude. Values of $m$ range from $-l$ to $l$, and thus there are $2l+1$ modes for each multiplet of degree $l$. Figure \ref{spher_harm} illustrates the appearance of the $l\!=\!3$ octupole modes on a stellar surface. Modes are additionally characterized by the radial order $n$, which is related to the number of radial nodes.

\figuremacroW{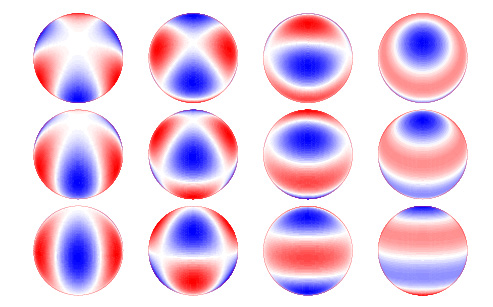}{Freeze-frame of the radial component of the $l\!=\!3$ octupole modes}{Rows display the same modes although with different inclination angles of the polar axis with respect to the line of sight: $30^\circ$ (top row), $60^\circ$ (middle row), and $90^\circ$ (bottom row). White bands represent the nodal surface lines. Red and blue sections represent portions of the stellar surface that are moving in and out, respectively. The rightmost column displays the axisymmetric (i.e., with $m\!=\!0$) mode $(l\!=\!3,m\!=\!0)$. From right to left, the middle columns display the tesseral (i.e., with $0\!<\!|m|\!<\!l$) modes $(l\!=\!3,m\!=\!\pm1)$ and $(l\!=\!3,m\!=\!\pm2)$. The leftmost column displays the sectoral (i.e., with $|m|\!=\!l$) mode $(l\!=\!3,m\!=\!\pm3)$. Figure courtesy of Conny Aerts.}{1}

As an example of an eigenfunction, I introduce the radial component of displacement which may be expressed as 
\begin{equation}
\label{displacement}
\xi_r(r,\theta,\phi;t)=\Re\left\{a(r) \, Y_l^m(\theta,\phi) \, \exp(-{\rm{i}}\,2\pi \nu t)\right\} \, ,
\end{equation}
where $a(r)$ is an amplitude function, and $\nu$ is the (cyclic) frequency of oscillation. For a spherically symmetric star the frequency of oscillation depends only on $n$ and $l$, i.e., $\nu\!=\!\nu_{nl}$. The spherical harmonic $Y_l^m(\theta,\phi)$ is expressed as
\begin{equation}
\label{spharm}
Y_l^m(\theta,\phi)=(-1)^m \, c_{lm} \, P_l^m(\cos \theta) \, \exp({\rm{i}}\,m\phi) \, ,
\end{equation}
where $P_l^m$ is an associated Legendre function given by
\begin{equation}
P_l^m(\cos \theta)=\frac{1}{2^l l!} \, (1-\cos^2 \theta)^{m/2} \, \frac{{\rm{d}}^{l+m}}{{\rm{d}}\cos^{l+m} \theta} (\cos^2 \theta -1)^l \, ,
\end{equation}
and the normalization constant $c_{lm}$ is determined by
\begin{equation}
c_{lm}^2=\frac{(2l+1)(l-m)!}{4\pi (l+m)!} \, ,
\end{equation}
such that the integral of $|Y_l^m|^2$ over the unit sphere is unity.

\subsection{Spatial filtering}\label{sect:spat_filt}
Unlike the case of the Sun, for which modes of very high degree $l$ can be observed, we have not yet reached the stage where we can resolve stellar surfaces using either velocity or intensity observations. In the stellar case our observations actually result from weighted averages of the pulsation amplitude over the stellar disk. Consequently, modes of moderate and high degree $l$, and hence of increasing complexity, tend to average out in what is known as partial cancellation or spatial filtering. Particularly for solar-like oscillations, whose intrinsic amplitudes are rather low, this means that only modes of the lowest degree, i.e., with $l\!\leq\!3$, are expected to be observed. Furthermore, in the case of velocity observations, the projection of the velocity onto the line of sight introduces an extra factor of $\cos \theta$ in the weighting function. This effectively gives more sensitivity to the center of the disk relative to the limb, ultimately resulting in a slightly larger response to modes of $l\!=\!3$ than for intensity observations. 

The preceding considerations can be supported by very simple calculations. Assuming the case of surface-integrated intensity of an axisymmetric mode over the stellar disk, while neglecting the effects of limb darkening and rotation, the spatial response function $S_l^{({\rm{I}})}$ is then given by
\begin{equation}
S_l^{({\rm{I}})}=2\sqrt{2l+1} \int_0^{\pi/2} P_l^0(\cos \theta) \cos \theta \sin \theta \, {\rm{d}}\theta \, .
\end{equation} 
A similar calculation can be carried out for the case of velocity observations and assuming that the velocity field is predominantly in the radial direction, giving
\begin{equation}
S_l^{({\rm{V}})}=2\sqrt{2l+1} \int_0^{\pi/2} P_l^0(\cos \theta) \cos^2 \theta \sin \theta \, {\rm{d}}\theta \, .
\end{equation}
The spatial response functions $S_l^{({\rm{I}})}$ and $S_l^{({\rm{V}})}$ are plotted in Fig.~\ref{spat_filt} as a function of $l$. 

\figuremacroW{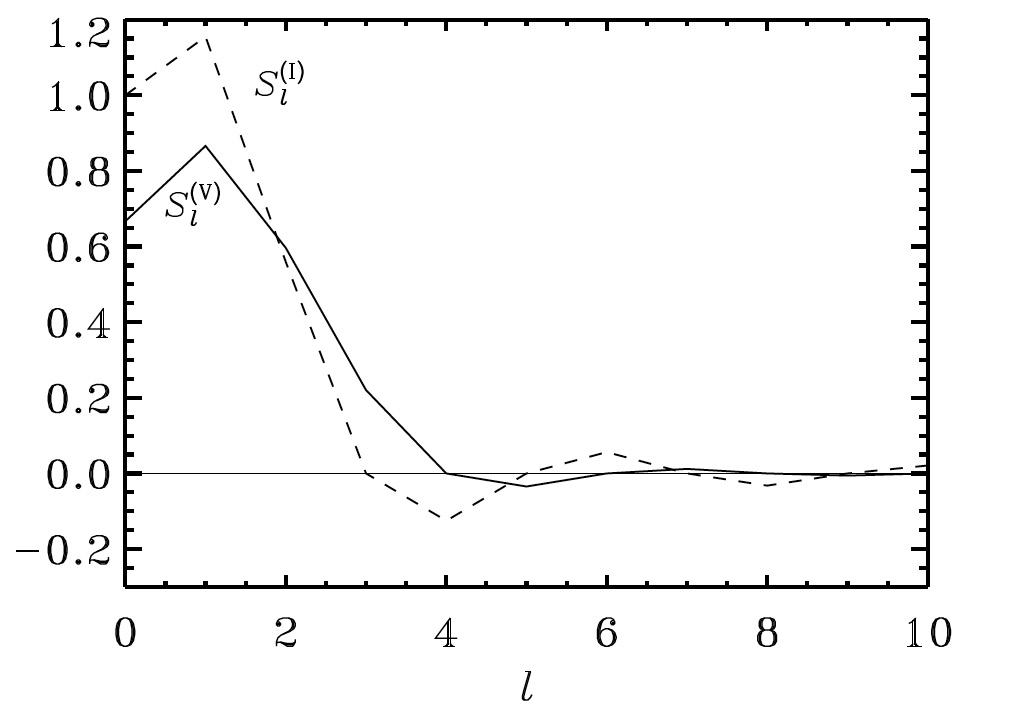}{Spatial response functions}{$S_l^{({\rm{I}})}$ and $S_l^{({\rm{V}})}$ for surface-integrated intensity and velocity, respectively, are plotted as a function of $l$. Negative values of either quantity mean that the oscillations will appear to have reversed phases. Figure courtesy of J.~Christensen-Dalsgaard.}{1.0}

\citet{Kjeldsen08} provide spatial responses to modes with degree $l$ relative to those with $l\!=\!0$ for a set of intensity and velocity observations (see their table 1). Radial modes make a sensible reference since they are not split by rotation, as will be seen in Sect.~\ref{sect:rot}. Those ratios were computed using the results of \citet{JCD89} and \citet{Bedding96}. Particularly useful is the latter work, where the authors provide approximate expressions for computing the spatial response functions taking into account the effect of limb darkening. Those expressions are, however, only valid in the case of a slow rotator. Recently, \citet{Salabert2011} provided precise estimates of mode visibilities for radial-velocity and photometric observations of the Sun-as-a-star (i.e., whole-disk observations of the Sun), further comparing them to theoretical predictions.

Table \ref{tab:spat_filt} displays the relative spatial response functions $S_l/S_0$, computed according to \citet{Bedding96}, for a number of present and upcoming instruments/missions used to measure solar-like oscillations. Those performing intensity measurements are the red channel of the VIRGO/SPM instrument \citep{VIRGO,VIRGO2} on board the \emph{SOHO} spacecraft, as well as the \emph{CoRoT} \citep{CoRoT} and NASA \emph{Kepler} \citep{Kepler1,Kepler2} space missions. On the other hand, radial-velocity measurements are performed by the HARPS spectrograph \citep{HARPS} and are the purpose of the forthcoming SONG network \citep{SONG,SONG2}.
\begin{table}[!ht]
	\centering
	\caption[Relative spatial response functions $S_l/S_0$]{\textbf{Relative spatial response functions $S_l/S_0$} - These are given for a number of present and upcoming instruments/missions.}
	\begin{tabular}{lcccccc}
	\hline\hline
	& \multicolumn{3}{c}{\bf Intensity} & & \multicolumn{2}{c}{\bf Velocity} \\
	& VIRGO/SPM & \emph{CoRoT} & \emph{Kepler} & & HARPS & SONG \\
	& (862 nm) & (660 nm) & (641 nm) & & (535 nm) & (550 nm) \\
	\hline
			$S_0/S_0$ &  1.00 &  1.00 &  1.00 & & 1.00 & 1.00 \\
			$S_1/S_0$ &  1.20 &  1.22 &  1.22 & & 1.35 & 1.35 \\
			$S_2/S_0$ &  0.67 &  0.70 &  0.71 & & 1.02 & 1.01 \\
			$S_3/S_0$ &  0.10 &  0.14 &  0.14 & & 0.48 & 0.47 \\
			$S_4/S_0$ & $-0.10$ & $-0.09$ & $-0.08$ & & 0.09 & 0.09 \\
	\hline
	\end{tabular}
	\label{tab:spat_filt}
\end{table}

\subsection{Understanding the behavior of mode eigenfunctions}\label{eigenf}
The diagnostic potential of the oscillation frequencies can be better understood through asymptotic analyses of the oscillation equations. This sort of approach approximates these equations to such an extent that they can be discussed analytically. The fact that a reasonable number of classes of pulsating stars display high-order acoustic or gravity modes justifies employing an asymptotic analysis. 

An approximate asymptotic description of the oscillation equations has been derived by D.~O.~Gough \citep{DeubnerGough,Gough86,Gough93}, on the basis of an earlier work by \citet{Lamb32}:
\begin{equation}
\frac{{\rm{d}}^2 X}{{\rm{d}} r^2} + K(r)X = 0 \, ,
\end{equation}
where
\begin{equation}
\label{eq:charfreqs}
K(r)=\frac{\omega^2}{c^2}\left[1 - \frac{\omega_{\rm{c}}^2}{\omega^2} - \frac{S_l^2}{\omega^2} \left(1 - \frac{N^2}{\omega^2}\right)\right] \, ,
\end{equation}
and
\begin{equation}
X=c^2 \rho^{1/2} \,\, {\rm{div}}\,{\bf{\delta r}} \, .
\end{equation}
The adiabatic sound speed, $c$, is given by $c^2\!=\!\Gamma_1\,p/\rho$, $p$ being pressure, and $\Gamma_1\!=\!(\partial\ln p/\partial\ln \rho)_{\rm{ad}}$ being the adiabatic exponent relating pressure and density; ${\bf{\delta r}}$ is the displacement vector in the last equation. The behavior of the eigenfunction of a mode is determined by three characteristic (angular) frequencies varying throughout the star: the acoustic ($S_l$), the buoyancy ($N$), and the acoustic cut-off ($\omega_{\rm{c}}$) frequencies (cf.~Eq.~\ref{eq:charfreqs}). Figure \ref{charact_freqs} displays the three characteristic frequencies as a function of fractional radius for a set of selected stellar models.

The acoustic (or Lamb) frequency{\footnote{Note that both the Lamb frequency and the spatial response function are represented by $S_l$. However, with attention to context, this should not result in confusion.}} $S_l$ is determined by
\begin{equation}
S_l^2=\frac{l(l+1)\,c^2}{r^2} \, ,
\end{equation}
being interpreted as the frequency of a sound wave traveling horizontally with local wave number $k_{\rm{h}}\!=\!\sqrt{l(l+1)}/r$. 

The buoyancy (or Brunt-V\"{a}is\"{a}l\"{a}) frequency $N$ is determined by
\begin{eqnarray}
\label{buoyancy}
N^2&=&g \left(\frac{1}{\Gamma_1} \frac{{\rm{d}}\ln p}{{\rm{d}}r} - \frac{{\rm{d}}\ln \rho}{{\rm{d}}r}\right) \nonumber \\
&\approx& \frac{g^2 \rho}{p} \left(\nabla_{\rm{ad}} - \nabla + \nabla_\mu \right) \, ,
\end{eqnarray}
where $g$ is the local gravitational acceleration. To obtain the second equality, the gas has been regarded as a fully-ionized ideal gas and the effects of degeneracy and radiation pressure, as well as of Coulomb interactions, have been neglected. This constitutes a fairly good approximation in much of the interior of the majority of stars. The resulting simple equation of state, $p\!=\!\rho \, k_{\rm{B}} T/\mu m_{\rm{u}}$, where $k_{\rm{B}}$ is Boltzmann's constant, $\mu$ is the mean molecular weight, and $m_{\rm{u}}$ is the atomic mass unit, then leads to
\begin{equation}
\label{eq:soundspeed}
c=\sqrt{\frac{\Gamma_1 \, k_{\rm{B}} T}{\mu \, m_{\rm{u}}}} \, ,
\end{equation}
with the sound speed depending on the temperature and chemical composition of the gas. Moreover,
\begin{equation}
\nabla=\frac{{\rm{d}}\ln T}{{\rm{d}}\ln p} \, , \;\;\; \nabla_{\rm{ad}}=\left(\frac{\partial \ln T}{\partial \ln p}\right)_{\rm{ad}} \, , \;\;\; \nabla_\mu=\frac{{\rm{d}}\ln \mu}{{\rm{d}}\ln p} \, .
\end{equation}
For $N^2\!>\!0$, $N$ can be interpreted as the frequency of a gas element of reduced horizontal extent which oscillates due to buoyancy. Conversely, regions for which $N^2\!<\!0$ satisfy the Ledoux criterion of convective instability, i.e.,
\begin{equation}
\nabla>\nabla_{\rm{ad}} + \nabla_\mu \, .
\end{equation}
Gravity waves cannot, therefore, propagate in convective regions.

The acoustic cut-off frequency $\omega_{\rm{c}}$ is determined by
\begin{equation}
\label{f_ac}
\omega_{\rm{c}}^2=\frac{c^2}{4 H_{\rho}^2} \left(1 - 2\frac{{\rm{d}}H_{\rho}}{{\rm{d}}r}\right) \, ,
\end{equation} 
where $H_{\rho}\!=\!-({\rm{d}}\ln \rho/{\rm{d}}r)^{-1}$ is the density scale height. In an isothermal atmosphere, $H_{\rho}$ is constant, and thus $\omega_{\rm{c}}\!=\!c/(2H_{\rho})$. In the solar atmosphere, $H_{\rho}\!\approx\!120\:{\rm{km}}$, corresponding to a (cyclic) cut-off frequency of about $5\:{\rm{mHz}}$, or a period of 3 minutes. A useful relation, describing the behavior of the acoustic cut-off frequency (in units of $\omega_{\rm{dyn}}\!=\!2\pi/t_{\rm{dyn}}$) as a function of the stellar parameters, is given by
\begin{equation}
\label{f_cut}
\frac{\omega_{\rm{c}}}{\omega_{\rm{dyn}}} \propto \, \left(\frac{M}{{\rm{M}}_\odot}\right)^{1/2} \, \left(\frac{L}{{\rm{L}}_\odot}\right)^{-1/4} \, \left(\frac{T_{\rm{eff}}}{{\rm{T}}_{{\rm{eff}},\odot}}\right)^{1/2} \, .
\end{equation}

\figuremacroW{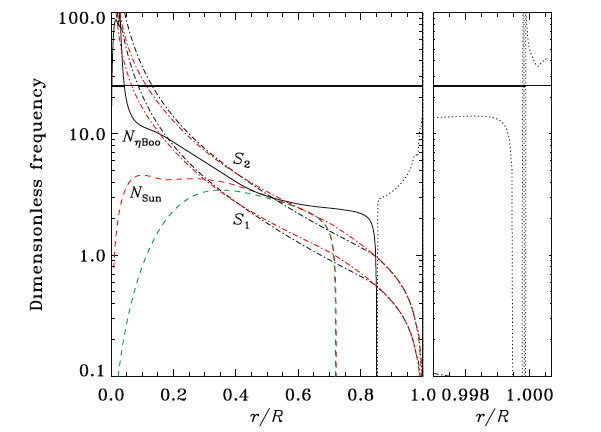}{Dimensionless characteristic frequencies}{The characteristic frequencies are given, in units of $\omega_{\rm{dyn}}$ and as a function of the fractional stellar radius, for a 1-${\rm{M}}_\odot$ ZAMS model, a model of the present Sun and a model of $\eta$ Boo (a 1.63-${\rm{M}}_\odot$ subgiant in the shell hydrogen-burning phase). Dot-dashed lines represent $S_l$, for $l\!=\!1$ and $l\!=\!2$, in the solar model (red) and the model of $\eta$ Boo (black), being barely indistinguishable. The buoyancy frequency $N$ is represented -- except for the atmosphere -- by a dashed line in the ZAMS model (green) and the solar model (red), and by a solid line in the model of $\eta$ Boo. The dotted line represents $\omega_{\rm{c}}$ in the model of $\eta$ Boo from the base of the convective envelope outward. The horizontal line represents the frequency of a stochastically-excited $l\!=\!1$ mode in the model of $\eta$ Boo, being thicker in regions where the mode propagates (see discussion in the text). From \citet{Cunhareview07}.}{1}

The eigenfunction of a mode oscillates as a function of $r$ in regions satisfying $K(r)\!>\!0$, where it is said to be propagating. Conversely, in regions satisfying $K(r)\!<\!0$, the eigenfunction behaves exponentially, and it is said to be evanescent. Finally, the location of the turning points of the eigenfunction are determined by $K(r)\!=\!0$. Typically, the eigenfunction has large amplitude in just one, dominant, propagating region, with the solution decaying exponentially away from it. This region, where the mode is said to be trapped, will then determine the eigenfrequency according to suitable phase relations at its boundaries. 

Let us start off with the superficial layers. Here, typically $\omega\!\gg\!S_l$ and the behavior of the eigenfunction is thus controlled by $\omega_{\rm{c}}$; the role of $\omega_{\rm{c}}$ is, nonetheless, minor in the remaining of the star, where the properties of the eigenfunction are effectively controlled by $S_l$ and $N$. Modes with frequency below the atmospheric value of $\omega_{\rm{c}}$ or, equivalently, with wavelength exceeding the density scale height, decay exponentially in the atmosphere, being reflected back and hence ending up trapped inside the star.

In unevolved stars (e.g., the Sun) the buoyancy frequency $N$ remains at relatively low values throughout the star, in which case the behavior of a high-frequency mode is mostly controlled by $S_l$. The eigenfunction of such a mode will be trapped between the near-surface reflection determined by $\omega\!=\!\omega_{\rm{c}}$ and an inner turning point located where $S_l(r_{\rm{t}})\!=\!\omega$, or
\begin{equation}
\label{turning}
\frac{c^2(r_{\rm{t}})}{r^2_{\rm{t}}}=\frac{\omega^2}{l(l+1)} \, ,
\end{equation}
with $r_{\rm{t}}$ being determined by $l$ and $\omega$. These are p modes, and so are the solar five-minute oscillations. For p modes, typically $\omega\!\gg\!N$, and $K$ may thus be approximated by
\begin{equation}
K(r) \approx \frac{1}{c^2} (\omega^2 - S_l^2) \, .
\end{equation}
In this approximation, the dynamics of the p modes is therefore solely determined by the variation of the sound speed with $r$. From Eq.~(\ref{turning}) it turns out that, for low-degree modes, $r_{\rm{t}}$ is small, meaning that those modes will sample most of the stellar interior. Radial p modes, in particular, travel all the way to the center of the star. Figure \ref{rays} illustrates the propagation of acoustic waves in a so-called ray plot.

\figuremacroW{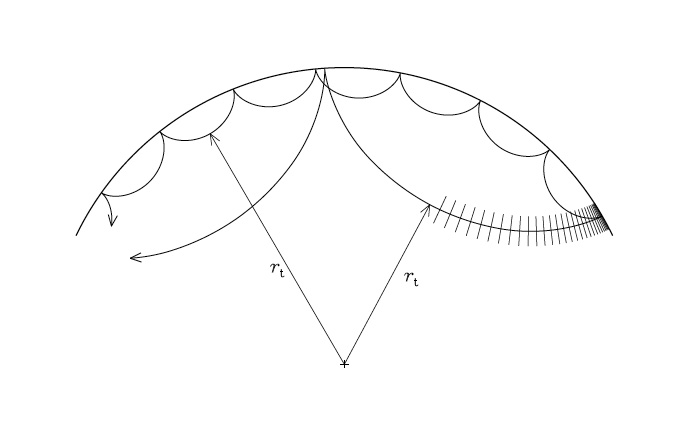}{Acoustic-ray propagation in a cross-section of the solar interior}{A mode with $l\!=\!30$ and $\nu\!=\!3\:{\rm{mHz}}$ penetrates deeper into the Sun than a mode with $l\!=\!100$ and $\nu\!=\!3\:{\rm{mHz}}$. The lines perpendicular to the ray path of the $l\!=\!30$ mode represent wave fronts. As the acoustic wave propagates into the star, the deeper wave fronts experience a higher sound speed. As a consequence, the ray path is bent away from the radial direction. At the inner turning point waves travel horizontally and undergo total internal refraction. At the surface, the acoustic waves are reflected due to a sudden decrease in density. Figure courtesy of J.~Christensen-Dalsgaard.}{0.95}

Let us continue looking at the case of an unevolved star. Low-frequency modes satisfy $\omega\!\ll\!S_l$ throughout most of the stellar radius. Under these circumstances the eingenfunction of a mode oscillates in a region approximately determined by $\omega\!<\!N$, and thus to great extent independent of the degree $l$. These are g modes, having one turning point very near the center of the star and a second one just below the base of the convection zone. For g modes in general $\omega^2\!\ll\!S_l^2$, and $K$ may then be approximated by
\begin{equation}
K(r) \approx \frac{1}{\omega^2} (N^2-\omega^2) \, \frac{l(l+1)}{r^2} \, .
\end{equation}
It is now obvious that the dynamics is controlled by the variation of $N$ with $r$.

However, it is evident from Eq.~(\ref{buoyancy}) and Fig.~\ref{charact_freqs} that $N$ may attain very large values in the core of an evolved star. This comes as a result of an increase of the local gravitational acceleration $g$ due to the contraction of the core. Furthermore, strong gradients in the hydrogen abundance may enhance that effect by causing $\nabla_\mu$ to become a large positive number. Consequently, even at high frequencies close to the atmospheric value of $\omega_{\rm{c}}$, relevant for stochastic excitation, $K$ may have a positive value both in the envelope where $\omega\!>\!S_l,N$ (p-mode behavior), and in the deep interior where $\omega\!<\!S_l,N$ (g-mode behavior). This interchangeable physical nature is illustrated in Fig.~\ref{charact_freqs} for a stochastically-excited $l\!=\!1$ mixed mode in the model of $\eta$ Boo. A more detailed discussion on these so-called mixed modes will be presented in Sect.~\ref{mixedmodes}.

\subsection{p modes and g modes in the Sun}

At this stage it is instructive to have a quick look at the eigenfrequencies computed for a model of the present Sun. These are displayed in Fig.~\ref{pgSun} as a function of degree $l$. Two distinct, although slightly overlapping, families of modes are obvious, viz., p and g modes. The frequencies of p modes are seen to increase with radial order $n$ and degree $l$. The frequencies of g modes -- also increasing with $l$ -- are now seen to decrease with overtone (i.e., with the number of radial nodes $|n|$), while increasing with $n${\footnote{Here, I adopt the convention that $n$ is negative for g modes, with $|n|$ corresponding to the number of radial nodes in the eigenfunction. On the other hand, p modes are assigned positive values of $n$ corresponding to the number of radial nodes.}}. Since buoyancy demands gas motions that are primarily horizontal, there are no radial (i.e., $l\!=\!0$) g modes. 

A third family of modes, labeled with $n\!=\!0$, although similar in behavior to the p modes, are in fact physically distinct. They are surface gravity waves and are known as f modes.     

\figuremacroW{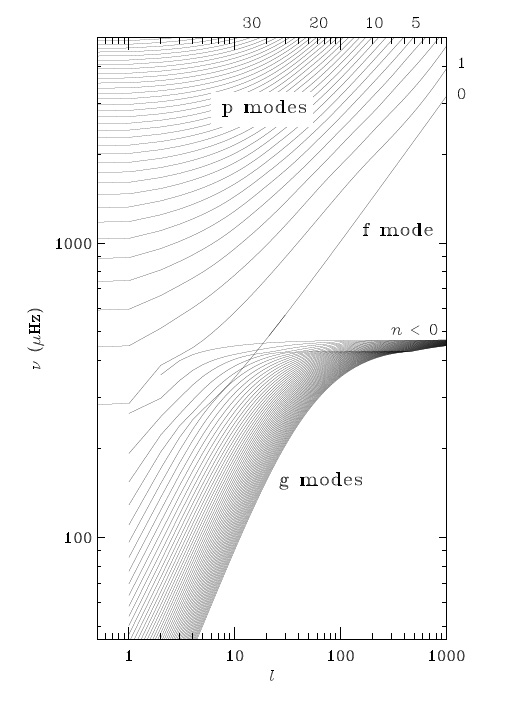}{Computed eigenfrequencies for a model of the present Sun}{Although only integral $l$ have physical meaning, continuous lines are shown for clarity. Selected values of the radial order $n$ are indicated for the p modes. Close inspection tells us that the solar five-minute oscillations are indeed standing acoustic waves, generally of high radial order. Figure courtesy of J.~Christensen-Dalsgaard.}{0.68}

\subsection{The effect of rotation}\label{sect:rot}
The dependence of the oscillations on the azimuthal order $m$ has been so far neglected. From Eqs.~(\ref{displacement}) and (\ref{spharm}) it can be seen that, for $m\!\neq\!0$, the exponentials in both equations combine to give $\exp[-{\rm{i}}\,(2\pi\nu t-m\phi)]$. The extra phase in this time-dependent term means that modes with $m\!\neq\!0$ are in fact traveling waves; modes formed by waves moving with the rotation of the star are called prograde modes (positive $m$), while those formed by waves traveling against the rotation of the star are called retrograde modes (negative $m$).

As already stated, there are $2l+1$ modes for each multiplet of degree $l$. Moreover, in the case of a spherically symmetric star, their frequencies will be the same. However, this frequency degeneracy is lifted by departures from spherical symmetry, of which the most notorious physical cause is rotation. Rotation introduces a dependence of the mode frequencies on $m$, with prograde (retrograde) modes having frequencies slightly higher (lower) than the axisymmetric mode in the observer's frame of reference. For the radial modes one simply cannot see this rotational signature. 

When the angular velocity of the star, $\Omega$, is small, as it is expected for most solar-like pulsators, the effect of rotation can be treated following a perturbative analysis. In the case of rigid-body rotation (i.e., $\Omega\!=\!\Omega(r)$), the frequency $\omega_{nlm}$ of a mode, as observed in an inertial frame, can be expressed to a first order of approximation as \citep{Ledoux51}:
\begin{equation}
\label{splitting}
\omega_{nlm}=\omega_{nl0}+m\langle\Omega\rangle_{nl}\,(1-C_{nl}) \, ,
\end{equation}
where $\langle\Omega\rangle_{nl}$ is an average of $\Omega$ over the stellar interior that depends on the properties of the eigenfunction in the non-rotating star. The kinematic splitting $m\langle\Omega\rangle_{nl}$ is corrected for the effect of the Coriolis force through the dimensionless Ledoux constant, $C_{nl}$. For high-order acoustic modes $C_{nl}\!\ll\!1$, and the rotational splitting is thus dominated by advection and given by the average angular velocity. With access only to low-degree acoustic modes, limited information can be achieved on the profile of rotation throughout the star. In that instance one would, however, still expect to obtain a measure of the average internal angular velocity. Finally, for high-order g modes $C_{nl}\!\simeq\!1/[l(l+1)]$.

To a second order of approximation, centrifugal effects that disrupt the equilibrium structure of the star are taken into account through an additional frequency perturbation that is independent of the sign of $m$. This perturbation scales as the ratio of the centrifugal to the gravitational forces at the stellar surface. Although negligible in the Sun, these effects may be significant for faster solar-like rotators. \citet{Ballot10} alerts to the need of considering second-order effects and, based on the work of \citet{Saio81}, presents an alternative description to that given in Eq.~(\ref{splitting}). Rotation is not the only physical cause behind a departure from spherical symmetry. Other agents, such as large-scale magnetic fields, may introduce additional corrections to the oscillation frequencies.

A way of measuring the inclination angle, $i$, between the direction of the rotation axis of a solar-like pulsator and the line of sight, is provided by asteroseismology. A knowledge of $i$ is not only important for obtaining improved stellar parameters, but also for determining the true masses of extrasolar planets that have been detected from periodic Doppler shifts seen in the spectra of their host stars. Assuming energy equipartition between multiplet components with different azimuthal order, the dependence of mode power on $m$ is given by \citep{GizonSolanki}:
\begin{equation}
\label{mheights}
\mathscr{E}_{l m}(i) = \frac{(l-|m|)!}{(l+|m|)!} \left[P_l^{|m|}(\cos i)\right]^2 \, .
\end{equation}
\citet{HandCamp} present explicit expressions for the computation of $\mathscr{E}_{l m}(i)$ with $l$ ranging from 0 to 4. Sensitivity to multiplet components with different $m$ is essentially a geometrical effect, mainly linked to the limb-darkening function. However, for velocity observations, the rotational shift of the spectral lines across the stellar disk may induce a departure from the description adopted above \citep{Brookes78,JCD89,Broomhall09}. I will mention this point again in Sect.~\ref{modelPS}.

According to Eq.~(\ref{mheights}), when the rotation axis points toward the observer (i.e., $i\!=\!0^\circ$), only the axisymmetric mode is visible and no inference can thus be made of the rotation. In the case of the Sun, on the other hand, whose rotation axis is approximately in the plane of the sky (i.e., $i\!\approx\!90^\circ$), whole-disk observations are essentially sensible only to modes with even $|l-m|$. Figure \ref{rotinc} displays the limit power spectra (for a definition see Sect.~\ref{pspec}) of dipole and quadrupole multiplets as a function of the inclination $i$.

\section{Stochastic excitation of oscillations}
Intrinsically stable oscillations, such as the ones present in stars on the cool side of the Cepheid instability strip, of which the Sun is an example, are thought to be stochastically excited by the vigorous near-surface convection \citep[e.g.,][]{GoldreichKeeley}. In these stars, the turbulent convective motion near the surface reaches speeds close to the speed of sound, and consequently acts as an efficient source of acoustic radiation that will excite the normal modes of the star. \citet{Houdek06} provides a recent review of the process of stochastic excitation in solar-like pulsators.

\figuremacroW{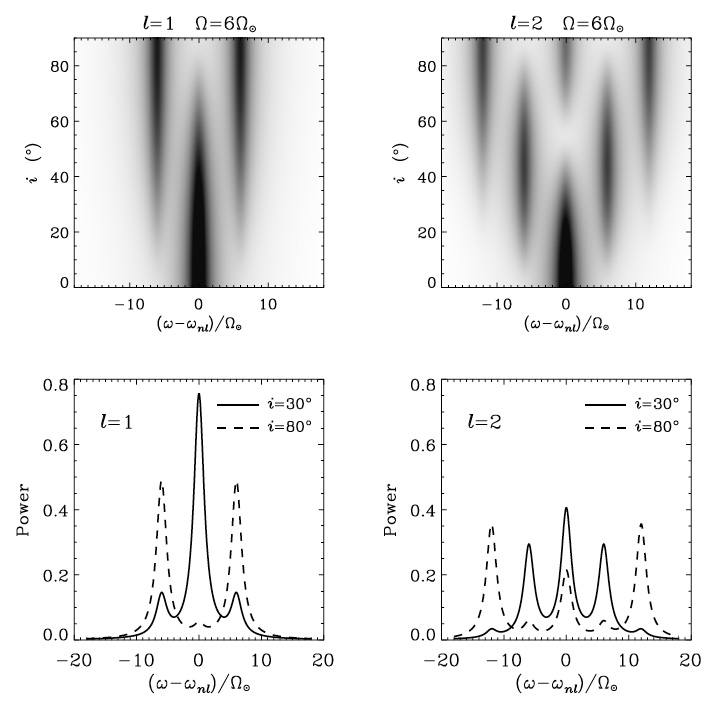}{Power spectra of dipole and quadrupole multiplets as a function of the inclination $i$}{Limit power spectra, with no background noise added, are displayed in the left-hand panels for a dipole (i.e., $l\!=\!1$) multiplet and in the right-hand panels for a quadrupole (i.e., $l\!=\!2$) multiplet. Mode linewidths were assigned solar values ($\Gamma_\odot\!\approx\!1\:{\rm{\mu Hz}}$; for a definition see Sect.~\ref{sect:height_width}) and the angular velocity is six times solar (where $\Omega_\odot/2\pi\!\approx\!0.5\:{\rm{\mu Hz}}$). From \citet{GizonSolanki}.}{0.85}

\subsection{Power spectrum of a solar-like oscillator}\label{pspec}
Understanding the characteristics of the power spectrum of a solar-like oscillator is fundamental in order to extract information on the physics of the modes. \citet{Batchelor} treated the general problem of the stochastic driving of a damped oscillator. 

Such a system can be described by
\begin{equation}
\label{oscillator}
\frac{{\rm{d}}^2}{{\rm{d}}t^2}y(t) + 2\eta\,\frac{{\rm{d}}}{{\rm{d}}t}y(t) + \omega_0^2\,y(t)=f(t) \, ,
\end{equation}
where $y(t)$ is the amplitude of the oscillator, $\eta$ is the linear damping rate, $\omega_0$ is the frequency of the undamped oscillator, and $f(t)$ is a random forcing function. By introducing the Fourier transforms of $y$ and $f$ as
\begin{equation}
Y(\omega)=\int y(t)\,{\rm{e}}^{{\rm{i}}\,\omega t} {\rm{d}}t \, , \;\;\; F(\omega)=\int f(t)\,{\rm{e}}^{{\rm{i}}\,\omega t} {\rm{d}}t \, , 
\end{equation}
the Fourier transform of Eq.~(\ref{oscillator}) is then expressed as
\begin{equation}
-\omega^2\,Y(\omega)-{\rm{i}}\,2\eta\omega\,Y(\omega) + \omega_0^2\,Y(\omega)=F(\omega).
\end{equation}

When a finite realization of the process described by $y(t)$ is observed for a given period of time, long enough so as to fully resolve the resonance, an estimate of the power spectrum (see Fig.~\ref{lorentz}) is then given by
\begin{equation}
P(\omega)=|Y(\omega)|^2=\frac{|F(\omega)|^2}{(\omega_0^2-\omega^2)^2 + 4\,\eta^2\omega^2} \, .
\end{equation}
The power spectrum of the random forcing function, $|F(\omega)|^2$, is uncorrelated at (angular) frequency separations of $2\pi/T_{\rm{obs}}$, with $T_{\rm{obs}}$ being the total observational span. Furthermore, at a fixed frequency, the spectra of different realizations take values that obey a $\chi^2$ distribution with 2 degrees of freedom{\footnote{The statistics of the power spectrum of a pure noise signal is derived in Sect.~\ref{PSstat}.}} (or $\chi^2_2$). \citet{Woodard84} showed that solar oscillation data are consistent with this distribution. 

In the limit of taking the ensemble average of an infinite number of realizations, and further considering that the damping rate is generally very small compared to the frequency of oscillation, one obtains near the resonance (i.e., for $|\omega-\omega_0|\!\ll\!\omega_0$) the following expression for the expectation value of the power spectrum (also called limit spectrum; see Fig.~\ref{lorentz}):
\begin{equation}
\label{lorentzprof}
\langle P(\omega)\rangle \simeq \frac{1}{4\,\omega_0^2} \frac{\langle P_f(\omega)\rangle}{(\omega-\omega_0)^2+\eta^2} \, .
\end{equation}  
The average power spectrum of the random forcing function, $\langle P_f(\omega)\rangle$, is expected to be a slowly-varying function of frequency. The result will thus be a Lorentzian profile, characterized by the central frequency $\omega_0$ and a width determined by the linear damping rate $\eta$. 

However, the fact that the dominant contributions to the driving of the oscillations are restricted to a region of small radial extent, will lead to asymmetries in the mode profiles \citep[e.g.,][]{Duvall93,AbramsKumar}. These asymmetries are determined by the relative location of the region responsible for the driving with respect to the resonant cavity. The detection of these asymmetries in the case of the Sun made it possible to estimate the location of the dominant source of mode excitation \citep[e.g.,][]{ChapAppour}. Moreover, the sign of the asymmetry depends upon the observable, an aspect that is related to the different correlation, found in velocity and intensity measurements, between the background stellar noise and the convective excitation of the oscillations \citep[e.g.,][]{Nigam98}. A simple expression for describing an asymmetrical mode profile is given by \citet{NigKos98}. In principle, similar detections are possible for other stars through the analysis of long and continuous observations with overall high signal-to-noise ratio (SNR).  

Very often, one implicitly assumes that the background stellar noise and the convective excitation of the oscillations are statistically uncorrelated stationary processes. In that case, the overall power spectrum is simply given by the sum of the separate power spectra. This is usually a fairly good approximation for solar-like oscillations, meaning that we end up neglecting any asymmetries in the mode profiles. If, however, one assumes that the processes are correlated but that their correlation is stationary, then we should take into account profile asymmetry. Such correlations have been studied by \citet{Severino} in the helioseismic context.  

\figuremacroW{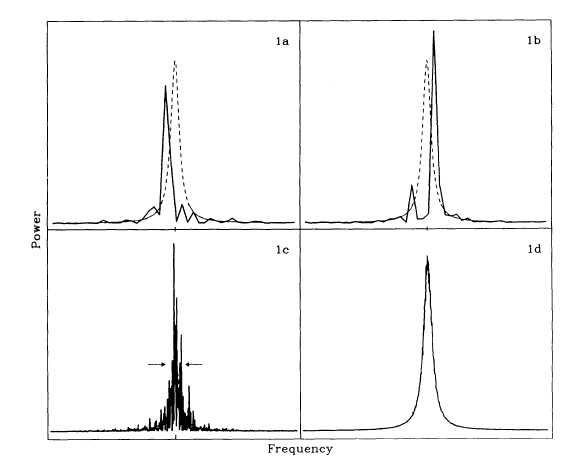}{Power spectrum of a solar-like oscillator}{Panels (a) and (b) display two realizations of the same limit spectrum (shown as a dotted curve). Both power spectra appear as an erratic function concealing the subjacent Lorentzian profile. Panel (c) displays a realization of the same limit spectrum, although with a resolution twenty times higher. Increasing the total observational span, and hence the resolution, did nothing to reduce the erratic behavior of the power spectrum. This panel is scaled differently to the remaining panels, with arrows marking the maximum value of the limit spectrum. Panel (d) displays the ensemble average of a large number of realizations with the same resolution as in (c). From \citet{Anderson90}.}{0.85}

\subsection{Mode excitation}\label{sect:mexcit}
I start by introducing two useful global properties of a mode, namely, its normalized inertia $E$ and its mode mass $M_{\rm{mode}}$:
\begin{equation}
\label{norminert}
E=\frac{M_{{\rm{mode}}}}{M} \equiv \frac{\int_V \rho|{\bf{\delta r}}|^2\,{\rm{d}}V}{M|{\bf{\delta r}}|^2_{\rm{ph}}} \, ,
\end{equation}
where the integration is over the volume $V$ of the star, and $|{\bf{\delta r}}|^2_{\rm{ph}}$ is the squared norm of the displacement vector at the photosphere. Based on the definition of mode inertia, one would thus expect modes trapped in the deep stellar interior to have large values of $E$. Mode inertia relates the photospheric rms velocity, $V_{\rm{rms}}$, to the kinetic energy of the mode, $\mathcal{E}_{\rm{kin}}$, through: 
\begin{equation}
\label{menergy}
\mathcal{E}_{\rm{kin}}=\frac{1}{2} M_{\rm{mode}} V_{\rm{rms}}^2 = \frac{1}{2} M E V_{\rm{rms}}^2 \, .
\end{equation}

Based on a detailed description of the stochastic mechanism of mode excitation, \citet{Chaplin05} obtained the following result for the expected mode amplitude{\footnote{This expression predicts the theoretical mode amplitude. In order to predict the observed quantity, one should multiply this expression by a factor $K_{\theta,\phi}$ accounting for the spatial filter of real observations.}}:
\begin{equation}
\label{expectamp}
V_{\rm{rms}}^2=\frac{1}{\eta E} \frac{\tilde{\mathcal{P}}}{E} \approx \frac{\mathcal{F}(\omega)}{E} \, ,
\end{equation} 
where $\tilde{\mathcal{P}}$ is a measure of the acoustic energy input. Both terms $\tilde{\mathcal{P}}$ and $\eta E$ depend on the properties of the eigenfunction in the near-surface region of the star and are thus predominantly functions of frequency, which justifies the second equality in the last equation. It follows from Eq.~(\ref{expectamp}) that, for a given frequency, mode amplitude essentially scales as $E^{-1/2}$; also, mode energy (cf.~Eq.~\ref{menergy}) is predominantly a function of frequency. 

The stochastic process gives rise to the excitation of all modes in a substantial range of frequencies, with an amplitude modulation that reflects the slow frequency dependence of the energy input and damping rate. The properties of the mode eigenfunctions in the near-surface region of vigorous convection also play an important role in determining the frequency dependence of the mode amplitudes. Low-frequency modes tend to be evanescent near the surface, hence leading to inefficient excitation and small mode amplitudes. High-frequency modes, on the other hand, see their amplitudes reduced due to a decrease of the convective energy at the timescale of the oscillations, combined with an increase in the damping rate. The driving is ultimately most efficient for those modes whose periods match the relevant timescales of near-surface convection, from 5 to 10 minutes in the solar case. The frequency of maximum amplitude, $\nu_{\rm{max}}$, is supposed to scale with the acoustic cut-off frequency \citep{Brown91,KjeldsenBedding95,BeddingKjeldsen03,Chaplin08,Belkacem11}, which when determined for an isothermal stellar atmosphere gives a scaling relation in terms of mass, radius, and effective temperature of (cf.~Eqs.~\ref{t_dyn} and \ref{f_cut})
\begin{equation}
\label{numaxscaling}
\nu_{\rm{max}} \propto M \, R^{-2} \, T_{\rm{eff}}^{-1/2} \, .
\end{equation}

The overall result is a characteristic amplitude distribution with frequency (very often modulated by a bell-shaped envelope), which constitutes a signature of the presence of solar-like oscillations (see Fig.~\ref{solarspec}). Our ability to theoretically predict the amplitudes of stochastically-excited modes, combined with a complete set of observed modes over a broad frequency range, substantially simplify the process of mode identification and hence the comparison with stellar models, ultimately exponentiating the asteroseismic diagnostic potential of solar-like oscillations. This comes in great contrast to heat-engine excitation, for which the mechanism determining the final amplitudes of the modes is not well understood.

\citet{JCDFrandsen83} provided rough estimates of the oscillation amplitudes in main-sequence stars and cool giants from model calculations. Based on those results, \citet{KjeldsenBedding95} found that mode amplitudes given in terms of surface velocities scale approximately as 
\begin{equation}
\label{scalerel}
V_{\rm{rms}} \propto \left(\frac{L}{M}\right)^{s} \, ,
\end{equation}
with $s\!=\!1$. They further argued that the oscillation amplitudes, $A_{\rm{rms}}$, observed in photometry at a wavelength $\lambda$, are related to the velocity amplitudes according to
\begin{equation}
\label{ampconv}
A_{\rm{rms}}=({\rm{d}}L/L)_{\lambda} \propto \frac{V_{\rm{rms}}}{\lambda\,T_{\rm{eff}}^r} \, ,
\end{equation}
or, in terms of bolometric amplitudes,
\begin{equation}
\label{scalerel2}
A_{\rm{rms}}^{\rm{bol}} \propto \frac{V_{\rm{rms}}}{T_{\rm{eff}}^{r-1}} \, .
\end{equation}
The exponent $s$ has since been revised both theoretically \citep{Houdek99,Houdek06,Samadi07}, as well as observationally using red-giant stars \citep{Gilliland08,Dziem10,MosserRG,StelloNGC6819}, main-sequence stars \citep{Vernercomparison}, and an ensemble of main-sequence and red-giant stars \citep{Baudin11}. As a result, its value is now seen to reside roughly within the range from 0.7 to 1.5. The value of $r$ is chosen to be either $r\!=\!1.5$ (assuming adiabatic oscillations) or $r\!=\!2$ \citep[following a fit to observational data in][]{KjeldsenBedding95}. 

Accordingly, amplitudes are predicted to increase with increasing luminosity along the main sequence and relatively large amplitudes are expected for red giants. Such predictions are now being increasingly tested against observations. Reasonable agreement is apparently found between predicted and observed amplitudes for stars cooler or as hot as the Sun, while for hotter stars predictions considerably exceed the observed values. As an example of the type of discrepancy just mentioned, early \emph{CoRoT} results, based on the analysis of the light curves of three main-sequence F stars fairly hotter than the Sun, showed that the oscillation amplitudes of those solar-like pulsators are about 25\% below the theoretical predictions \citep{Michel08}.

Recently, \citet{KB11} argued that amplitudes of oscillations in velocity should scale in proportion to velocity fluctuations due to granulation, since the physical motion of convective cells is what drives the oscillations. Therefore, they proposed a revised scaling relation for the velocities:
\begin{equation}
\label{scalerelrev}
V_{\rm{rms}} \propto \frac{L\,\tau^{0.5}}{M^{1.5}\,T_{\rm{eff}}^{2.25}} \, ,
\end{equation}
where $\tau$ is the e-folding mode lifetime. Compared to Eq.~(\ref{scalerel}), this revised scaling relation now incorporates a strong temperature dependence and also a weak dependence on mode lifetime. Stars with shorter mode lifetimes will show lower amplitudes, all other parameters remaining unaltered. Note that simple scaling relations exist in the literature for $\tau$ \citep{Chaplintau,Baudin11,Appourchauxlinewidths}. A revised scaling relation for (narrowband) intensity amplitudes is then given by
\begin{equation}
\label{ampconv2}
A_{\rm{rms}} \propto \frac{L\,\tau^{0.5}}{\lambda\,M^{1.5}\,T_{\rm{eff}}^{2.25+r}} \, ,
\end{equation}
or, in terms of bolometric amplitudes,
\begin{equation}
A_{\rm{rms}}^{\rm{bol}} \propto \frac{L\,\tau^{0.5}}{M^{1.5}\,T_{\rm{eff}}^{1.25+r}} \, .
\end{equation}
This revised scaling relation can now be extensively tested with observations from \emph{CoRoT} and \emph{Kepler} \citep[e.g.,][]{MulderScully}.

\figuremacroW{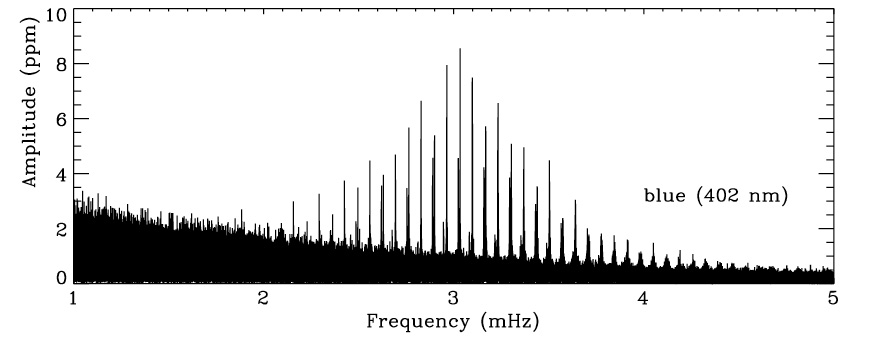}{p-mode amplitude spectrum of the Sun}{The spectrum is based on whole-disk observations of the solar irradiance using the blue channel of the VIRGO/SPM instrument on board the \emph{SOHO} spacecraft. The observations have been smoothed and rescaled to show the spectrum corresponding to 30 days. Adapted from \citet{BeddingKjeldsen03}.}{1}

\subsection{Mode height and mode linewidth}\label{sect:height_width}
It is not the integrated power $V_{\rm{rms}}^2$ (or $A_{\rm{rms}}^2$) that is observed directly in the power spectrum, but instead the power spectral density. If the total observational span is long enough in order to resolve a mode peak in the power spectrum (i.e., $T_{\rm{obs}}\!\gg\!2\tau$, where the mode lifetime is given by $\tau\!=\!1/\eta$), then the mode height (or maximum power spectral density) is given by \citep{Chaplin03,Chaplin05}: 
\begin{equation}
\label{regime1}
H\approx\frac{2\,V_{\rm{rms}}^2}{\eta}=\frac{2\,V_{\rm{rms}}^2}{\pi \Gamma} \, ,
\end{equation}
where $\Gamma\!=\!(\pi\tau)^{-1}$ is the full width at half maximum of the mode peak, being commonly called the mode linewidth{\footnote{The dominant modes in the Sun have linewidths of 1--$2\:{\rm{\mu Hz}}$ and hence lifetimes of 2--4 days \citep[e.g.,][]{solarlinewidths}.}}. In this regime, $H$ is independent of $E$, since both $V_{\rm{rms}}^2$ and $\eta$ scale as $E^{-1}$ at fixed frequency. Conversely, when $T_{\rm{obs}}\!\ll\!2\tau$ the mode peak is not resolved, and power is essentially confined in one bin of the power spectrum, meaning that
\begin{equation}
\label{regime2}
H\sim V_{\rm{rms}}^2 T_{\rm{obs}} \, ,
\end{equation}
and $H$ is thus proportional to $E^{-1}$. A proper description of $H$, covering these two extreme regimes as well as the intermediate regime, is given by \citep{Fletcher06,Chaplin08}
\begin{equation}
\label{regimegeneral}
H=\frac{2\,V_{\rm{rms}}^2T_{\rm{obs}}}{\pi\Gamma\,T_{\rm{obs}} + 2} \, .
\end{equation}

\subsection{Statistical properties of the oscillators}
\label{sectstatprop}
Let us bear in mind Eq.~(\ref{oscillator}) describing a damped linear oscillator forced by a random function. Since a very large number of convective elements is responsible for exciting the oscillations, it is reasonable to assume that $f(t)$ is a white-noise process with a Gaussian distribution. Furthermore, the energy of an oscillator in stochastic equilibrium can be interpreted in terms of the distance from the starting point of a two-dimensional random walk with a variable step size in the phase plane. The forcing function $f(t)$ being random means that each step is independent of the previous one. Once a large number of steps has been taken, the displacement $y(t)$ and velocity ${\rm{d}}y(t)/{\rm{d}}t$ will both be normally distributed. Therefore, the total energy of the oscillator, given by the sum of its kinetic energy, $[{\rm{d}}y(t)/{\rm{d}}t]^2$, and potential energy, $\omega_0^2\,y^2(t)$, follows by definition a $\chi^2_2$ distribution, i.e., a Boltzmann distribution:
\begin{equation}
p(\mathcal{E})=\frac{1}{\langle\mathcal{E}\rangle} \exp\left(-\frac{\mathcal{E}}{\langle\mathcal{E}\rangle}\right) \, ,
\end{equation} 
where $\langle\mathcal{E}\rangle$ is the mean energy. This relation will only hold if the mode energy can be measured over time intervals much smaller than the damping time \citep{Kumar88}. The observed solar oscillations amply satisfy this relation \citep[e.g.,][]{Chaplin97}. 

In order to obtain the amplitude distribution, one takes into account that $p(\mathcal{E}){\rm{d}}\mathcal{E}=p(A){\rm{d}}A$, and that the energy $\mathcal{E}$ is proportional to the square of the amplitude, $A^2$. As a result, the amplitude distribution turns out to be a Rayleigh-type distribution:
\begin{equation}
p(A)=\frac{2A}{\langle A^2\rangle} \exp\left(-\frac{A^2}{\langle A^2\rangle}\right) \, ,
\end{equation}
where $\langle A^2\rangle$ is the mean-square amplitude. Interestingly, the mean $\langle A\rangle$ and standard deviation $\sigma(A)$ of this distribution obey the following relation:
\begin{equation}
\sigma(A)=\left(\frac{4}{\pi}-1\right)^{1/2} \langle A\rangle \approx 0.52 \, \langle A\rangle \, , 
\end{equation}
which holds true as long as the oscillator is in stochastic equilibrium.

Therefore, one expects stars whose oscillations are stochastically excited to verify $\sigma(A) \approx 0.52 \langle A\rangle$. \citet{JCD01} noticed that observed amplitudes of semi-regular variables on the asymptotic giant branch approximately followed this relation. They argued that such variability might be due to stochastically-excited oscillations with mode lifetimes ranging from years to decades, a result later confirmed observationally by \citet{BeddingSRV}. This result, first obtained from amateur astronomer data from the American Association of Variable Star Observers (AAVSO), was later also confirmed by \citet{KissBedding} using data from the OGLE-II microlensing project. Furthermore, the regime $\sigma(A) < 0.52 \langle A\rangle$ is expected to hold for oscillations excited by thermal overstability. Most of the oscillations excited by the $\kappa$ mechanism, such as in subdwarf B stars \citep{PereiraLopes05}, are expected to be found in this regime. Finally, the regime $\sigma(A) > 0.52 \langle A\rangle$ corresponds to stochastic oscillations that are not in stochastic equilibrium, a type of oscillatory behavior yet to be observed. 

Based on the ratio $\sigma(A)/\langle A\rangle$, a simple diagnostic method has been established by \citet{PereiraLopes05} that probes the excitation mechanism of stellar pulsations through the analysis of the temporal variation of the amplitude of oscillation modes (see Fig.~\ref{diagnose}). Numerical simulations and the application to the $\gamma$ Dor star HD~22702 served as a test to this method \citep{PereiraLopes07}. The same method has also been applied by \citet{Campantehybrid} to the \emph{CoRoT} hybrid $\gamma$ Dor/$\delta$ Sct star HD~49434 in order to investigate the mechanism responsible for the excitation of the observed intermediate-order g modes (see Sect.~\ref{sechybrid}). 

\figuremacroW{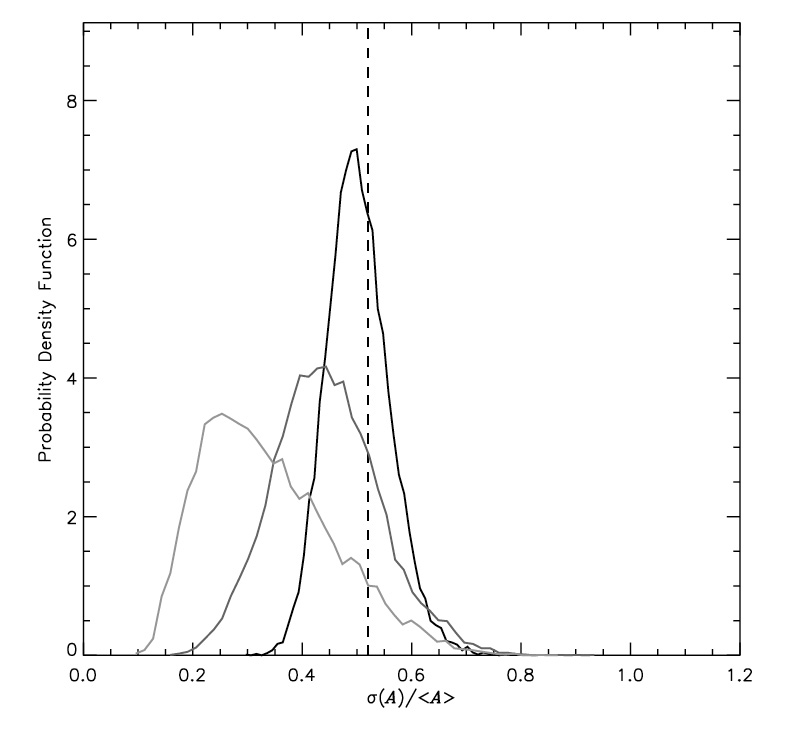}{Probing the excitation mechanism of stellar pulsations}{Probability density function of the statistic $\sigma(A)/\langle A\rangle$ computed by means of Monte Carlo simulations. From left to right, the curves correspond to samples of 40, 200 and 1000 amplitude measurements of a stochastic mode. These measurements come from contiguous time series considerably shorter than the damping time of the mode. The vertical dashed line corresponds to $\sigma(A)/\langle A\rangle\!=\!0.52$. As the number of measurements increases, the distribution gradually acquires a Gaussian profile with the location of its peak converging to 0.52. Moreover, the width of the distribution is substantially reduced.}{0.8}

Recent detection claims, based on \emph{CoRoT} observations, of solar-like oscillations in the massive star V1449 Aql \citep{Belkacem09}, previously known to be a $\beta$ Cep pulsator, constituted the first plausible evidence of simultaneous classical and solar-like oscillations, thus providing additional modeling constraints if we bear in mind that these modes probe different layers of the star. \emph{Kepler} observations of similar stars, however, have so far failed to confirm stochastically-excited oscillations \citep{Balona11}. This is clearly a domain that may benefit from the application of the aforementioned diagnostic method. This has certainly been the case of the very first detection of solar-like oscillations in a $\delta$ Sct star by \citet{AntociNature}, where its application proved decisive to set the claims on firm ground. Notice that solar-like oscillations in $\delta$ Sct stars had already been predicted by theory \citep{Houdek99,Samadi02}. The scientific relevance and implications of this groundbreaking work together with my substantial contribution to its completion, to be specific, in testing the stochastic origin of the oscillatory signal, led me to place the resulting article as a supplement in Appendix \ref{vichinature}.

\subsection{Near-surface effects on computed oscillation frequencies}
The effects of near-surface convection on the computation of oscillation frequencies are a very delicate matter. Computation of oscillation frequencies of stellar models usually assumes adiabaticity, a valid approximation in much of the stellar interior, viz., in regions where the thermal timescale is considerably longer than the period of the oscillations. This is certainly not the case in the near-surface region. Moreover, the dynamical effects of convection are usually neglected. This improper modeling of the near-surface layers gives rise to an offset between observed and computed oscillation frequencies. \citet{Grignonad} address the problem of frequency precision in non-adiabatic models using a time-dependent treatment of convection.

Near-surface effects are essentially independent of $l$ for low-degree modes. Moreover, these effects are predominantly functions of frequency, rapidly increasing with $\omega/\omega_{\rm{c}}$ \citep[e.g.,][]{JCDGough80}, and thus of vital importance if we are to correctly interpret the high-order acoustic modes. \citet{Kjeldsen08corrections} devised an empirical correction for the near-surface offset in the form of a power law:
\begin{equation}
\label{surfcorr}
\delta^{(\rm{surf})}\nu=a\,(\nu/\nu_0)^b \, ,
\end{equation} 
where $\nu_0$ is a suitably chosen reference frequency, and the amplitude $a$ and exponent $b$ are obtained from a fit to solar frequencies of radial modes. They extended this correction to the stellar case with reasonable success by adopting the solar value of the exponent $b$ and using the frequencies of a reference stellar model. Nevertheless, this correction is calibrated with respect to the Sun and thus needs to be thoroughly tested to assess its validity when applied to other solar-like oscillators. \citet{IsaHyi} show that, after applying this correction to the case of $\beta$ Hyi, the observed modes are well reproduced, including those that have mixed-mode character. On the other hand, application of this same correction to the case of Procyon has led to mixed success \citep{DoganProcyon}. One hopes that \emph{Kepler} observations of a broad sample of solar-like pulsators will yield insight into the dependence of these effects on stellar parameters.

In order to account for the minor dependence of the near-surface effects on mode inertia, we may want to rewrite Eq.~(\ref{surfcorr}) as 
\begin{equation}
\delta^{(\rm{surf})}\nu_{nl}=Q_{nl}^{-1}a\,(\nu_{nl}/\nu_0)^b \, ,
\end{equation} 
where the inertia ratio $Q_{nl}$ is given by
\begin{equation}
Q_{nl}=\frac{E_{nl}}{\overline{E}_0(\nu_{nl})} \, ,
\end{equation}
that is to say, the ratio of mode inertia to the interpolated inertia of radial modes. The power-law correction is based on a fit to the radial modes and, relative to these, the effect on the non-radial modes is reduced by a factor proportional to the mode inertia. $Q_{nl}$ also accounts for the presence of mixed modes, for which the near-surface effect is smaller due to the higher inertia ratio $Q_{nl}$.

\section{Asteroseismic inference}

\subsection{Asymptotic signatures}

\subsubsection{Asymptotic relation for p modes}\label{sect:pasympt}
The observed modes of solar-like oscillations are typically high-order acoustic modes. If interaction with a g mode can be neglected, linear, adiabatic, high-order acoustic modes, in a spherically symmetric star, satisfy an asymptotic relation for the frequencies \citep{Vandakurov67,Tassoul80}:
\begin{equation}
\label{asympt}
\nu_{nl} \simeq \left(n+\frac{l}{2}+\epsilon\right)\Delta\nu_0-\left\{\frac{l(l+1)}{4\pi^2\,\Delta\nu_0}\left[\frac{c(R)}{R}-\int_0^R\frac{{\rm{d}}c}{{\rm{d}}r}\frac{{\rm{d}}r}{r}\right]-\delta\right\}\frac{{\Delta\nu_0}^2}{\nu_{nl}} \, ,
\end{equation}
where
\begin{equation}
\Delta\nu_0 = \left(2\int_0^R\frac{{\rm{d}}r}{c}\right)^{-1}
\end{equation}
is the inverse of the sound-travel time across the stellar diameter; additionally, the term $\epsilon\!=\!\epsilon(\nu)$ is determined by the reflection properties of the surface layers, as is the small correction term $\delta$. 

To leading order, Eq.~(\ref{asympt}) predicts that modes should occur in groups corresponding to degree of the same parity (i.e., either even or odd degree) such that $n+l/2$ are the same, and being further uniformly spaced with a separation given by $\Delta\nu_0$. This degeneracy is lifted by considering the second-order term in Eq.~(\ref{asympt}). The spectrum is then characterized both by the large frequency separation
\begin{equation}
\Delta\nu_{nl}=\nu_{n+1\,l}-\nu_{nl}\approx\Delta\nu_0 \, ,
\end{equation}
a quantity depending both on frequency and on mode degree $l$, and by the small frequency separation
\begin{equation}
\label{eq:ssep1}
\delta\nu_{nl}=\nu_{nl}-\nu_{n-1\,l+2}\approx-(4l+6)\frac{\Delta\nu_0}{4\pi^2\,\nu_{nl}}\int_0^R\frac{{\rm{d}}c}{{\rm{d}}r}\frac{{\rm{d}}r}{r} \, ,
\end{equation}
also a frequency- and degree-dependent quantity (I have here neglected the term in the surface sound speed appearing in Eq.~\ref{asympt}). It should be noted that the small frequency separation may become a negative quantity during stellar evolution. This apparent violation of Eq.~(\ref{eq:ssep1}) is associated with the presence of a convective helium-rich core \citep{SorianoVauclair}. 

Both the large and the small frequency separations are shown in Fig.~\ref{solarspec2} in the case of the acoustic amplitude spectrum of the Sun. The quasi-regularity of the spectrum of high-order p modes, along with the characteristic amplitude distribution with frequency discussed in Sect.~\ref{sect:mexcit}, constitute the main signatures of the presence of solar-like oscillations. The large and small frequency separations are extremely valuable diagnostic tools for asteroseismic studies of solar-like oscillators. In fact, these two quantities can be measured with considerable precision, even in the case of low-SNR observations where a determination of individual oscillation frequencies is hindered.

\figuremacroW{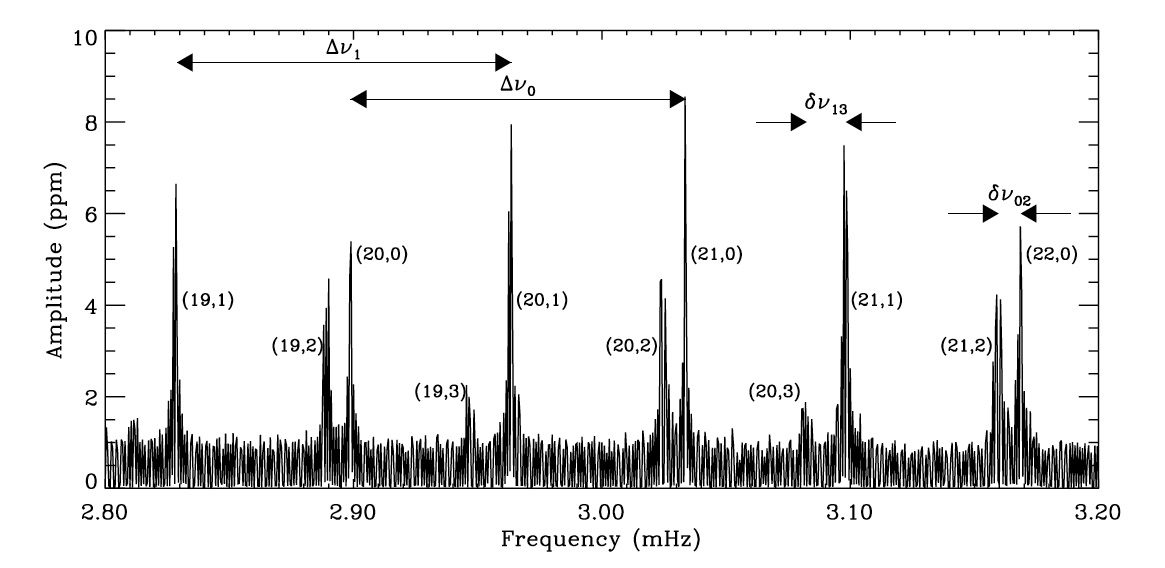}{Close-up of the p-mode amplitude spectrum of the Sun}{This is a close-up of the p-mode spectrum displayed in Fig.~\ref{solarspec}. Mode peaks are tagged with the corresponding ($n,l$) values, which were determined by comparison with theoretical models. The large and small frequency separations are indicated with a slightly different nomenclature than the one used in the text, which should, however, still be clear. From \citet{BeddingKjeldsen03}.}{1}

Moreover, it may be also convenient to consider small separations that take into account modes with adjacent degree:
\begin{equation}
\label{eq:ssep2}
\delta^{(1)}\nu_{nl}=\nu_{nl}-\frac{1}{2}(\nu_{n-1\,l+1}+\nu_{n\,l+1})\approx-(2l+2)\frac{\Delta\nu_0}{4\pi^2\,\nu_{nl}}\int_0^R\frac{{\rm{d}}c}{{\rm{d}}r}\frac{{\rm{d}}r}{r} \, ,
\end{equation}
viz., the amount by which modes with degree $l$ are offset from the midpoint between the $l+1$ modes on either side.

A recurrent way of visualizing the asymptotic properties of the acoustic spectrum is to build an \'echelle diagram \citep[e.g.,][]{Grec83}, for which one starts by expressing the frequencies as
\begin{equation}
\nu_{nl}=\nu_0+k\langle\Delta\nu\rangle+\tilde{\nu}_{nl} \, ,
\end{equation}
where $\nu_0$ is a reference frequency, $\langle\Delta\nu\rangle$ is a suitable average of the large frequency separation $\Delta\nu_{nl}$, and $k$ is an integer such that $\tilde{\nu}_{nl}$ takes a value between 0 and $\langle\Delta\nu\rangle$. Finally, the diagram is built by plotting $\tilde{\nu}_{nl}$ on the abscissa and $\nu_0+k\langle\Delta\nu\rangle$ on the ordinate, the graphical equivalent to slicing the spectrum into segments of length $\langle\Delta\nu\rangle$ and stacking them one on top of the other. Figure \ref{echelle} displays a scaled \'echelle diagram \citep{scaledechelle} where the p-mode frequencies of three main-sequence stars are simultaneously plotted. If the frequencies of these stars were to strictly obey the asymptotic relation in Eq.~(\ref{asympt}), then they would exhibit essentially vertical ridges in the \'echelle diagram. However, departures from regularity are clearly present: variations in the large separation with frequency are seen to introduce a curvature in the ridges, while variations in the small separation with frequency appear as a convergence or divergence of the relevant ridges.   

\figuremacroW{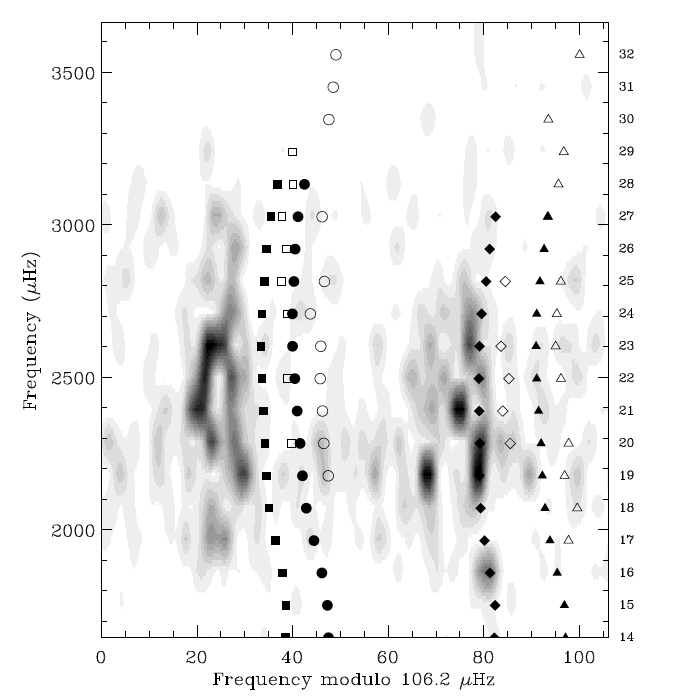}{Scaled \'echelle diagram}{Three main-sequence stars are simultaneously compared: the power spectrum of $\alpha$ Cen A is given in grayscale, filled symbols represent scaled solar frequencies, and open symbols represent scaled frequencies for $\alpha$ Cen B. Scaling factors for the frequencies were computed based on a homology relation (cf.~Eq.~\ref{homology}) and fine-tuned so as to match the slopes of the ridges for the three stars (making them vertical), which means matching $\Delta\nu_0$. Symbol shapes indicate mode degree: $l\!=\!0$ (circles), $l\!=\!1$ (triangles), $l\!=\!2$ (squares), and $l\!=\!3$ (diamonds). From \citet{scaledechelle}.}{0.86}

The small frequency separation is mostly sensitive to conditions in the stellar core, where the eigenfunctions of modes of similar frequency but of different degree mainly differ. As stellar evolution takes its course, hydrogen is burned into helium in the core leading to an increase of the mean molecular weight. Bearing in mind Eq.~(\ref{eq:soundspeed}) for the sound speed in an ideal gas, and taking into consideration that the central temperature will not vary significantly during the phase of hydrogen burning, the sound speed in the core will thus decrease as the star becomes more evolved, such decrease being more intense at the center and becoming more pronounced with increasing age. Therefore, the resulting positive sound-speed gradient ${\rm{d}}c/{\rm{d}}r$ in the core causes a gradual reduction of $\delta\nu_{nl}$ and $\delta^{(1)}\nu_{nl}$ with increasing stellar age (cf.~Eqs.~\ref{eq:ssep1} and \ref{eq:ssep2}, respectively). In conclusion, the small frequency separation can be seen as a diagnostic tool of the evolutionary stage of a (main-sequence) star. 

On the other hand, the large frequency separation provides a more global measure of the properties of a star, essentially scaling as $t_{\rm{dyn}}^{-1}$ (cf.~Eq.~\ref{t_dyn}):
\begin{equation}
\label{homology}
\Delta\nu_0 \propto \left(\frac{M}{R^3}\right)^{1/2}\propto\sqrt{\overline{\rho}} \, .
\end{equation}
Besides providing a measure of the mean stellar density, it may be taken as a measure of stellar mass for stars residing on the main sequence. Based on theoretical models, \citet{theodiagramsWhite} suggested that the scaling relation of $\Delta\nu_0$ with density may be improved by including a function of $T_{\rm{eff}}$. 

The preceding considerations suggest presenting the large and small frequency separations in a two-dimensional diagram, which can be thought of as an asteroseismic H-R diagram, known as the C-D diagram \citep[e.g.,][]{CDdiagramori,CDdiagram}. An example of such a diagram is displayed in Fig.~\ref{CDdiag}. Assuming that the remaining stellar parameters (e.g., the chemical composition) are known, the location of a star in this diagram would then determine its mass and age. \citet{Monteiro02} present an interesting analysis of the uncertainties associated with the use of this diagram due to the sensitivity to several model parameters.

For subgiants as well as for red giants, however, the small separation is approximately a fixed fraction of the large separation, regardless of mass or evolutionary state \citep{BeddingLetter10,obsdiagramsWhite}. As a consequence, the distribution of evolved stars in the C-D diagram becomes highly degenerate as evolutionary stellar tracks converge. Based on models extending from the zero-age main sequence to the tip of the red-giant branch, \citet{theodiagramsWhite} revived the diagnostic potential of an alternative asteroseismic diagram relating $\epsilon$ (cf.~Eq.~\ref{asympt}) to the large separation (see Fig.~\ref{epsilondiag} for an example). They found that evolutionary tracks in this so-called $\epsilon$ diagram \citep[originally introduced by][]{CDdiagramori} are more sensitive to the mass and age of evolved stars than in the C-D diagram. They have also shown that $\epsilon$ is mostly determined by $T_{\rm{eff}}$ and that it could thus be useful for addressing the problem of mode identification in F stars (see also Sect.~\ref{SectMLE}), as previously suggested by \citet{scaledechelle}.

\figuremacroW{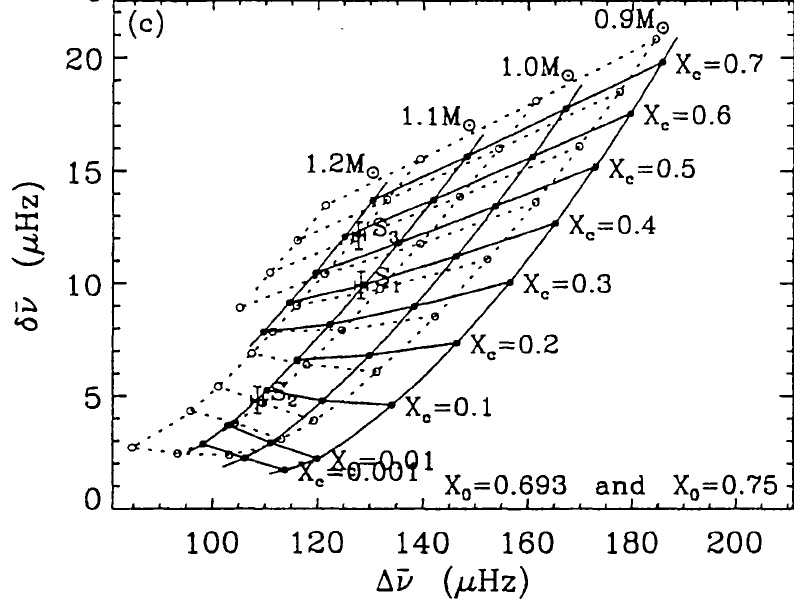}{Example of a C-D diagram}{Evolutionary sequences of stars with 0.9, 1.0, 1.1, and $1.2\:{\rm{M}}_\odot$ are displayed, together with curves of constant central hydrogen abundance, $X_{\rm{c}}$. Stellar age increases as one moves downward along an evolutionary sequence of given mass. Sequences corresponding to models with an initial hydrogen abundance $X_0\!=\!0.75$ appear as solid curves, whereas sequences corresponding to models with $X_0\!=\!0.693$ are displayed as dotted curves. Adapted from \citet{Monteiro02}.}{0.9}

Finally, the small frequency separation still retains a residual sensitivity to the properties of the stellar envelope. \citet{RoxVor03} showed that ratios such as
\begin{equation}
\label{r02}
r_{02}=\frac{\nu_{n0}-\nu_{n-1\,2}}{\nu_{n1}-\nu_{n-1\,1}} \, ,
\end{equation}
and
\begin{equation}
\label{r10}
r_{10}=\frac{-\frac{1}{8}(\nu_{n-1\,1}-4\nu_{n0}+6\nu_{n1}-4\nu_{n+1\,0}+\nu_{n+1\,1})}{\nu_{n+1\,0}-\nu_{n0}} \, ,
\end{equation}
between small and large frequency separations, are largely independent of the surface layers and provide a reliable measure of the core properties.

\figuremacroW{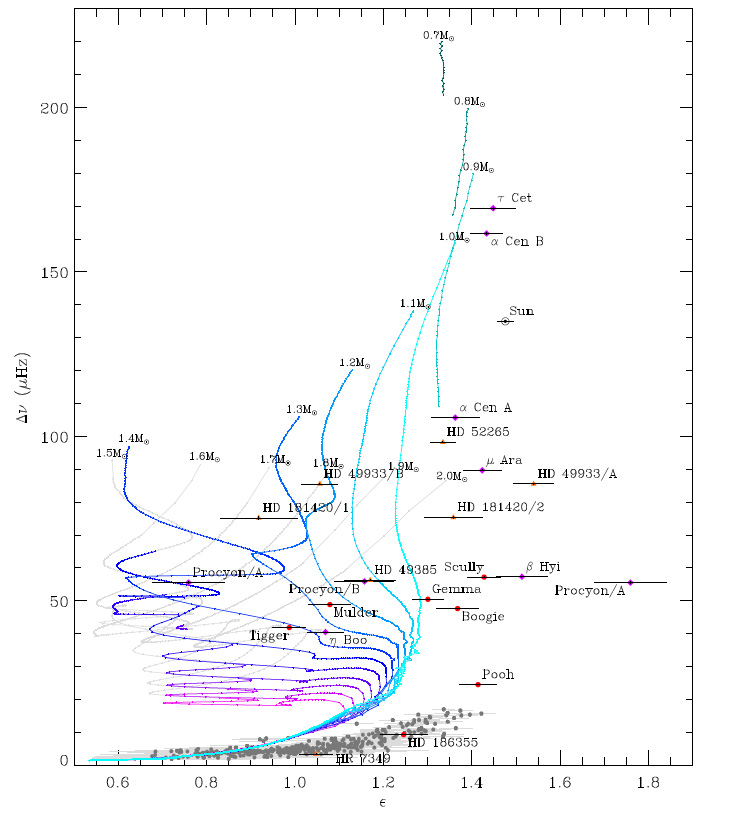}{Example of an $\epsilon$ diagram}{$\epsilon$ diagram with model tracks for near-solar metallicity ($Z_0\!=\!0.017$). Tracks increase in mass from 0.7 to $2.0\:{\rm{M}}_\odot$ (green to magenta lines) in steps of $0.1\:{\rm{M}}_\odot$. The sections of the evolutionary tracks in which the models are hotter than the approximate cool edge of the classical instability strip are shown in gray. Stars shown, as labeled, were observed by either \emph{CoRoT} (orange triangles), \emph{Kepler} (red circles) or from the ground (purple diamonds). Gray circles are \emph{Kepler} red giants \citep{HuberRG10}. The Sun is marked by its usual symbol. From \citet{theodiagramsWhite}.}{1}

\subsubsection{Asymptotic relation for g modes}
High-order, low-degree g modes obey the following first-order asymptotic expression for the periods \citep{Vandakurov67,Smeyers68,Tassoul80}:
\begin{equation}
\label{gasympt1}
\Pi_{nl}=\frac{1}{\nu_{nl}}\simeq\frac{\Pi_0}{\sqrt{l(l+1)}}\,\left(n+\epsilon_{\rm{g}}\right) \, ,
\end{equation}
where
\begin{equation}
\label{gasympt2}
\Pi_0=2\pi^2\left(\int\frac{N}{r}{\rm{d}}r\right)^{-1} \, ;
\end{equation}
the phase term $\epsilon_{\rm{g}}$ depends on the details of the boundaries of the g-mode trapping region and the integral is computed over that same region. The periods are now nearly uniformly spaced, and not the frequencies, as was the case for p modes (cf.~Eq.~\ref{asympt}). Furthermore, the period spacings depend on degree $l$. Departures from the simple asymptotic relation given in Eq.~(\ref{gasympt1}) are used as a means of diagnosing the stratification inside stars (e.g., inside white dwarfs), since the magnitude of these departures is very sensitive to strong abundance gradients and their effect on the buoyancy frequency. 

\subsection{Effects of sharp features}
Oscillation frequencies contain a greater deal of information besides what is suggested by the simple asymptotic description. In fact, sharp features (i.e., features varying more rapidly than the scale of the eigenfunction) in the internal structure of a star are known to give rise to oscillatory signals in observable seismic parameters \citep[e.g.,][]{Monteiro00,Ballot04,Basu04,Vernerglitches,HoudekGough07}. This oscillatory behavior is a function of frequency and arises from the varying phase of the oscillation at the location of the sharp feature, ultimately causing departures from the asymptotic description. In particular, these oscillatory signals can be found in the frequencies themselves, in the large frequency separation, and in higher-order differences. The second difference, defined as 
\begin{equation}
\Delta_2\nu_{nl}=\nu_{n-1\,l}-2\,\nu_{nl}+\nu_{n+1\,l} \, ,
\end{equation} 
is the most widely exploited parameter. Other diagnostics from which to extract such signatures are frequency differences that make use of the $l\!=\!0$ and $l\!=\!1$ modes \citep{Roxburgh09}.  

The modulation of the seismic parameters with frequency may be written in the form
\begin{equation}
A(\omega) \, \cos[2(\omega\,\tau_{\rm{d}}+\phi)] \, ,
\end{equation}
where $A(\omega)$ is an amplitude, $\tau_{\rm{d}}$ is the acoustic depth of the feature, and $\phi$ is a surface phase. The frequency dependence of the amplitude $A(\omega)$ is determined by the physical properties of the feature. The acoustic depth $\tau_{\rm{d}}$ is defined as
\begin{equation}
\tau_{\rm{d}}=\int_{r_{\rm{d}}}^R \frac{{\rm{d}}r}{c} \, ,
\end{equation} 
where $r_{\rm{d}}$ is the acoustic radius of the feature.

These sharp features are associated with abrupt variations of the sound speed and thus are also called acoustic glitches. The two main sources behind an abrupt variation of the sound speed are the border of a convection zone and the ionization of a dominant element. The former source is related to the sharp transition of the temperature gradient from being radiative to becoming adiabatic, which causes a discontinuity in the second derivative of the sound speed. Moreover, convective overshoot may produce a discontinuity in the temperature gradient with a consequent discontinuity in the first derivative of the sound speed, ultimately leading to a stronger oscillatory signal. Determination of the lower boundary of a convective envelope is a very important matter, since this region is believed to play a key role in stellar dynamos. The latter source is related to a rapid variation of the adiabatic exponent $\Gamma_1$ (and hence of the local sound speed) associated with the ionization of an abundant element, e.g., arising from the second ionization of helium. Extraction of the helium signature allows tight constraints to be placed on the helium abundance in stellar envelopes, otherwise not possible when dealing with such cool stars (since ionization temperatures are too high to yield usable photospheric lines for spectroscopy in these stars). 

The effects of sharp features are detectable from the analysis of the frequencies of low-degree modes. Therefore, one expects to be able to conduct such analyses in the stellar case once frequency precision is high enough. \citet{Monteiro00} conducted a seismic study that aimed at determining the characteristics of the convective envelopes of low-mass stars, namely, measuring the acoustic depth of the base of the convection zone and constraining the properties of an overshoot layer at the base of such an envelope. Using frequencies of low-degree modes (up to $l\!=\!2$) they concluded that the signal in the frequencies (see Fig.~\ref{glitches}) could be measured if the precision in frequency determination was $0.1\:{\rm{\mu Hz}}$ or better.\figuremacroW{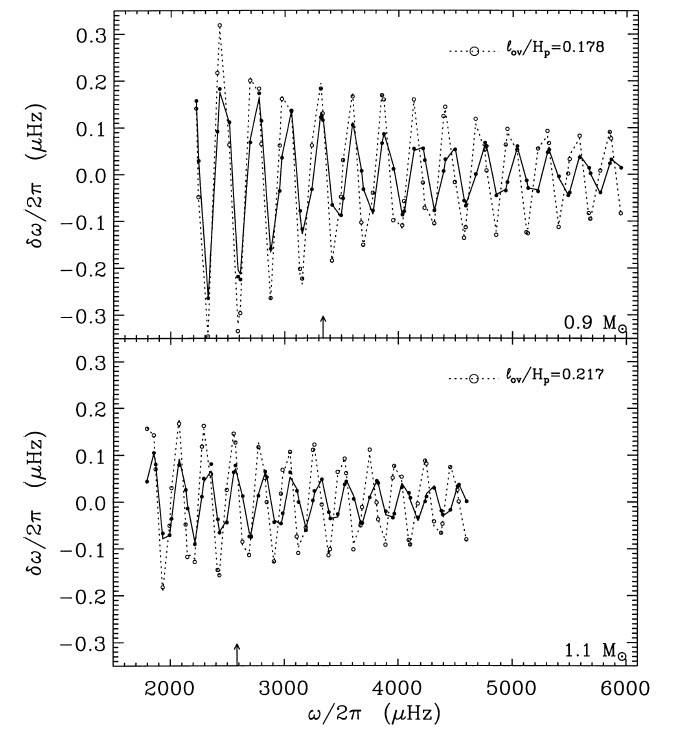}{Oscillatory signal from the base of the convection zone}{These signals are present in the frequencies of zero-age main-sequence models of a 0.9-${\rm{M}}_\odot$ (top panel) and a 1.1-${\rm{M}}_\odot$ (bottom panel) stars. Model frequencies are represented by circles: open circles correspond to models incorporating overshoot and filled circles correspond to models without overshoot. Dotted lines indicate the fit to the former groups of points, whereas solid lines indicate the fit to the latter groups of points. The size of the overshoot layer, $\ell_{\rm{ov}}$, is indicated in both panels in units of the pressure scale height, $H_p$. It can be clearly seen that the amplitude of the signal reflects the presence or not of convective overshoot. From \citet{Monteiro00}.}{0.95} \citet{Ballot04} conducted a detailed investigation on the seismic extraction of the convective extent in solar-like stars, again using low-degree data. Their analysis was mainly based on the use of the second difference $\Delta_2\nu_{nl}$, after having asserted that this seismic parameter constitutes the best compromise between enhancing the oscillatory signal while keeping the errors acceptably low. They concluded that an observational span of at least 150 days is necessary if we are to reliably extract the signature of the base of the convection zone for a large sample of solar-like stars. Very recently, \citet{Miglio10} found evidence of the seismic signature of a sharp transition in the internal structure of the \emph{CoRoT} red-giant star HR~7349. Through comparison with stellar models they were led to conclude that this feature is associated with the helium second-ionization region. In another very recent work, \citet{Mazumdar10} claim to have determined the acoustic depth of both the base of the convection zone and the helium second-ionization region of HD~49933 with a precision of 10\% by means of the second difference. 

Finally, the sharp transition associated with the edge of a convective core -- found in solar-type stars that are slightly more massive than the Sun -- also produces an effect on the oscillation frequencies \citep[e.g.,][]{CunhaMetcalfe07}. In the case of a main-sequence solar-like oscillator harboring a convective core, its edge will be situated near the inner turning point of low-degree p modes and, as a result, the signal will no longer be periodic. Measurement of the frequency dependence of suitable frequency separations of low-degree modes provides a diagnostic tool of both the presence and size of a convective core \citep[e.g.,][]{CunhaBrandao,Victorconvcore}. Determining the sizes of convective cores and the overshoot of the corresponding convective motions can provide an accurate calibration of the ages of such stars \citep[e.g.,][]{Mazumdar06}.

\subsection{Mixed modes}\label{mixedmodes}
I ended Sect.~\ref{eigenf} by mentioning that modes with mixed p- and g-mode character may occur in evolved stars. This comes as a result of the large magnitude attained by the buoyancy frequency in the stellar core, which reaches frequency values relevant for stochastic excitation. Hereafter, an illustration of the signatures of mixed modes is provided, based on a model{\footnote{This is not the same model as considered in Fig.~\ref{charact_freqs}. The general properties of the characteristic frequencies of both models are, however, very similar.}} of the subgiant $\eta$ Boo having a mass of $1.7\:{\rm{M}}_\odot$ and a heavy-element abundance of $Z\!=\!0.04$ \citep{DiMauro03,JCDHoudek10}. Interestingly, $\eta$ Boo is the first star other than the Sun for which definite frequencies of solar-like oscillations have been identified \citep{Kjeldsen95etaBoo}; these would be later confirmed by \citet{Kjeldsen03etaBoo} and \citet{Carrier05etaBoo} (see Sect.~\ref{obstatus} for a more detailed account).

In the course of its evolution along the subgiant branch, the star expands at roughly constant luminosity and consequently its effective temperature drops. In addition, as a result of this expansion, the eigenfrequencies tend to decrease (cf.~Eq.~\ref{t_dyn}). On the other hand, the increasing central condensation leads to an increase of the buoyancy frequency in the deep interior, which in turn tends to augment the frequencies of the g modes. Panel (a) of Fig.~\ref{avoided1} displays the evolution with age or, equivalently, decreasing effective temperature, of the frequencies of selected radial ($l\!=\!0$) and dipole ($l\!=\!1$) modes of the model of $\eta$ Boo being considered. The frequencies of the purely acoustic $l\!=\!0$ modes are seen to decrease monotonically in accordance with Eq.~(\ref{t_dyn}). Also, the frequencies of the predominantly acoustic $l\!=\!1$ modes -- with their values roughly halfway between those of the radial modes on either side -- follow the same general behavior. However, also evident, is a branch of increasing frequency that corresponds to a $l\!=\!1$ mode whose predominant character is that of a g mode. Where this mode meets a predominantly acoustic mode, their frequencies undergo what is called an avoided crossing \citep{Osaki75,Aizenman77}, i.e., closely approaching without actually crossing. At the point of closest approach these modes have a mixed character, with considerable amplitudes both in the g- and p-mode trapping regions. The important role of mixed modes as diagnostic tools resides here, namely, in the fact that their sensitivity to the properties of stellar cores is greatly enhanced when compared to purely acoustic modes. 

The changing nature of the modes can also be traced by means of the behavior of their normalized inertia $E$ (cf.~Eq.~\ref{norminert}), as depicted in panel (b) of Fig.~\ref{avoided1} for a couple of $l\!=\!1$ modes undergoing an avoided crossing. When their character is predominantly acoustic their inertia is similar to that of a neighboring (purely acoustic) radial mode. As they approach the g-mode branch, however, their inertia modestly rises above what would be expected for a purely acoustic mode. The two modes are seen to exchange nature during the avoided crossing. Furthermore, at the point of closest approach (near the vertical line) the two modes have essentially the same inertia.
 
\figuremacroW{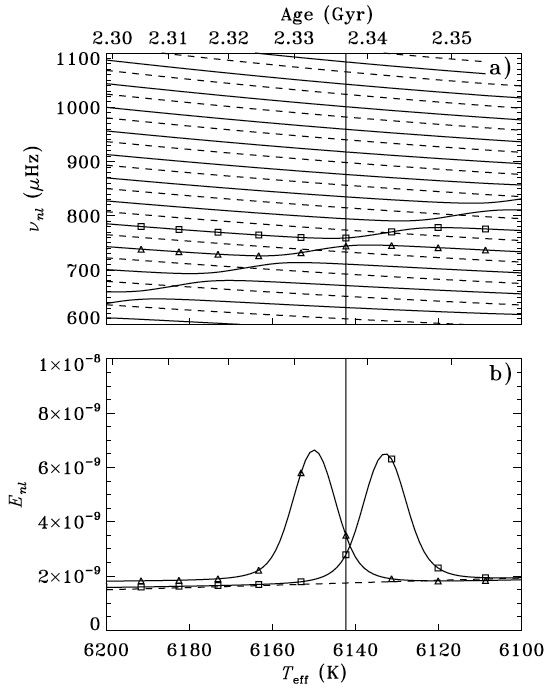}{Evolution of mode properties in a model of $\eta$ Boo}{Panel (a) displays the evolution of the frequencies of selected radial (dashed curves) and dipole (solid curves) modes. Panel (b) depicts the evolution of the normalized inertia for a couple of $l\!=\!1$ modes (solid curves) marked by triangles and squares in panel (a). The normalized inertia of a neighboring radial mode is also depicted (dashed curve), seen to vary slowly with age. The vertical line indicates the location of the specific model being considered and whose oscillation frequencies are displayed in Fig.~\ref{avoided2}. From \citet{JCDHoudek10}.}{0.85}

Inspection of Fig.~\ref{charact_freqs} tells us that the evanescent region is narrower for $l\!=\!1$ modes than for $l\!=\!2$ modes. Consequently, discrimination between g- and p-mode behavior is effectively blended for the dipole modes, as suggested by the gradual nature of the avoided crossings in panel (a) of Fig.~\ref{avoided1}. On the other hand, for modes with $l\!>\!1$, the evanescent region is broader. This gives rise to a better discrimination between the two types of behavior by reducing the coupling between the g- and p-mode regions, as well as to a decrease in the likelihood of finding a mode in a mixed state. Also, this is accompanied by a plummeting increase of the mode inertia while approaching a g-mode branch, which can become higher by several orders of magnitude than for a purely acoustic mode.

An issue not yet addressed is whether or not such mixed modes are expected to be excited to observable amplitudes. It follows from Eq.~(\ref{expectamp}) that, for a given frequency, the amplitudes of stochastically-excited modes will scale inversely to the mode inertia, more specifically, as $E^{-1/2}$. Based on this and the above considerations, one may predict that $l\!=\!1$ mixed modes are likely to be excited to observable amplitudes, whereas this is less likely to happen for mixed modes of higher degree. 

The \'echelle diagram in Fig.~\ref{avoided2} shows computed and observed frequencies for $\eta$ Boo. The jagged appearance of the $l\!=\!1$ ridge -- an obvious departure from the asymptotic description -- is a trademark of the presence of an avoided crossing and hence of the evolved nature of a star. This is also known as mode bumping, meaning that mode frequencies are shifted from their regular spacing.\figuremacroW{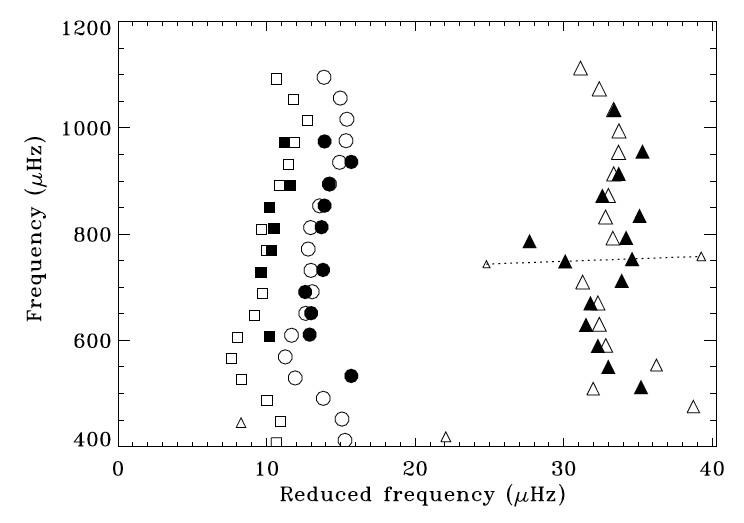}{\'Echelle diagram showing computed and observed oscillation frequencies for $\eta$ Boo}{Symbol shapes indicate mode degree: $l\!=\!0$ (circles), $l\!=\!1$ (triangles), and $l\!=\!2$ (squares). Open symbols represent computed frequencies corresponding to the model being considered in the current discussion and that has been indicated with a vertical line in Fig.~\ref{avoided1}. Filled symbols represent a combined set of observed frequencies based on the works of \citet{Kjeldsen03etaBoo} and \citet{Carrier05etaBoo}. Symbol size reflects the amplitude of a mode relative to that of a (purely acoustic) radial mode of the same frequency. The dotted line connects the two modes undergoing an avoided crossing as depicted in panel (b) of Fig.~\ref{avoided1}. Also visible, although only for the computed frequencies, is a second avoided crossing at the low-frequency end. From \citet{JCDHoudek10}.}{0.88} These same features have also been seen in the cases of ground-based observations of $\beta$ Hyi \citep{Bedding07betaHyi} and possibly Procyon \citep{BeddingProcyon}, as well as in the cases of the \emph{CoRoT} target HD~49385 \citep{Deheuvels10} and of a few \emph{Kepler} targets \citep[e.g.,][]{MetcalfeGemma,MulderScully,BoogieTigger}. The diagnostic potential of mixed modes is also being recognized by the red-giant community: detected g-mode period spacings in red giants \citep{Beckmixed} are used to discriminate between hydrogen- and helium-burning red-giant stars \citep{BeddingRGmixed,MosserRGmixed}. 

\citet{BeddingWS} suggested a new asteroseismic diagram -- inspired by the C-D diagram -- in which the frequencies of the avoided crossings (i.e., the frequencies of the pure g modes in the core cavity) for a number of stars are plotted against the large separation of the p modes \citep[e.g.,][]{MulderScully}. This so-called p-g diagram could prove to be an instructive way to display results of many stars and to allow for a first comparison with theoretical models. Much of the diagnostic potential of mixed modes can be captured in this way, since their overall pattern is determined by the mode bumping at each avoided crossing, which in turn is determined by the g modes trapped in the core. 

In his \emph{Lecture Notes on Stellar Oscillations}, J.~Christensen-Dalsgaard proposed a simple analogy based on coupled oscillators to describe an avoided crossing between two modes. Very recently, \citet{DeheuvelsMichel} presented an extension of that analogy to the case of more than two modes, having shown that the presence of an avoided crossing will induce a characteristic distortion of the ridge of degree $l$, an effect that is most prominent for $l\!=\!1$ modes. Based on the behavior of the eigenfrequencies, they concluded that HD~49385 should be in the post-main-sequence phase.   

\section{Observational aspects}

\subsection{Techniques}
There are two main observational techniques used in asteroseismology: (i) photometric observations of variability in the stellar flux resulting from the intrinsic pulsation of the star and (ii) spectroscopic observations of velocity variations due to the motion of elements on the stellar surface. It should be noted that intensity and velocity (or Doppler) observations sample the same properties  of the pulsations, although not in exactly the same way. In fact, we have already seen in Sect.~\ref{sect:spat_filt} that the response of velocity observations to modes of moderate degree is larger compared to intensity observations. Such difference in response can in principle be used in the process of mode identification, a particularly complicated issue in the case of classical pulsators.

Velocity shifts of spectral lines are measured using high-dispersion spectrographs with stable reference sources, mounted on ground-based telescope facilities. The oscillation amplitudes in solar-like pulsators are nonetheless extremely small, particularly for main-sequence stars (see Fig.~\ref{amplitudes}): e.g., for the Sun the velocity amplitude per mode is typically less than $15\:{\rm{cm\,s^{-1}}}$, while the corresponding amplitude in broadband intensity is around $4\:{\rm{ppm}}$. However, the last decade or so has seen a rampant increase in the achievable precision of radial-velocity measurements promoted by the detection of extrasolar planets. Presently, radial-velocity determination has reached a precision of only a few tens of ${\rm{cm\,s^{-1}}}$ per exposure. It should be noted that this method is strongly biased toward low-effective temperature and slowly-rotating stars, since velocity observations require the analysis of many narrow spectral lines, as well as toward subgiant and giant stars, as a result of their larger intrinsic amplitudes.  

\figuremacroW{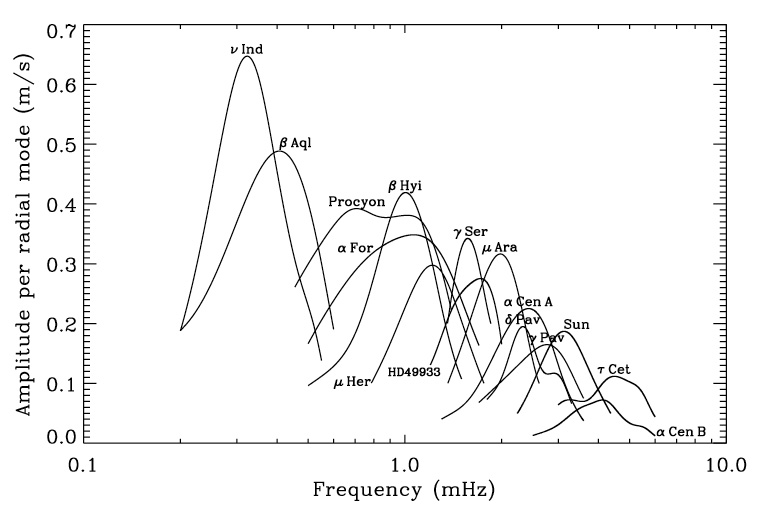}{Amplitudes of solar-like oscillations in a variety of stars}{Amplitudes have been normalized so as to correspond to the amplitude per radial mode. The location of $\nu_{\rm{max}}$ occurs at lower frequencies for more evolved stars, while the velocity amplitude generally increases with increasing $L/M$. In order to allow for a direct comparison, the solar data were obtained by observing the blue sky with a technique corresponding to the stellar observations. From \citet{ArentoftProcyon}.}{1}

Most of the pre-\emph{CoRoT} detections have indeed come from high-precision radial-velocity measurements using spectrographs such as: CORALIE at the 1.2-m Euler telescope at ESO La Silla in Chile, HARPS at the ESO La Silla 3.6-m telescope, UCLES at the 3.9-m AAT in Australia, UVES at the 8.2-m UT2 of the VLT at ESO Paranal in Chile, SARG at the 3.6-m TNG in La Palma, etc. Besides weather instabilities, a serious limitation when it comes to ground-based asteroseismic observations of solar-like pulsators is the lack of dedicated facilities, either in the format of a network of telescopes at low and/or intermediate latitudes or as an asteroseismic telescope in Antarctica. This considerably undermines both the duration and continuity of the available observations due to diurnal interruptions and the annual motion of the Earth. In an attempt to augment the duty cycle of ground-based observations, several successful double-site campaigns have been coordinated that used the instruments in Chile and at the AAT \citep[e.g.,][]{Butler04,Kjeldsen05}. These coordination efforts culminated in the realization of a multi-site campaign to measure oscillations in the F5IV star Procyon, which has been the most extensive campaign so far organized on any solar-like pulsator \citep{ArentoftProcyon}. Furthermore, a dedicated network of 1-m telescopes equipped with iodine-stabilized spectrographs is the goal of the SONG project \citep{SONG,SONG2}, while the SIAMOIS project \citep{SIAMOIS} has plans to install an instrument performing spectrometric observations at the Concordia station at Dome C in Antarctica.

On the other hand, performing intensity measurements presents two significant advantages: the instrumentation is rather simple and the flux is measured within bands of medium to low spectral resolution. Nevertheless, the intrinsic stellar background signal arising from non-oscillatory fluctuations associated with granulation, activity, etc., is substantially higher for photometric measurements than for spectroscopic measurements, meaning that velocity observations have higher SNR at low frequencies \citep[e.g.,][]{Harvey88}. The fact that photometric measurements are primarily sensitive to temperature variations -- caused by the compression and expansion of the stellar atmosphere during the oscillation cycle -- explains their higher sensitivity to granulation (the sloping background seen in Fig.~\ref{solarspec} is actually due to granulation). Figure \ref{phvsvel} compares the solar spectra as measured in velocity with GOLF \citep{GOLF} and in intensity using the green channel of the VIRGO/SPM, both instruments being on board the \emph{SOHO} spacecraft.

\figuremacroW{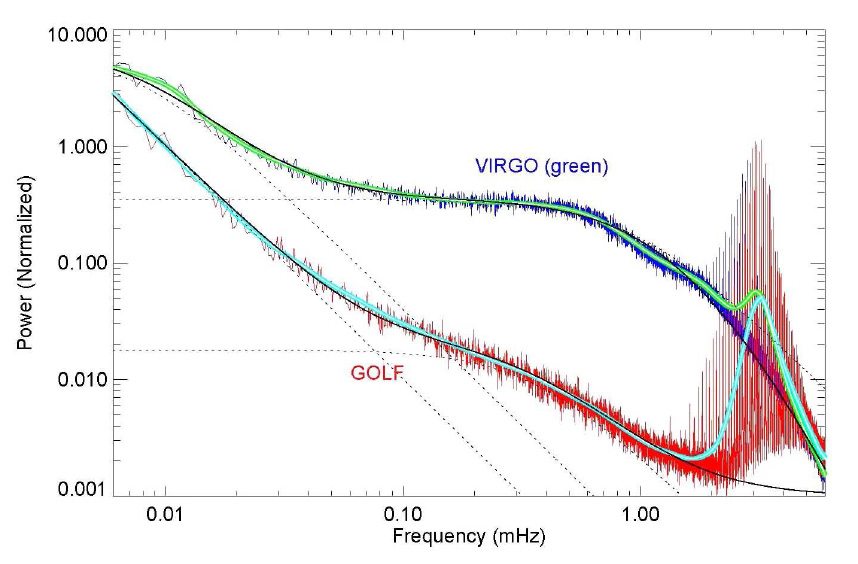}{Comparison of solar spectra computed from velocity and intensity observations}{Velocity observations were performed with GOLF, while intensity observations are from the green channel of the VIRGO/SPM. Both instruments are on board the \emph{SOHO} spacecraft. The stellar background signal has contributions both from activity and from granulation. A simple Harvey-like profile (cf.~Eq.~\ref{harveyprof}) has been used to describe the background signal with dashed curves representing the different components of such profile. The smoothed spectra are shown as solid thick curves. From \citet{Grundahl07}.}{0.98}

When envisaging the study of classical pulsators, characterized by the relatively high amplitudes of their oscillations, the use of ground-based photometry is ubiquitous, owing to the fact that most small and medium-sized telescopes have the necessary instrumentation to carry out absolute or relative photometry in a variety of photometric systems. However, scintillation from the Earth's atmosphere strongly limits the achievable precision of ground-based photometry. Until now, all ground-based attempts to detect solar-like oscillations in stars near the main sequence through intensity measurements have failed. The most ambitious enterprise consisted of a multi-site campaign -- using differential photometry and employing most of the world's then-largest telescopes -- on the open cluster M67 that failed, however, to detect any oscillations \citep{Gilliland93}. In some cases upper limits for oscillation amplitudes were well below the theoretical predictions, which are now known to have been overestimated. It should be noted that the observed stars in M67 consisted of turn-off stars somewhat hotter than the Sun. Lately, two ground-based photometric campaigns have attempted to detect solar-like oscillations in red-giant cluster members \citep{Frandsen07,Stello07}. Having been the only successful of the two campaigns, \citet{Stello07} were able to make the first test of relevant scaling relations with an homogeneous ensemble of stars and to detect excess power consistent with the expected signal from stellar oscillations.

The development of space-based asteroseismology using the technique of ultra-high-precision photometry has been a major breakthrough. It finally provided the possibility of carrying out long and almost uninterrupted observations of the same targets. Moreover, using moderate apertures, space photometry is capable of detecting oscillations whose amplitudes are about $1\:{\rm{ppm}}$. Non-dedicated instruments, such as the three Fine Guidance Sensors on the \emph{Hubble Space Telescope} \citep[e.g.,][]{Zwintz99,GillilandHST}, the 52-mm star camera on the \emph{WIRE} satellite \citep[e.g.,][]{BuzasiWIRE,Bruntt05} and the SMEI experiment on board the \emph{Coriolis} satellite \citep[e.g.,][]{Tarrant07,Tarrant08}, have been used to acquire high-precision photometric data for asteroseismic studies. 

The Canadian microsat \emph{MOST} \citep[e.g.,][]{WalkerMOST}, the first dedicated asteroseismology mission to be launched successfully (in June 2003), is able to measure intensity variations of relatively bright stars by producing light curves with time spans of the order of a few weeks. It is not expected, however, to reach the low-noise levels required for the detection of solar-like oscillations in main-sequence stars. The French-led \emph{CoRoT} mission, launched in December 2006, produced a major leap in the domain of space-based asteroseismology. Photometric variations of a few hundred asteroseismic targets are being monitored in the course of the mission, although the number of observed solar-like pulsators residing on or near the main sequence is quite reduced (just over ten). A major strength of this mission resides in the possibility of conducting nearly continuous observations, extending over five months, of the same field of view. A further advance is provided by the asteroseismic program of the \emph{Kepler} mission, launched in March 2009, whereby thousands of stars have been monitored during a survey phase, after which there is the possibility of conducting long-term follow-ups of a selection of those stars. \emph{Kepler} will lead to a revolution in the field of solar-like oscillations, since the number of known solar-like pulsators is expected to increase by several orders of magnitude. 

If selected for funding following its present design study, the ESA \emph{PLATO} mission{\footnote{Unfortunately, shortly before the submission of this dissertation, the news came that ESA had not selected \emph{PLATO} in the context of its Cosmic Vision program.}} \citep{PLATO} will provide very extensive and high-quality asteroseismic data as an integral part of its goal to study extrasolar planetary systems. A more distant milestone should be reached when imaging of stellar surfaces will be made possible. The Stellar Imager is a planned space-based interferometer, designed to enable 0.1-milliarcsecond spectral imaging of stellar surfaces and capable of probing flows and structures in stellar interiors through asteroseismology \citep[][]{SI}.

\subsection{The observational status}\label{obstatus}
The information conveyed by solar-like oscillations can be used to determine fundamental stellar properties such as mass, radius, and age. Furthermore, the internal stellar structure can be constrained to unprecedented levels provided that individual mode parameters are measured. We thus believe that asteroseismology will produce significant improvements on the theories related to stellar structure and evolution, on topics as diverse as energy generation and transport, rotation and stellar cycles. The interdisciplinary aspect of the field should not be neglected, the best example of which is probably its potential use to characterize exoplanet-host stars, thus providing key information for understanding the formation and evolution of planetary systems, as well as for constraining the location of habitable zones based on a knowledge of the stellar magnetic activity. 

The search for solar-like oscillations in stars other than the Sun has started some thirty years ago. The first hint of a hump of excess power with a frequency dependence similar to the one observed in the solar case was obtained by \citet{Brown91} from radial-velocity observations of Procyon ($\alpha$ CMi). The first plausible detection of individual oscillation frequencies and a large frequency separation is, however, attributed to \citet{Kjeldsen95etaBoo}, who observed the G0IV star $\eta$ Boo by employing a novel technique that involved measuring fluctuations in the equivalent widths of the temperature-sensitive Balmer lines using low-resolution spectroscopy (see Fig.~\ref{etaBoo} for a power spectrum of the observations).\figuremacroW{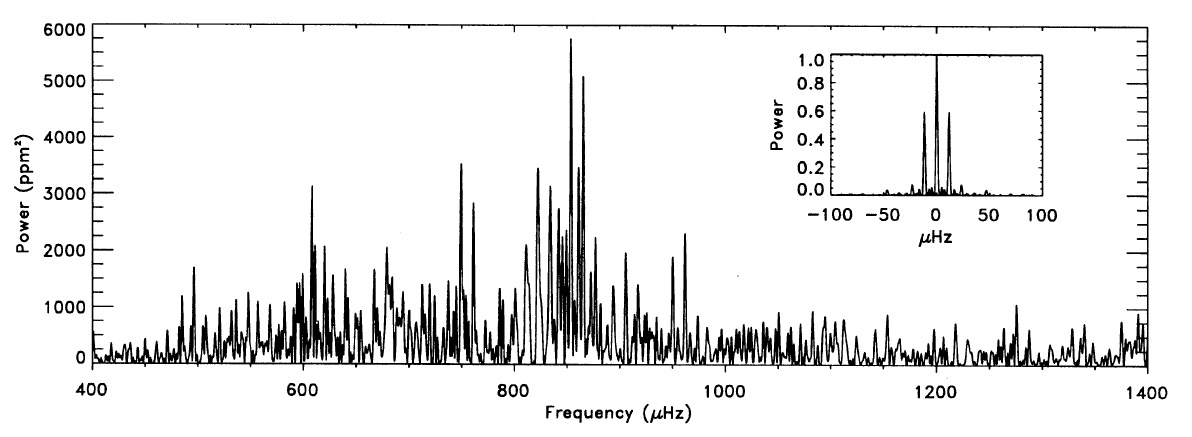}{Power spectrum of $\eta$ Boo from equivalent-width measurements}{There is a clear hump of excess power centered at about $850\:{\rm{\mu Hz}}$ due to solar-like oscillations. The inset depicts the power spectrum of the window function (see Sect.~\ref{Nyquist} for a definition). The observations were made over six nights using the 2.5-m Nordic Optical Telescope (single-site observations). Owing to the daily gaps in the data, strong sidelobes appear in the spectral window at splittings of $\pm1\:{\rm{cycle/day}}$ or, equivalently, $\pm11.57\:{\rm{\mu Hz}}$. These sidelobes greatly complicate the determination of individual frequencies. From \citet{Kjeldsen95etaBoo}.}{1} Subsequently, \citet{Brown97} were unable to confirm the presence of oscillations in $\eta$ Boo using radial-velocity measurements, which would be later confirmed by further equivalent-width and radial-velocity measurements by \citet{Kjeldsen03etaBoo}, as well as by independent radial-velocity measurements carried out by \citet{Carrier05etaBoo}. Space-based observations of this star, conducted with the \emph{MOST} satellite, led to considerable controversy after claims that a series of low-overtone p modes had been identified \citep{Guenther05}. The fact that the orbital frequency of the spacecraft is nearly a multiple of the large separation means that caution is needed in the interpretation of those low-frequency peaks.

Coming back to the case of Procyon, it is interesting to note that not until 1999 were solar-like oscillations definitely confirmed in this star \citep{Martic99}. Being the eighth brightest star in the night sky, Procyon has indeed been a long-time favorite for the search of oscillations, with several independent radial-velocity studies, mostly single-site, reporting an excess in the power spectrum. Controversy arised when photometric observations obtained with \emph{MOST} failed to detect oscillations \citep{MatthewsProcyon}, leading \citet{BeddingvsMatthews} shortly thereafter to argue that such non-detection was consistent with the ground-based data. All these efforts culminated in the realization of an extensive multi-site campaign carried out in January 2007, whereby high-precision velocity observations over more than three weeks were obtained with eleven telescopes at eight observatories. The analysis of these observations has been presented in \citet{ArentoftProcyon} and \citet{BeddingProcyon}. Given my extensive contribution to the latter paper, I have decided to place it as a supplement in Appendix \ref{beddingprocyon}. In the latest work dedicated to Procyon, \citet{HuberProcyon} compared a new and more accurate 2007 set of \emph{MOST} data with the simultaneous data acquired during the multi-site campaign, concluding that the \emph{MOST} power spectrum shows clear evidence of individual oscillation frequencies and thus refuting the renewed non-detection claims made by \citet{Walker08}.

The year 2001 came and oscillations were found in the G2IV star $\beta$ Hyi and in the G2V star $\alpha$ Cen A. The following year would also bring the first firm establishment of solar-like oscillations in a giant star \citep[$\xi$ Hya;][]{FrandsenHya}, based on a continuous 1-month monitoring with CORALIE. 

\citet{Bedding01betaHyi} and \citet{Carrier01betaHyi} confirmed the presence of solar-like oscillations in $\beta$ Hyi, but were unable to identify individual mode frequencies. Subsequently, \citet{Bedding07betaHyi} observed this star during more than a week with HARPS and UCLES, being able to identify 28 oscillation modes that included some mixed modes of degree $l\!=\!1$. Frequently regarded as an older Sun, $\beta$ Hyi is for that reason a particularly interesting object of study. 

During the last decade, the visual binary system $\alpha$ Cen has been a preferred asteroseismic target due to its proximity and to the similarity of its components to the Sun. The unambiguous detection of solar-like oscillations in $\alpha$ Cen A by \citet{BouchyCarrier01,BouchyCarrier02}, based on 13 nights of single-site observations with CORALIE, is hailed as a milestone in the field, confirming the earlier claimed detection made by \citet{SchouBuzasi01} with the \emph{WIRE} satellite. Simultaneously with the CORALIE observations, a double-site campaign was being devoted to this star employing UVES and UCLES \citep{Butler04,Bedding04}. In the meantime, a re-analysis of the \emph{WIRE} data \citep{Fletcher06} and ground-based spectroscopy with HARPS \citep{Bazot07} have been conducted. Very recently, \citet{deMeulenaer} combined and analysed the radial-velocity time series obtained in May 2001 with CORALIE, UVES, and UCLES (see Sect.~\ref{secalphaCenA}).

These early discoveries paved the way for the detection of solar-like oscillations in a number of stars. A thorough, although pre-\emph{CoRoT}, observational review is provided by \citet{BeddingKjeldsen08}, referring not only to main-sequence and subgiant stars, but also to G and K giants, semi-regular variables, and red supergiants. Although ground-based spectroscopic campaigns seem to be loosing momentum as we move deep into the era of space asteroseismology, it should be noted that they provide the ultimate precision for asteroseismic investigations, at least for bright and/or nearby stars. Not to mention the ability to cover the whole sky and to target stars whose parallaxes and other parameters are accurately known. While waiting for the advent of projects such as SONG and SIAMOIS, it is still rewarding to explore the use of existing facilities for short (1--2 weeks) ground-based campaigns devoted to carefully selected solar-like pulsators. A striking example was the first application of asteroseismology (8 nights of observations with HARPS) to an exoplanet-host star \citep[$\mu$ Ara;][]{BouchyAra,BazotAra}. Another remarkable example is the recent work of \citet{18Sco}, who have employed asteroseismology (12 nights with HARPS) and long-baseline interferometry (with the PAVO beam-combiner at the CHARA array) in order to derive the radius and mass of the solar twin 18 Sco (see Sect.~\ref{sec18Sco}).    

Asteroseismology has definitely entered a new and golden era with the advent of the \emph{CoRoT} and \emph{Kepler} space missions. Together, they are up to date responsible for the detection of solar-like oscillations in several hundred main-sequence and subgiant stars, as well as in several thousand stars residing on the red-giant branch. These numbers are indeed stratospheric. An overview of the most important results obtained so far with both these space missions is provided by \citet{Garciareview}.

The \emph{CoRoT} satellite has been the first asteroseismic mission to be able to perform ultra-high-precision, wide-field photometry for extended and continuous periods of time (up to 150 days) on the same targets. Launched into an inertial polar orbit at an altitude of $897\:{\rm{km}}$, it carries a 4-CCD array fed by a 27-cm telescope and is able to measure stellar brightnesses to $\mu$mag precision \citep{CoRoT}. The mission comprises two main scientific programs -- asteroseismology and the detection of exoplanets by the transit method -- simultaneously working on adjacent fields in the sky. 

In the asteroseismic context, \emph{CoRoT} has been successful in providing observations that allowed a seismic analysis of a few late-type main-sequence and post-main-sequence stars displaying solar-like oscillations \citep{Appourchaux08,Michel08,Barban09,Benomar09,Garcia09,Mosser09,Deheuvels10,Mathur10}. These have been joined by a study of the exoplanet-host star HD~52265 \citep[see Sect.~\ref{secHD52265};][]{HD52265}. The availability of long time-series data on solar-type stars presents good prospects for probing stellar cycles with asteroseismology \citep[e.g.,][]{Chaplincycle,Metcalfecycle,Karoffcycle,Chaplinactreview}. \citet{Garciaactivity} recently uncovered the first evidence of global changes in the oscillation frequencies and mode amplitudes associated with a stellar activity cycle in another solar-type star, to be specific, the \emph{CoRoT} target HD~49933 (see Fig.~\ref{activity}). The results on HD~49933 (with a period of the stellar cycle probably between 1 and 2 years) seem to be consistent with the paradigm that stars divide into two distinct sequences in terms of activity (the active sequence and the inactive sequence), with stars along each sequence displaying a similar number of rotational periods per activity cycle \citep[e.g.,][]{BV}, meaning that solar-type stars with short rotational periods -- HD~49933 has a surface rotational period of about 3.4 days -- tend to have short activity cycles. 

Notable success came from the asteroseismic study of several hundreds of red giants observed in the exoplanet channel of \emph{CoRoT}. These data made it possible to establish new seismic scaling relations \citep[e.g.,][]{HekkerRG09,MosserRG}, to unambiguously detect for the first time non-radial modes in red giants \citep{DeRidder09,CarrierRG} and to directly estimate stellar masses and radii from scaling relations \citep{KallingerRGCorot}. The studies are being extended to make inference on the red-giant population -- dominated by red-clump stars in the \emph{CoRoT} ensemble -- and by that means to test population-synthesis models of the evolution of the Galaxy \citep{MiglioRG09}. The establishment of a universal red-giant oscillation pattern \citep{MosserRGuni} and the analysis of mixed modes in these stars to determine their evolutionary status \citep{MosserRGmixed} constitute the latest developments. 

\figuremacroW{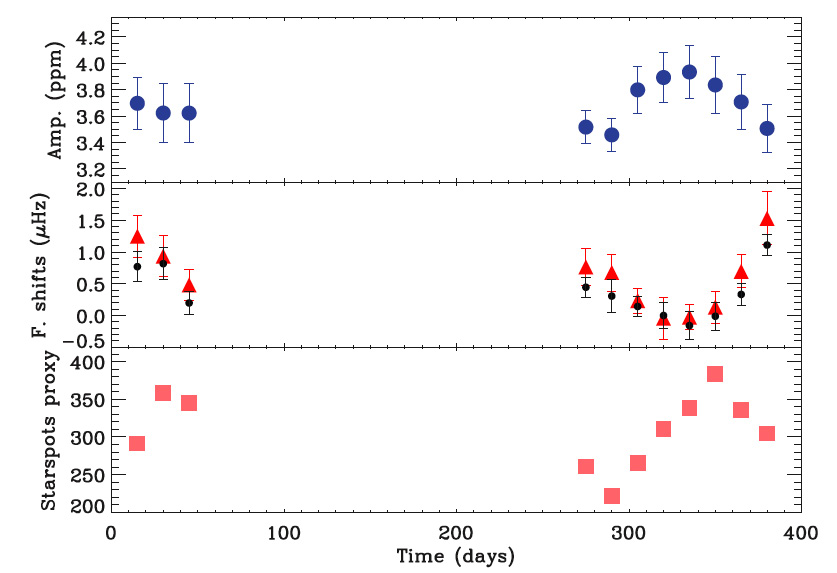}{Evidence of a magnetic activity cycle in the Sun-like star HD~49933}{Temporal evolution of the mode amplitudes (top panel), the frequency shifts using two different methods (middle panel), and a starspot proxy (bottom panel). The seismic indicators (top and middle panels) are anticorrelated in time -- as observed for the Sun -- and reveal a modulation in the second epoch that suggests a period of at least 120 days related to the internal magnetic activity. Moreover, the starspot signature confirms the existence of an activity cycle, which seems to be temporally shifted compared with the seismic indicators. From \citet{Garciaactivity}.}{0.93}

The NASA \emph{Kepler} mission was designed with the intent of detecting -- using the transit photometry method -- Earth-like planets in and near the habitable zones of late-type main-sequence stars \citep{Kepler1,Kepler2}. The satellite, which operates in an Earth-trailing heliocentric orbit, consists of a 95-cm aperture photometer with a CCD array capable of producing photometric observations with a precision of a few ppm during a period of 4--6 years. The high-quality data provided by \emph{Kepler} are also well suited for conducting asteroseismic studies of stars as part of the \emph{Kepler} Asteroseismic Investigation \citep[KAI;][]{KAI}. Photometry of the vast majority of these stars is conducted in long-cadence mode \citep[29.4 minutes; see][]{JenkinsLC}, whereas a revolving selection of up to 512 stars are monitored in short-cadence mode \citep[58.85 seconds; see][]{GillilandSC}. 

\figuremacroW{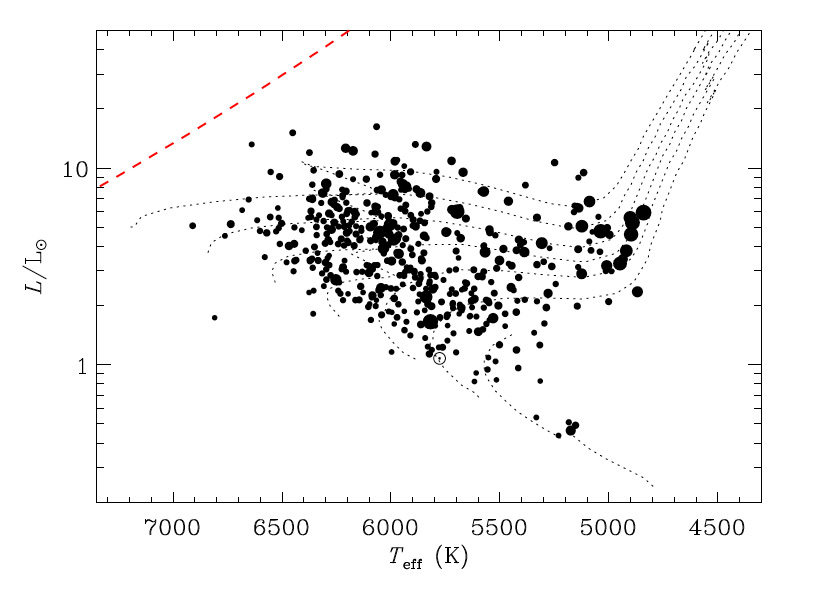}{H-R diagram of solar-type stars displaying solar-like oscillations from seven months of \emph{Kepler} survey data}{Symbol size is proportional to the SNR in the acoustic spectrum. The location of the Sun is indicated with the usual solar symbol. Dotted curves represent evolutionary tracks for solar composition, computed for masses ranging from 0.7 to $1.5\:{\rm{M}}_\odot$. The dashed red line marks the location of the red edge of the Cepheid instability strip. Adapted from \citet{Chaplin11detect}.}{0.93}

Short-cadence data are necessary in order to investigate solar-like oscillations in main-sequence and subgiant stars, whose dominant periods are of the order of several minutes \citep[e.g.,][]{ChaplinLetter10,MetcalfeGemma}. \citet{JCDExoHost10} reported the first application of asteroseismology to known exoplanet-host stars in the \emph{Kepler} field. During the first seven months of \emph{Kepler} science operations, an asteroseismic survey of solar-type stars -- each star being observed for one month at a time in short-cadence mode -- made it possible to detect solar-like oscillations in about 500 targets (see Fig.~\ref{ensemble}). This constitutes an increase of one order of magnitude in the number of such stars with confirmed oscillations \citep{Chaplin11detect}. This large, homogeneous data sample opens the possibility of, for the first time, conducting ensemble asteroseismology on a population of solar-type field stars (see Sect.~\ref{secensemble}). A statistical survey of trends in relevant seismic parameters will allow tests of basic scaling relations, comparisons with trends predicted from models, and lead to crucial insights on the detailed modeling of stars \citep[e.g.,][]{Chaplinactivity,ChaplinSci,Huberscaling,Vernercomparison,obsdiagramsWhite}. Performing what might be called differential (or comparative) seismology is also made possible by picking from a large ensemble pairs, small groups, or sequences of stars sharing common stellar properties such as mass, composition, or surface gravity \citep[e.g.,][]{Victordifferential}. This allows eliminating any dependence of the results on the common property, or properties. Since the start of the mission, a selection of survey stars have been continuously monitored in short-cadence mode to test and validate the time-series photometry, five of which show evidence of solar-like oscillations. The analysis of two of these solar-like oscillators, namely, KIC~10273246 and KIC~10920273, is presented in \citet{MulderScully}. This article can be found as a supplement in Appendix \ref{campantekepler}. Two other such oscillators, namely, KIC~11395018 and KIC~11234888, are analysed in a companion paper \citep[see Sect.~\ref{secMathurfurry};][]{BoogieTigger}.

Studies of large samples of long-cadence G and K giants, extending in luminosity from the red clump down to the bottom of the giant branch, have shown clear evidence of the presence of $l\!=\!3$ modes \citep{BeddingLetter10}, while confirming theoretical scaling laws \citep{HuberRG10,HekkerRG11} and allowing the computation of asteroseismic fundamental parameters \citep{KallingerRG10}. Photometric data of red giants in the open cluster NGC~6819 allowed a first clear detection of solar-like oscillations in cluster stars to be made, and provided additional tests for cluster membership based on the analysis of the asteroseismic parameters \citep{StelloNGC6819}. Other seismic studies of red giants in open clusters ensued \citep[e.g.,][]{BasuRG,HekkerRGclusters}.

\chapter{Data analysis in asteroseismology}\label{Ch2}

\ifpdf
    \graphicspath{{2/figures/PNG/}{2/figures/PDF/}{2/figures/}}
\else
    \graphicspath{{2/figures/EPS/}{2/figures/}}
\fi

This chapter is intended as a practical guide into some of the main data analysis methods and techniques employed contemporarily in asteroseismology of solar-like stars, and of which I have made recurrent use. The goal has been, since the very beginning, to produce a text that could prove useful to initiate graduate students, by providing them with a solid background in data analysis in asteroseismology, a mere starting point from which they can follow their own paths. From my own personal experience, I must say that this sort of manuals are indeed hard to come across with and increasingly so for the initiate student. The interested reader may also want to consult Hans Kjeldsen's notes on \emph{Time Series Analysis in Astrophysics}\footnote{\url{http://owww.phys.au.dk/~hans/tidsserie/}} and the very useful crash course in data analysis presented in \citet{Appourchauxguide}. 

The way matters are presented in this chapter does not deviate much from the structure adopted in the latter work. The subjects of digital signal processing and spectral analysis are treated first. These concern the acquisition of continuous physical signals to be subsequently digitally analysed. Notice that the instrumentation, although being itself an integrant part of the data analysis process, is beyond the scope of this dissertation and is thus not explicitly discussed here. The subjects of hypothesis testing and parameter estimation are discussed next both from the competing Bayesian and frequentist points of view. Finally, the implementation of a pipeline for mode parameter analysis of \emph{Kepler} data is described.      

\section{Digital signal processing and spectral analysis}

\subsection{Nyquist sampling theorem and aliasing}\label{Nyquist}
Let us assume that $s(\tau)$ is the time-averaged value of some continuous signal $x(t)$ around $t\!=\!\tau$:
\begin{equation}
\label{taverage}
s(\tau)=\frac{\int_{-\infty}^{+\infty}w(t-\tau)x(t)\,{\rm{d}}t}{\int_{-\infty}^{+\infty}w(t-\tau)\,{\rm{d}}t} \, ,
\end{equation}
where $w(t)$ is a suitable weighting function. Ideally, $w(t)\!=\!\delta(t)$ (the impulse or Dirac delta function) and, in that particular case, the result of sampling a continuous signal at uniform intervals separated by $\Delta t$ is represented by the product of the signal and a set of impulse functions regularly spaced in time. Bearing in mind that the Fourier transform of such a set of impulse functions is another set of impulse functions with separation $1/\Delta t$ in the frequency domain, one can use the convolution theorem to show that the transform of a properly sampled band-limited signal $x(t)$ is periodic, with each of its periods being equal to (within a constant) the transform of the continuous signal:
\begin{equation}
\label{sampling}
x(t) \sum_{n=-\infty}^{+\infty} \delta\left(t-n\,\Delta t\right) \Longleftrightarrow X(\nu) \ast \frac{1}{\Delta t} \sum_{n=-\infty}^{+\infty} \delta\left(\nu-\frac{n}{\Delta t}\right) \, ,
\end{equation}
where $X(\nu)$ is the Fourier transform of $x(t)$, the symbol ``$\Longleftrightarrow$'' indicates a Fourier pair, and the symbol ``$\ast$'' denotes convolution. Therefore, information is not lost about the original continuous signal $x(t)$, which can be identically reconstructed by filtering a single undistorted period out of the transform in Eq.~(\ref{sampling}) and then taking its inverse Fourier transform. In practice, however, the original signal can only be approximately recovered since we observe for a finite amount of time. 

At this point, I find it appropriate to introduce a very important concept, to be specific, that of the spectral window. For that purpose use is made of Eq.~(\ref{sampling}). In this idealized example, the window function (i.e., the observational window) is given by $\sum_{n=-\infty}^{+\infty} \delta\left(t-n\,\Delta t\right)$. The spectral window is simply the Fourier transform of the window function and is thus given by $\frac{1}{\Delta t} \sum_{n=-\infty}^{+\infty} \delta\left(\nu-\frac{n}{\Delta t}\right)$ in the current example. It is important to retain that the Fourier transform of a windowed signal is given by the convolution of the spectral window with the transform of the continuous and uninterrupted version of the signal (see Eq.~\ref{convoltruespec} below for the case of finite and discrete sampling).      

But what is meant by a properly sampled signal, as mentioned above? The answer to this question is given by the Nyquist sampling theorem \citep[also known as the Nyquist-Shannon sampling theorem;][]{Nyquist,Shannon}: It states that if the Fourier transform of a continuous signal $x(t)$ is band-limited{\footnote{Conversely, if $x(t)$ is time-limited, i.e., is zero for all $|t|\!\ge\!T/2$, then $x(t)$ can be uniquely reconstructed from samples of its transform, $X(\nu)$, at frequency intervals of $1/T$.}}, i.e., is zero for all $|\nu|\!\ge\!\nu_{\rm{lim}}$, then $x(t)$ can be uniquely reconstructed from a knowledge of its sampled values at uniform intervals of $\Delta t\!\leq\!1/(2\,\nu_{\rm{lim}})$ (see Fig.~\ref{Nyq1}). For a given uniform sampling interval $\Delta t$, the Nyquist frequency is defined as $\nu_{\rm{Nyq}}\!=\!1/(2\Delta t)$. In case the continuous signal being sampled contains frequency components above the Nyquist frequency, these will give rise to an effect known as aliasing, whereby the original spectrum will be distorted due to spectral leaking. The signal is then said to be undersampled and can no longer be uniquely recovered (see Fig.~\ref{Nyq2}). The Nyquist frequency can be thought of as the highest useful frequency to search for in the spectrum, although some authors will argue that, based on astrophysical arguments, one can also accept frequencies above $\nu_{\rm{Nyq}}$. 

\figuremacroW{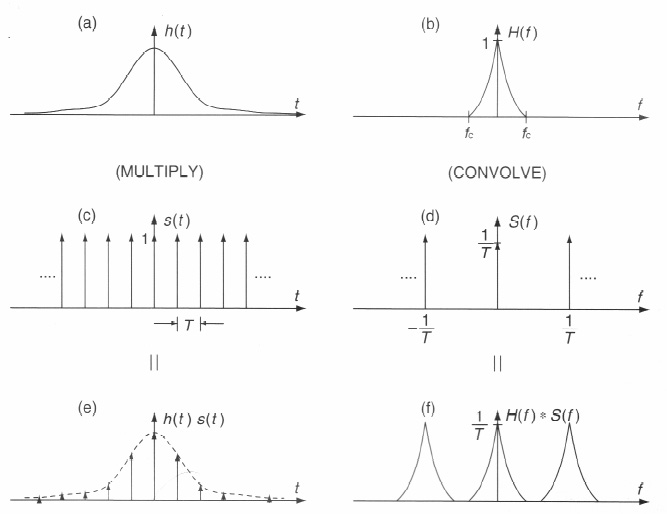}{Illustration of the Nyquist sampling theorem (proper sampling)}{Panels (a) and (b) respectively represent a continuous waveform, $h(t)$, and its band-limited Fourier transform, $H(f)$. Panels (c) and (d) respectively represent the window function, $s(t)$, and the spectral window, $S(f)$. Panels (e) and (f) respectively represent the sampled waveform, $h(t)s(t)$, and its Fourier transform, $H(f) \ast S(f)$. From \citet{GregoryBook}.}{1}

\figuremacroW{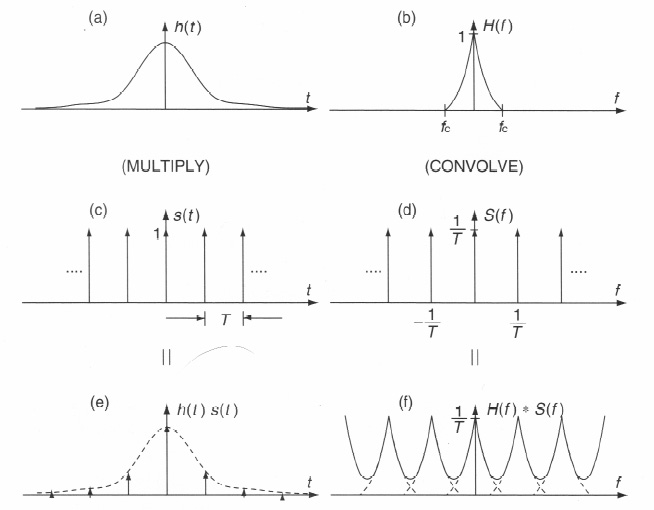}{Illustration of the Nyquist sampling theorem (undersampling)}{Similar to Fig.~\ref{Nyq1} but for the case of an undersampled signal. From \citet{GregoryBook}.}{1}

Going back to Eq.~(\ref{taverage}), a more realistic choice of $w(t)$ would be that of a rectangular pulse of width $\Delta t_{\rm{int}}\!\leq\!\Delta t$, thus mimicking the effect of integration of the detector, normally a CCD. This results in multiplication of the transform of the continuous signal by ${\rm{sinc}}(\nu\,\Delta t_{\rm{int}})$:
\begin{eqnarray}
\left[x(t) \ast \frac{1}{\Delta t_{\rm{int}}}\,\Pi\left(\frac{t-\Delta t_{\rm{int}}/2}{\Delta t_{\rm{int}}}\right)\right] \sum_{n=-\infty}^{+\infty} \delta\left(t-n\,\Delta t\right) \Longleftrightarrow \nonumber \\ \Longleftrightarrow \left[X(\nu)\,{\rm{sinc}}(\nu\,\Delta t_{\rm{int}})\,{\rm{e}}^{{\rm{i}}\,\pi\nu\Delta t_{\rm{int}}}\right] \ast \frac{1}{\Delta t} \sum_{n=-\infty}^{+\infty} \delta\left(\nu-\frac{n}{\Delta t}\right) \, ,
\end{eqnarray}
where $\Pi\left(\frac{t-\Delta t_{\rm{int}}/2}{\Delta t_{\rm{int}}}\right)$ is a boxcar function of width $\Delta t_{\rm{int}}$ centered at $\Delta t_{\rm{int}}/2$. Therefore, integration reduces the amplitude of the high-frequency noise in the time series, an effect that is maximal for a fill cycle of 100\%, i.e., for $\Delta t_{\rm{int}}\!=\!\Delta t$. In truth, the asteroseismic signal is not band-limited, meaning that the highest frequencies will not be properly sampled, ultimately resulting in an undersampled signal. In the solar case, the power of the background signal at high frequencies shows a drop-off proportional to $\nu^{-2}$, or possibly even more accentuated \citep[e.g.,][]{faculaeKaroff}. Therefore, it is critical to make sure that the fill cycle is as high as possible, thereby helping to reduce the aliasing effects. The VIRGO/SPM instrument on board \emph{SOHO}, for example, has a fill cycle of about 94\% \citep{VIRGO2}.  

When the available data are not uniformly sampled there is, strictly speaking, no Nyquist frequency, even though equivalent frequencies have been suggested in the literature \citep[e.g.,][]{Bretthorst2000}. Choosing an adequate value depends on the magnitude of the departure from uniformity of the sampling intervals. If large, it has been shown that $\nu_{\rm{Nyq}}$ could in fact be controlled by the greatest common divisor of the sampling intervals; if small, $\nu_{\rm{Nyq}}$ can be estimated using either the mean or the median sampling interval. 

Regular gaps in the time series due to diurnal interruptions and, for data sets spanning more than one year, caused by the annual motion of the Earth, are usually present in asteroseismic observations carried out from the ground. These regular gaps also give rise to frequency aliasing. The former kind of so-called daily aliases, appearing at splittings of $\pm1\:{\rm{cycle/day}}$ (or, equivalently, $\pm11.57\:{\rm{\mu Hz}}$) and their non-zero multiples (see Fig.~\ref{etaBoo}), are particularly problematic when observing solar-like oscillations since frequency separations of that same magnitude are common. A similar occurrence concerns the \emph{CoRoT} satellite. The extra noise associated with data collected by \emph{CoRoT} while passing across the South Atlantic anomaly \cite[e.g.,][]{SAA} gives rise to strong harmonics of the satellite's orbital frequency (see Fig.~\ref{SAA}). 

\figuremacroW{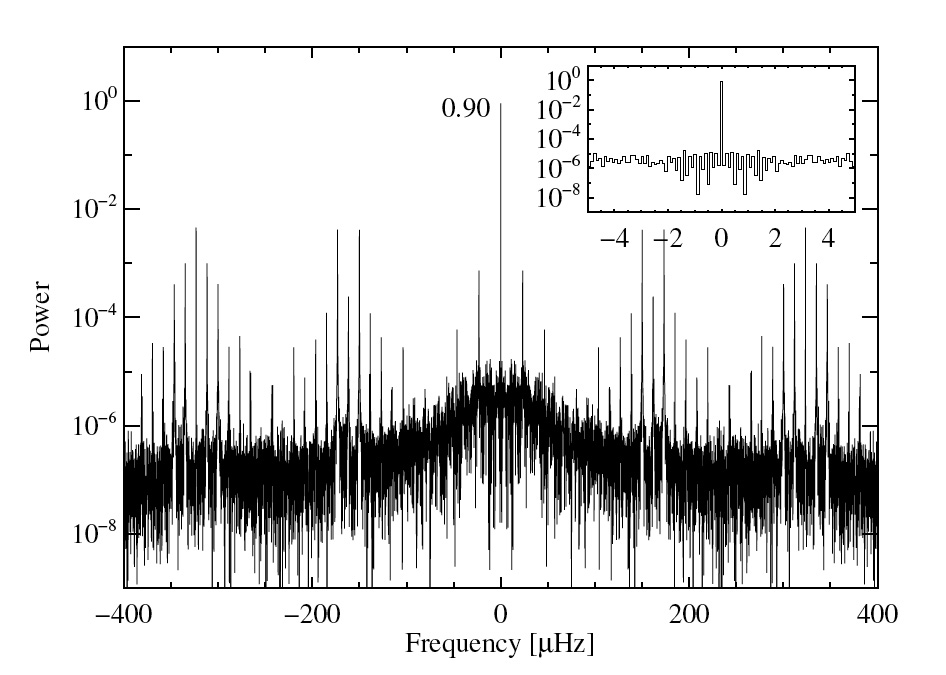}{Power spectral window for \emph{CoRoT} observations of HD~52265}{A forest of peaks corresponding to the orbital frequency ($161.7\:{\rm{\mu Hz}}$), its harmonics, and also daily aliases are clearly visible on a logarithmic scale. The inset provides a close-up around zero frequency. The power spectral window has been normalized so that the ordinate at zero frequency takes the value of the fractional duty cycle. From \citet{HD52265}.}{1}

The effect on the spectral window caused by the presence of regular gaps in the time series can be understood by means of a simple idealized example. Let us start by constructing a window function that is the result of the convolution of a boxcar function of width $T_0$ with a set of impulse functions with separation $T\!>\!T_0$. Such a window function then simply consists of a series of boxcar functions of width $T_0$ whose midpoints are separated at uniform intervals of $T$. Taking $T\!=\!1\:{\rm{day}}$ and $T_0\!=\!1/3\:{\rm{day}}$, for example, could very well correspond to the window function obtained from a single-site ground-based campaign where one observes during 8 hours every night. Let us compute the corresponding spectral window:
\begin{equation}
\frac{1}{T_0}\,\Pi\!\left(\frac{t}{T_0}\right) \ast \sum_{n=-\infty}^{+\infty} \! \delta\left(t-n\,T\right) \Longleftrightarrow {\rm{sinc}}(\nu T_0) \; \frac{1}{T} \sum_{n=-\infty}^{+\infty} \! \delta\left(\nu-\frac{n}{T}\right) \, .
\end{equation}
The spectral window thus results from the product of a sinc function and a bed of nails. As $T_0$ increases (i.e., increasing duty cycle) the central lobe of the sinc function becomes narrower with an accompanying reduction of the aliases at $\pm1/T$. In the limit when $T_0\!\to\!T$, i.e., the data recording is continuous and its length infinite, the spectral window is an impulse function and the aliases consequently vanish.

\subsection{Temporal filtering}
A problem commonly encountered while analysing asteroseismic time series is the presence of low-frequency drifts which can be either of instrumental origin or else intrinsic to the star. These low-frequency drifts introduce a background in the Fourier domain that will ultimately lead to a decrease of the SNR of the oscillation modes in the power spectrum. In order to prevent this from happening, high-pass filters are widely used, ideally reducing the effect of the drifts while preserving the relevant signals.  

Let us start by shedding some light on the process of smoothing of a time series, which can usually be interpreted as an application of a low-pass filter. Smoothing consists in convolving a signal $x(t)$ with a weighting function $w(t)$ (possibly complex) in the time domain:
\begin{equation}
\label{filter}
x_{\rm{low}}(t) = x(t) \ast w(t) \Longleftrightarrow X_{\rm{low}}(\nu) = X(\nu) \, W(\nu) \, ,
\end{equation}
where $X(\nu)$ and $W(\nu)$ are the transforms of $x(t)$ and $w(t)$, respectively. Conversely, a high-pass filter can be implemented by simply computing $x_{\rm{high}}(t)\!=\!x(t)-x_{\rm{low}}(t)$:
\begin{equation}
\label{filter2}
x_{\rm{high}}(t) \Longleftrightarrow X_{\rm{high}}(\nu) = X(\nu) \left[1-W(\nu)\right] \, .
\end{equation} 

Typical examples of the weighting function $w(t)$ are a boxcar function, a triangular function (equivalent to the convolution of two boxcar functions), and a bell-shaped function (equivalent to the convolution of four boxcar functions or two triangular functions). The transform of the simple boxcar function is the sinc function and thus leads to an excessive ringing (or Gibbs-like) effect in the Fourier domain. Multiple-boxcar smoothing is therefore advisable as a means to suppress this ringing effect. It is up to the data analyst, however, to carefully take into consideration the trade-off between ringing artifacts and frequency-domain sharpness.  

A commonly used high-pass filter in helioseismology is the backwards-difference filter \citep[e.g.,][]{backwards}:
\begin{equation}
x_{\rm{bd}}(t)=x(t)-x(t-t_0)=x(t)-\left[x(t)\ast\delta(t-t_0)\right] \, ,
\end{equation}
where a time shift $t_0$ has been considered, while becoming apparent that $w(t)\!=\!\delta(t-t_0)$ in Eq.~(\ref{filter}). Using Eq.~(\ref{filter2}), one can then determine the transfer function of the backwards-difference filter:
\begin{equation}
|1-W(\nu)|^2=\left[2 \sin\left(\frac{\pi}{2}\frac{\nu}{\nu_{\rm{c}}}\right)\right]^2 \, ,
\end{equation}  
where the cut-off frequency, $\nu_{\rm{c}}\!=\!1/(2\,t_0)$, has been introduced.

\subsection{Discrete Fourier Transform}\label{secDFT}
Attention is now drawn to the computation of an estimate of the Fourier transform of a function based on a finite number of samples. Suppose there are $N$ evenly spaced available samples $x(t_n)\!=\!x(n\Delta t)$ at intervals of $\Delta t$, with $n\!=\!0,1,2,\ldots,N\!-\!1$. Then according to the Nyquist sampling theorem (see Sect.~\ref{Nyquist}), useful frequency information is only obtainable for $|\nu|\leq\nu_{\rm{Nyq}}\!=\!1/(2\Delta t)$. The Discrete Fourier Transform (DFT) is defined as:
\begin{equation}
\label{DFT}
X_{\rm{DFT}}(\nu_p)=\sum_{n=0}^{N-1} x(t_n)\,{\rm{e}}^{{\rm{i}}\,2\pi\nu_p t_n} \;\;\;\; {\rm{for}} \:\: \nu_p=p/(N\Delta t) \, , \:\: p=0,1,2,\ldots,N-1 \, .
\end{equation} 
The transform $X_{\rm{DFT}}(\nu)$ has periodicity $1/\Delta t$ or twice the Nyquist frequency. Then $p\!=\!0$ corresponds to the DFT at zero frequency and $p\!=\!N/2$ to the value at $\pm\nu_{\rm{Nyq}}$. Values of $p$ between $N/2\!+\!1$ and $N\!-\!1$ correspond to values of the DFT for negative frequencies from $-(N/2\!-\!1)/(N\Delta t)$ to $-1/(N\Delta t)$. 

For deterministic processes (not necessarily periodic), the observed Fourier transform, $X_{\rm{DFT}}(\nu)$, results from the convolution of the true Fourier transform $X(\nu)$ with a spectral window $W_{\rm{DFT}}(\nu)$ \citep[e.g.,][]{Deeming}:
\begin{equation}
\label{convoltruespec}
X_{\rm{DFT}}(\nu)=X(\nu) \ast W_{\rm{DFT}}(\nu) \equiv \int_{-\infty}^{+\infty} X(\nu-\nu')\,W_{\rm{DFT}}(\nu') \, {\rm{d}}\nu' \, , 
\end{equation}
where the Fourier pair composed of the window function and spectral window is given by
\begin{equation}
\sum_{n=0}^{N-1} \delta(t-t_n) \Longleftrightarrow W_{\rm{DFT}}(\nu)=\sum_{n=0}^{N-1} {\rm{e}}^{{\rm{i}}\,2\pi\nu t_n} \, .
\end{equation}

\citet{FFT} introduced the Fast Fourier Transform (FFT), an efficient method of implementing the DFT that removes certain redundancies in the computation and speeds up the calculation (the speed enhancement is approximately given by $N\log_2\!N/N^2$).

\subsection{Power spectral density estimation}
As astrophysicists we are constantly being faced with the problem of having to analyse waveforms that are in fact random processes. Observations are affected by noise associated with the observed physical phenomenon and the instrumentation. Therefore, it is necessary to develop suitable statistical approaches to spectral estimation. 

Here, I present two common approaches to the estimation of the power spectral density of a random process denoted by $x(t)$. But first, a definition of the power spectral density (PSD) needs to be given:
\begin{equation}
P(\nu)=\lim_{T\to\infty} \frac{1}{T} \left|\int_{-T/2}^{T/2} x(t)\,{\rm{e}}^{{\rm{i}}\,2\pi\nu t}\,{\rm{d}}t\right|^2 \, ,
\end{equation}     
where $T$ is the length of the data set. 

A widely employed approach to the estimation of the PSD is to use the periodogram, also known as the Schuster periodogram and originally introduced in the field of meteorology \citep{Schuster}:
\begin{equation}
\label{Schuster}
\hat{P}_{\rm{p}}(\nu)=\frac{1}{T} \left|\int_{-T/2}^{T/2} x(t)\,{\rm{e}}^{{\rm{i}}\,2\pi\nu t}\,{\rm{d}}t\right|^2 \, .
\end{equation} 
In practice this usually involves the computation of a FFT.

Before presenting the second approach to spectral estimation, the Wiener-Khintchine theorem \citep{Wiener,Khintchine} is introduced, according to which the power spectral density, $P(\nu)$, and the autocorrelation function, $\phi(\tau)$, are a Fourier pair:
\begin{equation}
\label{WK}
\phi(\tau)=\int_{-\infty}^{+\infty} P(\nu)\,{\rm{e}}^{-{\rm{i}}\,2\pi\nu\tau}\,{\rm{d}}\nu \Longleftrightarrow P(\nu)=\int_{-\infty}^{+\infty} \phi(\tau)\,{\rm{e}}^{{\rm{i}}\,2\pi\nu\tau}\,{\rm{d}}\tau \, ,
\end{equation}  
where 
\begin{equation}
\phi(\tau)=\lim_{T\to\infty} \frac{1}{T} \int_{-T/2}^{T/2} x(t)x(t+\tau)\,{\rm{d}}t \, .
\end{equation}
If we further assume that $x(t)$ is a stationary and ergodic process, one then has 
\begin{equation}
\label{statergo}
\phi(\tau)={\rm{E}}\left[x(t)x(t+\tau)\right] \, .
\end{equation}
The Wiener-Khintchine theorem is absolutely crucial to the understanding of the spectral analysis of random processes. It straightforwardly explains, for instance, why white noise, whose autocorrelation function is the Dirac delta function, has constant power spectral density.

Since the random process $x(t)$ is observed over a finite time interval, an estimator of $\phi(\tau)$ has to be searched for:
\begin{equation}
\hat{\phi}(\tau)=\frac{1}{T-|\tau|} \int_0^{T-|\tau|} x(t)x(t+|\tau|)\,{\rm{d}}t \, , \:\: |\tau|<T \, .
\end{equation}
A second approach to the estimation of the PSD is the so-called correlation or lagged-product estimator \citep[commonly referred to as the Blackman-Tukey procedure;][]{BlackTukey1,BlackTukey2}, which can be simply arrived at by employing the Wiener-Khintchine theorem (cf.~Eq.~\ref{WK}):
\begin{equation}
\label{corrspec}
\hat{P}_{\rm{c}}(\nu)=\int_{-\infty}^{+\infty} w(\tau)\hat{\phi}(\tau)\,{\rm{e}}^{{\rm{i}}\,2\pi\nu\tau}\,{\rm{d}}\tau \, ,
\end{equation}
where the window function $w(\tau)$ is unit for $|\tau|\!<\!T$ and zero elsewhere.

Finally, it can be shown \citep{JenkinsWatts,Deeming} that the mean (expectation) value of both the periodogram (cf.~Eq.~\ref{Schuster}) and the correlation (cf.~Eq.~\ref{corrspec}) estimators is given by the convolution of the true power spectral density $P(\nu)$ with the power spectral window, i.e.,
\begin{equation}
{\rm{E}}\left[\hat{P}_{\rm{p}}(\nu)\right]={\rm{E}}\left[\hat{P}_{\rm{c}}(\nu)\right]=P(\nu) \ast W(\nu) \, ,
\end{equation}  
meaning that the mean value of the estimator will be biased unless, of course, the data recording has infinite duration, in which case $W(\nu)$ is the Dirac delta function. This relationship is essentially the same as that for deterministic processes presented in Eq.~(\ref{convoltruespec}).

\subsubsection{The discrete case}
The discrete form of the PSD is now presented based on the use of Parseval's theorem \citep{Parseval}. This theorem can be seen as an alternative formulation of the energy conservation principle, stating that the energy in a signal $x(t)$ computed in the time domain equals the energy as computed in the frequency domain:
\begin{equation}
\int_{-\infty}^{+\infty} x^2(t)\,{\rm{d}}t = \int_{-\infty}^{+\infty} \left|X(\nu)\right|^2\,{\rm{d}}\nu \, ,
\end{equation}  
where $X(\nu)$ is the Fourier transform of $x(t)$. One immediately notices that $\left|X(\nu)\right|^2$ is an energy spectral density. Recovering the same terminology that has been employed in Sect.~\ref{secDFT}, the discrete form of Parseval's theorem is introduced as:
\begin{equation}
\label{dParseval}
\sum_{n=0}^{N-1} x^2(t_n)\,\Delta t = \sum_{p=0}^{N-1} \left|\Delta t\,X_{\rm{DFT}}(\nu_p)\right|^2\,\Delta\nu \, ,
\end{equation}
where $\Delta\nu\!=\!1/(N\Delta t)$. By analogy with the continuous case, it is obvious that $\left|\Delta t\,X_{\rm{DFT}}(\nu_p)\right|^2$ is a discrete energy spectral density.

From Eq.~(\ref{dParseval}) it can be easily shown that the average signal power, i.e., the ratio of signal energy to its duration, is given by its mean-square amplitude:
\begin{eqnarray}
\label{normalization}
\frac{1}{N\Delta t} \sum_{n=0}^{N-1} x^2(t_n)\,\Delta t & = & \frac{1}{N} \sum_{n=0}^{N-1} x^2(t_n) \nonumber \\
& = & \sum_{p=0}^{N-1} \frac{\left|X_{\rm{DFT}}(\nu_p)\right|^2\,\Delta t}{N}\,\Delta\nu \, ,
\end{eqnarray}
where $\left|X_{\rm{DFT}}(\nu_p)\right|^2 \Delta t/N$ can be identified with the two-sided discrete PSD in units of power per unit of bandwidth{\footnote{To convert to units of power per bin one simply has to multiply by $\Delta \nu$.}}. It should be noticed that this result can also be derived from Eqs.~(\ref{WK}) and (\ref{statergo}) by setting $\tau\!=\!0$: 
\begin{equation}
\phi(0)=\int_{-\infty}^{+\infty} P(\nu)\,{\rm{d}}\nu={\rm{E}}\left[x^2(t)\right] \, .
\end{equation}

Finally, I introduce the one-sided discrete PSD, $P(\nu_q)$, defined only for nonnegative frequencies (with $q\!=\!0,1,\ldots,N/2$):
\begin{eqnarray}
P(\nu_0) & = & \frac{\Delta t}{N} \left|X_{\rm{DFT}}(\nu_0)\right|^2 \, , \nonumber \\\nonumber \\
P(\nu_q) & = & \frac{\Delta t}{N} \left[\left|X_{\rm{DFT}}(\nu_p)\right|^2 + \left|X_{\rm{DFT}}(\nu_{N-p})\right|^2\right] \, , \:\: p=1,2,\ldots,N/2-1 \, , \\\nonumber \\
P(\nu_{N/2}) & = & \frac{\Delta t}{N} \left|X_{\rm{DFT}}(\nu_{N/2})\right|^2 \, , \nonumber
\end{eqnarray} 
where $\nu_{N/2}\!=\!1/(2\Delta t)$, viz., the Nyquist frequency. Furthermore, it follows from Eq.~(\ref{normalization}) that $P(\nu_q)$ is normalized according to
\begin{equation}
\sum_{q=0}^{N/2} P(\nu_q)\,\Delta\nu = \frac{1}{N} \sum_{n=0}^{N-1} x^2(t_n) \, .
\end{equation}

\subsection{Statistics of the power spectrum}\label{PSstat}
In the following I consider the statistics of the power spectrum of a pure noise signal. Let $x(t)$ represent a random process from which a finite number of samples denoted by $x(t_n)$ have been drawn. These samples are assumed to be independent and identically distributed{\footnote{This would be the case for white Gaussian noise, although here I consider the more general case of an unknown distribution.}} (i.i.d.), with ${\rm{E}}\left[x(t_n)\right]\!=\!0$ and ${\rm{E}}\left[x^2(t_n)\right]\!=\!\sigma_0^2$ for all $n$ (i.e., the random process is further assumed to be stationary). The DFT of the set $x(t_n)$ (cf.~Eq.~\ref{DFT}) may be decomposed into its real and imaginary parts as:
\begin{eqnarray}
X_{\rm{DFT}}(\nu_p) & = & X_{\rm{DFT}}^{\rm{Re}}(\nu_p) + {\rm{i}}\,X_{\rm{DFT}}^{\rm{Im}}(\nu_p) \nonumber \\
& = & \sum_{n=0}^{N-1} x(t_n)\cos(2\pi\nu_pt_n) + {\rm{i}}\,\sum_{n=0}^{N-1} x(t_n)\sin(2\pi\nu_pt_n) \, . 
\end{eqnarray}    
It then follows from the Central Limit theorem that for a large number of samples (i.e., for large $N$) both $X_{\rm{DFT}}^{\rm{Re}}(\nu_p)$ and $X_{\rm{DFT}}^{\rm{Im}}(\nu_p)$ are normally distributed. Finally, given the fact that $X_{\rm{DFT}}^{\rm{Re}}(\nu_p)$ and $X_{\rm{DFT}}^{\rm{Im}}(\nu_p)$ are independent and have the same normal distribution, the power spectrum, $\left|X_{\rm{DFT}}(\nu_p)\right|^2$, of a pure noise signal then has by definition a $\chi^2$ distribution with 2 degrees of freedom. 

Bearing in mind that
\begin{equation}
\forall p: \; {\rm{E}}\left[\left(X_{\rm{DFT}}^{\rm{Re}}(\nu_p)\right)^2\right]={\rm{E}}\left[\left(X_{\rm{DFT}}^{\rm{Im}}(\nu_p)\right)^2\right]=\frac{N}{2}\sigma_0^2 \, ,
\end{equation} 
and adopting $\left|X_{\rm{DFT}}(\nu_p)\right|^2 \Delta t/N$ as our normalization of the power spectrum, yields a constant power spectral density for the noise given by $\sigma_0^2\Delta t$. Consequently, at a fixed frequency bin, the probability density, $p(z)$, that the observed power spectrum takes a particular value $z$, is given by
\begin{equation}
\label{pdfz}
p(z)=\frac{1}{\sigma_0^2\Delta t}\,\exp\left(-\frac{z}{\sigma_0^2\Delta t}\right) \, ,
\end{equation} 
having mean $\langle z \rangle\!=\!\sigma_0^2\Delta t$ and variance $\sigma_z^2\!\equiv\!\langle z^2 \rangle\!-\!\langle z \rangle^2\!=\!\left(\sigma_0^2\Delta t\right)^2$. The probability density function of an exponential distribution with unit mean and unit variance is displayed in the right-hand panel of Fig.~\ref{H0test}. 

Equation (\ref{pdfz}) enables one to derive the probability that the power in one bin is greater than $m$ times the mean level of the continuum, $\langle z \rangle$:
\begin{equation}
F(m)={\rm{e}}^{-m} \, .
\end{equation} 
For a frequency band containing $M$ bins, the probability that at least one bin has a normalized power greater than $m$ is then \citep[e.g.,][]{ChaplinBiSON}:
\begin{equation}
\label{probwin}
F_M(m)=1-(1-{\rm{e}}^{-m})^M \, ,
\end{equation}  
which approximates to $F_M(m)\!=\!M{\rm{e}}^{-m}$ for ${\rm{e}}^{-m}\!\ll\!1$. Based on Monte Carlo simulations, \citet{Gabriel02} found that, to account for the modified statistics resulting from oversampling{\footnote{This is achieved by zero-padding the time series, i.e., adding zeros onto the end of the series. Zero-padding is frequently used to accurately determine the amplitudes of intrinsically narrow spectral features.}} the spectrum, the number of original bins $M$ in the frequency interval being considered should be multiplied by a factor $\zeta$, which is a function of the oversampling factor (see their table 1):
\begin{equation}
F_M(m)=1-(1-{\rm{e}}^{-m})^{M\zeta} \, .
\end{equation}   

Surprisingly, as $N$ tends to infinity by sampling a longer stretch of data, the variance ($\sigma_z^2$) in the power spectrum remains unchanged. The information resulting from the introduction of additional sampled points goes instead into producing estimates at a greater number of discrete frequencies $\nu_p$. This rather frustrating property can be seen in Fig.~\ref{lorentz} by comparing panel (c) with panels (a) and (b). There are, however, ways of reducing the variance in the power spectrum. A very simple technique consists in partitioning the original time series into $k$ segments of equal length, computing their separate power spectra and finally averaging them. Despite the spectral resolution of the average spectrum being $k$ times lower than that of the original spectrum, its variance has now decreased by the same factor $k$. It should be noted that the statistics of the average power spectrum now follows a $\chi^2$ distribution with $2k$ degrees of freedom \citep{Appourchaux2003}. The effect of reducing the variance in the power spectrum can be seen in panel (d) of Fig.~\ref{lorentz}. There, the fact that the random excitation arises from a system in statistical equilibrium means that averaging the power spectrum over a large number of realizations will reduce the erratic behavior while preserving the spectral feature.       

Although stationarity is generally satisfied by the random processes under consideration, the same is not true about the property of their samples being i.i.d., since these processes may in fact have a memory. Nonetheless, it has been shown \citep{PeligradWu} that even in that event both components of the Fourier transform are normally distributed with zero mean and the same frequency-dependent variance.

A very interesting observational property of stochastically-excited modes can now be tackled. Let us recall Eqs.~(\ref{regime1}) and (\ref{regime2}), respectively describing the mode height $H$ of a resolved and an unresolved peak in the power spectrum. We define the noise-to-signal ratio, $\beta$, of a given mode as:
\begin{equation}
\label{betaproperty}
\beta \equiv B/H = \left\{ \begin{array}{ll}
         (\pi\sigma_0^2\Gamma\Delta t)/(2V^2_{\rm{rms}}) & \mbox{for $T \gg 2\tau$} \, ; \\
         \\
         \sim(\sigma_0^2\Delta t)/(V^2_{\rm{rms}}T) & \mbox{for $T \ll 2\tau$} \, .\end{array} \right.
\end{equation}   
Here, $B\!=\!\langle z\rangle\!=\!\sigma_0^2\Delta t$, represents the mean noise level. Therefore, given fixed noise and mode characteristics, and a chosen observational cadence, the noise-to-signal ratio cannot be improved with time once a mode is resolved, i.e., for $T\!\gg\!2\tau$ \citep{Chaplin03}.   

\subsection{Generalized Lomb-Scargle periodogram}\label{genLS}
In astrophysics it is very common to deal with unevenly sampled time series. In that event, an existing frequentist statistic known as the Lomb-Scargle periodogram \citep{Lomb76,Scargle82,Scargle89} is widely used as a replacement for the Schuster periodogram, the latter being only suitable in the case of uniform sampling. In the following I introduce the generalized Lomb-Scargle periodogram \citep{Bretthorst2001II}, which reduces to the Lomb-Scargle periodogram for a real stationary sinusoid.

Let us assume the general case of quadrature data sampling, i.e., involving the measurement of the real and imaginary parts of a complex signal. I let $x_{\rm{Re}}(t_i)$ denote the set of $N_{\rm{Re}}$ real data samples taken at times $t_i$ and $x_{\rm{Im}}(t'_j)$ denote the set of $N_{\rm{Im}}$ imaginary data samples taken at times $t'_j$ (total of $N\!=\!N_{\rm{Re}}\!+\!N_{\rm{Im}}$ data samples). The sampling is allowed to be non-uniform and it is not required that the $t_i$ and $t'_j$ are simultaneous. Furthermore, the adopted model for both signals is that of a non-stationary single sinusoid, and so one has:
\begin{equation}
x_{\rm{Re}}(t_i)=A\cos(2\pi\nu\,t_i-\theta) Z(t_i) + B\sin(2\pi\nu\,t_i-\theta) Z(t_i) + {\rm{error}}
\end{equation}
and
\begin{equation} 
x_{\rm{Im}}(t'_j)=B\cos(2\pi\nu\,t'_j-\theta) Z(t'_j) - A\sin(2\pi\nu\,t'_j-\theta) Z(t'_j) + {\rm{error}} \, ,
\end{equation}  
where ``error'' denotes the misfit between the data and the model, and $Z(t)$ describes an arbitrary amplitude modulation (e.g., exponential decay) whose time dependence and parameters (if any) are fully specified. $Z(t)$ can be thought of as a weighting or apodizing function. The condition that the overall complex model be orthogonal leads to the following definition of the angle $\theta${\footnote{For simultaneous sampling this condition is automatically satisfied and $\theta$ is defined to be zero.}}:
\begin{equation}
\theta=\frac{1}{2} \tan^{-1}\left[\frac{\sum_{i=1}^{N_{\rm{Re}}} \sin(4\pi\nu\,t_i)Z^2(t_i) - \sum_{j=1}^{N_{\rm{Im}}} \sin(4\pi\nu\,t'_j)Z^2(t'_j)}{\sum_{i=1}^{N_{\rm{Re}}} \cos(4\pi\nu\,t_i)Z^2(t_i) - \sum_{j=1}^{N_{\rm{Im}}} \cos(4\pi\nu\,t'_j)Z^2(t'_j)}\right] \, .
\end{equation}
Finally, having adopted the same nomenclature as used by \citet{Bretthorst2001II}, the generalized Lomb-Scargle periodogram is given by
\begin{equation}
\label{gLS}
\overline{h^2}=\frac{R^2(\nu)}{C(\nu)}+\frac{I^2(\nu)}{S(\nu)} \, ,
\end{equation}     
where
\begin{eqnarray}
\label{gLS1}
R(\nu)&\equiv&\sum_{i=1}^{N_{\rm{Re}}} x_{\rm{Re}}(t_i)\cos(2\pi\nu\,t_i-\theta)\,Z(t_i) \nonumber \\ &-& \sum_{j=1}^{N_{\rm{Im}}} x_{\rm{Im}}(t'_j)\sin(2\pi\nu\,t'_j-\theta)\,Z(t'_j) \, , \\ \nonumber\\
I(\nu)&\equiv&\sum_{i=1}^{N_{\rm{Re}}} x_{\rm{Re}}(t_i)\sin(2\pi\nu\,t_i-\theta)\,Z(t_i) \nonumber \\ &+& \sum_{j=1}^{N_{\rm{Im}}} x_{\rm{Im}}(t'_j)\cos(2\pi\nu\,t'_j-\theta)\,Z(t'_j) \, , \\ \nonumber\\
C(\nu)&\equiv&\sum_{i=1}^{N_{\rm{Re}}} \cos^2(2\pi\nu\,t_i-\theta)\,Z^2(t_i) + \sum_{j=1}^{N_{\rm{Im}}} \sin^2(2\pi\nu\,t'_j-\theta)\,Z^2(t'_j) \, , \\ \nonumber\\
\label{gLS2}
S(\nu)&\equiv&\sum_{i=1}^{N_{\rm{Re}}} \sin^2(2\pi\nu\,t_i-\theta)\,Z^2(t_i) + \sum_{j=1}^{N_{\rm{Im}}} \cos^2(2\pi\nu\,t'_j-\theta)\,Z^2(t'_j) \, .
\end{eqnarray}

In the case of a real stationary sinusoidal signal (i.e., $N_{\rm{Im}}\!=\!0$ and constant $Z(t)$, respectively), Eqs.~(\ref{gLS1}) to (\ref{gLS2}) greatly simplify and the generalized Lomb-Scargle periodogram (cf.~Eq.~\ref{gLS}) reduces to the Lomb-Scargle periodogram. Furthermore, for uniformly sampled quadrature data and a stationary sinusoid it reduces to the Schuster periodogram. The term generalized Lomb-Scargle has also been used in the literature by \citet{Zechmeister} to denote the generalization to a full sine-wave fit, including an offset and statistical weights based on the measurement errors. The two approaches are indeed similar and I invite the interested reader to consult both works and references therein.   

I still would like to make a few remarks on the Lomb-Scargle periodogram. The Lomb-Scargle periodogram is actually equivalent to a Least-Squares spectrum \citep[e.g.,][]{HorneBaliunas}, viz., the solution obtained by linear least-squares fitting to the model 
\begin{equation}
\sum_{i=1}^{N} \left[x(t_i)-a\cos(2\pi\nu\,t_i)-b\sin(2\pi\nu\,t_i)\right]^2 \, .
\end{equation}   
For each test frequency $\nu$, solutions are then obtained for the unknowns $a$ and $b$ that are given by $R(\nu)/C(\nu)$ and $I(\nu)/S(\nu)$, respectively, after setting $N_{\rm{Im}}\!=\!0$ and $Z(t)\!=\!1$. In addition, the Lomb-Scargle periodogram has the attractive property of retaining the $\chi^2_2$ statistics \citep{Scargle82}. Fast computation of the periodogram is achieved using the algorithm presented in \citet{PressRybicki}, whose trick is to carry out extirpolation{\footnote{Extirpolation is the process by which a function value at an arbitrary point is replaced by several function values on a regular mesh.}} of the data onto a regular mesh and subsequently employ the FFT, thus obtaining an accurate approximation of the periodogram.

\subsection{Gapped time series and bin correlations in the Fourier spectrum}
The presence of gaps in the time series introduces frequency correlations in complex Fourier space through convolution of the observable with the spectral window \citep{Gabriel94}. These correlations should be taken into account, especially when performing a fit to the power spectrum (see Sect.~\ref{parest}). To that end, \citet{StahnGizon} describe and implement a rather general method to retrieve maximum likelihood estimates of the oscillation parameters which accounts for the proper statistics of the spectrum. Here, however, I will merely summarize an important result derived in \citet{Appourchauxguide} that proves to be very useful for computing bin correlations in complex Fourier space.

Let us assume that a random process denoted by $x(t)$ is observed through a window function $w(t)$. Then the Fourier transform of the windowed signal is given by
\begin{equation}
\label{windsig}
\tilde{X}(\nu)=\int_{-\infty}^{+\infty} x(t)w(t)\,{\rm{e}}^{{\rm{i}}\,2\pi\nu t} {\rm{d}}t = X(\nu) \ast W(\nu) \, ,
\end{equation}   
where $X(\nu)$ and $W(\nu)$ are the transforms of $x(t)$ and $w(t)$, respectively. In his work, \citet{Appourchauxguide} elegantly arrives at the following useful expression for the mean correlation between any two frequency bins in complex Fourier space:
\begin{equation}
\label{bincorr}
{\rm{E}}\left[\tilde{X}(\nu_1)\tilde{X}^\ast(\nu_2)\right]=2\int_{-\infty}^{+\infty} {\rm{E}}\left[X_{\rm{Re}}^2(\nu)\right] W(\nu_1-\nu) W(\nu-\nu_2)\,{\rm{d}}\nu \, ,
\end{equation}  
where $X_{\rm{Re}}(\nu)\!\equiv\!\Re\{X(\nu)\}$ and the superscript asterisk denotes the complex conjugate. An alternative way to express this correlation, not provided in the aforementioned work, is to use Eq.~(\ref{windsig}) to write:
\begin{equation}
{\rm{E}}\left[\tilde{X}(\nu_1)\tilde{X}^\ast(\nu_2)\right]=\int\int {\rm{E}}\left[x(t)x(t')\right] {\rm{e}}^{\rm{i}\,2\pi(\nu_1t-\nu_2t')} w(t)w(t') \,{\rm{d}}t\,{\rm{d}}t' \, .
\end{equation}  
Further assuming that the process is stationary and ergodic, use of Eq.~(\ref{statergo}) then leads to the following alternative formulation:
\begin{equation}
{\rm{E}}\left[\tilde{X}(\nu_1)\tilde{X}^\ast(\nu_2)\right]=\int\int \phi(\tau) {\rm{e}}^{\rm{i}\,2\pi(\nu_1\tau+(\nu_1-\nu_2)t')} w(\tau+t')w(t') \,{\rm{d}}\tau\,{\rm{d}}t' \, ,
\end{equation} 
where $\tau\!=\!t-t'$.

Let us investigate what happens in the case of a continuous and infinite window function, for which we know the spectral window to be the impulse function $\delta(\nu)$. In this case one has:
\begin{equation}
\label{bincorr2}
{\rm{E}}\left[\tilde{X}(\nu_1)\tilde{X}^\ast(\nu_2)\right]=2\,{\rm{E}}\left[X_{\rm{Re}}^2(\nu_1)\right] \delta(\nu_1-\nu_2) \, ,
\end{equation}
where use was made of Eq.~(\ref{bincorr}). From Eq.~(\ref{bincorr2}) it becomes apparent that a given bin is uncorrelated with any other bin in the spectrum.

Furthermore, for slowly varying power spectra, i.e., for which the variations of ${\rm{E}}\left[X_{\rm{Re}}^2(\nu)\right]$ are slow with respect to $W(\nu)$, Eq.~(\ref{bincorr}) may be rewritten as
\begin{eqnarray}
\label{bincorr3}
{\rm{E}}\left[\tilde{X}(\nu_1)\tilde{X}^\ast(\nu_2)\right] & \approx & 2\,{\rm{E}}\left[X_{\rm{Re}}^2(\nu_1)\right] \int_{-\infty}^{+\infty} W(\nu_1-\nu) W(\nu-\nu_2)\,{\rm{d}}\nu \nonumber \\
& = & 2\,{\rm{E}}\left[X_{\rm{Re}}^2(\nu_1)\right] \int_{-\infty}^{+\infty} w^2(t) {\rm{e}}^{{\rm{i}}\,2\pi(\nu_1-\nu_2)t}\,{\rm{d}}\nu \nonumber \\
& = & 2\,{\rm{E}}\left[X_{\rm{Re}}^2(\nu_1)\right] W_{\rm{sq}}(\nu_1-\nu_2) \, ,
\end{eqnarray}
where the integral, denoted by $W_{\rm{sq}}(\nu_1-\nu_2)$, is simply the Fourier transform of the square of the window function. When observing white noise, ${\rm{E}}\left[X_{\rm{Re}}^2(\nu)\right]$ is constant and will be denoted by $\sigma^2$. The equality in Eq.~(\ref{bincorr3}) then holds exactly. 

Let us see what happens when the window function is rectangular with width $T$ and centered at $t\!=\!0$:
\begin{equation}
{\rm{E}}\left[\tilde{X}(\nu_1)\tilde{X}^\ast(\nu_2)\right] = 2\sigma^2\,\frac{\sin\left(\pi T(\nu_1-\nu_2)\right)}{\pi(\nu_1-\nu_2)} \, .
\end{equation}
Frequency bins are thus uncorrelated at frequency separations that are non-zero multiples of $1/T$, the so-called Rayleigh resolution. And if one would instead opt for apodizing (i.e., weighting) the data, such that the square of the window function becomes a triangular (Barlett) window while keeping $T$ as the length of the observation? The answer can be easily computed:
\begin{equation}
{\rm{E}}\left[\tilde{X}(\nu_1)\tilde{X}^\ast(\nu_2)\right] = 2\sigma^2\,\left[\frac{\sin\left(\pi T(\nu_1-\nu_2)/2\right)}{\pi(\nu_1-\nu_2)}\right]^2 \, .
\end{equation}
This would in fact reduce the spectral leakage at high frequencies while increasing it at the low-frequency end. Frequency bins are now uncorrelated at frequency separations that are non-zero multiples of $2/T$. 

These simple calculations exemplify how the introduction of weights on the data may induce modifications to bin correlations in Fourier space. Weights are commonly used when analysing ground-based observations of stellar oscillations as a way of taking into account the significant variations in data quality during a typical observing campaign, especially when two or more telescopes are involved. This has certainly been the case in the analysis of the time series of velocity observations obtained for Procyon over 25 days using 11 telescopes at eight observatories \citep{BeddingProcyon}. In that work use was made of two alternative weighting schemes, namely, noise-optimized weights and sidelobe-optimized weights \citep[see also][]{Arentoftweights}. Proper account was taken of the weighting by using Eq.~(\ref{bincorr3}) to determine the effective frequency resolution prior to the computation of the power spectrum.    

\subsection{Bayesian insight into the periodogram}
The Schuster periodogram was originally introduced, rather intuitively, in order to detect a periodicity and estimate its frequency. \citet{Jaynes87}, however, demonstrated that the periodogram follows naturally from Bayesian probability theory{\footnote{The topic of Bayesian inference will be dealt with throughout Sect.~\ref{statinf}.}}. Considering the analysis of a time series known to contain a single sine wave with frequency $\nu_0$, which is additionally contaminated with additive independent Gaussian noise with variance $\sigma^2$, \citet{Jaynes87} showed that the posterior probability of the frequency of the periodic sinusoidal signal is approximately given by
\begin{equation}
\label{postpdf}
p(\nu|D,I) \propto \exp\left\{\frac{\hat{P}_{\rm{p}}(\nu)}{\sigma^2}\right\} \, ,
\end{equation}  
where $D$ and $I$ respectively represent the observed data and the prior information, and $\hat{P}_{\rm{p}}(\nu)$ is the Schuster periodogram (cf.~Eq.~\ref{Schuster}). The periodogram is defined here as the squared magnitude of the FFT times the reciprocal of the number $N$ of samples drawn. If the noise variance is not a known quantity, then the resulting posterior probability may instead be expressed in the form of a Student's $t$ distribution and is given approximately by \citep{Bretthorst88}
\begin{equation}
\label{postpdf2}
p(\nu|D,I) \propto \left[1-\frac{2\,\hat{P}_{\rm{p}}(\nu)}{N\overline{x^2}}\right]^{\frac{2-N}{2}} \, ,
\end{equation}
where $\overline{x^2}$ is the mean-square amplitude of the data values. Equations (\ref{postpdf}) and (\ref{postpdf2}) do not require the data to be uniformly sampled provided that several specific conditions are met. 

By computing the first and second moments of the distribution in Eq.~(\ref{postpdf}), \citet{Jaynes87} also showed that $\langle\nu\rangle\!=\!\nu_0$ and that, for a sine wave of amplitude $A$, the rms error on the determination of the frequency is given by
\begin{eqnarray}
\label{sigmafreq}
\sigma_\nu&=&\frac{\sqrt{6}}{\pi} \frac{\sigma}{A} \frac{1}{T\sqrt{N}} \nonumber \\
&=&\frac{\sqrt{6}}{\pi} \frac{\sigma}{A} \frac{\sqrt{\Delta t}}{T^{3/2}} \, ,
\end{eqnarray} 
where $T$ is the length of the observation and the second equality assumes the data to be uniformly sampled at intervals of $\Delta t$ \citep[a formula also given in][]{Cuypers87,Koen99,MontDon}. Frequency precision is therefore determined by the SNR in the time domain, the length of the observation, and the number of data points. Notice that, for high SNR, frequency precision can greatly surpass the Rayleigh resolution. 

\citet{Bretthorst2000,Bretthorst2001I,Bretthorst2001II} generalized this Bayesian approach to the periodogram to a broader range of single-frequency estimation problems (single sinusoid with arbitrary decay as well as periodic but nonsinusoidal functions) and sampling conditions. An exact Bayesian expression for $p(\nu|D,I)$ was derived that involves a nonlinear processing of the generalized Lomb-Scargle periodogram analogous to the nonlinear processing of the Schuster periodogram in Eq.~(\ref{postpdf2}):
\begin{equation}
p(\nu|D,I) \propto \frac{1}{\sqrt{C(\nu)S(\nu)}} \left[N\overline{x^2}-\overline{h^2}\right]^{\frac{2-N}{2}} \, .
\end{equation}  
Here, I have recovered the terminology introduced in Sect.~\ref{genLS}, where $\overline{h^2}$ represents the generalized Lomb-Scargle periodogram, and the mean-square amplitude of the data values is defined as
\begin{equation}
\overline{x^2}=\frac{1}{N} \left[\sum_{i=1}^{N_{\rm{Re}}}x^2_{\rm{Re}}(t_i) + \sum_{j=1}^{N_{\rm{Im}}}x^2_{\rm{Im}}(t'_j)\right] \, .
\end{equation}

The latter formalism is not applicable to stochastic oscillators due to the multiplicative nature of the associated noise. In fact, separation between the deterministic signal and the noise is not possible in the time domain for stochastically-excited oscillators, but only in the frequency domain. This same formalism is, however, certainly relevant to the case of classical pulsators, whose oscillations are periodic but not necessarily sinusoidal. 

\subsection{Multisine estimation}\label{multisine}
Fourier-based methods are well suited to the analysis of sinusoidal oscillations. Even though approaches such as the Lomb-Scargle periodogram \citep{Lomb76,Scargle82,Scargle89} and the date-compensated DFT \citep{DCDFT} take into account sampling irregularities in the time series, they are only statistically valid in the case of a single sinusoid present in the data \citep{Foster96}. If, on the other hand, multiple sinusoids are present then these are required to be well separated in frequency \citep{Bretthorst88}, which is not always the case. So how does one proceed when searching for multiple sinusoids in the data? 

Multisine estimation is usually performed by employing sequential methods. These methods iteratively remove sinusoidal components from the data while taking into account the effect of the window function in a rather straightforward way. Such components are identified as the maxima of the Fourier spectrum of the residuals after a prewhitening step, i.e., after the contributions of previous estimated frequencies have been removed. This approach is commonly known in the literature as Iterative Sine-Wave Fitting (ISWF). However, due to the presence of noise peaks and sampling artifacts in the Fourier spectrum, these maxima may not correspond to genuine frequencies, thus leading to fatal error propagation. Several refinements of the basic ISWF procedure have been posteriorly developed that aim at improving its efficiency, most notoriously the CLEAN algorithm \citep{Roberts87}, originally introduced in radio astronomy \citep{Hogbom74}. The CLEAN algorithm as described by \citet{Roberts87} introduces a clean gain whereby only a fraction of the sinusoidal component is removed at each iteration, in an attempt to prevent the propagation of errors resulting from false detections. Furthermore, since the number of oscillation modes present in the data is unknown a priori, a stopping rule is needed that associates a confidence level to the amplitude of each extracted sinusoidal component \citep[e.g.,][]{Breger93,Reegen04}. Although these methods are better suited to the analysis of classical pulsators, they have been successfully employed in the analysis of a number of ground-based observations of solar-like stars \citep[e.g.,][]{Bedding04,Kjeldsen05,Bedding07betaHyi}, always under the assumption that the characteristic mode lifetime is (much) longer than the length of the time series (i.e., that the modes are unresolved). These methods have also been extensively tested on solar-like artificial data \citep[][]{White10}.

Several other methods exist for multisine estimation. One such method, based on the framework of sparse representations, is suggestively named SparSpec \citep{SparSpec}. SparSpec addresses multisine estimation by reconstructing a high-dimensional vector of spectral amplitudes corresponding to a discrete frequency grid. A sparse representation of the spectrum is desired, i.e., one with the fewest non-zero spectral amplitudes, which can be accomplished by minimizing a convex criterion. In a second article (in preparation) devoted to the solar twin 18 Sco, we compare several different approaches to frequency estimation, including CLEAN, SparSpec, and also Lorentzian-profile fitting techniques (see Sect.~\ref{parest} for more on the last-mentioned techniques).

\subsection{Wavelet analysis}
I cannot come up with a better way of finishing the current section than to present a tool performing time-frequency analysis, namely, the wavelet transform. The following discussion is based on the work by \citet{TorrenceCompo}, who supply a useful guide to wavelet analysis containing examples of its application in the field of geophysics and a link to open-source software. In short, a wavelet-based analysis decomposes a time series into time-frequency space allowing one to determine both the dominant modes of oscillation and their temporal variability. For us, asteroseismologists, I perceive at least two novel domains of application for this sort of tool: (i) in assessing whether or not a given set of modes are stochastically excited \citep[e.g.,][]{Belkacem09,AntociNature} and (ii) in determining the stellar rotational period from the modulation of a light curve caused by photospheric spots \citep[e.g.,][]{Mathur10,HD52265,Garciafast}.

As a preamble, I should mention the windowed Fourier transform, by which the Fourier transform of a given time series is performed on a running segment of length $T$. This analysis tool is widely used for time-frequency localization, although in a rather inaccurate and inefficient way, since it imposes a scale $T$ on the analysis. When analysing a signal possibly containing non-stationary power over a wide range of frequencies, one should opt instead for a method of time-frequency localization that is independent of scale, the wavelet transform being such an approach.

Let us start by considering a time series that has been uniformly sampled at intervals of $\Delta t$, i.e., $x(t_n)\!=\!x(n\Delta t)$, with $n\!=\!0,1,2,\ldots,N\!-\!1$. We also consider a wavelet function, $\psi_0(\eta)$, which depends on the nondimensional time parameter $\eta$, having zero mean and being localized both in time and in frequency space. A common example of such a wavelet function is the Morlet wavelet, assumed hereafter, and which consists of a plane wave modulated by a Gaussian. Finally, the continuous wavelet transform of the set $x(t_n)$ is defined as the convolution of $x(t_n)$ with a scaled and translated version of $\psi_0(\eta)$ that has further been normalized (and so the subscript in $\psi_0$ is dropped):
\begin{equation}
W_n(s)=\sum_{n'=0}^{N-1} x(t_{n'}) \, \psi^\ast\left[\frac{(n'-n)\Delta t}{s}\right] \, .
\end{equation}  
Therefore, by varying the wavelet scale{\footnote{Notice that, for the Morlet wavelet, the equivalent Fourier period is given by $1.03s$.}} $s$ and translating along the time index $n$, one may generate a diffuse two-dimensional time-frequency image of the signal amplitude. Computation of the wavelet transform is, however, notably faster if done in Fourier space:
\begin{equation}
\label{wavelet}
W_n(s)=\sum_{p=0}^{N-1} \hat{x}_p\,\hat{\psi}^\ast(s\,\omega_p)\,{\rm{e}}^{{\rm{i}}\,\omega_p n \Delta t} \, ,
\end{equation} 
where $\hat{x}_p$ is the DFT of $x(t_n)$ and $\hat{\psi}(s\,\omega)$ is the Fourier transform of $\psi(t/s)$. The angular frequency $\omega_p$ is in turn defined as
\begin{equation}
\omega_p = \left\{ \begin{array}{ll}
         \frac{2\pi p}{N \Delta t} & \mbox{for $0 \leq p \leq N/2$} \, ; \\
         \\
         -\frac{2\pi p}{N \Delta t} & \mbox{for $N/2 < p \leq N-1$} \, .\end{array} \right.
\end{equation} 
Moreover, a properly normalized wavelet transform satisfies, for each scale $s$, the following equality (cf.~Eq.~\ref{wavelet}):
\begin{equation}
\label{normwavelet}
\sum_{p=0}^{N-1} |\hat{\psi}(s\,\omega_p)|^2 = N \, .
\end{equation}
The wavelet transform is, therefore, weighted solely by the amplitude of the Fourier coefficients $\hat{x}_p$ and not by the wavelet function itself.
 
The wavelet power spectrum is then simply given by $|W_n(s)|^2$ (see Fig.~\ref{waveletrot} for an example). In order to compare wavelet power spectra arising from different time series, one is obviously interested in finding a common normalization factor. This can be easily done. From Eqs.~(\ref{wavelet}) and (\ref{normwavelet}), the expectation value of $|W_n(s)|^2$ is $N$ times that of $|\hat{x}_p|^2$. Assuming a white-noise process, the expectation value of $|\hat{x}_p|^2$ is given by{\footnote{In the present discussion, a normalization factor of $1/N$ is applied to the definition of the DFT in Eq.~(\ref{DFT}).}} $\sigma_0^2/N$, with $\sigma_0^2$ representing the variance in the time series. Finally, the expectation value of $|W_n(s)|^2$ is given by $\sigma_0^2$, and normalization of the wavelet power spectrum by $1/\sigma_0^2$ will then provide a measure of the power relative to white noise.

\figuremacroW{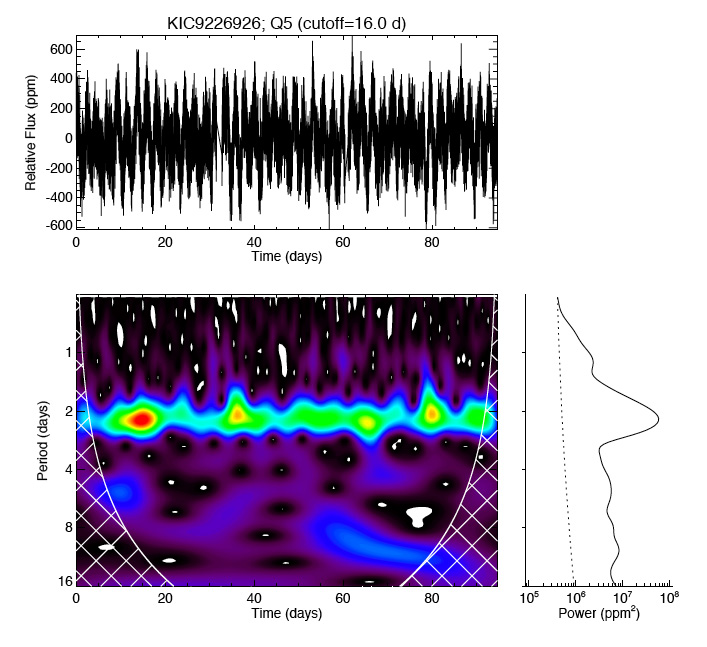}{Wavelet power spectrum of KIC~9226926 from three months of \emph{Kepler} photometry}{The top panel displays the light curve of KIC~9226926 obtained during Quarter 5 (Q5) after being corrected for instrumental perturbations according to \citet{Garciacorrections}. The light curve has further been high-pass filtered with a cutoff of 16 days. The corresponding wavelet spectrum is shown on the bottom left-hand panel for periods ranging from 0.5 to 16 days. A signature of the stellar rotational period (first harmonic) can be immediately recognized at a period of approximately 2.2 days, being nearly stationary over the entire observational span. The cone of influence (see discussion in the text) is represented by the hatched area. The bottom right-hand panel depicts the so-called global wavelet spectrum (solid line), i.e., the time-averaged wavelet spectrum. The dotted line represents the 99\% confidence level assuming a red-noise background.}{1}

We should not forget that we will always be dealing with finite-length time series and, since the inverse Fourier transform in Eq.~(\ref{wavelet}) assumes the data are cyclic{\footnote{Note that sampling in the frequency domain results in a periodic version of the signal in the time domain.}}, this means that the wavelet power spectrum will be subject to errors at both ends of the time interval. A way of reducing these edge effects consists in padding the time series with enough zeros. This will, of course, introduce discontinuities at both endpoints and, as we move to larger scales, the amplitude decrease near the edges will, as a result of the larger number of included zeros, be more conspicuous. Hence we define the so-called cone of influence, viz., the region of the wavelet power spectrum where edge effects are significant, underneath which any reading is dubious. The cone of influence is determined by the e-folding time of the autocorrelation of wavelet power at each scale, given by $\sqrt{2}s$ for the Morlet wavelet.  

Significance levels can be easily computed for wavelet power spectra based on hypothesis testing (these are, however, not displayed in Fig.~\ref{waveletrot}). As a first step one needs to choose an appropriate mean background for the local wavelet power spectrum, defined as a vertical slice through the bottom left-hand panel of Fig.~\ref{waveletrot}. This theoretical background spectrum is usually chosen as being either a white- (constant spectral density) or a red-noise (spectral density inversely proportional to frequency squared) spectrum. Moreover, it can be shown that the wavelet power spectrum, $|W_n(s)|^2$, is distributed as $\chi^2_2$. Finally, in order to compute the, say, 95\% confidence level for $|W_n(s)|^2/\sigma_0^2$, one simply has to multiply the theoretical background spectrum by the 95th percentile for $\frac{1}{2}\,\chi^2_2$ at each scale (the $\frac{1}{2}$ factor accounts for the degrees of freedom in $\chi^2_2$).

\section{Statistical inference}\label{statinf}

\subsection{Setting the scene: Are you a Bayesian or a frequentist?}
This question was posed by Thierry Appourchaux to a few of us, young asteroseismologists, during a doctoral school that took place in Tenerife in the autumn of 2010. The audience fell inconveniently silent as if no one had ever given it any thought. 

It was a controversy between Ronald Fisher and Harold Jeffreys, dating back to the first half of the last century, that marked the beginning of two schools of thought with different views on probability and statistics, namely, that of the frequentists and that of the Bayesians. The debate between frequentists and Bayesians relates to the topic of objective versus subjective probabilities. For a frequentist, to whom the laws of physics are deterministic, the probability of an event is identified with the long-run relative frequency with which that event occurs in identical repeats of the experiment or observation. In the frequentist approach probabilities are only assigned to propositions about random variables. The Bayesian, on the other hand, regards the laws of physics as being operational. The Bayesian has recognized that the mathematical rules of probability are not only suitable for calculating relative frequencies of random variables, but can also be interpreted as valid principles of logic for directly computing the probability of any proposition or hypothesis of interest, based on our current state of knowledge. An excellent presentation of the topic of Bayesian logical data analysis is given in the book by \citet{GregoryBook}.

I leave the skeptical frequentist (presumably) reader with Edwin T.~Jaynes' words on the topic of objectivity versus subjectivity: ``The only thing objectivity requires of a scientific approach is that experimenters with the same state of knowledge reach the same conclusion.''

\subsection{Hypothesis testing}
Inferring the truth of one or more hypotheses related to some physical phenomenon is one of the main goals in science. As a further matter, the information that we have available is always incomplete, meaning that our knowledge of nature is probabilistic. Consequently, the funny thing about it is that, due to our state of incomplete information, we can never prove any hypothesis is true. Bayesian inference comes to the rescue by allowing us to directly compute the probabilities of two or more competing hypotheses based on the current state of knowledge. On the other hand, the frequentist approach to hypothesis testing is rather indirect. The present discussion on hypothesis testing is adapted from \citet{AppourchauxReviewg}. 

\subsubsection{Frequentist hypothesis testing}\label{Freqhyptest}
In the frequentist approach, the argument of a probability is restricted to a random variable. Since a given hypothesis cannot be considered a random variable, the truth of the hypothesis must be indirectly inferred. Frequentist hypothesis testing thus involves considering each hypothesis individually and deciding whether to reject that hypothesis or fail to reject it, based on the computed value of a suitable choice of statistic. In our case, we will be interested in testing the following two hypotheses:
\begin{itemize}
\item $H_0$ or null hypothesis: We observe pure noise;
\item $H_1$ or alternative hypothesis: We observe a signal embedded in noise. 
\end{itemize}
The test based on the $H_0$ hypothesis consists in determining a significance level for which peaks in the spectrum have a low probability of being due to noise. It thus tests for the presence of a signal in the data. The test based on the $H_1$ hypothesis, on the other hand, allows computing the probability that a signal can be detected given its characteristics. 

For a mode with a lifetime much longer than the length of the observation, i.e., a long-lived mode, it is assumed that we search for a peak that is restricted to a single frequency bin in the power spectrum. An analytical example based on the detection of a long-lived mode in a power spectrum (e.g., a g mode in the power spectrum of the Sun or an unresolved solar/stellar p mode) will guide the reader throughout the discussion taking place in this and the following section.  

Let us start by considering the $H_0$ hypothesis. The statistics of the power spectrum of a pure noise signal is assumed to be known and taken to be $\chi^2$ with 2 degrees of freedom. I denote by $Z$ the random variable representing the power level in a given bin, which is observed to take a particular value $z$ (cf.~Eq.~\ref{pdfz}). Our choice of statistic is simply that of $Z${\footnote{Such a choice of statistic is not done ad hoc, but can be systematically derived using the Neyman-Pearson lemma \citep{NeymanPearson}.}}. Next we set the false alarm probability or $p$-value \citep{Scargle82}, thus defining the detection threshold $z_{\rm{det}}$:
\begin{equation}
\label{detectionlevel}
p=p(Z\ge z_{\rm{det}}|H_0)={\rm{e}}^{-\frac{z_{\rm{det}}}{\langle z\rangle}} \, .
\end{equation}      
For instance, a confidence level of 95\% or, equivalently, a false alarm probability of 5\%, leads to a threshold given by $z_{\rm{det}}\!\approx\!3\langle z\rangle$. Finally, the probability of having a value of the statistic at least as extreme as the one observed is computed (the so-called detection significance). We are now in a position to make a decision: If the observed power level $z_{\rm{obs}}$ is greater than $z_{\rm{det}}$, then the $H_0$ hypothesis is rejected; otherwise we fail to reject the null hypothesis. In the former case, the detection significance, given by $p_{\rm{obs}}\!=\!{\rm{e}}^{-\frac{z_{\rm{obs}}}{\langle z\rangle}}$, is quoted. A test based on the $H_0$ hypothesis was used in helioseismology, for example, by \citet{Appourchaux2000}, to impose an upper limit on g-mode amplitudes. A range of similar tests have been applied by \citet{ChaplinBiSON} in the search for low-degree, low-frequency solar p modes making use of 9 years of BiSON data. Figure \ref{H0test} illustrates the application of a test based on the $H_0$ hypothesis to the detection of an unresolved solar p mode. Figure \ref{probechelle}, on the other hand, illustrates the application of a test based on the $H_0$ hypothesis to the detection of short-lived stellar p modes.

\figuremacroW{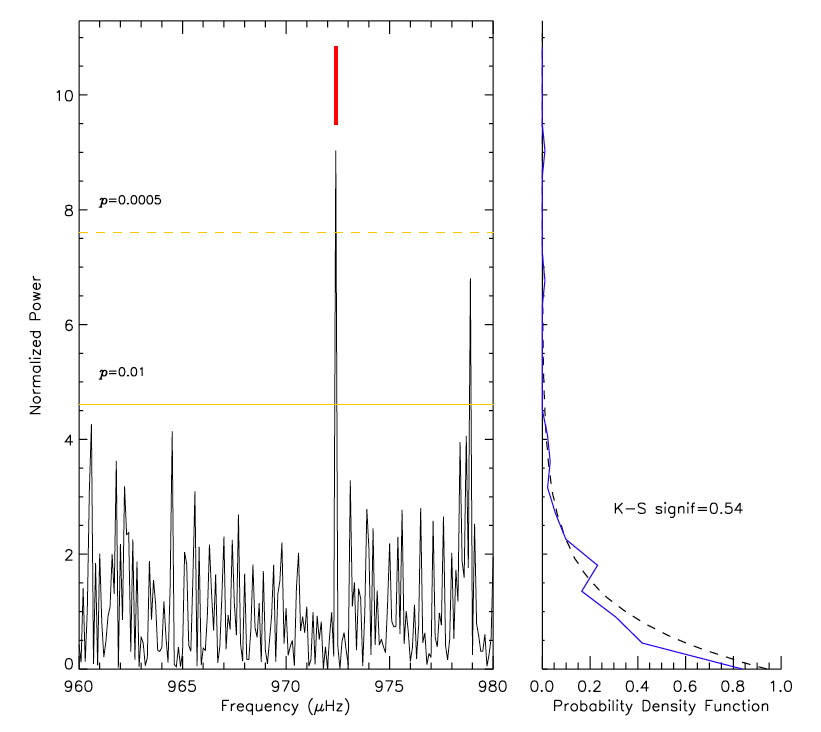}{Application of a test based on the $H_0$ hypothesis to the detection of an unresolved solar p mode}{The left-hand panel displays the simulated power spectrum of an isolated low-frequency solar $l\!=\!0$ mode (its true location being marked by a vertical red dash) normalized to the local background. The mode linewidth was taken to be one tenth of the bin width and its relative height was rather optimistically set to 10 for illustrative purposes only. The solid horizontal line corresponds to the detection threshold obtained by setting $p\!=\!0.01$ in Eq.~(\ref{detectionlevel}). According to Eq.~(\ref{probwin}), the chance of finding at least one noise spike within the displayed window (containing 200 bins) at or above this detection threshhold is then of about 87\% (with a total of 2 such spikes, on average, expected to be found). A more conservative approach would be to set to, say, 10\%, the probability of finding at least one spike within the considered window (or a total of 0.1 spikes on average). This results in $p\!=\!0.0005$, corresponding to the dashed horizontal line. The right-hand panel displays the probability density function of the simulated spectral noise (solid curve) and the exponential distribution with unit mean and unit variance (dashed curve) from which noise samples have been drawn. The significance level of the Kolmogorov-Smirnov statistic is indicated, telling us how similar the corresponding cumulative distribution functions are.}{0.84}

\figuremacroW{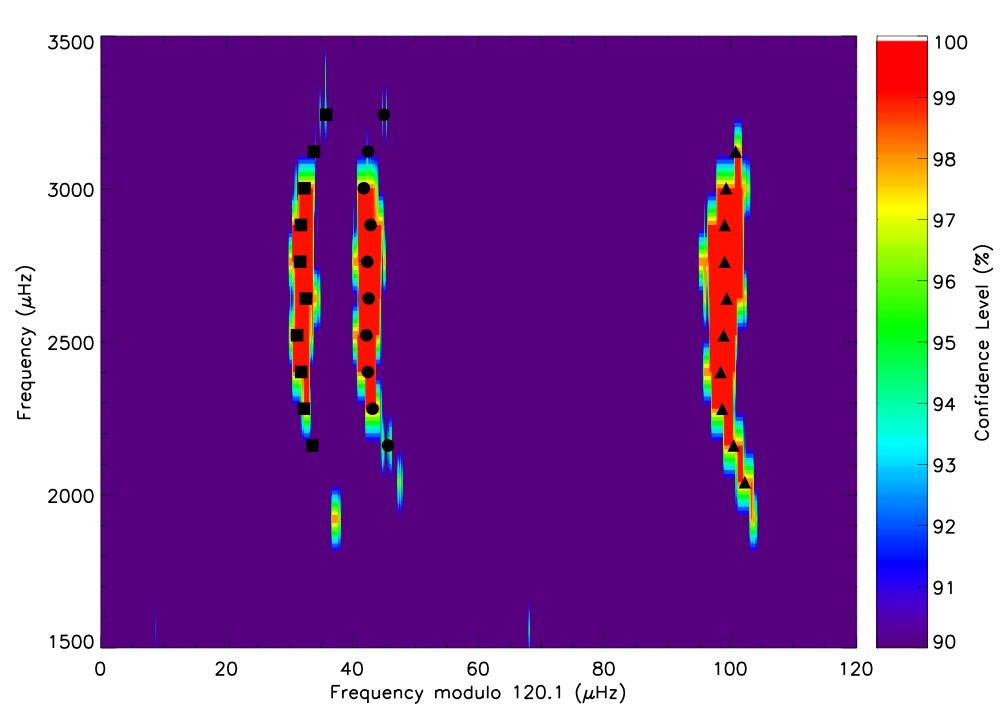}{Application of a test based on the $H_0$ hypothesis to the detection of short-lived stellar p modes}{In order to keep the discussion as concise as possible, the problem of detecting short-lived modes in a binned power spectrum has not been addressed. Nevertheless, an example is given here of the application of a test based on the $H_0$ hypothesis to the detection of short-lived modes \citep[cf.][]{Appourchaux04}. What one may call a probabilistic \'echelle diagram (i.e., displaying probability instead of power) is plotted based on a 18-month-long time series of KIC~3427720. The displayed frequencies come from the analysis of a subseries of only 9 months long with symbol shapes indicating mode degree: $l\!=\!0$ (circles), $l\!=\!1$ (triangles), and $l\!=\!2$ (squares). A confidence level of, say, 95\%, corresponds to a detection threshold such that the probability of finding at least one noise spike within the whole plotted window is 5\%. Confidence levels have been clipped to a minimum of 90\%. The inclusion of an extra 9 months of photometry allows detecting a few additional modes at the low-frequency range.}{0.95}

For the $H_1$ hypothesis, the characteristics of both the noise and the sought-after signal are assumed to be known. A significance level is then set that defines the acceptance or rejection of that hypothesis. It was based on the $H_1$ hypothesis that \citet{Gabriel02} provided the probability of detecting a pure sine wave with given amplitude in GOLF data. \citet{Appourchaux04} provides tests based both on the $H_0$ and $H_1$ hypotheses. In that work, a false alarm test for detecting short-lived p modes is defined and the probability of detecting such modes is given subject to a set of assumptions about their characteristics.

Decisions based on hypothesis testing are prone to errors. For example, it may happen that the $H_0$ hypothesis be rejected when true (false positive or type I error) or accepted when false (false negative or type II error). A type I error is considered to be more serious{\footnote{A courtroom analogy would be that the possibility of convicting an innocent party is considered worse than the possibility of acquitting a guilty party.}}. It is not possible, however, to minimize both the type I and type II errors. The usual procedure is then to state the maximum size of the type I error that is tolerable and construct a test procedure that minimizes the type II error. Assuming, for instance, that one is willing to accept a maximum type I error of 5\%, then setting $p\!=\!0.05$ will minimize the occurrence of a type II error.

The detection significance $p_{\rm{obs}}$ is often incorrectly regarded as the probability that $H_0$ is true. The point here is that any particular value of the detection significance may arise even in the event that the alternative hypothesis is true. The detection significance instead tells us how likely the observed data are given the null hypothesis and provided we could repeat our experiment ad infinitum. In the framework of Bayesian inference, one is not interested in the detection significance but in the posterior probability of $H_0$, viz., the likelihood that $H_0$ is true given the observed data, $p(H_0|z)$. This is the quantity that we really want to know and its computation will be the subject of the next section.

\subsubsection{Bayesian hypothesis testing}\label{Bayeshyptest}
A preliminary step should be to formally introduce Bayes' theorem \citep{Bayestheorem}. Let us consider a set of competing hypotheses, $\{H_i\}$, assumed to be mutually exclusive. One should be able to assign a probability, $p(H_i|D,I)$, to each hypothesis, taking into account the observed data, $D$, and any available prior information, $I$, arising from theoretical considerations and/or previous observations. This is done through Bayes' theorem:
\begin{equation}
\label{BayesTheorem}
p(H_i|D,I) = \frac{p(H_i|I)\,p(D|H_i,I)}{p(D|I)} \, .
\end{equation}
The probability of the hypothesis $H_i$ in the absence of $D$ is called the \emph{prior probability}, $p(H_i|I)$, whereas the probability including $D$ is called the \emph{posterior probability}, $p(H_i|D,I)$. The quantity $p(D|H_i,I)$ is called the \emph{likelihood} of $H_i$, $p(D|I)$ being the \emph{global likelihood} for the entire class of hypotheses. Bayesian inference thus encodes our current state of knowledge into a posterior probability concerning each member of the hypothesis space of interest. Moreover, the sum of the posterior probabilities over the hypothesis space of interest is unity, and hence one has:
\begin{equation}
\label{GlobalLikelihood}
p(D|I)=\sum_i p(H_i|I)\,p(D|H_i,I) \, .
\end{equation}

It now becomes a trivial exercise to derive an expression for $p(H_0|z)$ as a function of $p(z|H_0)$ and $p(z|H_1)$ \citep{BergerSellke}:
\begin{equation}
p(H_0|z)=\frac{p(H_0)\,p(z|H_0)}{p(H_0)\,p(z|H_0)+p(H_1)\,p(z|H_1)} \, ,
\end{equation}
where I believe the nomenclature to be self-explanatory. By setting the prior on $H_0$ as $p(H_0)\!=\!p_0$, one then necessarily has $p(H_1)\!=\!1\!-\!p_0$. The adopted value for $p_0$ thus reflects our prejudice on which hypothesis is more likely to be true. Finally, we obtain
\begin{equation}
\label{postH0}
p(H_0|z)=\left(1+\frac{1-p_0}{p_0}\,\mathscr{L}\right)^{-1} \, ,
\end{equation} 
where the likelihood ratio $\mathscr{L}$ is defined as:
\begin{equation}
\label{likeratio}
\mathscr{L}=\frac{p(z|H_1)}{p(z|H_0)} \, .
\end{equation}
Hereafter, we set $p_0\!=\!0.5$. The prescription of \citet{Berger97} should then be used when performing hypothesis testing:
\begin{itemize}
\item If $\mathscr{L}\!>\!1$, reject $H_0$ and report $p(H_0|z)\!=\!1/(1+\mathscr{L})$;
\item If $\mathscr{L}\!\le\!1$, accept $H_0$ and report $p(H_1|z)\!=\!1/(1+\mathscr{L}^{-1})$.
\end{itemize}
Furthermore, \citet{Sellke2001} found that $p(H_0|z)$ has a lower bound given by
\begin{equation}
\label{lbound}
p(H_0|z) \ge \left(1-\frac{1}{{\rm{e}}\,p\,\ln p}\right)^{-1} \, ,
\end{equation}
where the previously introduced detection significance, given by{\footnote{From now on we drop the subscript in $p_{\rm{obs}}$.}} $p\!=\!{\rm{e}}^{-z}$, has been incorporated.

We now recover our analytical example based on the detection of a long-lived mode in a power spectrum. The likelihood of $H_0$ is simply given by
\begin{equation}
p(z|H_0)={\rm{e}}^{-z} \, ,
\end{equation} 
where we have conveniently set the mean noise level to unity in Eq.~(\ref{pdfz}), i.e., $\langle z\rangle\!=\!1$. In order to define the likelihood of $H_1$, we assume that there is a peak corresponding to a long-lived mode. The mode is further assumed to be stochastically excited, being characterized by a known height $H$. This is a strong prior. The more realistic case of a mode with an unknown height is briefly mentioned below. Since the mean noise level is unity, $H$ can be regarded as the signal-to-noise ratio. We are dealing with multiplicative noise and a proper expression for $p(z|H_1)$ can be found in \citet{Moreira2005}:
\begin{equation}
p(z|H_1)=\frac{1}{1+H} {\rm{e}}^{-z/(1+H)} \, .
\end{equation}
Finally, after some manipulation, we rewrite Eq.~(\ref{postH0}) as
\begin{equation}
\label{postH0_2}
p(H_0|z)=\left(1+\frac{1}{1+H}\,p^{-H/(1+H)}\right)^{-1} \, .
\end{equation}
It can be easily shown that Eq.~(\ref{postH0_2}) has a minimum at $H\!=\!-(\ln p+1)$, where it takes the value:
\begin{equation}
\label{postH0min}
p_{\rm{min}}(H_0|z)=\left(1-\frac{1}{{\rm{e}}\,p\,\ln p}\right)^{-1} \, ,
\end{equation}
coinciding with the lower bound given in Eq.~(\ref{lbound}). Take a detection significance of 1\% (i.e., $p\!=\!0.01$). The odds against $H_0$ will then at most be of about $9\!:\!1$ (cf.~Eq.~\ref{postH0min}). This means that the likelihood of wrongly rejecting $H_0$ (type I error) is considerably higher than that suggested by the detection significance. This simple example brings to light a pitfall of frequentist hypothesis testing: the detection significance when rejecting the $H_0$ hypothesis can lead to the incorrect conclusion that the null hypothesis is unlike to occur at that level of significance \citep{AppourchauxBayesTest}. In conclusion, it should be stated that the posterior probability of $H_0$ (Bayesian approach) provides a more conservative quantification of a detection when compared to the use of the detection significance (frequentist approach). 

In the more realistic case of a mode with an unknown height, the posterior probability of $H_0$ is instead given by \citep{AppourchauxBayesTest}:
\begin{equation}
\label{postH0_3}
p(H_0|z)=\left(1+\frac{1}{H_{\rm{u}}} \int_0^{H_{\rm{u}}} \frac{1}{1+H'}\,p^{-H'/(1+H')}\,{\rm{d}}H'\right)^{-1} \, ,
\end{equation}
where a uniform prior for the mode height in the range $[0,H_{\rm{u}}]$ has been considered. Figure \ref{hypothesis} displays the behavior of $p(H_1|z)$, i.e., $1\!-\!p(H_0|z)$, as a function of signal-to-noise ratio and detection significance $p$, for modes both with known and with unknown height. One immediately notices that, for a given $p$, assuming an unknown mode height intuitively causes the probability of detecting the mode to decrease for low to moderate SNR. This effect is reversed at high SNR, when $p(H_1|z)$ enters a downward trend, such trend being more accentuated for a mode with known height. This downward trend is contrary to our intuition and can be easily explained for the case of a mode with known height. In fact, our prior assumption of a high SNR is incorrect, viz., it does not find support in the data. As a consequence, $p(H_1|z)$ is penalized. Finally, the likelihood of wrongly rejecting $H_0$ (type I error) is considerably higher than that suggested by the value of $p$ (recall discussion above), by a factor that progressively increases as the detection significance increases (i.e., as the magnitude of $p$ decreases).

\figuremacroW{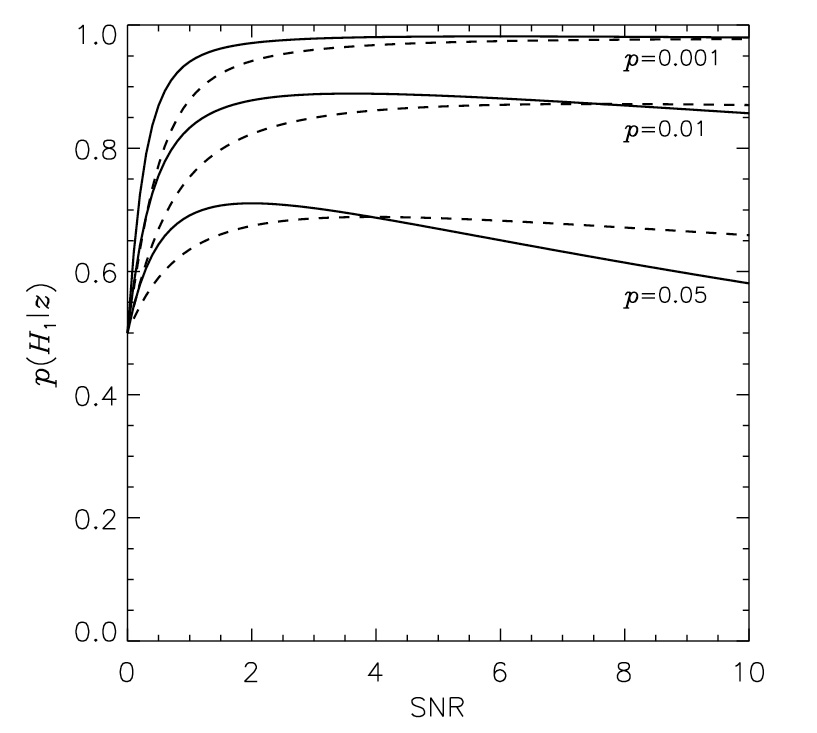}{Posterior probability of detecting a stochastically-excited long-lived mode}{The posterior probability of $H_1$, i.e., $1\!-\!p(H_0|z)$, is shown as a function of signal-to-noise ratio and detection significance $p$, for modes both with known (via Eq.~\ref{postH0_2}; solid lines) and with unknown height (via Eq.~\ref{postH0_3}; dashed lines).}{0.8}

\citet{Broomhall2010} provide a comparison of frequentist and Bayesian approaches in the search for low-frequency p modes and g modes in Sun-as-a-star data. A Bayesian approach has also been used by \citet{Deheuvels10} who, based upon posterior probability estimates, made decisions concerning the presence of $l\!=\!3$ modes and mixed modes in the power spectrum of the \emph{CoRoT} target HD~49385.

\subsection{Parameter estimation}\label{parest}

\subsubsection{Modeling the power spectrum}\label{modelPS}
The very first step in such a parameter estimation problem consists in defining an appropriate model of the limit power spectrum. The model will be denoted by $\mathscr{P}(\nu;{\boldsymbol{\lambda}})$, being described by a set of parameters ${\boldsymbol{\lambda}}$ which contain the desired information on the physical processes at play. As I have already noted, the power spectrum is distributed around this limit (or mean) spectrum with an exponential distribution (see Sects.~\ref{pspec} and \ref{PSstat}, as well as Eq.~\ref{pdfspec} below).

Neglecting any asymmetries in the mode profiles, we have seen that the mean spectrum of a single stochastically-excited mode follows a standard Lorentzian profile near the resonance (cf.~Eq.~\ref{lorentzprof}):
\begin{equation}
\mathscr{M}(\nu;H,\nu_0,\Gamma)=\frac{H}{1+\left[\frac{2(\nu-\nu_0)}{\Gamma}\right]^2} \, ,
\end{equation} 
where $H$ is the mode height, $\nu_0$ is the mode frequency, and $\Gamma$ is the mode linewidth. Assuming energy equipartition between multiplet components with different azimuthal order, one may define the following overall profile for a $(n,l)$ multiplet:
\begin{equation}
\mathscr{M}_{nl}(\nu;H_{nl},\nu_{nlm},\Gamma_{nlm},i)=\sum_{m=-l}^l \mathscr{E}_{lm}(i)\,\mathscr{M}(\nu;H_{nl},\nu_{nlm},\Gamma_{nlm}) \, ,
\end{equation} 
where the $\mathscr{E}_{lm}(i)$ coefficients are given by Eq.~(\ref{mheights}). We have seen in Sect.~\ref{sect:rot} how departures from spherical symmetry, particularly when caused by rotation, will lift the degeneracy of the mode frequency $\nu_{nl}$.  

We are primarily interested in performing a so-called global fit \citep[e.g.,][]{Appourchaux08} to the observed power spectrum, whereby several radial orders are fitted simultaneously within a broad frequency range{\footnote{Conversely, pseudo-global (or local) fitting \citep[e.g.,][]{localfit} is an approach traditionally adopted for Sun-as-a-star data, whereby narrow frequency windows are considered at a time.}}. Therefore, one ends up modeling the limit acoustic power spectrum according to the following general relation:
\begin{equation}
\label{modelspec}
\mathscr{P}(\nu;{\boldsymbol{\lambda}})=\sum_{n}\sum_{l}\sum_{m=-l}^{l}\,\frac{\mathscr{E}_{lm}(i)\,H_{nl}}{1+\left[\frac{2(\nu-\nu_{nlm})}{\Gamma_{nlm}}\right]^2} + S_{\rm{bg}}(\nu) + B \, ,
\end{equation}
where $S_{\rm{bg}}(\nu)$ has been introduced to describe the background signal (of both instrumental and stellar origin), while $B$ is used to represent the photon shot noise. Depending on the model used, different assumptions are possible that concern the mode and background parameters, which may be extremely useful in order to reduce the dimension of parameter space \citep[e.g.,][]{HandCamp}. 

The stellar background signal results from the superposition of several components related to activity, different scales of granulation, and even faculae. It may be modeled as \citep[e.g.,][]{Harvey85,Harvey93,Aigrain94,faculaeKaroff}:
\begin{equation}
\label{harveyprof}
S_{\rm{bg}}(\nu)=\sum_{j} \frac{H_j}{1+(2\pi\nu\tau_j)^{a_j}} \, ,
\end{equation}    
where $H_j$ is the height in the power spectrum at $\nu\!=\!0$, $\tau_j$ is the characteristic time of the decaying autocorrelation function of the process, and $a_j$ determines the slope of this so-called Harvey-like profile. Such functional form is representative of a random non-harmonic field whose autocorrelation decays exponentially with time. The original model by \citet{Harvey85}, having $a_j\!=\!2$, failed to reproduce the observed solar background signal above the acoustic cut-off frequency. \citet{Harvey93} refined the original model by letting $a_j$ be a free parameter and thus solved the misfit at high frequencies. The exponent $a_j$ calibrates the amount of memory in the process and there is no physical reason why its value should be fixed or even the same for different components. Furthermore, \citet{faculaeKaroff} argues that the slope of a given component should be different at low and high frequencies based on physical arguments. The instrumental background signal, being also a $1/f$-noise process, may just as well be incorporated in Eq.~(\ref{harveyprof}). 

An important remark should be made at this stage on how we model the mode heights. Once again assuming energy equipartition between multiplet components with different azimuthal order, their heights may be expressed as:
\begin{equation}
\label{heightsnrad}
H_{nlm}=\mathscr{E}_{lm}(i)\,H_{nl}=\mathscr{E}_{lm}(i)\,S_l^2 \,\alpha_{nl} \, ,
\end{equation}
where $S_l$ is the spatial response function introduced in Sect.~\ref{sect:spat_filt}, and the factor $\alpha_{nl}\!\approx\!\alpha(\nu_{nl})$ depends mainly on the frequency. This relation, however, is only strictly valid under one assumption: When the stellar flux is integrated over the full apparent disk, one must assume that the weighting function giving the contribution of a surface element to the integral is a function of the distance to the disk center alone. In that case, the apparent mode amplitude can effectively be separated into two factors: $\mathscr{E}_{lm}(i)$ and $S_l^2$. This assumption holds very well in the case of intensity measurements, since the weighting function is then mainly linked to the limb darkening, whereas for velocity measurements departures might be observed due to asymmetries in the velocity field induced by rotation \citep[e.g.,][]{Ballot2006}. As a means to reduce the number of parameters in our model, the heights of non-radial modes are commonly defined based on the heights of radial modes according to Eq.~(\ref{heightsnrad}) and taking into account the $S_l/S_0$ ratios as given in Table \ref{tab:spat_filt}.

The final model should be given by the convolution of the model in Eq.~(\ref{modelspec}) with the power spectral window normalized to unit total area \citep{Anderson90}, therefore taking into account the redistribution of power caused by gaps in the data. Inclusion of this last step may result in a computationally demanding implementation of the fitting problem, especially when using Monte Carlo techniques, and therefore should be well pondered.

\subsubsection{Maximum Likelihood Estimation}\label{SectMLE}
\begin{sidewaysfigure}
	\centering
	\includegraphics[width=\textheight]{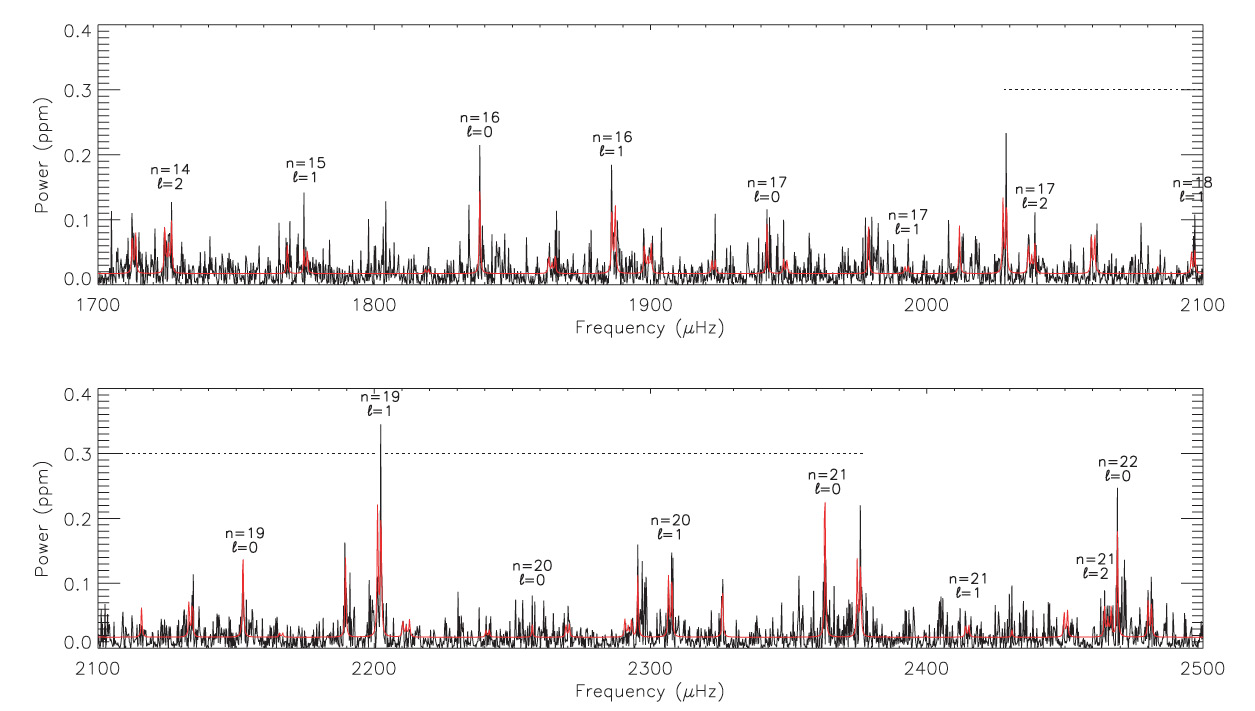}
	\caption[Fit to the \emph{WIRE} power spectrum of $\alpha$ Cen A using MLE]{\textbf{Fit to the \emph{WIRE} power spectrum of $\alpha$ Cen A using MLE} - The \emph{WIRE} spectrum is shown with the fitted model overlaid in red. Fitted modes are labeled with the radial order $n$ and degree $l$. The horizontal dotted lines, extending from the prominent peak at $\sim\!2200\:{\rm{\mu Hz}}$, indicate the spacing at which the associated sidelobes caused by the severe window function are placed. From \citet{Fletcher06}.}
	\label{fitMLE}
\end{sidewaysfigure}

In the case of a power spectrum of a solar-like oscillator, for which measurement errors are not normally distributed, one should employ the concept of Maximum Likelihood Estimation (MLE), thereby determining estimates of the parameters ${\boldsymbol\lambda}$ describing the model $\mathscr{P}(\nu;{\boldsymbol\lambda})$ that maximize the likelihood of the observed power spectrum. As an example, a fit to the power spectrum of $\alpha$ Cen A obtained with \emph{WIRE} \citep{Fletcher06} is displayed in Fig.~\ref{fitMLE}. It should be noted that this methodology is not equivalent to a traditional least-squares minimization problem, where it is implicitly assumed that the measurement errors are normally distributed.

The use of MLE for fitting solar power spectra was first mentioned by \citet{DuvallHarvey86}, being later applied and tested by \citet{Anderson90}. It has now been in use by helioseismologists for more than 20 years and has been applied in the analysis of low-degree modes as well as of medium- and high-degree modes \citep[e.g.,][]{Schou92,ToutainAppour94,Appourchaux98,solarflag}. Asteroseismologists, myself included, are now increasingly adopting this and other related approaches. As Sir Isaac Newton would say, we stand on the shoulders of giants.

I start by introducing the probability density function (pdf) at frequencies separated by $1/T$, $T$ being the length of the observation \citep{DuvallHarvey86}:
\begin{equation}
\label{pdfspec}
f(P_j;{\boldsymbol\lambda})=\frac{1}{\mathscr{P}(\nu_j;{\boldsymbol\lambda})} \exp\left[-\frac{P_j}{\mathscr{P}(\nu_j;{\boldsymbol\lambda})}\right] \, ,
\end{equation}   
where $P_j$ is the observed power spectrum at frequency channel $j$. This equation is nothing but the $\chi^2_2$ distribution scaled to the limit spectrum. As we have seen, an average power spectrum instead obeys a $\chi^2$ probability distribution with $2k$ degrees of freedom, $k$ being the number of combined spectra. In that case one has \citep{Appourchaux2003}:
\begin{equation}
f(P_j;{\boldsymbol\lambda})=\frac{k^{k-1}}{(k-1)!} \frac{P_j^{k-1}}{\mathscr{P}^k(\nu_j;{\boldsymbol\lambda})} \exp\left[-\frac{k P_j}{\mathscr{P}(\nu_j;{\boldsymbol\lambda})}\right] \, .
\end{equation}

The next step is to specify the likelihood function, i.e., the joint pdf for the data sample $\{P_j\}$. Assuming that the frequency bins are uncorrelated, the joint pdf is simply given by the product of $f(P_j;{\boldsymbol\lambda})$ over some frequency interval of interest spanned by $j$:
\begin{equation}
\label{eqlikelihood}
L({\boldsymbol\lambda})=\prod_{j=1}^{N} f(P_j;{\boldsymbol\lambda}) \, .
\end{equation}
In spite of Eq.~(\ref{eqlikelihood}) being valid for an uninterrupted data set, the same is not true when gaps are present in the time series. In that event, \citet{StahnGizon} have derived an expression for the joint pdf of solar-like oscillations in complex Fourier space, in agreement with the earlier work of \citet{Gabriel94}. The last-mentioned pdf explicitly takes into account frequency correlations introduced by the convolution with the spectral window.

The basic idea behind MLE is to determine estimates ${\boldsymbol{\tilde{\lambda}}}$ so as to maximize the likelihood function of the observed power spectrum. Due to improved numerical stability, however, it is more convenient to work with logarithmic probabilities. In practice, one ends up minimizing
\begin{eqnarray}
\label{eqloglikelihood}
\ell({\boldsymbol\lambda}) & \equiv & -\ln L({\boldsymbol\lambda}) \nonumber \\
& = & \sum_{j=1}^{N} \left\{\ln\mathscr{P}(\nu_j;{\boldsymbol\lambda}) + \frac{P_j}{\mathscr{P}(\nu_j;{\boldsymbol\lambda})}\right\} \, ,
\end{eqnarray}
where Eq.~(\ref{pdfspec}) has been used in conjunction with Eq.~(\ref{eqlikelihood}). Symbolically, one thus has:
\begin{equation}
\label{symbolicMLE}
{\boldsymbol{\tilde{\lambda}}}=\arg \min_{\boldsymbol\lambda}\left\{\ell({\boldsymbol\lambda})\right\} \, .
\end{equation} 
Finding the parameters ${\boldsymbol{\tilde{\lambda}}}$ that minimize $\ell({\boldsymbol\lambda})$ may in principle be accomplished by employing, for instance, Powell's algorithm \citep{Powell}. However, things are not that simple. Owing to the nonlinearity of the fitting problem in hand and to the large number of model parameters, one expects the likelihood to be multimodal. The risk of attaining a local minimum is real and further enhanced by the proneness of most standard algorithms to the initial guess parameters. Nevertheless, hope is not lost, since more sophisticated algorithms, such as Markov chain Monte Carlo techniques (see next section), are available.   

The method of MLE possesses a number of attractive asymptotic properties \citep[e.g.,][]{KendallStuart}. Accordingly, in the limit of a very large sample (i.e., $N\!\to\!\infty$), the estimator ${\boldsymbol{\tilde{\lambda}}}$ is unbiased:
\begin{equation}
\lim_{N\to\infty} {\rm{E}}\left[{\boldsymbol{\tilde{\lambda}}}\right]={\boldsymbol{\lambda_0}} \, ,
\end{equation}
where ${\boldsymbol{\lambda_0}}$ is the true parameter value. Moreover, the Cram\'er-Rao lower bound \citep{Rao,Cramer} is reached:
\begin{equation}
\lim_{N\to\infty} {\rm{cov}}\left[{\boldsymbol{\tilde{\lambda}}}\right]=\mathsf{I}^{-1}({\boldsymbol{\lambda_0}}) \, ,
\end{equation} 
where $\mathsf{I}({\boldsymbol{\lambda}})$ is the Fisher information matrix whose elements are given by
\begin{equation}
\mathsf{i}_{uv}={\rm{E}}\left[\frac{\partial^2\ell({\boldsymbol\lambda})}{\partial\lambda_u\partial\lambda_v}\right] \, .
\end{equation}
This means that no other asymptotically unbiased estimator has lower variance (minimum-variance unbiased estimator). Finally, the estimator ${\boldsymbol{\tilde{\lambda}}}$ is asymptotically normally distributed with mean ${\boldsymbol{\lambda_0}}$ and covariance matrix $\mathsf{I}^{-1}({\boldsymbol{\lambda_0}})$. 

Nonetheless, the world we live in is finite and hence there is no guarantee that the estimator ${\boldsymbol{\tilde{\lambda}}}$ will be normally distributed or even unbiased. Assuming that ${\boldsymbol{\tilde{\lambda}}}$ is approximately normally distributed, the covariance matrix is then given by the inverse of the Hessian matrix, $\mathsf{H}^{-1}({\boldsymbol{\lambda}})$, computed at ${\boldsymbol{\tilde{\lambda}}}$. The elements of $\mathsf{H}({\boldsymbol{\lambda}})$ are given by: 
\begin{equation}
\mathsf{h}_{uv}=\frac{\partial^2\ell({\boldsymbol\lambda})}{\partial\lambda_u\partial\lambda_v} \, .
\end{equation}
The so-called formal error bars on ${\boldsymbol{\tilde{\lambda}}}$ are given by the diagonal elements of the inverse of the Hessian matrix. Furthermore, the covariance matrix obeys the following inequality:
\begin{equation}
{\rm{cov}}\left[{\boldsymbol{\tilde{\lambda}}}\right]\!=\!\mathsf{H}^{-1}({\boldsymbol{\tilde{\lambda}}})\,\,\ge\,\,\mathsf{I}^{-1}({\boldsymbol{\lambda_0}}) \, ,
\end{equation} 
meaning that the precision to which we can estimate ${\boldsymbol{\tilde{\lambda}}}$ is fundamentally limited by the  Fisher information matrix. 

{\textbf{Monte Carlo simulations.}} The considerations in the previous paragraph suggest that one should run Monte Carlo simulations on synthetic data as a means of validating the method \citep[e.g.,][]{Anderson90,SchouBrown94,ToutainAppour94,Appourchaux98,GizonSolanki,Ballot2006,Ballot2008}. This is accomplished by fitting a large number of simulated realizations of a power spectrum in order to obtain the distributions of the fitted model parameters. Monte Carlo simulations enable us not only to determine the bias and precision associated with each parameter but also to assess the correlations between parameters. Besides, a matter of the utmost importance is the calibration of error bars. The procedure consists in calibrating the formal error bars such that the standard deviation of a given fitted parameter as obtained from Monte Carlo simulations equals the mean of the formal errors returned by the inverse Hessian, i.e., $\mathsf{H}({\boldsymbol{\tilde{\lambda}}})\approx\mathsf{I}({\boldsymbol{\lambda_0}})$.   

Having defined a theoretical Hessian, \citet{ToutainAppour94} derived the following analytical expression for the frequency precision of a single p mode:
\begin{eqnarray}
\label{sigmafreq2}
\sigma_{\nu}&=&f(\beta) \sqrt{\frac{\Gamma}{4\pi T}} \nonumber \\
&=&\sqrt{\sqrt{\beta+1}\left(\sqrt{\beta+1}+\sqrt{\beta}\right)^3} \sqrt{\frac{\Gamma}{4\pi T}} \, ,
\end{eqnarray} 
where $\beta$ is the noise-to-signal ratio as before, and $f(\beta)$ is close to unity when $\beta$ approaches 0. This is the same result as obtained in \citet{Libbrecht92}. Equation (\ref{sigmafreq2}) is to be compared with Eq.~(\ref{sigmafreq}) for a pure sine wave. Assuming white noise, $\sqrt{\beta}\!\sim\!\sigma/(A\sqrt{N})$ in the latter equation. Consequently, as the amplitude increases, the frequency precision of long-lived modes increases while that of short-lived modes reaches a limit. From Eq.~(\ref{sigmafreq2}) it becomes apparent that once a mode is resolved, and thus $\beta$ cannot be improved with time (recall Eq.~\ref{betaproperty}), frequency precision is expected to scale with the square root of time, i.e., $\sigma_\nu\!\propto\!T^{-1/2}$. This differs from the relation $\sigma_\nu\!\propto\!T^{-3/2}$ found in Eq.~(\ref{sigmafreq}). Figure \ref{precision} displays the behavior of the frequency precision achievable for a stochastically-excited p mode as the length of the observation is increased. A transition between the unresolved ($T\!\ll\!2\tau$) and resolved ($T\!\gg\!2\tau$) regimes can be clearly seen.  

\citet{ToutainAppour94} have also computed theoretical error bars for other parameters. To validate the approach, the theoretical precisions derived in their work were compared with those obtained from Monte Carlo simulations. Just like \citet{Anderson90} before them, the authors realized that the amplitude, the linewidth, and the background noise tend to have a log-normal distribution. They suggest fitting the natural logarithm of these parameters instead, in order to derive meaningful error bars from the inverse of the Hessian.  

\figuremacroW{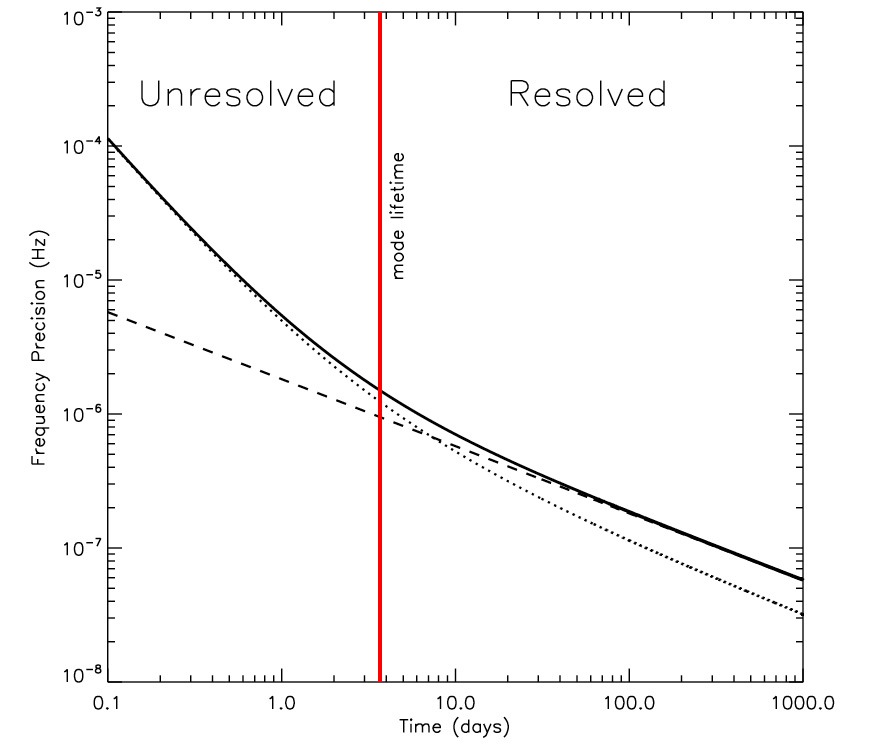}{Frequency precision of a stochastically-excited p mode as a function of the length of the observation}{Use has been made of Eq.~(\ref{sigmafreq2}) to compute the frequency precision. The definition of $\beta$ entering that equation depends on whether or not the mode is resolved and is taken from Eq.~(\ref{betaproperty}). The dotted and dashed curves respectively represent the behavior of frequency precision in the unresolved and resolved regimes. Moreover, a more general expression, covering these two extreme regimes as well as the intermediate regime, has been computed for the frequency precision via Eq.~(\ref{regimegeneral}) and is shown as a solid curve. The rms noise, $\sigma_0$, per $\Delta t\!=\!58.85\:{\rm{s}}$ integration, was computed according to the empirical minimal-term model for the noise presented in \citet{GillilandSC} assuming a \emph{Kepler} apparent magnitude of $Kp\!=\!10$. The mode has a typical solar linewidth of $1\:{\rm{\mu Hz}}$ and a rms amplitude $A_{\rm{rms}}\!=\!2.68\:{\rm{ppm}}$, which corresponds to the maximum solar rms velocity amplitude of $0.16\:{\rm{m\,s^{-1}}}$ according to Eq.~(\ref{ampconv}).}{0.95}

I would also like to highlight a series of very instructive articles that have devoted particular attention to the issue of fitting the inclination of the rotation axis, $i$, and the rotational splitting{\footnote{In the asymptotic regime one has $\nu_{\rm{s}}\!\simeq\!\langle \Omega\rangle_{nl}/(2\pi)$ in Eq.~(\ref{splitting}).}}, $\nu_{\rm{s}}$, of solar-like stars. Making use of extensive Monte Carlo simulations, \citet{GizonSolanki} estimated the precision of the measurement of $i$. They found that the inclination angle can be retrieved accurately when $i\!\gtrsim\!30^\circ$ for stars rotating at least twice as fast as the Sun that have been observed for 6 months, having further assumed solar linewidths. An extension of this analysis is provided by the same authors in \citet{GizonSolanki04}, where they conclude that information can be obtained about the latitudinal differential rotation from observations of dipole and quadrupole modes, depending on the value of the mean rotation and on the inclination of the rotation axis. \citet{Ballot2006,Ballot2008} emphasize the difficulties found when fitting these two quantities in the case of slow rotators, for which the splitting is comparable to the mode linewidth, thus leading to blending of the multiplet components. They explore in depth the correlations between the two parameters (see Fig.~\ref{incsplitcorr} for an example).

\figuremacroW{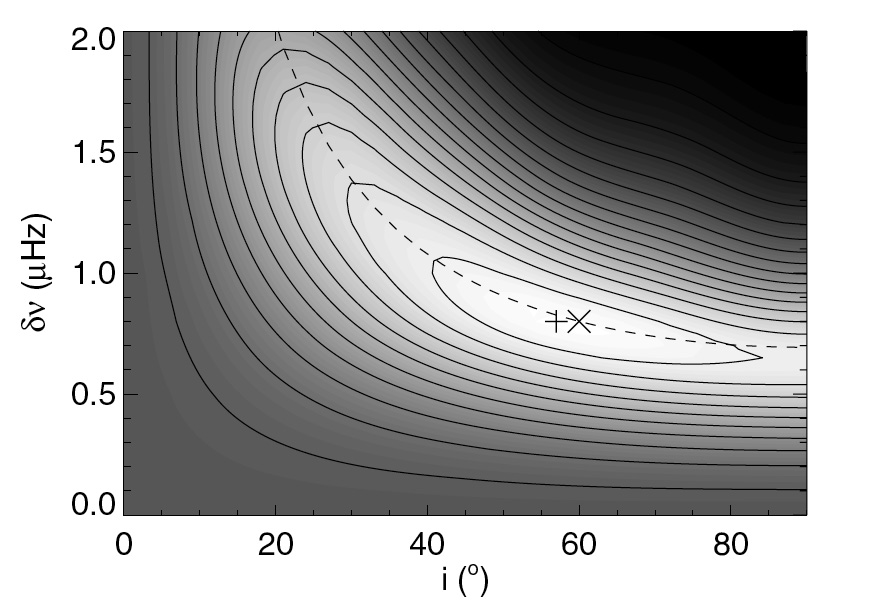}{Correlation between maximum likelihood estimates of the inclination $i$ and splitting $\nu_{\rm{s}}$}{The likelihood function for a simulated spectrum is shown in the plane $(i,\nu_{\rm{s}})$ of the space of parameters (notice that a different nomenclature is used for the splitting than the one in the text). Lighter tones correspond to higher likelihood and solid curves represent contours of constant likelihood. The cross ($\times$) marks the input pair $(i_0,\nu_{{\rm{s}}0})$ while the plus sign ($+$) marks the maximum of the likelihood. A clear correlation is seen between the two parameters whose values are organized along the dashed curve defined by $\nu_{\rm{s}}\sin i\!=\!\nu_{{\rm{s}}0}\sin i_0$. From \citet{Ballot2006}.}{1}

{\textbf{Parameter significance and the likelihood ratio.}} The discussion about MLE is not yet complete as one question still remains unanswered: How can one test the statistical significance of the fitted parameters? This can be done by employing the so-called likelihood ratio test \citep{Appourchaux98}. First, one starts by maximizing the likelihood $L({\boldsymbol{\lambda}}_p)$ of the observed spectrum, where $p$ parameters were used to construct the model. If one intends to model the same observed spectrum with $n$ additional parameters, then the likelihood $L({\boldsymbol{\lambda}}_{p+n})$ has to be maximized. The natural logarithm of the likelihood ratio $\mathscr{L}$ (cf.~Eq.~\ref{likeratio}) is then given by:
\begin{eqnarray}
\ln \mathscr{L} & = & \ln L({\boldsymbol{\lambda}}_p) - \ln L({\boldsymbol{\lambda}}_{p+n}) \nonumber \\
& = & \ell({\boldsymbol{\lambda}}_{p+n}) - \ell({\boldsymbol{\lambda}}_p) \, .
\end{eqnarray} 
Clearly, when $\mathscr{L}\!\ll\!1$ one has reasons to believe that the additional parameters are indeed significant. To assess how significant they are, an hypothesis test called likelihood ratio test may be performed. The null hypothesis is stated as: The $n$ additional parameters are not needed to model the observed spectrum. Taking comfort in the knowledge that the statistical test $-2\ln\mathscr{L}$ approximately follows a $\chi^2$ distribution with $n$ degrees of freedom under the null hypothesis \citep{Wilks38}, one then proceeds with the computation of the detection significance (recall Sect.~\ref{Freqhyptest} for its definition). 

\citet{faculaeKaroff} employs the likelihood ratio test to assess the statistical significance of a facular component in power spectra of the Sun, having made use of 13 years of observations carried out by the VIRGO/SPM instrument. Figure \ref{faculae} portrays the case of KIC~6603624, a main-sequence star exhibiting solar-like oscillations, for which the statistical significance of a facular component is over the $99.9\%$ level according to the likelihood ratio test. 

A rather interesting work on the reliability of the likelihood ratio when applied to mode identification (i.e., tagging of modes by degree $l$) in solar-like stars is presented in \citet{Salabertlikeratio}. It was also based on the likelihood ratio that \citet{Appourchaux08} computed the posterior probabilities of the two competing mode identification scenarios in a study of the \emph{CoRoT} F5 main-sequence star HD~49933. Their calculations were done in a way similar to the procedure outlined in Sect.~\ref{Bayeshyptest}, hence providing the statistical significance of either scenario. Mode identification was made difficult due to the conjugation of two factors, namely, a mode linewidth larger than solar together with a small frequency separation narrower than in the Sun. As a result, the $l\!=\!0,2$ and $l\!=\!1$ ridges were indistinguishable from one another. This has been coined the ``F star problem'' \citep{BeddingWS}, which is also a problem in, e.g., Procyon \citep{BeddingProcyon,scaledechelle} and $\theta$ Cyg (see Fig.~\ref{thetacyg}). Notoriously, the preferred mode identification of \citet{Appourchaux08} was proved to be wrong as soon as a longer time series became available for analysis \citep[see Sect.~\ref{secHD49933};][]{Benomar09}. To be fair, it should be stressed that \citet{Appourchaux08} were already envisaging the inclusion of prior knowledge through a full Bayesian approach as the next logical step to take. This would have saved the authors from all the controversy \citep{BenAppourBaud}. Bayesian parameter estimation, including the analogous problem of model comparison, are the subjects of the next section.     

\figuremacroW{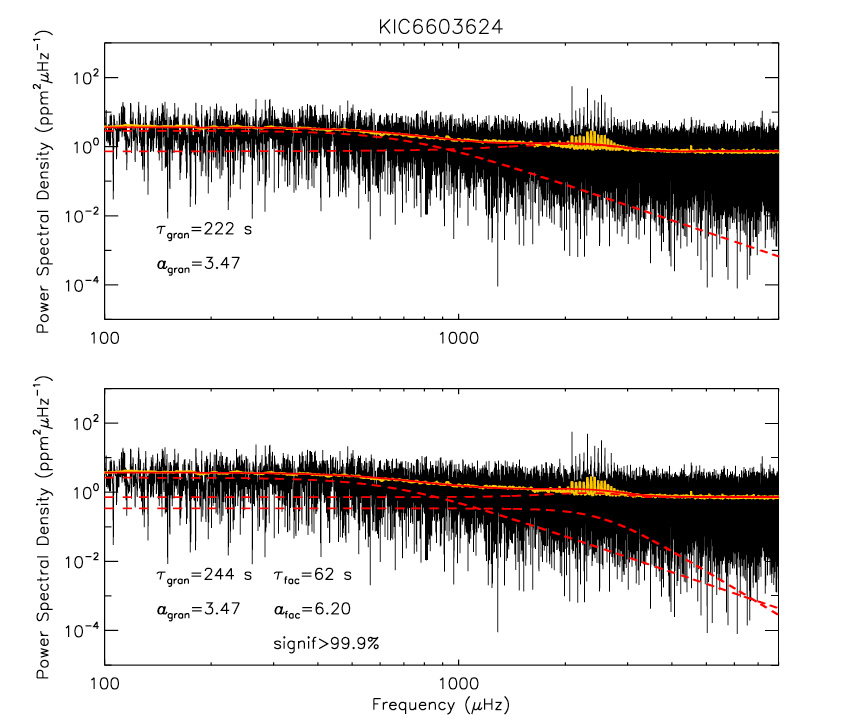}{Presence of a facular signal in the power spectrum of KIC~6603624}{The power spectrum (black) has been computed from twelve months of \emph{Kepler} photometry (Q5 to Q8). A smoothed version of the power spectrum (yellow) is also shown in order to visually enhance the acoustic signal centered at about $2.5\:{\rm{mHz}}$. Besides a Gaussian component used to describe the p-mode power-excess hump \citep[e.g.,][]{KallingerRGCorot} and a flat component representing white noise, the background model also included both a granular and a facular signal (bottom panel) or else only a granular signal (top panel). The characteristic timescales ($\tau_{\rm{gran}}$ and $\tau_{\rm{fac}}$) and slopes ($a_{\rm{gran}}$ and $a_{\rm{fac}}$) of the corresponding Harvey-like profiles are indicated (cf.~Eq.~\ref{harveyprof}). The latter have been fixed to their solar values according to \citet{faculaeKaroff}.}{1.05}

\figuremacroW{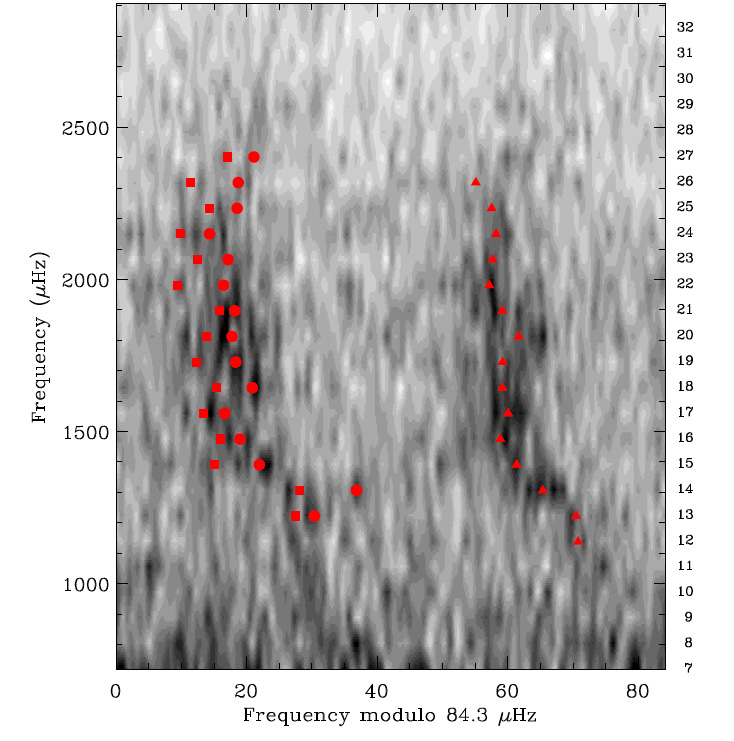}{\'Echelle diagram of the power density spectrum of $\theta$ Cyg}{$\theta$ Cyg is a very bright F-type dwarf. It has been observed with \emph{Kepler} throughout Quarter 8 (Q8) using a special photometric mask. The resulting power density spectrum is shown in grayscale. Symbols in red represent the individual frequencies obtained from a provisional fit to the power spectrum using a regularized version of the MLE algorithm (or MAP; see discussion in the next section). Symbol shapes indicate mode degree: $l\!=\!0$ (circles), $l\!=\!1$ (triangles), and $l\!=\!2$ (squares). Identification of the ridges from a simple visual inspection is far from evident. Another fit was performed assuming the alternative mode identification scenario. The plotted scenario was found to be statistically more likely based on the computation of the likelihood ratio, although not in a decisive way (only by a factor of $\sim\!3.7$). At the time of writing a debate is still ongoing as to which is the correct scenario.}{0.95}

\subsubsection{Bayesian parameter estimation using MCMC}
The first attempt at applying a Bayesian approach to parameter estimation in asteroseismology is due to \citet{Brewer07}. However, their use of a time-based model consisting of pure sine waves incorrectly models the stochastic excitation of an harmonic oscillator. They have since improved their approach by considering damped oscillations and by modeling the asteroseismic data as a Gaussian process \citep{Brewer09}. It would then be the controversy surrounding the early work of \citet{Appourchaux08} on HD~49933 to trigger the development of several Bayesian data analysis tools \citep{BenAppourBaud,Gruberbauer09,HandCamp}.

Herein, I highlight the main features of a Bayesian peak-bagging{\footnote{The term ``peak-bagging'', coined by Jesper Schou, a keen mountain climber, refers in the present context to the analysis of individual oscillation peaks in the power spectrum.}} tool that employs Markov chain Monte Carlo (MCMC) techniques. The following discussion is adapted from \citet{HandCamp}, where the reader will find a comprehensive guide to the implementation of such a peak-bagging tool. This is unquestionably one of my main contributions to the field and, for that reason, the aforementioned article can be found as a supplement in Appendix \ref{handcamp}. 

This tool is to be applied to the power spectra of solar-like oscillators and used as a means to infer both individual oscillation mode parameters and parameters describing non-resonant features. Besides making it possible to incorporate relevant prior information through Bayes' theorem, this tool also allows obtaining the marginal probability density function for each of the model parameters. Moreover, it provides larger error bars on the parameters than does MLE, thus making it a more conservative approach. \citet{HandCamp} apply this tool to a couple of recent asteroseismic data sets, namely, to \emph{CoRoT} observations of HD~49933 and to ground-based observations of Procyon. 

Having set up the model of the power spectrum in Sect.~\ref{modelPS}, I will start by introducing the Bayesian statistical framework to be used for estimating the model parameters and for comparing the merits of competing models. In the meantime, the reader should recall Eqs.~(\ref{BayesTheorem}) and (\ref{GlobalLikelihood}), where Bayes' theorem was formally introduced and a posterior probability $p(H_i|D,I)$ was assigned to each member of the hypothesis space of interest, denoted by $\{H_i\}$. 

{\textbf{Parameter estimation.}} Very often, a particular hypothesis, i.e., a given model $M$ of the power spectrum, is assumed true and the hypothesis space of interest then concerns the values taken by the model parameters ${\boldsymbol{\lambda}}$. These parameters are continuous, meaning that one will be interested in their probability density functions. The global likelihood of model $M$ is then given by the continuous counterpart of Eq.~(\ref{GlobalLikelihood}):
\begin{equation}
\label{globallike}
p(D|I)=\int p({\boldsymbol{\lambda}}|I) \,p(D|{\boldsymbol{\lambda}},I) \,{\rm{d}}{\boldsymbol{\lambda}} \, .
\end{equation} 
Computation of the global likelihood is rather complex, but can be achieved rather straightforwardly by using the Metropolis-Hastings algorithm under parallel tempering \citep[e.g.,][]{GregoryBook,HandCamp}. 

We may also want to restate Bayes' theorem in order to account for this new formalism:
\begin{equation}
\label{BayesTheorem2}
p({\boldsymbol{\lambda}}|D,I)=\frac{p({\boldsymbol{\lambda}}|I)\,p(D|{\boldsymbol{\lambda}},I)}{p(D|I)} \, .
\end{equation}
The terms entering this equation have exactly the same meaning as the corresponding terms entering Eq.~(\ref{BayesTheorem}). As a result, $p(D|{\boldsymbol{\lambda}},I)$ is nothing but the likelihood function $L({\boldsymbol{\lambda}})$ as given in Eq.~(\ref{eqlikelihood}). We are ultimately interested in using MCMC techniques to map the posterior pdf given by Eq.~(\ref{BayesTheorem2}). This approach is exceedingly more powerful than the Bayesian point-estimation method of Maximum A Posteriori \citep[MAP; e.g.,][]{MAP}. The MAP approach can in fact be regarded as a regularized version of the frequentist MLE, also a point-estimation method. Drawing a parallelism with Eq.~(\ref{symbolicMLE}), one has symbolically:
\begin{equation}
\label{symbolicMAP}
{\boldsymbol{\tilde{\lambda}}}_{\rm{MAP}}=\arg \max_{\boldsymbol\lambda}\left\{p({\boldsymbol{\lambda}}|I)\,L({\boldsymbol{\lambda}})\right\} \, .
\end{equation} 

One of the main advantages of a Bayesian approach when compared to a frequentist approach resides in the fact that the posterior pdf, $p({\boldsymbol{\lambda}}|D,I)$, is directly accessed, and not only the likelihood function. Moreover, the use of MCMC techniques to map the posterior pdf clearly supersedes the MLE approach, which only provides the location of the maximum of the likelihood. Another main advantage provided by the Bayesian framework is the ability to incorporate relevant prior information through Bayes' theorem and evaluate its effect on our conclusions. When basing the fitting/detection problem upon a priori theoretical knowledge, its outcome will effectively be restricted to what one can imagine or conceive \citep{AppourBayes08}. This is particularly useful when fitting a model to an acoustic power spectrum exhibiting low SNR in the p modes, a scenario where frequentist approaches tend to break down. \citet{GregoryBook} provides useful insight on the use and effect of the prior in the framework of Bayesian data analysis.

The procedure of marginalization makes it possible to compute the marginal posterior pdf for a subset of parameters ${\boldsymbol{\lambda}}_{\rm{A}}$ by integrating over the remaining parameters ${\boldsymbol{\lambda}}_{\rm{B}}$, the so-called nuisance parameters:
\begin{equation}
\label{marginalization}
p({\boldsymbol{\lambda}}_{\rm{A}}|D,I)=\int p({\boldsymbol{\lambda}}_{\rm{A}},{\boldsymbol{\lambda}}_{\rm{B}}|D,I)\,{\rm{d}}{\boldsymbol{\lambda}}_{\rm{B}} \, .
\end{equation}   
Furthermore, assuming that the prior on ${\boldsymbol{\lambda}}_{\rm{A}}$ is independent of the prior on ${\boldsymbol{\lambda}}_{\rm{B}}$, then by applying the product rule one has:
\begin{equation}
p({\boldsymbol{\lambda}}_{\rm{A}},{\boldsymbol{\lambda}}_{\rm{B}}|I)=p({\boldsymbol{\lambda}}_{\rm{A}}|I)\,p({\boldsymbol{\lambda}}_{\rm{B}}|{\boldsymbol{\lambda}}_{\rm{A}},I)=p({\boldsymbol{\lambda}}_{\rm{A}}|I)\,p({\boldsymbol{\lambda}}_{\rm{B}}|I) \, .
\end{equation}

In practice, one will be working with logarithmic probabilities. The global likelihood of the model plays the role of a normalization constant and we rewrite Eq.~(\ref{BayesTheorem2}) as follows:
\begin{equation}
\ln p({\boldsymbol{\lambda}}|D,I)={\rm{const.}}+\ln p({\boldsymbol{\lambda}}|I)-\ell({\boldsymbol{\lambda}}) \, ,
\end{equation}
where $\ell({\boldsymbol{\lambda}})$ is given by Eq.~(\ref{eqloglikelihood}).

{\textbf{Model comparison.}} As we will see below, the problem of model comparison is analogous to that of parameter estimation. When facing a situation in which several parameterized models are available for describing the same physical phenomenon, one expects Bayes' theorem to allow for a statistical comparison between such models. Bayesian model comparison has a built-in Occam's razor, a principle also known as lex parsimoniae, by which a complex model is automatically penalized, unless the available data justify its additional complexity. Notice that these competing models may be either intrinsically different models or else similar but with varying number of parameters (i.e., nested models), or even the same model with different priors affecting its parameters. 

Given two or more competing models and our prior information, $I$, being in the current context that one and only one of the models is true, we can assign individual probabilities similarly to what has been done in Eq.~(\ref{BayesTheorem}), after replacing $H_i$ by $M_i$:
\begin{equation}
\label{BayesTheorem3}
p(M_i|D,I) = \frac{p(M_i|I)\,p(D|M_i,I)}{p(D|I)} \, ,
\end{equation}
where the global likelihood of model $M_i$, $p(D|M_i,I)$, also called the evidence of the model, is given by Eq.~(\ref{globallike}). As stated above, after comparison of Eqs.~(\ref{BayesTheorem2}) and (\ref{BayesTheorem3}), one realizes that the problem of model comparison is indeed analogous to the problem of parameter estimation. 

Of particular interest is the computation of the ratio of the probabilities of two competing models:
\begin{equation}
\label{oddsratio}
O_{ij}\equiv\frac{p(M_i|D,I)}{p(M_j|D,I)}=\frac{p(M_i|I)\,p(D|M_i,I)}{p(M_j|I)\,p(D|M_j,I)}=\frac{p(M_i|I)}{p(M_j|I)}B_{ij} \, ,
\end{equation}   
where $O_{ij}$ is the odds ratio in favor of model $M_i$ over model $M_j$, $B_{ij}$ is the so-called Bayes' factor, and $p(M_i|I)/p(M_j|I)$ is the prior odds ratio. The Bayesian odds ratio closely resembles the frequentist likelihood ratio discussed in the previous section. However, the former ratio is the product of the ratio of the prior probabilities of the models and the ratio of their global likelihoods, in contrast to the ratio of point-like probability estimates in the latter case. Computation of $O_{ij}$ is thus a means of assessing, for instance, which of two mode identification scenarios is statistically more likely \citep[e.g.,][]{BenAppourBaud,BeddingProcyon,HandCamp}. It is usual to assume that one has no prior information impelling us to prefer one model over the other, and in that case one sets $p(M_i|I)/p(M_j|I)\!=\!1$. This assumption should, however, be used with great care, especially when dealing with nested models which is often the case \citep[e.g.,][]{ScottBerger}. One is now in need of a scale by which to judge the ratio of the evidences of two competing models. The usual scale employed is Jeffreys' scale \citep{Jeffreys61}, which is displayed in Table \ref{scaleJeffreys} for convenience.

\begin{table}[!t]
	\centering
	\caption[Jeffreys' scale]{\textbf{Jeffreys' scale} - The usual scale by which to judge the evidences of two competing models.}
	\begin{tabular}{ll}
	\hline\hline
	$\ln B_{ij}$ & Strength of evidence \\
	\hline
			$<1$ & Not worth more than a bare mention \\
			1--2.5 & Significant \\
			2.5--5 & Strong to very strong \\
			$>5$ & Decisive \\
	\hline
	\end{tabular}
	\label{scaleJeffreys}
\end{table}

Moreover, the Bayesian framework makes it possible to extract parameter constraints even in the presence of model uncertainty, i.e., when the implementation of model selection has not been successful. This is done by simply combining the probability distribution of the parameters within each individual model, weighted by the model probability. This procedure, called Bayesian model averaging \citep[e.g.,][]{Liddle09}, is an analog of the superposition of eigenstates of an observable in quantum mechanics.

{\textbf{Markov chain Monte Carlo.}} After inspection of Eq.~(\ref{marginalization}), the need becomes clear for a mathematical tool that is able to efficiently evaluate the multidimensional integrals required in the computation of the marginal distributions. This constitutes the rationale behind the method known as Markov chain Monte Carlo, first introduced in the early 1950s by statistical physicists and nowadays extensively used in all areas of science and economics.      

The aim is to draw samples from the target distribution, $p({\boldsymbol{\lambda}}|D,I)$, by constructing a pseudo-random walk in parameter space such that the number of samples drawn from a particular region is proportional to its posterior density. Such a pseudo-random walk is attained by generating a Markov chain, whereby a new sample, ${\boldsymbol{\lambda}}_{t+1}$, depends on the previous sample, ${\boldsymbol{\lambda}}_{t}$, according to a time-independent quantity called the transition kernel, $p({\boldsymbol{\lambda}}_{t+1}|{\boldsymbol{\lambda}}_{t})$. After a burn-in phase, $p({\boldsymbol{\lambda}}_{t+1}|{\boldsymbol{\lambda}}_{t})$ is able to generate samples of ${\boldsymbol{\lambda}}$ with a probability density converging on the target distribution.

An algorithm widely employed to generate a Markov chain was initially proposed by \citet{Metropolis}, and subsequently generalized by \citet{Hastings}, this latter version being commonly referred to as the Metropolis-Hastings algorithm. It works in the following way: Suppose the current sample, at some instant denoted by $t$, is represented by ${\boldsymbol{\lambda}}_t$. We would like to steer the Markov chain toward the next sampling state, ${\boldsymbol{\lambda}}_{t+1}$, by first proposing a new sample to be drawn, ${\boldsymbol{\xi}}$, from a proposal distribution, $q({\boldsymbol{\xi}}|{\boldsymbol{\lambda}}_t)$, that can have almost any form. Here, I specifically treat $q({\boldsymbol{\xi}}|{\boldsymbol{\lambda}}_t)$ as being a multivariate normal distribution with covariance matrix $\mathsf{\Sigma}$. The proposal distributions for the individual parameters are further assumed to be independent, meaning that $\mathsf{\Sigma}$ is diagonal. The proposed sample is then accepted with a probability given by:
\begin{equation}
\alpha({\boldsymbol{\lambda}}_t,{\boldsymbol{\xi}}) = \min(1,r) = \min\left[1,\frac{p({\boldsymbol{\xi}}|D,I)}{p({\boldsymbol{\lambda}}_t|D,I)} \frac{q({\boldsymbol{\lambda}}_t|{\boldsymbol{\xi}})}{q({\boldsymbol{\xi}}|{\boldsymbol{\lambda}}_t)}\right] \, ,
\end{equation}
where $\alpha({\boldsymbol{\lambda}}_t,{\boldsymbol{\xi}})$ is the acceptance probability and $r$ is called the Metropolis ratio. In the present case $q({\boldsymbol{\lambda}}_t|{\boldsymbol{\xi}})\!=\!q({\boldsymbol{\xi}}|{\boldsymbol{\lambda}}_t)$, since the proposal distribution is symmetric. As a result, if the posterior density for the proposed sample is greater than or equal to that of the current sample, i.e., $p({\boldsymbol{\xi}}|D,I)\!\geq\!p({\boldsymbol{\lambda}}_t|D,I)$, then the proposal will be accepted, otherwise it will be accepted with a probability given by the ratio of the posterior densities. If ${\boldsymbol{\xi}}$ is not accepted, then the chain will keep the current sampling state, i.e., ${\boldsymbol{\lambda}}_{t+1}\!=\!{\boldsymbol{\lambda}}_t$. The procedure just described is repeated for a predefined number of iterations or, alternatively, for a number of iterations determined by a convergence test applied to the Markov chain. 

Once the posterior pdf, $p({\boldsymbol{\lambda}}|D,I)$, has been mapped, the procedure of marginalization becomes trivial. The marginal posterior distribution of a given parameter $\lambda$, $p(\lambda|D,I)$, is then simply obtained by collecting its samples in an histogram and further normalizing it. An estimate of the $k$th moment of $\lambda$ about the origin is then given by
\begin{equation}
\langle \lambda^k\rangle \equiv \int \lambda^k \,p(\lambda|D,I) \,{\rm{d}}\lambda \approx \frac{1}{N_{\rm{it}}} \sum_{t} \lambda_{t}^{k} \, , 
\end{equation}
where $N_{\rm{it}}$ is the total number of samples.

{\textbf{Automated MCMC and parallel tempering.}} The basic Metropolis-Hastings algorithm outlined above can be refined by incorporating a statistical control system that allows to automatically fine-tune the proposal distribution during the burn-in phase. Moreover, inclusion of parallel tempering will increase the mixing properties of the Markov chain and consequently reduce the risk of the algorithm becoming stuck in a local mode of the target distribution. The reader is referred to \citet{HandCamp} for how to implement these two features. Here, I only give a brief summary.

The basic Metropolis-Hastings algorithm runs a serious risk of becoming stuck in a local mode of the target distribution, thus failing to fully explore all regions of parameter space containing significant probability. A way of overcoming this difficulty is to employ parallel tempering \citep[e.g.,][]{EarlDeem}, whereby a discrete set of progressively flatter versions of the target distribution is created by introducing a temperature parameter, $\mathcal{T}$. In practice, use is made of its reciprocal, $\gamma\!=\!1/\mathcal{T}$, referred to as the tempering parameter. We modify Eq.~(\ref{BayesTheorem2}) to generate the tempered distributions as follows:
\begin{equation}
p({\boldsymbol{\lambda}}|D,\gamma,I)=C\,p({\boldsymbol{\lambda}}|I)\,p(D|{\boldsymbol{\lambda}},I)^\gamma \, , \:\: 0 < \gamma \leq 1 \, ,
\end{equation}
where $C$ is a constant. For $\gamma\!=\!1$, we retrieve the target distribution, also called the cold sampler, while for $\gamma\!<\!1$, the hotter distributions are effectively flatter versions of the target distribution. By running such a set of chains in parallel ($n_\gamma$ in total) and further allowing the swap of their respective parameter states, we enable the algorithm to sample the target distribution in a way that makes possible both the investigation of its overall features (low-$\gamma$ chains) and the examination of the fine details of a local mode (high-$\gamma$ chains). Figure \ref{MetroHast} provides a pseudocode version of the Metropolis-Hastings algorithm with the inclusion of parallel tempering.

Based on a statistical control system similar to the one described in \citet{GregoryBook}, we may automate the process of calibration of the Gaussian proposal $\sigma$ values, which specify the direction and step size in parameter space when proposing a new sample to be drawn. The optimal choice of $\{\sigma\}$ is closely related to the average rate at which proposed state changes are accepted, the so-called acceptance rate. The control system makes use of an error signal to steer the selection of the $\sigma$ values during the burn-in phase of a single parallel tempering MCMC run, acting independently on each of the tempered chains. The error signal is proportional to the difference between the current acceptance rate and a target acceptance rate. As soon as the error signal for each of the tempered chains is less than a measure of the statistical fluctuation expected for a zero-mean error, the control system is turned off and the algorithm switches to the standard parallel tempering MCMC.

\begin{figure}[!t]
         \centering
	\begin{framed}
	\begin{algorithmic}[1]
		\Procedure{Parallel Tempering Metropolis-Hastings}{}
			\State ${\boldsymbol{\lambda}}_{0,i} = {\boldsymbol{\lambda}}_0 \, , \; 1 \leq i \leq n_\gamma$
			\For{$t = 0,1,\ldots,N_{\rm{it}}-1$}
				\For{$i = 1,2,\ldots,n_\gamma$}
					\State Propose a new sample to be drawn from a
					\Statex \hspace{5em} proposal distribution: ${\boldsymbol{\xi}} \sim N({\boldsymbol{\lambda}}_{t,i};\mathsf{\Sigma}_i)$
					\State Compute the Metropolis ratio:
					\Statex \hspace{5em} $\ln r = \ln p({\boldsymbol{\xi}}|D,\gamma_i,I) - \ln p({\boldsymbol{\lambda}}_{t,i}|D,\gamma_i,I)$
					\State Sample a uniform random variable:
					\Statex \hspace{5em} $U_1\sim\textup{Uniform}(0,1)$
					\If{$\ln U_1 \leq \ln r$}
						\State ${\boldsymbol{\lambda}}_{t+1,i} = {\boldsymbol{\xi}}$
					\Else
						\State ${\boldsymbol{\lambda}}_{t+1,i} = {\boldsymbol{\lambda}}_{t,i}$
					\EndIf
				\EndFor
				\State $U_2 \sim\textup{Uniform}(0,1)$
				\If{$U_2 \leq 1/n_\textup{swap}$}
					\State Select random chain: 
					\Statex \hspace{5em} $i\sim \textup{UniformInt}(1,n_\gamma-1)$ 
					\State Compute $r_\textup{swap}$:
					\Statex \hspace{5em} $\ln r_\textup{swap} = \ln p({\boldsymbol{\lambda}}_{t,i+1}|D,\gamma_i,I) + \ln p({\boldsymbol{\lambda}}_{t,i}|D,\gamma_{i+1},I)$
					\Statex \hspace{6em} $- \ln p({\boldsymbol{\lambda}}_{t,i}|D,\gamma_i,I) - \ln p({\boldsymbol{\lambda}}_{t,i+1}|D,\gamma_{i+1},I)$
					\State $U_3 \sim\textup{Uniform}(0,1)$
					\If{$\ln U_3 \leq \ln r_\textup{swap}$}
						\State Swap parameter states of chains $i$ and $i+1$:
						\Statex \hspace{7em} ${\boldsymbol{\lambda}}_{t,i} \leftrightarrow {\boldsymbol{\lambda}}_{t,i+1}$
					\EndIf
				\EndIf
			\EndFor
			\State \Return ${\boldsymbol{\lambda}}_{t,i} \, , \; \forall t \, , \; i\!:\!\gamma_i\!=\!1$
		\EndProcedure
	\end{algorithmic}
	\end{framed}
	\caption[Metropolis-Hastings algorithm]{\textbf{Metropolis-Hastings algorithm} - A pseudocode version of the Metropolis-Hastings algorithm with the inclusion of parallel tempering is given. For a complete understanding of the implementation of parallel tempering (lines 14 to 22) I refer the reader to Appendix \ref{handcamp}.}
	\label{MetroHast}
\end{figure}

\section{Getting practical: a pipeline for \emph{Kepler}}\label{sectpipe}
In the past few years, considerable effort has been invested in making preparations for the mode parameter analysis of \emph{Kepler} data. This analysis involves the estimation of individual and global oscillation mode parameters, as well as estimation of parameters describing non-resonant signatures of convection and activity. An example of such an effort is the work conducted in the framework of the AsteroFLAG consortium \citep{asteroflag} which followed from the earlier work undertaken by the \emph{CoRoT} data analysis team \citep{DAT1,DAT2}. This naturally paved the way for the development of a number of automated pipelines to measure global asteroseismic parameters of solar-like oscillators \citep{Huberpipeline,Mosserpipeline,Roxburghpipeline,Campantepipeline,Hekkerpipeline,Karoffpipeline,Mathurpipeline,Vernerpipeline}. 

In order to fully characterize a star using asteroseismology, it is desirable to have accurate estimates of individual p-mode parameters. These include the frequencies, amplitudes, and lifetimes of a large number of modes for which the angular degree $l$ and radial order $n$ have been identified. However, this is only possible for data above a certain signal-to-noise level. Global asteroseismic parameters, indicative of the global structure, are on the other hand readily obtainable using automated analysis methods that can incorporate data with a lower SNR and for which a full peak-bagging analysis is not always possible. Furthermore, the automated nature of these pipelines is required if we are to efficiently exploit the plenitude of data made available by \emph{Kepler} on these targets. A thorough comparison of complementary analysis methods used to extract global asteroseismic parameters of main-sequence and subgiant solar-like oscillators is presented in \citet{Vernercomparison} (see Sect.~\ref{secpipecomp}). 

I have been personally involved in the development of two such automated pipelines, namely, the KAB and the AAU pipelines, as they are known internally. The KAB pipeline is fully described in \citet{Karoffpipeline}. It operates based on a novel algorithm for modeling and fitting the autocovariance of the power spectrum, which has in turn been developed by \citet{Campantepipeline}. This algorithm overcomes the problem of mode identification and thus suits the automated nature of the pipeline, since there is no longer the need to make subjective choices during the analysis process. The article describing this algorithm can be found as a supplement in Appendix \ref{campanteautocov}.

Regarding the AAU pipeline, no work has unfortunately been published describing its methodology. That is, however, what I intend to do herein. The AAU pipeline is an automated pipeline designed to measure global asteroseismic parameters of main-sequence and subgiant solar-like oscillators, which is accomplished by using exclusively the time-series data as input. This pipeline is based on a series of programming modules that were passed to me by William J.~Chaplin, hence explaining some of the similarities between the AAU and the Birmingham-Sheffield Hallam \citep{Hekkerpipeline} pipelines. The underlying methodology of the AAU pipeline is described below. The results obtained from the automated analysis -- carried out using the methods of nine independent research teams -- of 1948 main-sequence and subgiant \emph{Kepler} survey stars are presented in \citet{Vernercomparison}. The AAU and KAB pipelines have taken part in this collective effort.   

The AAU pipeline consists of a series of modules that aim at extracting the following information from the power spectra of the time-series data: 
\begin{enumerate}
\item Frequency range of the oscillations;
\item Parameterization of the stellar and instrumental background signals;
\item Average large frequency separation, $\Delta\nu$;
\item Maximum mode amplitude, $A_{\rm{max}}$;
\item Frequency of maximum amplitude, $\nu_{\rm{max}}$.
\end{enumerate}

\subsection{Range of oscillations}
We have seen in Sect.~\ref{sect:pasympt} that the quasi-regularity of the spectrum of high-order p modes constitutes one of the main signatures of the presence of solar-like oscillations. We thus look for a frequency range in the power spectrum in which peaks appear at nearly regular intervals. It should be noted that the assumption of quasi-regularity of the spectrum may be too strong in the case of subgiants due to the presence of g modes and mixed modes. Nevertheless, we start, from $100\:{\rm{\mu Hz}}$ up to the Nyquist frequency ($\sim\!8300\:{\rm{\mu Hz}}$ for short-cadence data), by partitioning the power spectrum into overlapping windows of variable width, $w$, which in turn depends on the value of the central frequency of the window, $\nu_{\rm{central}}$, with the successive $\nu_{\rm{central}}$ separated by $w/20$. The values of $\nu_{\rm{central}}$ are actually used as proxies for $\nu_{\rm{max}}$. Since the width of the p-mode hump roughly scales with $\nu_{\rm{max}}$ \citep[e.g.,][]{Stello07,MosserRG}, $w$ is defined as $w\!=\!(\nu_{\rm{central}}/\nu_{{\rm{max}},\odot})w_\odot$, with $\nu_{{\rm{max}},\odot}\!=\!3100\:{\rm{\mu Hz}}$ and $w_\odot\!=\!2000\:{\rm{\mu Hz}}$, the latter being the expected value were the Sun to be observed as a bright star with \emph{Kepler}.

The next step consists in computing the power spectrum of the power spectrum, PS$\otimes$PS, for each of the frequency windows. The presence of prominent features in the PS$\otimes$PS around the predicted values of $\Delta\nu/2$, $\Delta\nu/4$, and $\Delta\nu/6$ (the first, second, and third harmonics, respectively) is then examined. The predicted value of $\Delta\nu$ is computed according to the observed relation between $\Delta\nu$ and $\nu_{\rm{max}}$ presented in \citet{Stello09}, to be precise, $\Delta\nu\!\propto\!\nu_{\rm{central}}^{0.77}$. An hypothesis test is subsequently applied, whereby the presence of oscillations in a given window is established if the probability of the three above features being due to noise is less than 1\% (i.e., a confidence level of 99\% is required). The same considerations that allowed us to derive the statistics of the power spectrum in Sect.~\ref{PSstat} are also applicable here, thus meaning that the PS$\otimes$PS follows a $\chi_2^2$ distribution. Therefore, the probability that the peaks are due to noise is given by $(1-p)^{\zeta N}$, where $p$ is the probability that a random variable following a $\chi^2_6$ distribution is larger than the average normalized height of the three peaks in the PS$\otimes$PS, $\zeta$ is an empirical correction factor that accounts for the effect of oversampling \citep[$\zeta\!=\!3$ when oversampling by a factor of 10;][]{Gabriel02}, and $N$ is the number of independent bins in the frequency window being considered. Finally, the frequency range of the oscillations is determined based on the overall span of the windows with confirmed oscillations. 

In case the above strategy fails to detect oscillations, an alternative approach is used that no longer assumes $\Delta\nu$ to be constant with frequency, but instead takes into account its frequency dependency by slightly stretching (or compressing) the frequency axis of the power spectrum before computing the PS$\otimes$PS for each of the windows. This aims at producing a nearly regular pattern of peaks in the stretched (compressed) power spectrum in contrast with the original spectrum, consequently enhancing the SNR of the features associated with the large frequency separation in the PS$\otimes$PS. The details on how to implement this stretching can be found in \citet{Hekkerpipeline} and only a summary is provided here. In short, the stretched frequencies are given by:
\begin{equation}
\nu_{\rm{stretch}}=(\nu-\nu_{\rm{central}}) - j\,s_{\rm{max}} \left(\frac{\nu}{\nu_{\rm{central}}} - 1\right)^2 \, ,
\end{equation} 
where $j$ is an integer, and $s_{\rm{max}}$ is the maximum amount of stretching allowed. The optimal amount of stretching is then evaluated by looking for the value of $j$ that minimizes the probability of the features in the PS$\otimes$PS being due to noise.

\subsection{Background signal}
The model of the background signal is kept simple. We opt for a model merely containing a granulation component and white noise, and fit it to a power spectrum that has been previously smoothed. The smoothing implies Gaussian statistics and we therefore employ a nonlinear least-squares fit based on a gradient-expansion algorithm. The frequency range of the oscillations (see previous section) is excluded from the fitting window{\footnote{It should be noted, however, that a background fit is performed for all stars regardless of whether or not oscillations have been detected.}}. The fitting window starts at $100\:{\rm{\mu Hz}}$ to allow for the decay of any possible activity component, characterized by considerably longer timescales, and extends all the way up to the Nyquist frequency. The granulation component is represented by a Harvey-like profile (cf.~Eq.~\ref{harveyprof}) to which an offset $B$ is added containing mostly white noise, resulting in the following functional form of the model:
\begin{equation}
S_{\rm{bg}}(\nu)+B=\frac{H_{\rm{gran}}}{1+(2\pi\nu\tau_{\rm{gran}})^{a}}+B \, ,
\end{equation}     
where $a$ is left as a free parameter. An example of a fit to the background signal is displayed in the top panel of Fig.~\ref{pipeline}.

A careful choice of the initial guesses for each model parameter proves critical for the convergence of the fitting procedure. $H_{\rm{gran}}$ is initially set to one thousandth of the maximum power in the smoothed spectrum. The input for $\tau_{\rm{gran}}$ is computed on the assumption that it scales inversely with $\nu_{\rm{max}}$ \citep{Huberpipeline,KB11}. Since at this stage an estimate of $\nu_{\rm{max}}$ is not yet available, a proxy is used instead, namely, the midpoint of the frequency range of the oscillations. The exponent $a$ is initially set to 2 following the original model of \citet{Harvey85}. Finally, we choose as input for $B$ the mean power at high frequencies, well beyond the range of oscillations. 

The above might, however, not be enough to achieve convergence and hence a trap has been devised for the non-convergence of the fit. We start by randomly selecting an input parameter and make a random increment (or decrement) to it before performing the fit. This is repeated up to a maximum of a few hundred times until convergence is achieved. The fitting procedure is said to converge -- for a given set of input parameters -- when the relative decrease in $\chi^2$ is less than $1\!\times\!10^{-6}$ in one iteration of the procedure. Failure to converge means that no other pipeline modules will be run on that particular star. Finally, the standard deviations of the model parameters are used as errors.   

\figuremacroW{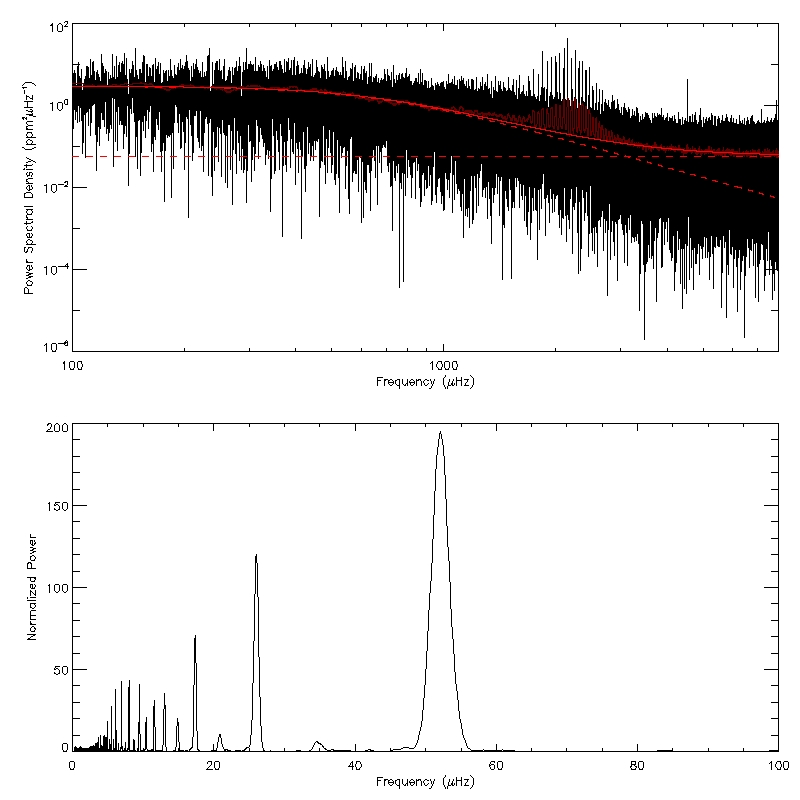}{Pipeline output from the analysis of the light curve of the bright G-type dwarf 16 Cyg A}{The solar analog 16 Cyg A is a member of a hierarchical triple system and has been observed with \emph{Kepler} throughout Quarter 7 (Q7) using a special photometric mask. The top panel displays the smoothed power density spectrum (in dark red), used in the fit to the background signal, atop the original spectrum (in black), on a log-log scale. The fit to the background signal (red solid line) and both its components (red dashed lines) are also shown (see text for details). The bottom panel displays the PS$\otimes$PS over the frequency range of the oscillations. The features at $\Delta\nu/2$ ($\sim\!52\:{\rm{\mu Hz}}$), $\Delta\nu/4$ ($\sim\!26\:{\rm{\mu Hz}}$) and $\Delta\nu/6$ ($\sim\!17\:{\rm{\mu Hz}}$) are conspicuous.}{1}

\subsection{$\Delta\nu$}
In order to estimate the average large frequency separation, $\Delta\nu$, we compute the PS$\otimes$PS over the frequency range of the oscillations (see bottom panel of Fig.~\ref{pipeline} for an example). Furthermore, the frequency dependence of $\Delta\nu$ is taken into account by computing the PS$\otimes$PS for a power spectrum that has been optimally stretched (compressed). The feature at $\Delta\nu/2$ (first harmonic) in the PS$\otimes$PS is then located and its power-weighted centroid computed to provide an estimate of $\Delta\nu$. The standard deviation of grouped data, given by $\sqrt{\left[\sum hx^2 - (\sum hx)^2/\sum h\right]\,/\,(\sum h - 1)}$, is also computed and used as the error on $\Delta\nu$, meaning that the feature in the PS$\otimes$PS is interpreted as an assembly of spectral heights ($h$) over a number of bins (with midpoint $x$).    

\subsection{$A_{\rm{max}}$ and $\nu_{\rm{max}}$}
We start by computing the power envelope for radial modes as a function of frequency according to \citet{Kjeldsen08}. First, we subtract the fit to the background signal from the power spectrum. The residual spectrum thus obtained is then heavily smoothed over the range occupied by the p modes by convolving it with a Gaussian having a full width at half maximum of $4\Delta\nu$. Finally, we multiply the smoothed, residual spectrum by $\Delta\nu/c$, where $c$, defined as $c\!=\!\sum (S_l/S_0)^2$ (cf.~Table \ref{tab:spat_filt}), is a factor{\footnote{The dependence of $c$ on limb darkening is ignored.}} measuring the effective number of modes per order and taken here to be 3.03. 

The frequency at which the power envelope attains its maximum value is used next as a proxy for $\nu_{\rm{max}}$ in the estimation of the maximum mode power. An estimate of the maximum mode power is computed by averaging the spectrum over a rectangular window of width $2\Delta\nu$ centered at this proxy and converting to power per radial mode as seen above. Computation of the associated uncertainty is then greatly simplified and is given by the standard deviation of the powers in the bins within the rectangular window. 

The amplitude envelope is obtained by taking the square root of the power envelope. This amplitude envelope is nothing but $A_{\rm{rms}}$ (see Eqs.~\ref{ampconv} and \ref{ampconv2}). Velocity amplitude envelopes are also displayed in Fig.~\ref{amplitudes} for a number of stars. Once again, notice that we are dealing with amplitudes scaled to be equivalent radial-mode amplitudes. It follows that the maximum mode amplitude, $A_{\rm{max}}$, is given by the square root of the estimated maximum mode power. Moreover, the fractional error on the maximum mode amplitude is half that on the maximum mode power.

Finally, a proper estimate of $\nu_{\rm{max}}$ is provided. We average the p-mode spectrum over adjacent (independent) rectangular windows of width $2\Delta\nu$ and convert to power per radial mode. An estimate of $\nu_{\rm{max}}$ is then given by the power-weighted centroid, with an associated uncertainty derived from the standard deviation of grouped data.

\chapter{Selected results}\label{Ch3}

\ifpdf
    \graphicspath{{3/figures/PNG/}{3/figures/PDF/}{3/figures/}}
\else
    \graphicspath{{3/figures/EPS/}{3/figures/}}
\fi

In the course of the two previous chapters, I have made reference to a series of published works that have benefited to a great extent from my contribution. For that reason, I have decided to compile them in Appendices \ref{vichinature} to \ref{campanteautocov}. In the current chapter, I present the reader with a series of additional published results to which I have also contributed, although in a somewhat less determinant (but still significant) way. These scientific results are divided into three categories according to the origin of the acquired data, namely, arising from radial-velocity measurements taken during ground-based campaigns, from \emph{CoRoT} photometry, or from \emph{Kepler} photometry. 

The above does not enclose my contribution to the field since, during the past three years, I have been involved (and continue to be) in a panoply of additional projects that have meanwhile been published or still await publication. However, I feel that the compendium of scientific results presented in this dissertation is fully representative of my work and technical expertise.

\section{Results from ground-based campaigns}
\subsection{An asteroseismic and interferometric study of the solar twin 18 Scorpii}\label{sec18Sco}
Solar twins, defined as having fundamental physical properties very similar or identical to solar \citep{solartwin}, are of great importance because they allow for a precise differential analysis relative to the Sun. In \citet{18Sco}, our goal has been to use asteroseismology and interferometry on a study of 18 Scorpii, the brightest among the known solar twins. These techniques had already been combined to study the bright subgiant $\beta$ Hyi \citep{North}, for which a mass was derived through the use of homology relations. In this study, we have applied a similar methodology to 18 Sco.

18 Scorpii, a 5th-magnitude star, has been observed for 12 nights with HARPS in May 2009 as part of the asteroseismic component of this study. Figure \ref{18ScoPS} displays the resulting power spectrum of radial-velocity measurements. An average large frequency separation of $134.4\pm0.3\:{\rm{\mu Hz}}$ has been estimated based on the autocorrelation of the time series. We have also performed long-baseline interferometry at visible wavelengths by using the PAVO beam-combiner \citep{PAVO} at the CHARA array \citep{CHARA}. An angular diameter of $0.6759\pm0.0062\:{\rm{mas}}$ has been estimated that, combined with the known parallax, leads to a radius of $1.010\pm0.009\:{\rm{R}}_\odot$. Using the homology relation given in Eq.~(\ref{homology}), one obtains a mass of $1.02\pm0.03\:{\rm{M}}_\odot$ for 18 Sco. This value for the mass is in good agreement with previously published estimates derived from indirect methods, such as comparison between spectroscopic or photometric observations and stellar evolutionary tracks.

\figuremacroW{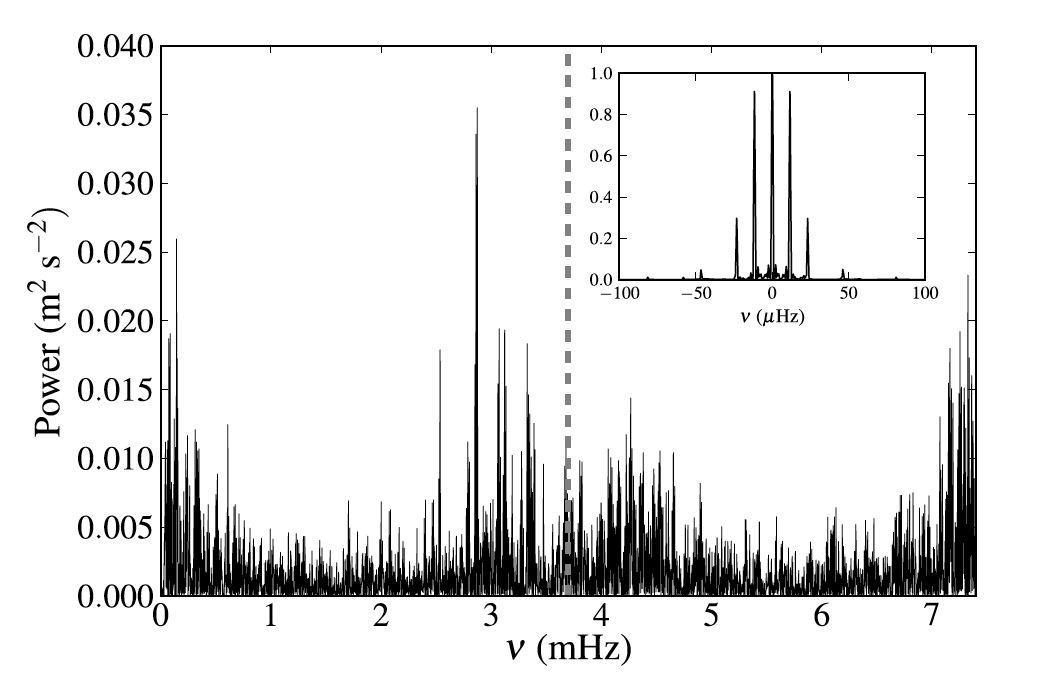}{Power spectrum from radial-velocity measurements of 18 Sco}{The vertical dashed line marks the location of the equivalent Nyquist frequency. The inset shows the spectral window, normalized to its maximum. The power spectrum shows a clear hump of excess power around $3\:{\rm{mHz}}$, reaching $\sim\!0.04\:{\rm{m^2\,s^{-2}}}$, which corresponds to amplitudes of $\sim\!20\:{\rm{cm\,s^{-1}}}$. From \citet{18Sco}.}{1}

This work shows the possibilities offered by the synergy between asteroseismology, even if from a ground-based single site, and interferometry \citep[for more on the combined application of these two techniques see][]{Cunhareview07}. Our results confirm that 18 Sco is remarkably similar to the Sun in terms of both radius and mass. The next steps to be taken involve measuring the individual oscillation frequencies and performing full modeling using all the available observations. 

\subsection{Probing the core properties of $\alpha$ Centauri A with asteroseismology}\label{secalphaCenA}
The proximity of the visual binary system $\alpha$ Cen, allied to the similarity of its components to the Sun, caused this system to become a preferred asteroseismic target{\footnote{An account is given in Sect.~\ref{obstatus} of the asteroseismic campaigns dedicated to $\alpha$ Cen A during the last decade.}}. Moreover, its primary component, $\alpha$ Cen A (G2V), has a mass of $1.105\pm0.007\:{\rm{M}}_\odot$ \citep{Pourbaix}, very close to the limit above which main-sequence stars keep the convective core developed during the pre-main-sequence phase. For this reason, a theoretical study of $\alpha$ Cen A is particularly valuable for testing the poorly-modeled treatment of convection and extra mixing in the central regions of low-mass stars.  

In \citet{deMeulenaer}, we have combined and analysed the radial-velocity time series obtained in May 2001 with CORALIE \citep{BouchyCarrier01,BouchyCarrier02}, UVES and UCLES \citep{Butler04,Bedding04}. The aim has been to derive a precise set of asteroseismic constraints to be compared to models, that would eventually allow us to improve the knowledge of the interior of $\alpha$ Cen A, namely, by determining the nature of its core.

While the combined time series is as long as the CORALIE time series (12.45 days), it contains almost five times more data points, significantly reducing the daily aliases in the spectral window due to an enhanced time coverage (see Fig.~\ref{combinedTS}). Three different weighting schemes have been used in the computation of a Lomb-Scargle periodogram from the combined time series, thus resulting in three different power spectra. Since we have employed Iterative Sine-Wave Fitting for frequency estimation (with all its known problems concerning false detections and fatal error propagation; recall Sect.~\ref{multisine}), we hoped to guarantee the genuineness of the detected frequencies by analysing three different power spectra. We have detected 44 modes with $l\!=\!0,1,2,3$, in overall good agreement with previous works, of which 14 showed possible rotational splittings. New average values have been derived for the large and small frequency separations.

\figuremacroW{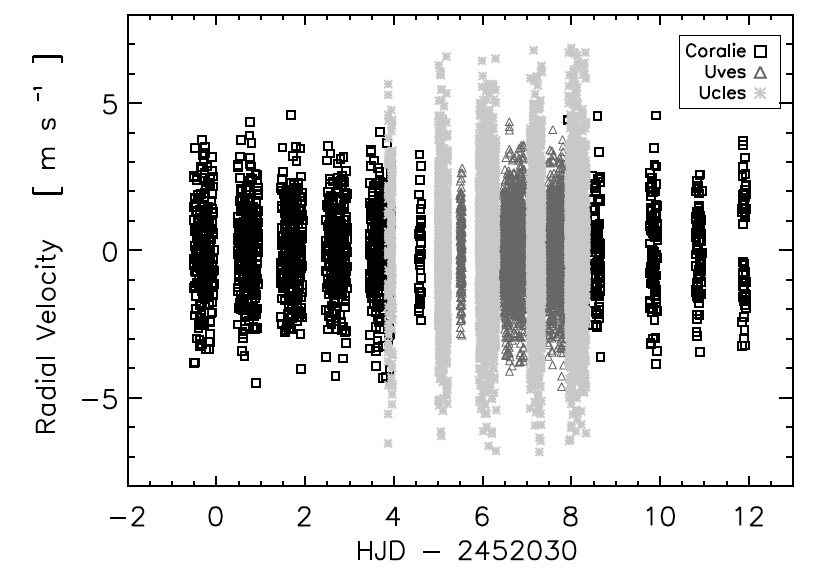}{Combined time series obtained for $\alpha$ Cen A with CORALIE, UVES, and UCLES}{Note that the UCLES data set, which has been obtained in Australia, fills several of the gaps in the time series left by the other two data sets, obtained in Chile. From \citet{deMeulenaer}.}{1}

\begin{sidewaysfigure}
	\centering
	\includegraphics[width=\textheight]{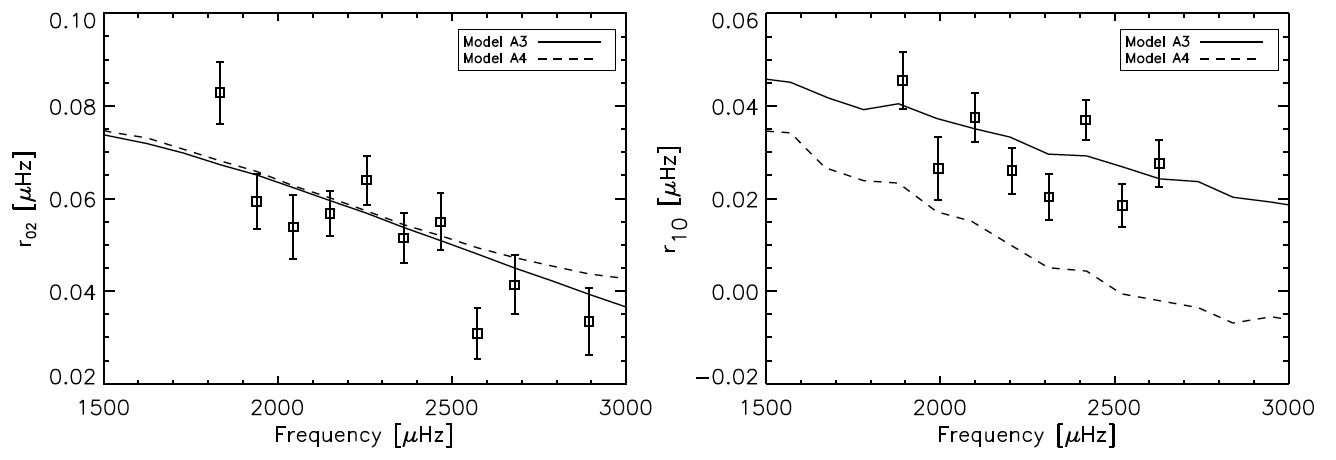}
	\caption[Comparison with models of $\alpha$ Cen A based on the ratios $r_{02}$ and $r_{10}$]{\textbf{Comparison with models of $\alpha$ Cen A based on the ratios $r_{02}$ and $r_{10}$} - The left-hand (right-hand) panel shows a comparison of the observed ratio $r_{02}$ ($r_{10}$) with that computed from a couple of models from \citet{MiglioMont} (see text for details). From \citet{deMeulenaer}.}
	\label{ratios}
\end{sidewaysfigure}

\citet{MiglioMont} have performed several calibrations of the $\alpha$ Cen system using both classical (photometric, spectroscopic, and astrometric) and asteroseismic constraints. They found that models with either a radiative or a convective core could equally well reproduce the classical constraints, and that an overshooting parameter $\alpha_{\rm{ov}}\!>\!0.15$ is sufficient for a convective core to persist once the star leaves the pre-main-sequence phase. They also realized that the structural differences in these two sets of models leave distinctive signatures on the oscillation frequencies, in particular on the ratios $r_{02}$ (cf.~Eq.~\ref{r02}) and $r_{10}$ (cf.~Eq.~\ref{r10}). In Fig.~\ref{ratios} we compare the observed ratios $r_{02}$ and $r_{10}$ -- derived from the frequencies estimated in our study -- with those computed from a couple of models from \citet{MiglioMont}, to be specific, their models A3 (radiative core and no overshooting) and A4 (convective core and $\alpha_{\rm{ov}}\!=\!0.2$). Although the ratio $r_{02}$ does not help discriminating between models, the ratio $r_{10}$ allows us to reject model A4. Therefore, current observational constraints seem not to be in favor of a convective core in $\alpha$ Cen A. Moreover, we are able to set an upper limit to the amount of convective-core overshooting needed to model stars of mass and metallicity similar to those of $\alpha$ Cen A.

\section{Results from \emph{CoRoT}}
\subsection{A fresh look at the seismic spectrum of HD~49933}\label{secHD49933}
In \citet{Benomar09}, we have analysed 180 days of \emph{CoRoT} photometry on the F5 main-sequence star HD~49933. In 2007, \emph{CoRoT} had already produced a 60-day time series on this target as a result of an initial short run. These initial data would be subsequently analysed by \citet{Appourchaux08}, thus providing the first asteroseismic results from \emph{CoRoT} on a star displaying solar-like oscillations. A new and longer run of 137 days then followed in 2008. Note also that this star had been observed during 10 nights with HARPS \citep{Mosser05} previous to the advent of \emph{CoRoT}.

The seismic spectrum of HD~49933 has proven very difficult to interpret. \citet{Mosser05} were unable to isolate individual p modes in the power spectrum, but could still measure the large frequency separation. The data analysed by \citet{Appourchaux08} clearly showed individual p-mode peaks in the power spectrum. However, they still had to face the problem of mode identification (recall the discussion on the ``F star problem'' in Sect.~\ref{SectMLE}), which was apparently solved based on the computation of a likelihood ratio. This same time series would be the object of further studies. For instance, \citet{BenAppourBaud}, based on the computation of a Bayesian odds ratio (cf.~Eq.~\ref{oddsratio}), found mode identification to be ambiguous for this star. Needless to say, controversy arised.

As already stated above, in this work we have analysed 180 days of photometric data. Three power spectra were computed: from the initial 60-day run and from two non-overlapping subsets of 60 days each belonging to the second longer run. These spectra were then averaged, hence reducing the variance in the resulting average power spectrum. This final power spectrum was distributed to a number of fitters, who then fitted a common model to the spectrum using different methods (MLE, MAP and a Bayesian approach using MCMC techniques) or the same method applied in an independent manner. The different methods yielded consistent results, allowing us to make a robust identification of the modes in terms of the degree $l$ (see Fig.~\ref{echelleHD49933}). Our preferred mode identification, opposite to the one advanced by \citet{Appourchaux08}, was established with a very high confidence level. Moreover, the precision with which mode parameters were estimated has significantly increased, with the rotational splitting remaining the only mode parameter that is poorly constrained.

\figuremacroW{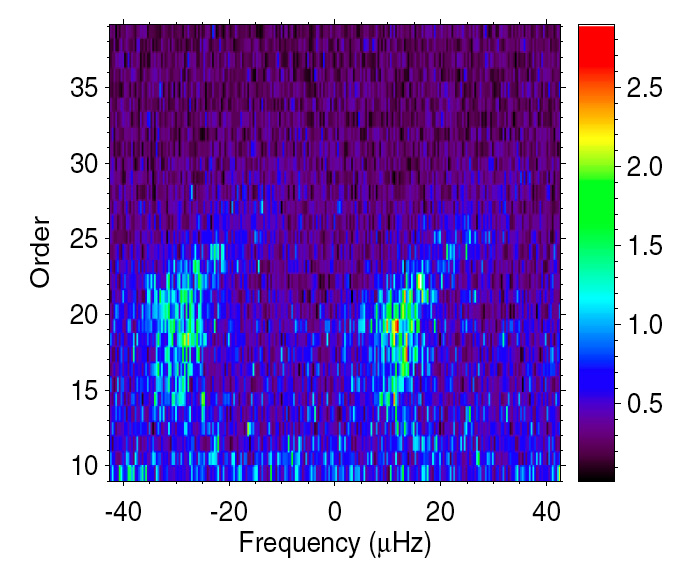}{Average spectrum of HD~49933 in \'echelle format}{This \'echelle diagram has been computed using a large frequency separation of $85\:{\rm{\mu Hz}}$ and then smoothed to $0.8\:{\rm{\mu Hz}}$ (4 bins). Identification of the ridges from a simple visual inspection is far from evident. In our preferred identification scenario, the left-hand ridge corresponds to $l\!=\!1$ and the right-hand ridge to $l\!=\!0,2$. From \citet{Benomar09}.}{0.95}

\subsection{The \emph{CoRoT} target HD~52265: a G0V metal-rich exoplanet-host star}\label{secHD52265}
HD~52265 is a G0V metal-rich exoplanet-host{\footnote{A planet orbiting HD~52265 was independently discovered in 2000 by \citet{Butlerplanet} and \citet{Naefplanet}.}} star observed in the seismology field of the \emph{CoRoT} space telescope from November 2008 to March 2009. The satellite collected 117 days of high-precision photometric data on this star with a duty cycle of 90\%, allowing for a clear detection of solar-like oscillations. Complementary ground-based observations were obtained with the Narval spectrograph at the Pic du Midi observatory in December 2008 and January 2009, i.e., simultaneously with \emph{CoRoT} observations. 

In \citet{HD52265}, our aim has been to characterize HD~52265 using both spectroscopic and seismic data. To date, only a handful of exoplanet-host stars have been the object of seismic studies: $\mu$ Ara \citep{BouchyAra}, $\iota$ Hor \citep{VauclairHor}, HD~46375 \citep{HD46375}, HAT-P-7, HAT-P-11 and TrES-2 \citep{JCDExoHost10}. The high-quality \emph{CoRoT} observations of HD~52265 have allowed us to determine its seismic properties with a precision never before obtained for any other exoplanet-host star.  

Precise fundamental stellar parameters have been obtained: $T_{\rm{eff}}\!=\!6100\pm60\:{\rm{K}}$, ${\rm{log}}\,g\!=\!4.35\pm0.09$, ${\rm{[M/H]}}\!=\!0.19\pm0.05$, as well as $v\sin i\!=\!3.6_{-1.0}^{+0.3}\:{\rm{km\,s^{-1}}}$. We have derived the granulation properties and have analysed the signature of stellar rotation arising from the modulation of the light curve due to photospheric magnetic activity (see Fig.~\ref{ballot1}). Thanks to spot-modeling of the light curve, we have found a mean rotational period of $P_{\rm{rot}}\!=\!12.3\pm0.15\:{\rm{days}}$.  

\figuremacroW{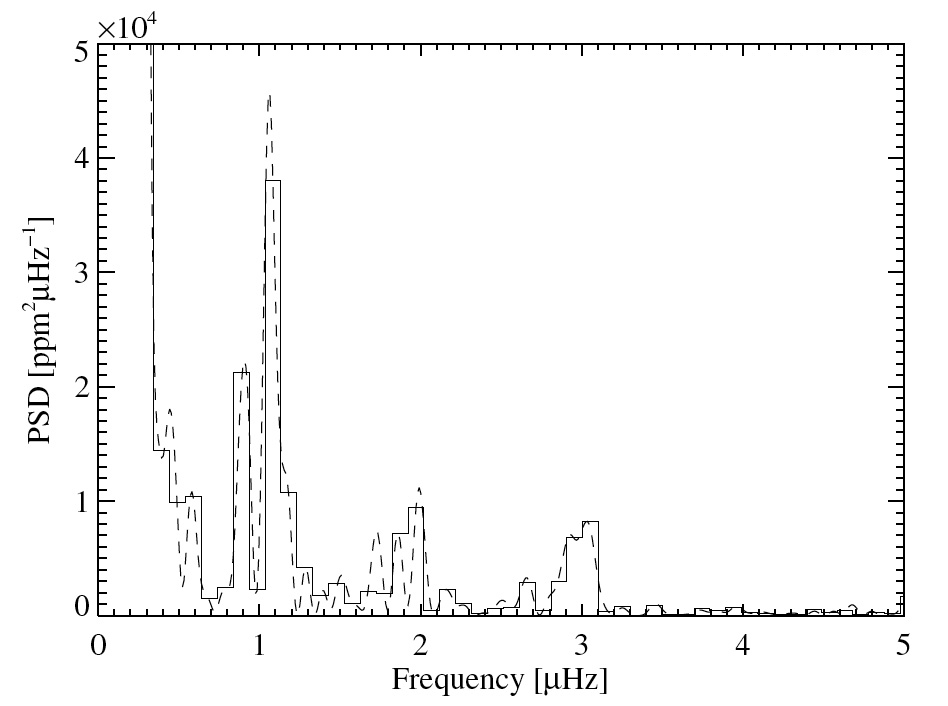}{Power density spectrum of HD~52265 at the low-frequency end}{The power density spectrum is represented by the solid line, while the dashed line corresponds to an oversampled version of the spectrum (by a factor of 10). The modulation of the light curve caused by photospheric spots produces a significant peak at $1.05\:{\rm{\mu Hz}}$ and another very close one at $0.91\:{\rm{\mu Hz}}$. We interpret this broadened structure as being the signature of differential rotation. Peaks around 2 and $3\:{\rm{\mu Hz}}$ are simply overtones of the rotational period. From \citet{HD52265}.}{0.97}

Parameters describing the observed p modes have been estimated using MLE. A global fit to the spectrum has been performed, over about ten radial orders, for degrees $l\!=\!0,1,2$. The frequencies of 31 modes are reported in the range 1500--$2550\:{\rm{\mu Hz}}$. The large separation exhibits a clear modulation around the mean value $\langle \Delta\nu \rangle\!=\!98.3\pm0.1\:{\rm{\mu Hz}}$, which we interpret as possibly being associated with the helium second-ionization region. Mode linewidths vary with frequency along a S-shaped curve with a local maximum around $1800\:{\rm{\mu Hz}}$ (see Fig.~\ref{ballot2}). Mode lifetimes range between 0.5 and 3 days, being shorter than solar, although significantly longer than those observed for F stars. Finally, amplitudes increase almost regularly until reaching a maximum around $2100\:{\rm{\mu Hz}}$, remaining close to that maximum before sharply dropping above $2450\:{\rm{\mu Hz}}$. The fitted maximum bolometric amplitude for radial modes is $3.96\pm0.24\:{\rm{ppm}}$.

Thanks to the precise estimates of mode frequencies (a precision of about $0.2\:{\rm{\mu Hz}}$ has been achieved for the highest-amplitude modes) and fundamental parameters, HD~52265 has become a promising object for stellar modeling. This seismic study will also help improving our knowledge of the planetary companion.    

\figuremacroW{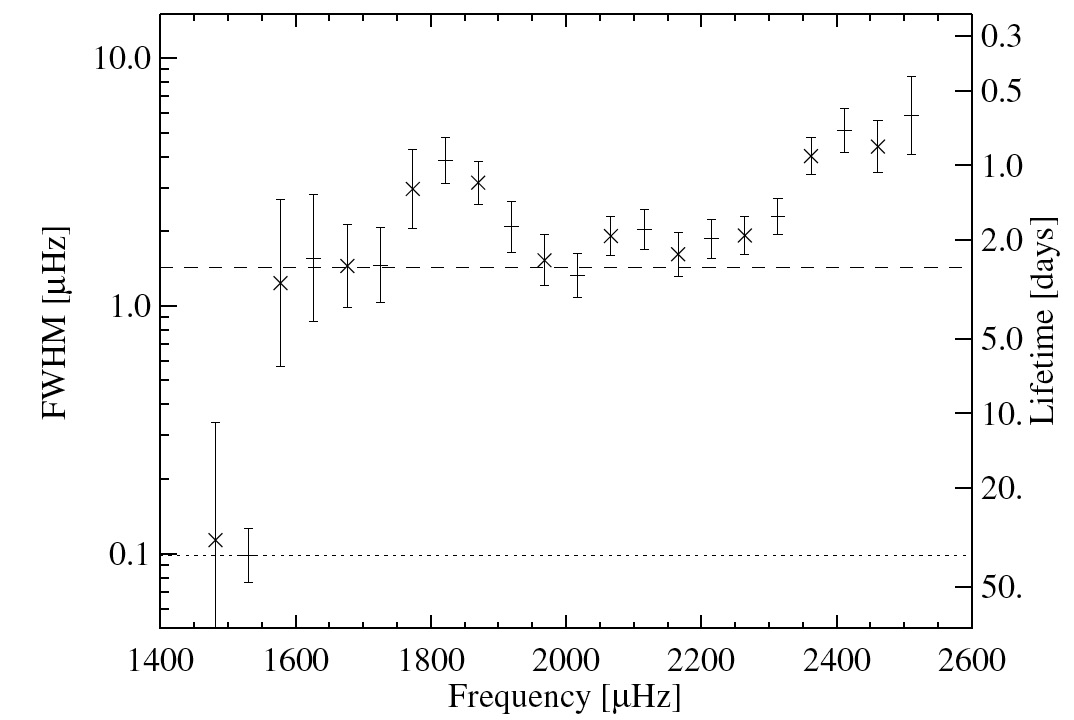}{Mode linewidths as a function of frequency for HD~52265}{Crosses correspond to values obtained by fitting a common width for modes with $(n-1,l\!=\!1)$ and $(n,l\!=\!0)$, whereas plus signs correspond to values obtained by fitting a common width for modes with $(n,l\!=\!1)$ and $(n,l\!=\!0)$. The dotted line indicates the spectral resolution. The dashed line shows the prediction of \citet{Chaplintau}. From \citet{HD52265}.}{1}

\subsection{On the origin of the intermediate-order g modes observed in the hybrid $\gamma$ Dor/$\delta$ Sct star HD~49434}\label{sechybrid}
$\gamma$ Doradus stars pulsate in high-order g modes with periods of order 1 day, driven by convective-flux blocking at the base of their convective envelopes. $\delta$ Scuti stars, on the other hand, pulsate in low-order p modes with periods of order 2 hours, driven by the $\kappa$ mechanism operating in the \ion{He}{ii} ionization zone. The two types of modes have their properties determined by different portions of the stellar interior. Therefore, hybrid $\gamma$ Dor/$\delta$ Sct pulsators are of great interest because they offer additional constraints on stellar structure and may be used to test theoretical models \citep{Ahmedhybrids}. Furthermore, in the same locus of the Hertzsprung-Russell diagram where the $\gamma$ Dor and $\delta$ Sct instability strips overlap, solar-like oscillations are also predicted to occur for $\delta$ Sct stars \citep{Houdek99,Samadi02}, having actually been confirmed observationally for the very first time by \citet{AntociNature}.

In \citet{Campantehybrid}{\footnote{This non-refereed proceedings paper resulted from a poster presentation at the Fourth HELAS International Conference, held in Lanzarote in February 2010.}}, we have analysed the frequency power spectrum of HD~49434 resulting from 136.9 days of \emph{CoRoT} photometry. HD~49434 (F1V) had been referenced as a candidate hybrid $\gamma$ Dor/$\delta$ Sct star by \citet{Uytter08}, following an extensive photometric and spectroscopic ground-based campaign, a classification that \citet{Chapellier} could, however, not confirm. A compelling feature of its frequency power spectrum is the presence of intermediate-order g modes between the simultaneously excited high-order g modes ($\gamma$ Dor regime) and low-order p modes ($\delta$ Sct regime). However, time-dependent convection models \citep{DupretTDC,AhmedTDC} predict the existence of a theoretical frequency gap that is stable to pulsations in the range 5--$15\:{\rm{d^{-1}}}$ (i.e., between the two regimes just mentioned) for low-degree modes. This raises the question as to which mechanism is responsible for the excitation of the observed intermediate-order g modes.

In this work, we have addressed the possibility that such modes are excited by a stochastic mechanism. A search for the signature of stochastic excitation in a selection of modes within the theoretical frequency gap was carried out according to the statistical method described in \citet{PereiraLopes05} (recall the discussion in Sect.~\ref{sectstatprop}). Figure \ref{exdia} displays the so-called excitation diagram for a selection of modes within the theoretical frequency gap. The fact that the observational results lie outside the confidence interval (for the two strongest modes) or just over the 1-$\sigma$ lower bound (for the two faintest modes) may tempt one to conclude that these modes are not stochastically excited. However, we need to be cautious and a new analysis should be carried out that uses a larger number of amplitude measurements (i.e., using shorter subseries or, alternatively, based on a longer time series), thus increasing the significance of the statistic $\sigma(A)/\langle A \rangle$.

\figuremacroW{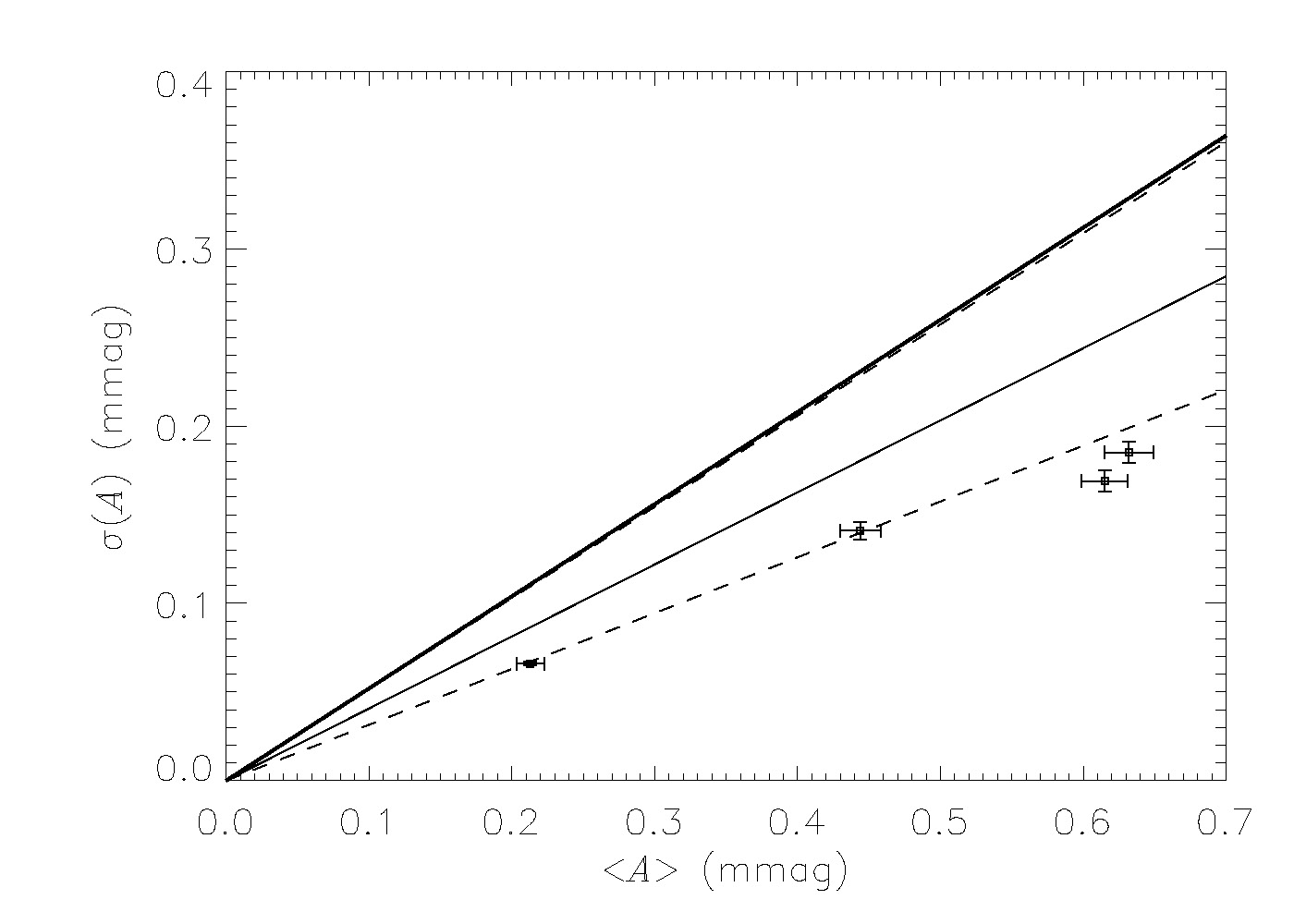}{Excitation diagram for 4 selected modes within the frequency gap}{Observational results for a selection of 4 modes within the gap are plotted with accompanying error bars. The thick solid line represents the theoretical relation $\sigma(A)\!=\!0.52\langle A \rangle$. The thin solid line represents the outcome of Monte Carlo simulations -- assuming stochastic excitation and the same sampling used to obtain the observational results -- which gave $\sigma(A)\!\approx\!0.41\langle A \rangle$, with the corresponding 1-$\sigma$ bounds represented by the dashed lines. From \citet{Campantehybrid}.}{0.96}

\section{Results from \emph{Kepler}}
\subsection{Global properties of solar-like oscillations observed by \emph{Kepler}: a comparison of complementary analysis methods}\label{secpipecomp}
In Sect.~\ref{sectpipe}, I emphasized the main reasons why automated pipelines have been used to analyse \emph{Kepler} data on main-sequence and subgiant solar-like oscillators. Firstly, global asteroseismic parameters are readily obtainable -- even from data with low SNR -- using automated analysis methods. Secondly, the automated nature of these pipelines is required to efficiently exploit the plenitude of data made available by \emph{Kepler} on these targets.

In \citet{Vernercomparison}, we present the asteroseismic analysis of 1948 F-, G-, and K-type main-sequence and subgiant stars observed by \emph{Kepler} during the first seven months of science operations. This incorporates all short-cadence observations of stars that have been identified as potentially showing solar-like oscillations. We have detected solar-like oscillations in 642 of these stars and have characterized them by their average large frequency separation ($\Delta\nu$), frequency of maximum amplitude ($\nu_{\rm{max}}$), and maximum mode amplitude ($A_{\rm{max}}$). This represents the largest cohort of main-sequence and subgiant solar-like oscillators observed to date. By combining results from the analysis methods of nine independent research teams (see Table \ref{9pipes}), we have verified the detections, rejected outliers, and devised a method to ensure the results are consistent within an accurate uncertainty range. It is apparent that the formal uncertainties returned from automated analysis methods are often inconsistent with the actual precision of the results. Obtaining an accurate uncertainty on the global asteroseismic parameters is essential when using such results to model stellar structure.

\ctable[
cap=Summary of the methods used by each automated pipeline,
caption={\textbf{Summary of the methods used by each automated pipeline} - Abbreviations used are SPS -- smoothed power spectrum, PSPS -- power spectrum of the power spectrum, TSACF -- time series autocorrelation function, PSACF -- power spectrum autocorrelation function, and K08 -- method based on \citet{Kjeldsen08}.},
label=9pipes,
pos=t,
notespar
]{llll}{
\tnote[a]{\citet{Mathurpipeline};}
\tnote[b]{see Sect.~\ref{sectpipe};}
\tnote[c]{\citet{Mosserpipeline};}
\tnote[d]{\citet{Hekkerpipeline};}
\tnote[e]{\citet{Bonannopipeline};}
\tnote[f]{\citet{Vernerpipeline};}
\tnote[g]{\citet{Campantepipeline};}
\tnote[h]{\citet{Karoffpipeline};}
\tnote[i]{\citet{Huberpipeline};}
\tnote[j]{only used for the analysis of simulated data.}
}{ 
\hline\hline
Pipeline & Method for $\Delta\nu$ & Method for $\nu_{\rm{max}}$ & Method for $A_{\rm{max}}$ \\
\hline
A2Z\tmark[a] & Peak of PSPS & Fit to SPS & Fit to SPS (K08) \\
AAU\tmark[b] & Peak of PSPS & Peak of SPS & Peak of SPS \\
COR\tmark[c] & Fit to TSACF & Fit to SPS & Fit to TSACF \\
IAS & Peak of TSACF & Fit to SPS & Fit to SPS (K08) \\
OCT\tmark[d] & 1: PSPS (full PS) & 1: Fit to SPS & 1: Peak of SPS \\
         & 2: PSPS (Bayesian on full PS)\tmark[j] & 2: Peak of SPS\tmark[j] & 2: Fit to SPS \\
         & 3: PSPS (small $\nu$ range)\tmark[j] & & 3: Fit to SPS (K08) \\ 
         & 4: PSPS (Bayesian on small $\nu$ range)\tmark[j] & & \\
ORK\tmark[e] & Comb response function & CLEAN algorithm & --- \\
QML\tmark[f] & 1: Peak of TSACF & Fit to PSACF & Fit to PSACF \\
          & 2: Fit to PSACF & & \\			  
KAB\tmark[g,h] & Fit to PSACF & 1: Fit to PSACF & Fit to SPS \\
         & & 2: Fit to SPS & \\
SYD\tmark[i] & Fit to PSACF & 1: Peak of SPS & 1: Peak of SPS (K08) \\
         & & 2: Fit to SPS & 2: Fit to SPS (K08) \\
\hline
}

We have correlated the fraction of stars for which we detected oscillations with the stellar parameters from the \emph{Kepler} Input Catalog \citep[KIC; e.g.,][]{KIC,KIC2}, and found a significant reduction in the proportion of solar-like oscillators with effective temperatures in the range $5300\:{\rm{K}}\!\lesssim\!T_{\rm{eff}}\!\lesssim\!5700\:{\rm{K}}$, viz., the temperature range that separates the distributions of main-sequence stars from subgiants. This has also been noted by \citet{Chaplin11detect}, who have suggested that this absence of oscillators may be due to evolutionary effects that cause an increase in surface magnetic activity, thus reducing the detectability of oscillations. Moreover, a drop-off has been found in the relative number of detected oscillators approaching the red edge of the classical instability strip. 

\begin{sidewaysfigure}
	\centering
	\includegraphics[width=\textheight]{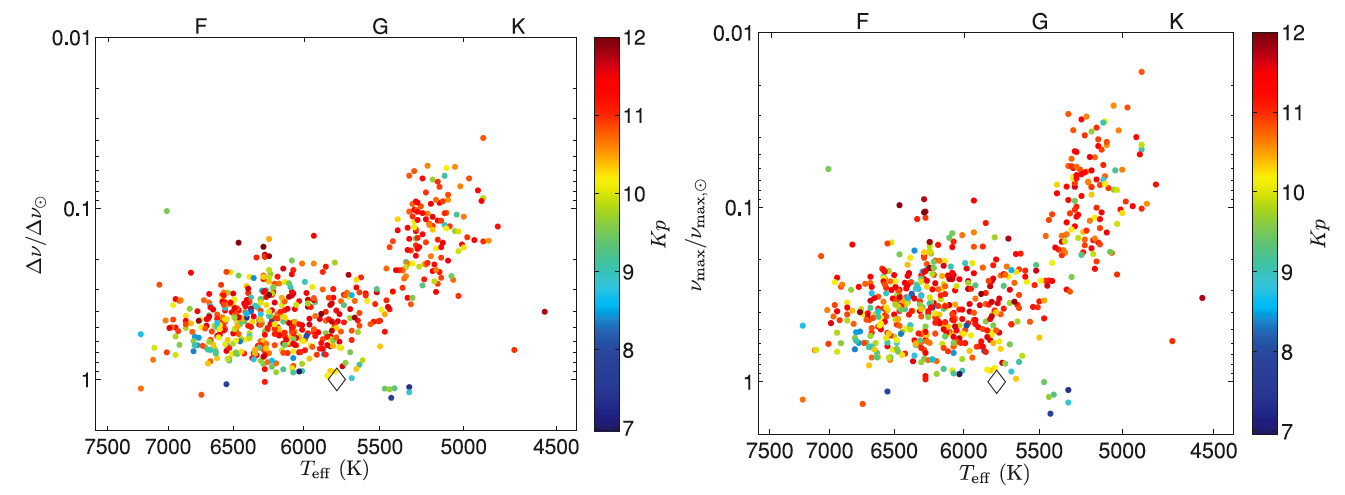}
	\caption[Asteroseismic population diagrams]{\textbf{Asteroseismic population diagrams} - These diagrams have been produced using the values of $\Delta\nu$ (left-hand panel) and $\nu_{\rm{max}}$ (right-hand panel). The subgiants are grouped in the upper right of the diagrams, with the main-sequence stars grouped in the lower left. The diamonds indicate the position of the Sun. Points are colored according to the corresponding apparent magnitude taken from the KIC. From \citet{Vernercomparison}.}
	\label{populations}
\end{sidewaysfigure}

By characterizing the stars by $\Delta\nu$ or $\nu_{\rm{max}}$, we have clearly identified the separate main-sequence and subgiant populations (see Fig.~\ref{populations}). The distributions of these global asteroseismic parameters for the observed main-sequence stars show that we typically find values of $\Delta\nu$ and $\nu_{\rm{max}}$ smaller than solar. This reflects the increased amplitude of oscillations in stars with lower $\nu_{\rm{max}}$ and the higher intrinsic luminosity of such stars. The stars with asteroseismic parameters closer to those of the Sun also tend to be the brighter stars in the set.

By combining the scaling relations in Eqs.~(\ref{numaxscaling}) and (\ref{homology}), \citet{Stello09} showed that the expected relationship between $\Delta\nu$ and $\nu_{\rm{max}}$ is almost independent of luminosity and only weakly dependent on mass and effective temperature. To a good approximation, we can assume
\begin{equation}
\frac{\Delta\nu}{\Delta\nu_\odot} \approx \left(\frac{\nu_{\rm{max}}}{\nu_{{\rm{max}},\odot}}\right)^a \, ,
\end{equation}
after scaling by solar values. We have determined the exponents of power-law relationships between the verified $\Delta\nu$ and $\nu_{\rm{max}}$ values and found that the scatter in the results is at a level greater than the formal uncertainties obtained for the individual methods. Taking a weighted average of the exponents, we have determined a power-law exponent of $a\!=\!0.795\pm0.007$, which is significantly higher than that found using data from a similar number of red giants: $a\!=\!0.784\pm0.003$ \citep{HekkerRG09}. This discrepancy can be explained by the different temperatures and evolutionary states of the stars in each study.

The maximum mode amplitudes obtained by each method have been used to determine the exponent $s$ (cf.~Eq.~\ref{scalerel}) relating the dependence of mode amplitude on $L/M$. We have used bolometric-corrected amplitudes and set $r\!=\!1.5$ (cf.~Eq.~\ref{scalerel2}) to find that a strong temperature dependence remains in $s$. This is not surprising, since setting $r\!=\!1.5$ in Eq.~(\ref{scalerel2}) relies on a very simple adiabatic description of the stellar atmosphere. We have constrained this dependence by a linear relationship, which gives a value of $s$ for stars in the red-giant regime that agrees with that determined in \citet{MosserRG}.

\subsection{Ensemble asteroseismology of solar-type stars with \emph{Kepler}}\label{secensemble}

During the first seven months of science operations, more than 2000 stars have been selected for observation at short cadence for a 1-month period each, as part of an asteroseismic survey of the solar-type population in the \emph{Kepler} field of view. These stars have magnitudes down to \emph{Kepler} apparent magnitude $Kp\!\approx\!12$ and have been selected as potential solar-type targets based upon parameters in the KIC. In \citet{ChaplinSci}, we report the detection of solar-like oscillations in 500 of these stars, an ensemble that is large enough to allow statistical studies of intrinsic stellar properties (such as mass, radius, and age) and to test theories of stellar evolution.

We have made use of the $\Delta\nu$ and $\nu_{\rm{max}}$ of the stars in the ensemble, together with estimates of $T_{\rm{eff}}$ derived from multicolor photometry available in the KIC, to estimate the masses and radii in a way that is independent of stellar evolutionary models. This so-called direct method of estimation uses the following relations \citep{KallingerRGCorot}:
\begin{equation}
\frac{R}{{\rm{R}}_\odot}\approx\frac{\nu_{\rm{max}}}{\nu_{{\rm{max}},\odot}}\left(\frac{\Delta\nu}{\Delta\nu_\odot}\right)^{-2}\left(\frac{T_{\rm{eff}}}{{\rm{T}}_{{\rm{eff}},\odot}}\right)^{0.5}
\end{equation} 
and
\begin{equation}
\frac{M}{{\rm{M}}_\odot}\approx\left(\frac{\nu_{\rm{max}}}{\nu_{{\rm{max}},\odot}}\right)^{3}\left(\frac{\Delta\nu}{\Delta\nu_\odot}\right)^{-4}\left(\frac{T_{\rm{eff}}}{{\rm{T}}_{{\rm{eff}},\odot}}\right)^{1.5} \, ,
\end{equation}
which can be arrived at by combining Eqs.~(\ref{numaxscaling}) and (\ref{homology}), and then solving for $R$ and $M$, respectively. We have obtained a median fractional uncertainty of just over 10\% in $M$ and about 5.5\% in $R$. The direct method gives larger uncertainties on $M$ and on $R$ than would be obtained from a grid-based method of estimation \citep[i.e., matching the observations to stellar evolutionary tracks; e.g.,][]{Stellogrid,Basugrid,Quiriongrid,Gaigrid}. However, the lack of precise independent constraints on the metallicities meant that the grid-based approach would be vulnerable to a systematic bias in the estimates of $M$ (although not $R$).

\figuremacroW{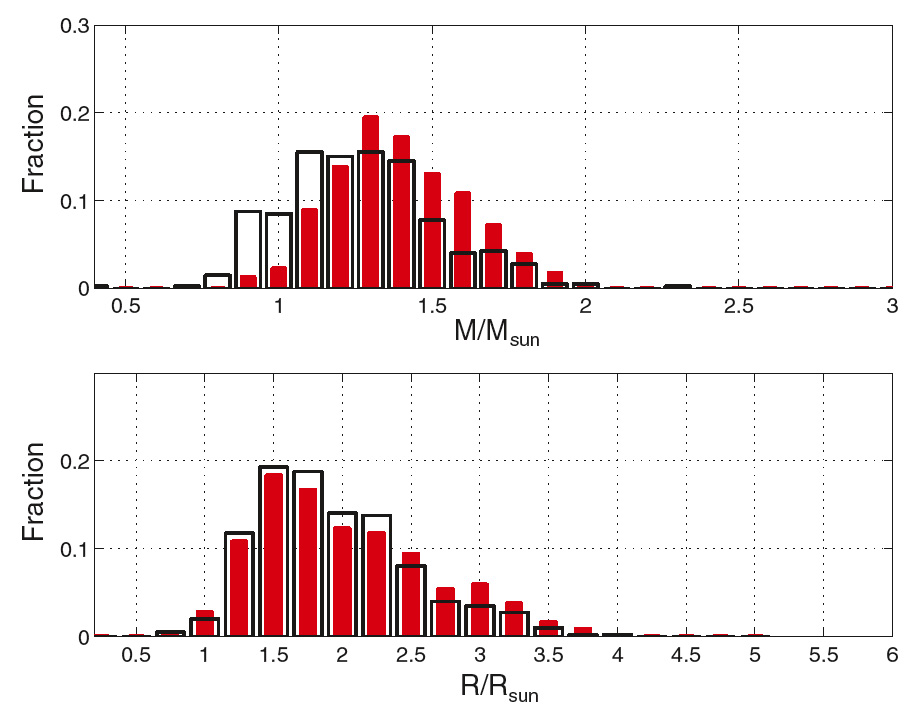}{Observed versus predicted distributions of fundamental stellar parameters}{Black lines show histograms of the observed distributions of masses (top panel) and radii (bottom panel). In red, the predicted distributions from population synthesis modeling, after correction for the effects of detection bias. From \citet{ChaplinSci}.}{1}

The observed distributions of stellar masses and radii have been compared with those predicted from synthetic stellar populations (see Fig.~\ref{synpop}). The synthetic populations have been calculated by modeling the formation and evolution of stars in the \emph{Kepler} field of view. This modeling requires descriptions of, for example, the star-formation history, the spatial density of stars in the disk of the Galaxy, and the rate at which the Galaxy is chemically enriched by stellar evolution. While the distributions of stellar radii are similar, the same cannot be said for the mass distributions. We have applied the Kolmogorov-Smirnov test in order to quantify differences between the observed and synthetic distributions. Differences in radius were found to be marginally significant at best. In contrast, those in mass were found to be highly significant ($>\!99.99\%$). The observed distribution of masses is wider at its peak than the predicted distribution and is offset toward slightly lower masses. On the assumption that the observed masses and radii are robust, this result may have implications for both the star-formation rate and the initial mass function of stars.  

\subsection{Solar-like oscillations in KIC~11395018 and KIC~11234888 from eight months of \emph{Kepler} data}\label{secMathurfurry}
Since the start of \emph{Kepler} science operations in May 2009, a selection of survey stars have been continuously monitored at short cadence to test and validate the time-series photometry, five of which show evidence of solar-like oscillations. Such continuous and long observations are unprecedented for solar-type stars other than the Sun. In \citet{BoogieTigger}, we have analysed 8-month-long time series with a duty cycle in excess of 90\% for two of these stars{\footnote{Two other stars, namely, KIC~10273246 and KIC~10920273, are analysed in \citet{MulderScully}.}}, namely, KIC~11395018 and KIC~11234888. The two stars selected for this study are relatively faint -- KIC~11395018 (G-type) and KIC~11234888 (late F-type) have \emph{Kepler} apparent magnitudes of $Kp\!=\!10.8$ and $Kp\!=\!11.9$, respectively -- and display low SNR in the p-mode peaks. The light curves used in the analysis have been corrected for instrumental effects in a manner independent of the \emph{Kepler} science pipeline \citep[for details see][]{Garciacorrections}.

Different fitting strategies have been employed to extract estimates of p-mode frequencies as well as of other individual mode parameters, from which we have selected frequency lists that will help constrain stellar models. A total of 22 and 16 modes of degree $l\!=\!0,1,2$ have been identified for KIC~11395018 (in the range 600--$1000\:{\rm{\mu Hz}}$) and KIC~11234888 (in the range 500--$900\:{\rm{\mu Hz}}$), respectively. Moreover, two avoided crossings ($l\!=\!1$ ridge) have been identified for KIC~11395018, while a more complex \'echelle spectrum has been found for KIC~11234888 displaying several avoided crossings (see Fig.~\ref{complexechelle}). Both stars are thus thought to have evolved off the main sequence. These results confirm previous expectations that asteroseismology of solar-type survey targets is possible down to apparent magnitudes of 11 and fainter, provided we work with a multi-month time series \citep[e.g.,][]{Stellogrid}. 

\figuremacroW{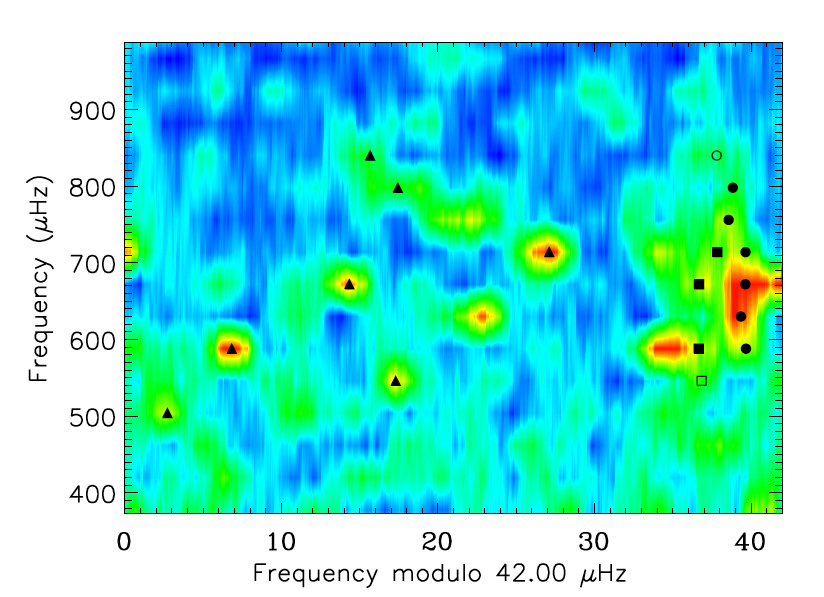}{\'Echelle diagram of the power density spectrum of KIC~11234888}{The minimal (filled symbols) and maximal (filled and open symbols) frequency sets are displayed \citep[for a definition see][]{MulderScully}. Symbol shapes indicate mode degree: $l\!=\!0$ (circles), $l\!=\!1$ (triangles), and $l\!=\!2$ (squares). From \citet{BoogieTigger}.}{1}

The global asteroseismic parameters reported for these stars, together with a detailed atmospheric analysis, should allow constraining their radii, masses and ages with considerable precision \citep{CreeveyFurry}. Further insight into the physics of these evolved solar-type stars -- based on detailed modeling and inversion techniques -- is now possible due to the high quality of the seismic parameters found.





\begin{multicols}{2} 
\begin{footnotesize}

\bibliographystyle{Latex/Classes/aa}
\renewcommand{\bibname}{Bibliography} 

\bibliography{9_backmatter/references} 

\begin{thebibliography}{334}
\expandafter\ifx\csname natexlab\endcsname\relax\def\natexlab#1{#1}\fi

\bibitem[{{Abrams} \& {Kumar}(1996)}]{AbramsKumar}
{Abrams}, D. \& {Kumar}, P. 1996, \apj, 472, 882

\bibitem[{{Aerts} {et~al.}(2010){Aerts}, {Christensen-Dalsgaard}, \&
  {Kurtz}}]{BookAstero}
{Aerts}, C., {Christensen-Dalsgaard}, J., \& {Kurtz}, D.~W. 2010,
  {Asteroseismology}, 1st edn. (Springer)

\bibitem[{{Aigrain} {et~al.}(2004){Aigrain}, {Favata}, \&
  {Gilmore}}]{Aigrain94}
{Aigrain}, S., {Favata}, F., \& {Gilmore}, G. 2004, \aap, 414, 1139

\bibitem[{{Aizenman} {et~al.}(1977){Aizenman}, {Smeyers}, \&
  {Weigert}}]{Aizenman77}
{Aizenman}, M., {Smeyers}, P., \& {Weigert}, A. 1977, \aap, 58, 41

\bibitem[{{Anderson} {et~al.}(1990){Anderson}, {Duvall}, \&
  {Jefferies}}]{Anderson90}
{Anderson}, E.~R., {Duvall}, Jr., T.~L., \& {Jefferies}, S.~M. 1990, \apj, 364,
  699

\bibitem[{{Antoci} {et~al.}(2011){Antoci}, {Handler}, {Campante}, {Thygesen},
  {Moya}, {Kallinger}, {Stello}, {Grigahc{\`e}ne}, {Kjeldsen}, {Bedding},
  {L{\"u}ftinger}, {Christensen-Dalsgaard}, {Catanzaro}, {Frasca}, {De Cat},
  {Uytterhoeven}, {Bruntt}, {Houdek}, {Kurtz}, {Lenz}, {Kaiser}, {van Cleve},
  {Allen}, \& {Clarke}}]{AntociNature}
{Antoci}, V., {Handler}, G., {Campante}, T.~L., {et~al.} 2011, \nat, 477, 570

\bibitem[{{Appourchaux}(2003)}]{Appourchaux2003}
{Appourchaux}, T. 2003, \aap, 412, 903

\bibitem[{{Appourchaux}(2004)}]{Appourchaux04}
{Appourchaux}, T. 2004, \aap, 428, 1039

\bibitem[{{Appourchaux}(2008)}]{AppourBayes08}
{Appourchaux}, T. 2008, Astronomische Nachrichten, 329, 485

\bibitem[{{Appourchaux}(2011)}]{Appourchauxguide}
{Appourchaux}, T. 2011, in Asteroseismology, Canary Islands Winter School of
  Astrophysics, ed. P.~L. {Pall{\'e}}, Vol. XXII (Cambridge University Press)

\bibitem[{{Appourchaux} {et~al.}(2010){Appourchaux}, {Belkacem}, {Broomhall},
  {Chaplin}, {Gough}, {Houdek}, {Provost}, {Baudin}, {Boumier}, {Elsworth},
  {Garc{\'{\i}}a}, {Andersen}, {Finsterle}, {Fr{\"o}hlich}, {Gabriel}, {Grec},
  {Jim{\'e}nez}, {Kosovichev}, {Sekii}, {Toutain}, \&
  {Turck-Chi{\`e}ze}}]{AppourchauxReviewg}
{Appourchaux}, T., {Belkacem}, K., {Broomhall}, A.-M., {et~al.} 2010, \aapr,
  18, 197

\bibitem[{{Appourchaux} {et~al.}(2012){Appourchaux}, {Benomar}, {Gruberbauer},
  {Chaplin}, {Garc{\'{\i}}a}, {Handberg}, {Verner}, {Antia}, {Campante},
  {Davies}, {Deheuvels}, {Hekker}, {Howe}, {Salabert}, {Bedding}, {White},
  {Houdek}, {Silva Aguirre}, {Elsworth}, {van Cleve}, {Clarke}, {Hall}, \&
  {Kjeldsen}}]{Appourchauxlinewidths}
{Appourchaux}, T., {Benomar}, O., {Gruberbauer}, M., {et~al.} 2012, \aap, 537,
  A134

\bibitem[{{Appourchaux} {et~al.}(2006{\natexlab{a}}){Appourchaux},
  {Berthomieu}, {Michel}, {Aerts}, {Ballot}, {Barban}, {Baudin}, {Boumier}, {De
  Ridder}, {Floquet}, {Garcia}, {Garrido}, {Goupil}, {Lambert}, {Lochard},
  {Neiner}, {Poretti}, {Provost}, {Roxburgh}, {Samadi}, \& {Toutain}}]{DAT1}
{Appourchaux}, T., {Berthomieu}, G., {Michel}, E., {et~al.} 2006{\natexlab{a}},
  in ESA Special Publication, Vol. 1306, Proceedings of The \emph{CoRoT}
  Mission Pre-Launch Status -- Stellar Seismology and Planet Finding, ed.
  {M.~Fridlund, A.~Baglin, J.~Lochard, \& L.~Conroy}, 377

\bibitem[{{Appourchaux} {et~al.}(2006{\natexlab{b}}){Appourchaux},
  {Berthomieu}, {Michel}, {Ballot}, {Barban}, {Baudin}, {Boumier}, {De Ridder},
  {Floquet}, {Garcia}, {Garrido}, {Goupil}, {Lambert}, {Lochard}, {Mazumdar},
  {Neiner}, {Poretti}, {Provost}, {Roxburgh}, {Samadi}, \& {Toutain}}]{DAT2}
{Appourchaux}, T., {Berthomieu}, G., {Michel}, E., {et~al.} 2006{\natexlab{b}},
  in ESA Special Publication, Vol. 1306, Proceedings of The \emph{CoRoT}
  Mission Pre-Launch Status -- Stellar Seismology and Planet Finding, ed.
  {M.~Fridlund, A.~Baglin, J.~Lochard, \& L.~Conroy}, 429

\bibitem[{{Appourchaux} {et~al.}(2000){Appourchaux}, {Fr{\"o}hlich},
  {Andersen}, {Berthomieu}, {Chaplin}, {Elsworth}, {Finsterle}, {Gough},
  {Hoeksema}, {Isaak}, {Kosovichev}, {Provost}, {Scherrer}, {Sekii}, \&
  {Toutain}}]{Appourchaux2000}
{Appourchaux}, T., {Fr{\"o}hlich}, C., {Andersen}, B.~N., {et~al.} 2000, \apj,
  538, 401

\bibitem[{{Appourchaux} {et~al.}(1998){Appourchaux}, {Gizon}, \&
  {Rabello-Soares}}]{Appourchaux98}
{Appourchaux}, T., {Gizon}, L., \& {Rabello-Soares}, M. 1998, \aaps, 132, 107

\bibitem[{{Appourchaux} {et~al.}(2008){Appourchaux}, {Michel}, {Auvergne},
  {Baglin}, {Toutain}, {Baudin}, {Benomar}, {Chaplin}, {Deheuvels}, {Samadi},
  {Verner}, {Boumier}, {Garc{\'{\i}}a}, {Mosser}, {Hulot}, {Ballot}, {Barban},
  {Elsworth}, {Jim{\'e}nez-Reyes}, {Kjeldsen}, {R{\'e}gulo}, \&
  {Roxburgh}}]{Appourchaux08}
{Appourchaux}, T., {Michel}, E., {Auvergne}, M., {et~al.} 2008, \aap, 488, 705

\bibitem[{{Appourchaux} {et~al.}(2009){Appourchaux}, {Samadi}, \&
  {Dupret}}]{AppourchauxBayesTest}
{Appourchaux}, T., {Samadi}, R., \& {Dupret}, M.-A. 2009, \aap, 506, 1

\bibitem[{{Arentoft} {et~al.}(2009){Arentoft}, {Kjeldsen}, \&
  {Bedding}}]{Arentoftweights}
{Arentoft}, T., {Kjeldsen}, H., \& {Bedding}, T.~R. 2009, in Astronomical
  Society of the Pacific Conference Series, ed. {M.~Dikpati, T.~Arentoft,
  I.~Gonz{\'a}lez Hern{\'a}ndez, C.~Lindsey, \& F.~Hill}, Vol. 416, 347

\bibitem[{{Arentoft} {et~al.}(2008){Arentoft}, {Kjeldsen}, {Bedding}, {Bazot},
  {Christensen-Dalsgaard}, {Dall}, {Karoff}, {Carrier}, {Eggenberger},
  {Sosnowska}, {Wittenmyer}, {Endl}, {Metcalfe}, {Hekker}, {Reffert}, {Butler},
  {Bruntt}, {Kiss}, {O'Toole}, {Kambe}, {Ando}, {Izumiura}, {Sato}, {Hartmann},
  {Hatzes}, {Bouchy}, {Mosser}, {Appourchaux}, {Barban}, {Berthomieu},
  {Garcia}, {Michel}, {Provost}, {Turck-Chi{\`e}ze}, {Marti{\'c}}, {Lebrun},
  {Schmitt}, {Bertaux}, {Bonanno}, {Benatti}, {Claudi}, {Cosentino}, {Leccia},
  {Frandsen}, {Brogaard}, {Glowienka}, {Grundahl}, \&
  {Stempels}}]{ArentoftProcyon}
{Arentoft}, T., {Kjeldsen}, H., {Bedding}, T.~R., {et~al.} 2008, \apj, 687,
  1180

\bibitem[{{Auvergne} {et~al.}(2009){Auvergne}, {Bodin}, {Boisnard}, {Buey},
  {Chaintreuil}, {Epstein}, {Jouret}, {Lam-Trong}, {Levacher}, {Magnan},
  {Perez}, {Plasson}, {Plesseria}, {Peter}, {Steller}, {Tiph{\`e}ne}, {Baglin},
  {Agogu{\'e}}, {Appourchaux}, {Barbet}, {Beaufort}, {Bellenger}, {Berlin},
  {Bernardi}, {Blouin}, {Boumier}, {Bonneau}, {Briet}, {Butler}, {Cautain},
  {Chiavassa}, {Costes}, {Cuvilho}, {Cunha-Parro}, {de Oliveira Fialho},
  {Decaudin}, {Defise}, {Djalal}, {Docclo}, {Drummond}, {Dupuis}, {Exil},
  {Faur{\'e}}, {Gaboriaud}, {Gamet}, {Gavalda}, {Grolleau}, {Gueguen},
  {Guivarc'h}, {Guterman}, {Hasiba}, {Huntzinger}, {Hustaix}, {Imbert},
  {Jeanville}, {Johlander}, {Jorda}, {Journoud}, {Karioty}, {Kerjean},
  {Lafond}, {Lapeyrere}, {Landiech}, {Larqu{\'e}}, {Laudet}, {Le Merrer},
  {Leporati}, {Leruyet}, {Levieuge}, {Llebaria}, {Martin}, {Mazy}, {Mesnager},
  {Michel}, {Moalic}, {Monjoin}, {Naudet}, {Neukirchner}, {Nguyen-Kim},
  {Ollivier}, {Orcesi}, {Ottacher}, {Oulali}, {Parisot}, {Perruchot},
  {Piacentino}, {Pinheiro da Silva}, {Platzer}, {Pontet}, {Pradines},
  {Quentin}, {Rohbeck}, {Rolland}, {Rollenhagen}, {Romagnan}, {Russ}, {Samadi},
  {Schmidt}, {Schwartz}, {Sebbag}, {Smit}, {Sunter}, {Tello}, {Toulouse},
  {Ulmer}, {Vandermarcq}, {Vergnault}, {Wallner}, {Waultier}, \&
  {Zanatta}}]{SAA}
{Auvergne}, M., {Bodin}, P., {Boisnard}, L., {et~al.} 2009, \aap, 506, 411

\bibitem[{{Baglin} {et~al.}(2006){Baglin}, {Michel}, {Auvergne}, \& {the
  \emph{CoRoT} Team}}]{CoRoT}
{Baglin}, A., {Michel}, E., {Auvergne}, M., \& {the \emph{CoRoT} Team}. 2006,
  in ESA Special Publication, Vol. 624, Proceedings of \emph{SOHO} 18/GONG
  2006/HELAS I, Beyond the spherical Sun

\bibitem[{{Ballot}(2010)}]{Ballot10}
{Ballot}, J. 2010, Astronomische Nachrichten, 331, 933

\bibitem[{{Ballot} {et~al.}(2008){Ballot}, {Appourchaux}, {Toutain}, \&
  {Guittet}}]{Ballot2008}
{Ballot}, J., {Appourchaux}, T., {Toutain}, T., \& {Guittet}, M. 2008, \aap,
  486, 867

\bibitem[{{Ballot} {et~al.}(2006){Ballot}, {Garc{\'{\i}}a}, \&
  {Lambert}}]{Ballot2006}
{Ballot}, J., {Garc{\'{\i}}a}, R.~A., \& {Lambert}, P. 2006, \mnras, 369, 1281

\bibitem[{{Ballot} {et~al.}(2011){Ballot}, {Gizon}, {Samadi}, {Vauclair},
  {Benomar}, {Bruntt}, {Mosser}, {Stahn}, {Verner}, {Campante},
  {Garc{\'{\i}}a}, {Mathur}, {Salabert}, {Gaulme}, {R{\'e}gulo}, {Roxburgh},
  {Appourchaux}, {Baudin}, {Catala}, {Chaplin}, {Deheuvels}, {Michel}, {Bazot},
  {Creevey}, {Dolez}, {Elsworth}, {Sato}, {Vauclair}, {Auvergne}, \&
  {Baglin}}]{HD52265}
{Ballot}, J., {Gizon}, L., {Samadi}, R., {et~al.} 2011, \aap, 530, A97

\bibitem[{{Ballot} {et~al.}(2004){Ballot}, {Turck-Chi{\`e}ze}, \&
  {Garc{\'{\i}}a}}]{Ballot04}
{Ballot}, J., {Turck-Chi{\`e}ze}, S., \& {Garc{\'{\i}}a}, R.~A. 2004, \aap,
  423, 1051

\bibitem[{{Balona} {et~al.}(2011){Balona}, {Pigulski}, {De Cat}, {Handler},
  {Guti{\'e}rrez-Soto}, {Engelbrecht}, {Frescura}, {Briquet}, {Cuypers},
  {Daszy{\'n}ska-Daszkiewicz}, {Degroote}, {Dukes}, {Garcia}, {Green}, {Heber},
  {Kawaler}, {Lehmann}, {Leroy}, {Molenda-{\.Z}aaowicz}, {Neiner}, {Noels},
  {Nuspl}, {{\O}stensen}, {Pricopi}, {Roxburgh}, {Salmon}, {Smith},
  {Su{\'a}rez}, {Suran}, {Szab{\'o}}, {Uytterhoeven}, {Christensen-Dalsgaard},
  {Kjeldsen}, {Caldwell}, {Girouard}, \& {Sanderfer}}]{Balona11}
{Balona}, L.~A., {Pigulski}, A., {De Cat}, P., {et~al.} 2011, \mnras, 413, 2403

\bibitem[{{Barban} {et~al.}(2009){Barban}, {Deheuvels}, {Baudin},
  {Appourchaux}, {Auvergne}, {Ballot}, {Boumier}, {Chaplin}, {Garc{\'{\i}}a},
  {Gaulme}, {Michel}, {Mosser}, {R{\'e}gulo}, {Roxburgh}, {Verner}, {Baglin},
  {Catala}, {Samadi}, {Bruntt}, {Elsworth}, \& {Mathur}}]{Barban09}
{Barban}, C., {Deheuvels}, S., {Baudin}, F., {et~al.} 2009, \aap, 506, 51

\bibitem[{{Basu} \& {Antia}(2008)}]{BasuAntia}
{Basu}, S. \& {Antia}, H.~M. 2008, \physrep, 457, 217

\bibitem[{{Basu} {et~al.}(2010){Basu}, {Chaplin}, \& {Elsworth}}]{Basugrid}
{Basu}, S., {Chaplin}, W.~J., \& {Elsworth}, Y. 2010, \apj, 710, 1596

\bibitem[{{Basu} {et~al.}(2011){Basu}, {Grundahl}, {Stello}, {Kallinger},
  {Hekker}, {Mosser}, {Garc{\'{\i}}a}, {Mathur}, {Brogaard}, {Bruntt},
  {Chaplin}, {Gai}, {Elsworth}, {Esch}, {Ballot}, {Bedding}, {Gruberbauer},
  {Huber}, {Miglio}, {Yildiz}, {Kjeldsen}, {Christensen-Dalsgaard},
  {Gilliland}, {Fanelli}, {Ibrahim}, \& {Smith}}]{BasuRG}
{Basu}, S., {Grundahl}, F., {Stello}, D., {et~al.} 2011, \apjl, 729, L10

\bibitem[{{Basu} {et~al.}(2004){Basu}, {Mazumdar}, {Antia}, \&
  {Demarque}}]{Basu04}
{Basu}, S., {Mazumdar}, A., {Antia}, H.~M., \& {Demarque}, P. 2004, \mnras,
  350, 277

\bibitem[{{Batchelor}(1953)}]{Batchelor}
{Batchelor}, G.~K. 1953, {The Theory of Homogeneous Turbulence}, 1st edn.
  (Cambridge University Press)

\bibitem[{{Baudin} {et~al.}(2011){Baudin}, {Barban}, {Belkacem}, {Hekker},
  {Morel}, {Samadi}, {Benomar}, {Goupil}, {Carrier}, {Ballot}, {Deheuvels}, {De
  Ridder}, {Hatzes}, {Kallinger}, \& {Weiss}}]{Baudin11}
{Baudin}, F., {Barban}, C., {Belkacem}, K., {et~al.} 2011, \aap, 529, A84

\bibitem[{{Bayes} \& {Price}(1763)}]{Bayestheorem}
{Bayes}, T. \& {Price}, R. 1763, Philosophical Transactions, 53, 370

\bibitem[{{Bazot} {et~al.}(2007){Bazot}, {Bouchy}, {Kjeldsen}, {Charpinet},
  {Laymand}, \& {Vauclair}}]{Bazot07}
{Bazot}, M., {Bouchy}, F., {Kjeldsen}, H., {et~al.} 2007, \aap, 470, 295

\bibitem[{{Bazot} {et~al.}(2011){Bazot}, {Ireland}, {Huber}, {Bedding},
  {Broomhall}, {Campante}, {Carfantan}, {Chaplin}, {Elsworth}, {Mel{\'e}ndez},
  {Petit}, {Th{\'e}ado}, {van Grootel}, {Arentoft}, {Asplund}, {Castro},
  {Christensen-Dalsgaard}, {Do Nascimento}, {Dintrans}, {Dumusque}, {Kjeldsen},
  {McAlister}, {Metcalfe}, {Monteiro}, {Santos}, {Sousa}, {Sturmann},
  {Sturmann}, {Ten Brummelaar}, {Turner}, \& {Vauclair}}]{18Sco}
{Bazot}, M., {Ireland}, M.~J., {Huber}, D., {et~al.} 2011, \aap, 526, L4

\bibitem[{{Bazot} {et~al.}(2005){Bazot}, {Vauclair}, {Bouchy}, \&
  {Santos}}]{BazotAra}
{Bazot}, M., {Vauclair}, S., {Bouchy}, F., \& {Santos}, N.~C. 2005, \aap, 440,
  615

\bibitem[{{Beck} {et~al.}(2011){Beck}, {Bedding}, {Mosser}, {Stello}, {Garcia},
  {Kallinger}, {Hekker}, {Elsworth}, {Frandsen}, {Carrier}, {De Ridder},
  {Aerts}, {White}, {Huber}, {Dupret}, {Montalb{\'a}n}, {Miglio}, {Noels},
  {Chaplin}, {Kjeldsen}, {Christensen-Dalsgaard}, {Gilliland}, {Brown},
  {Kawaler}, {Mathur}, \& {Jenkins}}]{Beckmixed}
{Beck}, P.~G., {Bedding}, T.~R., {Mosser}, B., {et~al.} 2011, Science, 332, 205

\bibitem[{{Bedding}(2003)}]{BeddingSRV}
{Bedding}, T.~R. 2003, \apss, 284, 61

\bibitem[{{Bedding}(2011)}]{BeddingWS}
{Bedding}, T.~R. 2011, in Asteroseismology, Canary Islands Winter School of
  Astrophysics, ed. P.~L. {Pall{\'e}}, Vol. XXII (Cambridge University Press)

\bibitem[{{Bedding} {et~al.}(2007{\natexlab{a}}){Bedding}, {Brun},
  {Christensen-Dalsgaard}, {Crouch}, {De Cat}, {Garc{\'{\i}}a}, {Gizon},
  {Hill}, {Kjeldsen}, {Leibacher}, {Maillard}, {Mathis}, {Rabello-Soares},
  {Rozelot}, {Rempel}, {Roxburgh}, {Samadi}, {Talon}, \&
  {Thompson}}]{Bedding07}
{Bedding}, T.~R., {Brun}, A.~S., {Christensen-Dalsgaard}, J., {et~al.}
  2007{\natexlab{a}}, Highlights of Astronomy, 14, 491

\bibitem[{{Bedding} {et~al.}(2001){Bedding}, {Butler}, {Kjeldsen}, {Baldry},
  {O'Toole}, {Tinney}, {Marcy}, {Kienzle}, \& {Carrier}}]{Bedding01betaHyi}
{Bedding}, T.~R., {Butler}, R.~P., {Kjeldsen}, H., {et~al.} 2001, \apjl, 549,
  L105

\bibitem[{{Bedding} {et~al.}(2010{\natexlab{a}}){Bedding}, {Huber}, {Stello},
  {Elsworth}, {Hekker}, {Kallinger}, {Mathur}, {Mosser}, {Preston}, {Ballot},
  {Barban}, {Broomhall}, {Buzasi}, {Chaplin}, {Garc{\'{\i}}a}, {Gruberbauer},
  {Hale}, {De Ridder}, {Frandsen}, {Borucki}, {Brown}, {Christensen-Dalsgaard},
  {Gilliland}, {Jenkins}, {Kjeldsen}, {Koch}, {Belkacem}, {Bildsten}, {Bruntt},
  {Campante}, {Deheuvels}, {Derekas}, {Dupret}, {Goupil}, {Hatzes}, {Houdek},
  {Ireland}, {Jiang}, {Karoff}, {Kiss}, {Lebreton}, {Miglio}, {Montalb{\'a}n},
  {Noels}, {Roxburgh}, {Sangaralingam}, {Stevens}, {Suran}, {Tarrant}, \&
  {Weiss}}]{BeddingLetter10}
{Bedding}, T.~R., {Huber}, D., {Stello}, D., {et~al.} 2010{\natexlab{a}},
  \apjl, 713, L176

\bibitem[{{Bedding} \& {Kjeldsen}(2003)}]{BeddingKjeldsen03}
{Bedding}, T.~R. \& {Kjeldsen}, H. 2003, \pasa, 20, 203

\bibitem[{{Bedding} \& {Kjeldsen}(2008)}]{BeddingKjeldsen08}
{Bedding}, T.~R. \& {Kjeldsen}, H. 2008, in Astronomical Society of the Pacific
  Conference Series, Vol. 384, 14th Cambridge Workshop on Cool Stars, Stellar
  Systems, and the Sun, ed. {G.~van Belle}, 21

\bibitem[{{Bedding} \& {Kjeldsen}(2010)}]{scaledechelle}
{Bedding}, T.~R. \& {Kjeldsen}, H. 2010, Communications in Asteroseismology,
  161, 3

\bibitem[{{Bedding} {et~al.}(2007{\natexlab{b}}){Bedding}, {Kjeldsen},
  {Arentoft}, {Bouchy}, {Brandbyge}, {Brewer}, {Butler},
  {Christensen-Dalsgaard}, {Dall}, {Frandsen}, {Karoff}, {Kiss}, {Monteiro},
  {Pijpers}, {Teixeira}, {Tinney}, {Baldry}, {Carrier}, \&
  {O'Toole}}]{Bedding07betaHyi}
{Bedding}, T.~R., {Kjeldsen}, H., {Arentoft}, T., {et~al.} 2007{\natexlab{b}},
  \apj, 663, 1315

\bibitem[{{Bedding} {et~al.}(2005){Bedding}, {Kjeldsen}, {Bouchy}, {Bruntt},
  {Butler}, {Buzasi}, {Christensen-Dalsgaard}, {Frandsen}, {Lebrun},
  {Marti{\'c}}, \& {Schou}}]{BeddingvsMatthews}
{Bedding}, T.~R., {Kjeldsen}, H., {Bouchy}, F., {et~al.} 2005, \aap, 432, L43

\bibitem[{{Bedding} {et~al.}(2004){Bedding}, {Kjeldsen}, {Butler}, {McCarthy},
  {Marcy}, {O'Toole}, {Tinney}, \& {Wright}}]{Bedding04}
{Bedding}, T.~R., {Kjeldsen}, H., {Butler}, R.~P., {et~al.} 2004, \apj, 614,
  380

\bibitem[{{Bedding} {et~al.}(2010{\natexlab{b}}){Bedding}, {Kjeldsen},
  {Campante}, {Appourchaux}, {Bonanno}, {Chaplin}, {Garcia}, {Marti{\'c}},
  {Mosser}, {Butler}, {Bruntt}, {Kiss}, {O'Toole}, {Kambe}, {Ando}, {Izumiura},
  {Sato}, {Hartmann}, {Hatzes}, {Barban}, {Berthomieu}, {Michel}, {Provost},
  {Turck-Chi{\`e}ze}, {Lebrun}, {Schmitt}, {Bertaux}, {Benatti}, {Claudi},
  {Cosentino}, {Leccia}, {Frandsen}, {Brogaard}, {Glowienka}, {Grundahl},
  {Stempels}, {Arentoft}, {Bazot}, {Christensen-Dalsgaard}, {Dall}, {Karoff},
  {Lundgreen-Nielsen}, {Carrier}, {Eggenberger}, {Sosnowska}, {Wittenmyer},
  {Endl}, {Metcalfe}, {Hekker}, \& {Reffert}}]{BeddingProcyon}
{Bedding}, T.~R., {Kjeldsen}, H., {Campante}, T.~L., {et~al.}
  2010{\natexlab{b}}, \apj, 713, 935

\bibitem[{{Bedding} {et~al.}(1996){Bedding}, {Kjeldsen}, {Reetz}, \&
  {Barbuy}}]{Bedding96}
{Bedding}, T.~R., {Kjeldsen}, H., {Reetz}, J., \& {Barbuy}, B. 1996, \mnras,
  280, 1155

\bibitem[{{Bedding} {et~al.}(2011){Bedding}, {Mosser}, {Huber},
  {Montalb{\'a}n}, {Beck}, {Christensen-Dalsgaard}, {Elsworth},
  {Garc{\'{\i}}a}, {Miglio}, {Stello}, {White}, {De Ridder}, {Hekker}, {Aerts},
  {Barban}, {Belkacem}, {Broomhall}, {Brown}, {Buzasi}, {Carrier}, {Chaplin},
  {di Mauro}, {Dupret}, {Frandsen}, {Gilliland}, {Goupil}, {Jenkins},
  {Kallinger}, {Kawaler}, {Kjeldsen}, {Mathur}, {Noels}, {Aguirre}, \&
  {Ventura}}]{BeddingRGmixed}
{Bedding}, T.~R., {Mosser}, B., {Huber}, D., {et~al.} 2011, \nat, 471, 608

\bibitem[{{Belkacem} {et~al.}(2011){Belkacem}, {Goupil}, {Dupret}, {Samadi},
  {Baudin}, {Noels}, \& {Mosser}}]{Belkacem11}
{Belkacem}, K., {Goupil}, M.-J., {Dupret}, M.-A., {et~al.} 2011, \aap, 530,
  A142

\bibitem[{{Belkacem} {et~al.}(2009){Belkacem}, {Samadi}, {Goupil},
  {Lef{\`e}vre}, {Baudin}, {Deheuvels}, {Dupret}, {Appourchaux}, {Scuflaire},
  {Auvergne}, {Catala}, {Michel}, {Miglio}, {Montalban}, {Thoul}, {Talon},
  {Baglin}, \& {Noels}}]{Belkacem09}
{Belkacem}, K., {Samadi}, R., {Goupil}, M.-J., {et~al.} 2009, Science, 324,
  1540

\bibitem[{{Benomar} {et~al.}(2009{\natexlab{a}}){Benomar}, {Appourchaux}, \&
  {Baudin}}]{BenAppourBaud}
{Benomar}, O., {Appourchaux}, T., \& {Baudin}, F. 2009{\natexlab{a}}, \aap,
  506, 15

\bibitem[{{Benomar} {et~al.}(2009{\natexlab{b}}){Benomar}, {Baudin},
  {Campante}, {Chaplin}, {Garc{\'{\i}}a}, {Gaulme}, {Toutain}, {Verner},
  {Appourchaux}, {Ballot}, {Barban}, {Elsworth}, {Mathur}, {Mosser},
  {R{\'e}gulo}, {Roxburgh}, {Auvergne}, {Baglin}, {Catala}, {Michel}, \&
  {Samadi}}]{Benomar09}
{Benomar}, O., {Baudin}, F., {Campante}, T.~L., {et~al.} 2009{\natexlab{b}},
  \aap, 507, L13

\bibitem[{{Berger} {et~al.}(1997){Berger}, {Boukai}, \& {Wang}}]{Berger97}
{Berger}, J.~O., {Boukai}, B., \& {Wang}, Y. 1997, Statistical Science, 12, 133

\bibitem[{{Berger} \& {Sellke}(1987)}]{BergerSellke}
{Berger}, J.~O. \& {Sellke}, T. 1987, Journal of the American Statistical
  Association, 82, 112

\bibitem[{{Blackman} \& {Tukey}(1958{\natexlab{a}})}]{BlackTukey1}
{Blackman}, R.~B. \& {Tukey}, J.~W. 1958{\natexlab{a}}, Bell System Technical
  Journal, 37, 185

\bibitem[{{Blackman} \& {Tukey}(1958{\natexlab{b}})}]{BlackTukey2}
{Blackman}, R.~B. \& {Tukey}, J.~W. 1958{\natexlab{b}}, Bell System Technical
  Journal, 37, 485

\bibitem[{{B{\"o}hm-Vitense}(2007)}]{BV}
{B{\"o}hm-Vitense}, E. 2007, \apj, 657, 486

\bibitem[{{Bonanno} {et~al.}(2008){Bonanno}, {Benatti}, {Claudi}, {Desidera},
  {Gratton}, {Leccia}, \& {Patern{\`o}}}]{Bonannopipeline}
{Bonanno}, A., {Benatti}, S., {Claudi}, R., {et~al.} 2008, \apj, 676, 1248

\bibitem[{{Borucki} {et~al.}(2010){Borucki}, {Koch}, {Basri}, {Batalha},
  {Brown}, {Caldwell}, {Caldwell}, {Christensen-Dalsgaard}, {Cochran},
  {DeVore}, {Dunham}, {Dupree}, {Gautier}, {Geary}, {Gilliland}, {Gould},
  {Howell}, {Jenkins}, {Kondo}, {Latham}, {Marcy}, {Meibom}, {Kjeldsen},
  {Lissauer}, {Monet}, {Morrison}, {Sasselov}, {Tarter}, {Boss}, {Brownlee},
  {Owen}, {Buzasi}, {Charbonneau}, {Doyle}, {Fortney}, {Ford}, {Holman},
  {Seager}, {Steffen}, {Welsh}, {Rowe}, {Anderson}, {Buchhave}, {Ciardi},
  {Walkowicz}, {Sherry}, {Horch}, {Isaacson}, {Everett}, {Fischer}, {Torres},
  {Johnson}, {Endl}, {MacQueen}, {Bryson}, {Dotson}, {Haas}, {Kolodziejczak},
  {Van Cleve}, {Chandrasekaran}, {Twicken}, {Quintana}, {Clarke}, {Allen},
  {Li}, {Wu}, {Tenenbaum}, {Verner}, {Bruhweiler}, {Barnes}, \&
  {Prsa}}]{Kepler1}
{Borucki}, W.~J., {Koch}, D., {Basri}, G., {et~al.} 2010, Science, 327, 977

\bibitem[{{Bouchy} {et~al.}(2005){Bouchy}, {Bazot}, {Santos}, {Vauclair}, \&
  {Sosnowska}}]{BouchyAra}
{Bouchy}, F., {Bazot}, M., {Santos}, N.~C., {Vauclair}, S., \& {Sosnowska}, D.
  2005, \aap, 440, 609

\bibitem[{{Bouchy} \& {Carrier}(2001)}]{BouchyCarrier01}
{Bouchy}, F. \& {Carrier}, F. 2001, \aap, 374, L5

\bibitem[{{Bouchy} \& {Carrier}(2002)}]{BouchyCarrier02}
{Bouchy}, F. \& {Carrier}, F. 2002, \aap, 390, 205

\bibitem[{{Bourguignon} {et~al.}(2007){Bourguignon}, {Carfantan}, \&
  {B{\"o}hm}}]{SparSpec}
{Bourguignon}, S., {Carfantan}, H., \& {B{\"o}hm}, T. 2007, \aap, 462, 379

\bibitem[{{Brand{\~a}o} {et~al.}(2011){Brand{\~a}o}, {Do{\u g}an},
  {Christensen-Dalsgaard}, {Cunha}, {Bedding}, {Metcalfe}, {Kjeldsen},
  {Bruntt}, \& {Arentoft}}]{IsaHyi}
{Brand{\~a}o}, I.~M., {Do{\u g}an}, G., {Christensen-Dalsgaard}, J., {et~al.}
  2011, \aap, 527, A37

\bibitem[{{Breger} {et~al.}(1993){Breger}, {Stich}, {Garrido}, {Martin},
  {Jiang}, {Li}, {Hube}, {Ostermann}, {Paparo}, \& {Scheck}}]{Breger93}
{Breger}, M., {Stich}, J., {Garrido}, R., {et~al.} 1993, \aap, 271, 482

\bibitem[{{Bretthorst}(1988)}]{Bretthorst88}
{Bretthorst}, G.~L. 1988, Bayesian Spectrum Analysis and Parameter Estimation
  (New York: Springer-Verlag)

\bibitem[{{Bretthorst}(2000)}]{Bretthorst2000}
{Bretthorst}, G.~L. 2000, in Bulletin of the American Astronomical Society,
  Vol.~32, American Astronomical Society Meeting Abstracts, 1438

\bibitem[{{Bretthorst}(2001{\natexlab{a}})}]{Bretthorst2001I}
{Bretthorst}, G.~L. 2001{\natexlab{a}}, in American Institute of Physics
  Conference Series, Vol. 568, Bayesian Inference and Maximum Entropy Methods
  in Science and Engineering, ed. {A.~Mohammad-Djafari}, 241--245

\bibitem[{{Bretthorst}(2001{\natexlab{b}})}]{Bretthorst2001II}
{Bretthorst}, G.~L. 2001{\natexlab{b}}, in American Institute of Physics
  Conference Series, Vol. 568, Bayesian Inference and Maximum Entropy Methods
  in Science and Engineering, ed. {A.~Mohammad-Djafari}, 246--251

\bibitem[{{Brewer} {et~al.}(2007){Brewer}, {Bedding}, {Kjeldsen}, \&
  {Stello}}]{Brewer07}
{Brewer}, B.~J., {Bedding}, T.~R., {Kjeldsen}, H., \& {Stello}, D. 2007, \apj,
  654, 551

\bibitem[{{Brewer} \& {Stello}(2009)}]{Brewer09}
{Brewer}, B.~J. \& {Stello}, D. 2009, \mnras, 395, 2226

\bibitem[{{Brookes} {et~al.}(1978){Brookes}, {Isaak}, \& {van der
  Raay}}]{Brookes78}
{Brookes}, J.~R., {Isaak}, G.~R., \& {van der Raay}, H.~B. 1978, \mnras, 185,
  19

\bibitem[{{Broomhall} {et~al.}(2010){Broomhall}, {Chaplin}, {Elsworth},
  {Appourchaux}, \& {New}}]{Broomhall2010}
{Broomhall}, A.-M., {Chaplin}, W.~J., {Elsworth}, Y., {Appourchaux}, T., \&
  {New}, R. 2010, \mnras, 406, 767

\bibitem[{{Broomhall} {et~al.}(2009){Broomhall}, {Chaplin}, {Elsworth}, \&
  {New}}]{Broomhall09}
{Broomhall}, A.-M., {Chaplin}, W.~J., {Elsworth}, Y., \& {New}, R. 2009, in
  Astronomical Society of the Pacific Conference Series, ed. {M.~Dikpati,
  T.~Arentoft, I.~Gonz{\'a}lez Hern{\'a}ndez, C.~Lindsey, \& F.~Hill}, Vol.
  416, 245

\bibitem[{{Brown} {et~al.}(1991){Brown}, {Gilliland}, {Noyes}, \&
  {Ramsey}}]{Brown91}
{Brown}, T.~M., {Gilliland}, R.~L., {Noyes}, R.~W., \& {Ramsey}, L.~W. 1991,
  \apj, 368, 599

\bibitem[{{Brown} {et~al.}(1997){Brown}, {Kennelly}, {Korzennik}, {Nisenson},
  {Noyes}, \& {Horner}}]{Brown97}
{Brown}, T.~M., {Kennelly}, E.~J., {Korzennik}, S.~G., {et~al.} 1997, \apj,
  475, 322

\bibitem[{{Brown} {et~al.}(2011){Brown}, {Latham}, {Everett}, \&
  {Esquerdo}}]{KIC}
{Brown}, T.~M., {Latham}, D.~W., {Everett}, M.~E., \& {Esquerdo}, G.~A. 2011,
  \aj, 142, 112

\bibitem[{{Bruntt} {et~al.}(2005){Bruntt}, {Kjeldsen}, {Buzasi}, \&
  {Bedding}}]{Bruntt05}
{Bruntt}, H., {Kjeldsen}, H., {Buzasi}, D.~L., \& {Bedding}, T.~R. 2005, \apj,
  633, 440

\bibitem[{{Butler} {et~al.}(2004){Butler}, {Bedding}, {Kjeldsen}, {McCarthy},
  {O'Toole}, {Tinney}, {Marcy}, \& {Wright}}]{Butler04}
{Butler}, R.~P., {Bedding}, T.~R., {Kjeldsen}, H., {et~al.} 2004, \apjl, 600,
  L75

\bibitem[{{Butler} {et~al.}(2000){Butler}, {Vogt}, {Marcy}, {Fischer}, {Henry},
  \& {Apps}}]{Butlerplanet}
{Butler}, R.~P., {Vogt}, S.~S., {Marcy}, G.~W., {et~al.} 2000, \apj, 545, 504

\bibitem[{{Buzasi}(2002)}]{BuzasiWIRE}
{Buzasi}, D.~L. 2002, in Astronomical Society of the Pacific Conference Series,
  Vol. 259, IAU Colloq.~185: Radial and Nonradial Pulsations as probes of
  Stellar Physics, ed. {C.~Aerts, T.~R.~Bedding, \& J.~Christensen-Dalsgaard},
  616

\bibitem[{{Campante} {et~al.}(2010{\natexlab{a}}){Campante}, {Grigahc{\`e}ne},
  {Su{\'a}rez}, \& {Monteiro}}]{Campantehybrid}
{Campante}, T.~L., {Grigahc{\`e}ne}, A., {Su{\'a}rez}, J.~C., \& {Monteiro},
  M.~J.~P.~F.~G. 2010{\natexlab{a}}, Astronomische Nachrichten, in press
  [arXiv:1003.4427v1]

\bibitem[{{Campante} {et~al.}(2011){Campante}, {Handberg}, {Mathur},
  {Appourchaux}, {Bedding}, {Chaplin}, {Garc{\'{\i}}a}, {Mosser}, {Benomar},
  {Bonanno}, {Corsaro}, {Fletcher}, {Gaulme}, {Hekker}, {Karoff}, {R{\'e}gulo},
  {Salabert}, {Verner}, {White}, {Houdek}, {Brand{\~a}o}, {Creevey}, {Do{\v
  g}an}, {Bazot}, {Christensen-Dalsgaard}, {Cunha}, {Elsworth}, {Huber},
  {Kjeldsen}, {Lundkvist}, {Molenda-{\.Z}akowicz}, {Monteiro}, {Stello},
  {Clarke}, {Girouard}, \& {Hall}}]{MulderScully}
{Campante}, T.~L., {Handberg}, R., {Mathur}, S., {et~al.} 2011, \aap, 534, A6

\bibitem[{{Campante} {et~al.}(2010{\natexlab{b}}){Campante}, {Karoff},
  {Chaplin}, {Elsworth}, {Handberg}, \& {Hekker}}]{Campantepipeline}
{Campante}, T.~L., {Karoff}, C., {Chaplin}, W.~J., {et~al.} 2010{\natexlab{b}},
  \mnras, 408, 542

\bibitem[{{Carrier} {et~al.}(2001){Carrier}, {Bouchy}, {Kienzle}, {Bedding},
  {Kjeldsen}, {Butler}, {Baldry}, {O'Toole}, {Tinney}, \&
  {Marcy}}]{Carrier01betaHyi}
{Carrier}, F., {Bouchy}, F., {Kienzle}, F., {et~al.} 2001, \aap, 378, 142

\bibitem[{{Carrier} {et~al.}(2010){Carrier}, {De Ridder}, {Baudin}, {Barban},
  {Hatzes}, {Hekker}, {Kallinger}, {Miglio}, {Montalb{\'a}n}, {Morel}, {Weiss},
  {Auvergne}, {Baglin}, {Catala}, {Michel}, \& {Samadi}}]{CarrierRG}
{Carrier}, F., {De Ridder}, J., {Baudin}, F., {et~al.} 2010, \aap, 509, A73

\bibitem[{{Carrier} {et~al.}(2005){Carrier}, {Eggenberger}, \&
  {Bouchy}}]{Carrier05etaBoo}
{Carrier}, F., {Eggenberger}, P., \& {Bouchy}, F. 2005, \aap, 434, 1085

\bibitem[{{Catala}(2009)}]{PLATO}
{Catala}, C. 2009, Experimental Astronomy, 23, 329

\bibitem[{{Cayrel de Strobel} {et~al.}(1981){Cayrel de Strobel}, {Knowles},
  {Hernandez}, \& {Bentolila}}]{solartwin}
{Cayrel de Strobel}, G., {Knowles}, N., {Hernandez}, G., \& {Bentolila}, C.
  1981, \aap, 94, 1

\bibitem[{{Chapellier} {et~al.}(2011){Chapellier}, {Rodr{\'{\i}}guez},
  {Auvergne}, {Uytterhoeven}, {Mathias}, {Bouabid}, {Poretti}, {Le Contel},
  {Mart{\'{\i}}n-Ru{\'{\i}}z}, {Amado}, {Garrido}, {Hareter}, {Rainer}, {Eyer},
  {Paparo}, {D{\'{\i}}az-Fraile}, {Baglin}, {Baudin}, {Catala}, {Michel}, \&
  {Samadi}}]{Chapellier}
{Chapellier}, E., {Rodr{\'{\i}}guez}, E., {Auvergne}, M., {et~al.} 2011, \aap,
  525, A23

\bibitem[{{Chaplin}(2011)}]{Chaplinactreview}
{Chaplin}, W.~J. 2011, in Asteroseismology, Canary Islands Winter School of
  Astrophysics, ed. P.~L. {Pall{\'e}}, Vol. XXII (Cambridge University Press)

\bibitem[{{Chaplin} \& {Appourchaux}(1999)}]{ChapAppour}
{Chaplin}, W.~J. \& {Appourchaux}, T. 1999, \mnras, 309, 761

\bibitem[{{Chaplin} {et~al.}(2008{\natexlab{a}}){Chaplin}, {Appourchaux},
  {Arentoft}, {Ballot}, {Christensen-Dalsgaard}, {Creevey}, {Elsworth},
  {Fletcher}, {Garc{\'{\i}}a}, {Houdek}, {Jim{\'e}nez-Reyes}, {Kjeldsen},
  {New}, {R{\'e}gulo}, {Salabert}, {Sekii}, {Sousa}, {Toutain}, \& {the rest of
  the asteroFLAG group}}]{asteroflag}
{Chaplin}, W.~J., {Appourchaux}, T., {Arentoft}, T., {et~al.}
  2008{\natexlab{a}}, Astronomische Nachrichten, 329, 549

\bibitem[{{Chaplin} {et~al.}(2006){Chaplin}, {Appourchaux}, {Baudin},
  {Boumier}, {Elsworth}, {Fletcher}, {Fossat}, {Garc{\'{\i}}a}, {Isaak},
  {Jim{\'e}nez}, {Jim{\'e}nez-Reyes}, {Lazrek}, {Leibacher}, {Lochard}, {New},
  {Pall{\'e}}, {R{\'e}gulo}, {Salabert}, {Seghouani}, {Toutain}, \&
  {Wachter}}]{solarflag}
{Chaplin}, W.~J., {Appourchaux}, T., {Baudin}, F., {et~al.} 2006, \mnras, 369,
  985

\bibitem[{{Chaplin} {et~al.}(2010){Chaplin}, {Appourchaux}, {Elsworth},
  {Garc{\'{\i}}a}, {Houdek}, {Karoff}, {Metcalfe}, {Molenda-{\.Z}akowicz},
  {Monteiro}, {Thompson}, {Brown}, {Christensen-Dalsgaard}, {Gilliland},
  {Kjeldsen}, {Borucki}, {Koch}, {Jenkins}, {Ballot}, {Basu}, {Bazot},
  {Bedding}, {Benomar}, {Bonanno}, {Brand{\~a}o}, {Bruntt}, {Campante},
  {Creevey}, {Di Mauro}, {Do{\u g}an}, {Dreizler}, {Eggenberger}, {Esch},
  {Fletcher}, {Frandsen}, {Gai}, {Gaulme}, {Handberg}, {Hekker}, {Howe},
  {Huber}, {Korzennik}, {Lebrun}, {Leccia}, {Martic}, {Mathur}, {Mosser},
  {New}, {Quirion}, {R{\'e}gulo}, {Roxburgh}, {Salabert}, {Schou}, {Sousa},
  {Stello}, {Verner}, {Arentoft}, {Barban}, {Belkacem}, {Benatti}, {Biazzo},
  {Boumier}, {Bradley}, {Broomhall}, {Buzasi}, {Claudi}, {Cunha}, {D'Antona},
  {Deheuvels}, {Derekas}, {Garc{\'{\i}}a Hern{\'a}ndez}, {Giampapa}, {Goupil},
  {Gruberbauer}, {Guzik}, {Hale}, {Ireland}, {Kiss}, {Kitiashvili},
  {Kolenberg}, {Korhonen}, {Kosovichev}, {Kupka}, {Lebreton}, {Leroy},
  {Ludwig}, {Mathis}, {Michel}, {Miglio}, {Montalb{\'a}n}, {Moya}, {Noels},
  {Noyes}, {Pall{\'e}}, {Piau}, {Preston}, {Roca Cort{\'e}s}, {Roth}, {Sato},
  {Schmitt}, {Serenelli}, {Silva Aguirre}, {Stevens}, {Su{\'a}rez}, {Suran},
  {Trampedach}, {Turck-Chi{\`e}ze}, {Uytterhoeven}, {Ventura}, \&
  {Wilson}}]{ChaplinLetter10}
{Chaplin}, W.~J., {Appourchaux}, T., {Elsworth}, Y., {et~al.} 2010, \apjl, 713,
  L169

\bibitem[{{Chaplin} \& {Basu}(2008)}]{ChaplinBasu}
{Chaplin}, W.~J. \& {Basu}, S. 2008, \solphys, 251, 53

\bibitem[{{Chaplin} {et~al.}(2011{\natexlab{a}}){Chaplin}, {Bedding},
  {Bonanno}, {Broomhall}, {Garc{\'{\i}}a}, {Hekker}, {Huber}, {Verner}, {Basu},
  {Elsworth}, {Houdek}, {Mathur}, {Mosser}, {New}, {Stevens}, {Appourchaux},
  {Karoff}, {Metcalfe}, {Molenda-{\.Z}akowicz}, {Monteiro}, {Thompson},
  {Christensen-Dalsgaard}, {Gilliland}, {Kawaler}, {Kjeldsen}, {Ballot},
  {Benomar}, {Corsaro}, {Campante}, {Gaulme}, {Hale}, {Handberg}, {Jarvis},
  {R{\'e}gulo}, {Roxburgh}, {Salabert}, {Stello}, {Mullally}, {Li}, \&
  {Wohler}}]{Chaplinactivity}
{Chaplin}, W.~J., {Bedding}, T.~R., {Bonanno}, A., {et~al.} 2011{\natexlab{a}},
  \apjl, 732, L5

\bibitem[{{Chaplin} {et~al.}(2007){Chaplin}, {Elsworth}, {Houdek}, \&
  {New}}]{Chaplincycle}
{Chaplin}, W.~J., {Elsworth}, Y., {Houdek}, G., \& {New}, R. 2007, \mnras, 377,
  17

\bibitem[{{Chaplin} {et~al.}(1997{\natexlab{a}}){Chaplin}, {Elsworth}, {Howe},
  {Isaak}, {McLeod}, {Miller}, \& {New}}]{Chaplin97}
{Chaplin}, W.~J., {Elsworth}, Y., {Howe}, R., {et~al.} 1997{\natexlab{a}},
  \mnras, 287, 51

\bibitem[{{Chaplin} {et~al.}(2002){Chaplin}, {Elsworth}, {Isaak}, {Marchenkov},
  {Miller}, {New}, {Pinter}, \& {Appourchaux}}]{ChaplinBiSON}
{Chaplin}, W.~J., {Elsworth}, Y., {Isaak}, G.~R., {et~al.} 2002, \mnras, 336,
  979

\bibitem[{{Chaplin} {et~al.}(1997{\natexlab{b}}){Chaplin}, {Elsworth}, {Isaak},
  {McLeod}, {Miller}, \& {New}}]{solarlinewidths}
{Chaplin}, W.~J., {Elsworth}, Y., {Isaak}, G.~R., {et~al.} 1997{\natexlab{b}},
  \mnras, 288, 623

\bibitem[{{Chaplin} {et~al.}(2003){Chaplin}, {Elsworth}, {Isaak}, {Miller},
  {New}, {Pint{\'e}r}, \& {Thiery}}]{Chaplin03}
{Chaplin}, W.~J., {Elsworth}, Y., {Isaak}, G.~R., {et~al.} 2003, \aap, 398, 305

\bibitem[{{Chaplin} {et~al.}(2008{\natexlab{b}}){Chaplin}, {Houdek},
  {Appourchaux}, {Elsworth}, {New}, \& {Toutain}}]{Chaplin08}
{Chaplin}, W.~J., {Houdek}, G., {Appourchaux}, T., {et~al.} 2008{\natexlab{b}},
  \aap, 485, 813

\bibitem[{{Chaplin} {et~al.}(2005){Chaplin}, {Houdek}, {Elsworth}, {Gough},
  {Isaak}, \& {New}}]{Chaplin05}
{Chaplin}, W.~J., {Houdek}, G., {Elsworth}, Y., {et~al.} 2005, \mnras, 360, 859

\bibitem[{{Chaplin} {et~al.}(2009){Chaplin}, {Houdek}, {Karoff}, {Elsworth}, \&
  {New}}]{Chaplintau}
{Chaplin}, W.~J., {Houdek}, G., {Karoff}, C., {Elsworth}, Y., \& {New}, R.
  2009, \aap, 500, L21

\bibitem[{{Chaplin} {et~al.}(2011{\natexlab{b}}){Chaplin}, {Kjeldsen},
  {Bedding}, {Christensen-Dalsgaard}, {Gilliland}, {Kawaler}, {Appourchaux},
  {Elsworth}, {Garc{\'{\i}}a}, {Houdek}, {Karoff}, {Metcalfe},
  {Molenda-{\.Z}akowicz}, {Monteiro}, {Thompson}, {Verner}, {Batalha},
  {Borucki}, {Brown}, {Bryson}, {Christiansen}, {Clarke}, {Jenkins}, {Klaus},
  {Koch}, {An}, {Ballot}, {Basu}, {Benomar}, {Bonanno}, {Broomhall},
  {Campante}, {Corsaro}, {Creevey}, {Esch}, {Gai}, {Gaulme}, {Hale},
  {Handberg}, {Hekker}, {Huber}, {Mathur}, {Mosser}, {New}, {Pinsonneault},
  {Pricopi}, {Quirion}, {R{\'e}gulo}, {Roxburgh}, {Salabert}, {Stello}, \&
  {Suran}}]{Chaplin11detect}
{Chaplin}, W.~J., {Kjeldsen}, H., {Bedding}, T.~R., {et~al.}
  2011{\natexlab{b}}, \apj, 732, 54

\bibitem[{{Chaplin} {et~al.}(2011{\natexlab{c}}){Chaplin}, {Kjeldsen},
  {Christensen-Dalsgaard}, {Basu}, {Miglio}, {Appourchaux}, {Bedding},
  {Elsworth}, {Garc{\'{\i}}a}, {Gilliland}, {Girardi}, {Houdek}, {Karoff},
  {Kawaler}, {Metcalfe}, {Molenda-{\.Z}akowicz}, {Monteiro}, {Thompson},
  {Verner}, {Ballot}, {Bonanno}, {Brand{\~a}o}, {Broomhall}, {Bruntt},
  {Campante}, {Corsaro}, {Creevey}, {Do{\u g}an}, {Esch}, {Gai}, {Gaulme},
  {Hale}, {Handberg}, {Hekker}, {Huber}, {Jim{\'e}nez}, {Mathur}, {Mazumdar},
  {Mosser}, {New}, {Pinsonneault}, {Pricopi}, {Quirion}, {R{\'e}gulo},
  {Salabert}, {Serenelli}, {Aguirre}, {Sousa}, {Stello}, {Stevens}, {Suran},
  {Uytterhoeven}, {White}, {Borucki}, {Brown}, {Jenkins}, {Kinemuchi}, {Van
  Cleve}, \& {Klaus}}]{ChaplinSci}
{Chaplin}, W.~J., {Kjeldsen}, H., {Christensen-Dalsgaard}, J., {et~al.}
  2011{\natexlab{c}}, Science, 332, 213

\bibitem[{{Christensen-Dalsgaard}(1984)}]{CDdiagramori}
{Christensen-Dalsgaard}, J. 1984, in Space Research in Stellar Activity and
  Variability, ed. {A.~Mangeney \& F.~Praderie}, 11

\bibitem[{{Christensen-Dalsgaard}(1988)}]{CDdiagram}
{Christensen-Dalsgaard}, J. 1988, in IAU Symposium, Vol. 123, Advances in
  Helio- and Asteroseismology, ed. {J.~Christensen-Dalsgaard \& S.~Frandsen},
  295

\bibitem[{{Christensen-Dalsgaard}(1989)}]{JCD89}
{Christensen-Dalsgaard}, J. 1989, \mnras, 239, 977

\bibitem[{{Christensen-Dalsgaard}(2002)}]{JCDhelio}
{Christensen-Dalsgaard}, J. 2002, Reviews of Modern Physics, 74, 1073

\bibitem[{{Christensen-Dalsgaard}(2004)}]{JCDreview04}
{Christensen-Dalsgaard}, J. 2004, \solphys, 220, 137

\bibitem[{{Christensen-Dalsgaard} {et~al.}(2011){Christensen-Dalsgaard},
  {Carpenter}, {Schrijver}, {Karovska}, \& {the SI Team}}]{SI}
{Christensen-Dalsgaard}, J., {Carpenter}, K.~G., {Schrijver}, C.~J.,
  {Karovska}, M., \& {the SI Team}. 2011, Journal of Physics Conference Series,
  271, 012085

\bibitem[{{Christensen-Dalsgaard} \& {Frandsen}(1983)}]{JCDFrandsen83}
{Christensen-Dalsgaard}, J. \& {Frandsen}, S. 1983, \solphys, 82, 469

\bibitem[{{Christensen-Dalsgaard} \& {Gough}(1980)}]{JCDGough80}
{Christensen-Dalsgaard}, J. \& {Gough}, D.~O. 1980, \nat, 288, 544

\bibitem[{{Christensen-Dalsgaard} \& {Houdek}(2010)}]{JCDHoudek10}
{Christensen-Dalsgaard}, J. \& {Houdek}, G. 2010, \apss, 328, 51

\bibitem[{{Christensen-Dalsgaard} {et~al.}(2010){Christensen-Dalsgaard},
  {Kjeldsen}, {Brown}, {Gilliland}, {Arentoft}, {Frandsen}, {Quirion},
  {Borucki}, {Koch}, \& {Jenkins}}]{JCDExoHost10}
{Christensen-Dalsgaard}, J., {Kjeldsen}, H., {Brown}, T.~M., {et~al.} 2010,
  \apjl, 713, L164

\bibitem[{{Christensen-Dalsgaard} {et~al.}(2001){Christensen-Dalsgaard},
  {Kjeldsen}, \& {Mattei}}]{JCD01}
{Christensen-Dalsgaard}, J., {Kjeldsen}, H., \& {Mattei}, J.~A. 2001, \apjl,
  562, L141

\bibitem[{{Claverie} {et~al.}(1979){Claverie}, {Isaak}, {McLeod}, {van der
  Raay}, \& {Roca Cort\'es}}]{Claverie}
{Claverie}, A., {Isaak}, G.~R., {McLeod}, C.~P., {van der Raay}, H.~B., \&
  {Roca Cort\'es}, T. 1979, \nat, 282, 591

\bibitem[{{Cooley} \& {Tukey}(1965)}]{FFT}
{Cooley}, J.~W. \& {Tukey}, J.~W. 1965, Mathematics of Computing, 19, 297

\bibitem[{{Cox} \& {Whitney}(1958)}]{CoxWhitney}
{Cox}, J.~P. \& {Whitney}, C. 1958, \apj, 127, 561

\bibitem[{{Cram\'er}(1946)}]{Cramer}
{Cram\'er}, H. 1946, Mathematical Methods of Statistics (Princeton University
  Press)

\bibitem[{{Creevey} {et~al.}(2012){Creevey}, {Do{\u g}an}, {Frasca},
  {Thygesen}, {Basu}, {Bhattacharya}, {Biazzo}, {Brand{\~a}o}, {Bruntt},
  {Mazumdar}, {Niemczura}, {Shrotriya}, {Sousa}, {Stello}, {Subramaniam},
  {Campante}, {Handberg}, {Mathur}, {Bedding}, {Garc{\'{\i}}a}, {R{\'e}gulo},
  {Salabert}, {Molenda-{\.Z}akowicz}, {Quirion}, {White}, {Bonanno}, {Chaplin},
  {Christensen-Dalsgaard}, {Christiansen}, {Elsworth}, {Fanelli}, {Karoff},
  {Kinemuchi}, {Kjeldsen}, {Gai}, {Monteiro}, \& {Su{\'a}rez}}]{CreeveyFurry}
{Creevey}, O.~L., {Do{\u g}an}, G., {Frasca}, A., {et~al.} 2012, \aap, 537,
  A111

\bibitem[{{Cunha} {et~al.}(2007){Cunha}, {Aerts}, {Christensen-Dalsgaard},
  {Baglin}, {Bigot}, {Brown}, {Catala}, {Creevey}, {Domiciano de Souza},
  {Eggenberger}, {Garcia}, {Grundahl}, {Kervella}, {Kurtz}, {Mathias},
  {Miglio}, {Monteiro}, {Perrin}, {Pijpers}, {Pourbaix}, {Quirrenbach},
  {Rousselet-Perraut}, {Teixeira}, {Th{\'e}venin}, \&
  {Thompson}}]{Cunhareview07}
{Cunha}, M.~S., {Aerts}, C., {Christensen-Dalsgaard}, J., {et~al.} 2007, \aapr,
  14, 217

\bibitem[{{Cunha} \& {Brand{\~a}o}(2011)}]{CunhaBrandao}
{Cunha}, M.~S. \& {Brand{\~a}o}, I.~M. 2011, \aap, 529, A10

\bibitem[{{Cunha} \& {Metcalfe}(2007)}]{CunhaMetcalfe07}
{Cunha}, M.~S. \& {Metcalfe}, T.~S. 2007, \apj, 666, 413

\bibitem[{{Cuypers}(1987)}]{Cuypers87}
{Cuypers}, J. 1987, Medelingen van de Koninklijke Academie voor Wetenschappen,
  Letteren en Schone Kunsten van Belg\"{i}e, 49, 1

\bibitem[{{de Meulenaer} {et~al.}(2010){de Meulenaer}, {Carrier}, {Miglio},
  {Bedding}, {Campante}, {Eggenberger}, {Kjeldsen}, \&
  {Montalb{\'a}n}}]{deMeulenaer}
{de Meulenaer}, P., {Carrier}, F., {Miglio}, A., {et~al.} 2010, \aap, 523, A54

\bibitem[{{De Ridder} {et~al.}(2009){De Ridder}, {Barban}, {Baudin}, {Carrier},
  {Hatzes}, {Hekker}, {Kallinger}, {Weiss}, {Baglin}, {Auvergne}, {Samadi},
  {Barge}, \& {Deleuil}}]{DeRidder09}
{De Ridder}, J., {Barban}, C., {Baudin}, F., {et~al.} 2009, \nat, 459, 398

\bibitem[{{Deeming}(1975)}]{Deeming}
{Deeming}, T.~J. 1975, \apss, 36, 137

\bibitem[{{Deheuvels} {et~al.}(2010){Deheuvels}, {Bruntt}, {Michel}, {Barban},
  {Verner}, {R{\'e}gulo}, {Mosser}, {Mathur}, {Gaulme}, {Garcia}, {Boumier},
  {Appourchaux}, {Samadi}, {Catala}, {Baudin}, {Baglin}, {Auvergne},
  {Roxburgh}, \& {P{\'e}rez Hern{\'a}ndez}}]{Deheuvels10}
{Deheuvels}, S., {Bruntt}, H., {Michel}, E., {et~al.} 2010, \aap, 515, A87

\bibitem[{{Deheuvels} \& {Michel}(2010)}]{DeheuvelsMichel}
{Deheuvels}, S. \& {Michel}, E. 2010, \apss, 328, 259

\bibitem[{{Deubner} \& {Gough}(1984)}]{DeubnerGough}
{Deubner}, F. \& {Gough}, D.~O. 1984, \araa, 22, 593

\bibitem[{{di Mauro} {et~al.}(2003){di Mauro}, {Christensen-Dalsgaard},
  {Kjeldsen}, {Bedding}, \& {Patern{\`o}}}]{DiMauro03}
{di Mauro}, M.~P., {Christensen-Dalsgaard}, J., {Kjeldsen}, H., {Bedding},
  T.~R., \& {Patern{\`o}}, L. 2003, \aap, 404, 341

\bibitem[{{Do{\u g}an} {et~al.}(2010){Do{\u g}an}, {Bonanno}, {Bedding},
  {Campante}, {Christensen-Dalsgaard}, {Kjeldsen}, \& {et al.}}]{DoganProcyon}
{Do{\u g}an}, G., {Bonanno}, A., {Bedding}, T.~R., {et~al.} 2010, Astronomische
  Nachrichten, 331, 949

\bibitem[{{Dupret} {et~al.}(2005){Dupret}, {Grigahc{\`e}ne}, {Garrido},
  {Gabriel}, \& {Scuflaire}}]{DupretTDC}
{Dupret}, M.-A., {Grigahc{\`e}ne}, A., {Garrido}, R., {Gabriel}, M., \&
  {Scuflaire}, R. 2005, \aap, 435, 927

\bibitem[{{Duvall} \& {Harvey}(1986)}]{DuvallHarvey86}
{Duvall}, Jr., T.~L. \& {Harvey}, J.~W. 1986, in NATO ASIC Proc.~169:
  Seismology of the Sun and the Distant Stars, ed. {D.~O.~Gough}, 105--116

\bibitem[{{Duvall} {et~al.}(1993){Duvall}, {Jefferies}, {Harvey}, {Osaki}, \&
  {Pomerantz}}]{Duvall93}
{Duvall}, Jr., T.~L., {Jefferies}, S.~M., {Harvey}, J.~W., {Osaki}, Y., \&
  {Pomerantz}, M.~A. 1993, \apj, 410, 829

\bibitem[{{Dziembowski} \& {Soszy{\'n}ski}(2010)}]{Dziem10}
{Dziembowski}, W.~A. \& {Soszy{\'n}ski}, I. 2010, \aap, 524, A88

\bibitem[{{Earl} \& {Deem}(2005)}]{EarlDeem}
{Earl}, D.~J. \& {Deem}, M.~W. 2005, Physical Chemistry Chemical Physics
  (Incorporating Faraday Transactions), 7, 3910

\bibitem[{{Eddington}(1926)}]{EddingtonBook}
{Eddington}, A.~S. 1926, {The Internal Constitution of the Stars} (Cambridge
  University Press)

\bibitem[{{Ferraz-Mello}(1981)}]{DCDFT}
{Ferraz-Mello}, S. 1981, \aj, 86, 619

\bibitem[{{Fletcher} {et~al.}(2006){Fletcher}, {Chaplin}, {Elsworth}, {Schou},
  \& {Buzasi}}]{Fletcher06}
{Fletcher}, S.~T., {Chaplin}, W.~J., {Elsworth}, Y., {Schou}, J., \& {Buzasi},
  D.~L. 2006, \mnras, 371, 935

\bibitem[{{Foster}(1996)}]{Foster96}
{Foster}, G. 1996, \aj, 111, 541

\bibitem[{{Frandsen} {et~al.}(2007){Frandsen}, {Bruntt}, {Grundahl}, {Kopacki},
  {Kjeldsen}, {Arentoft}, {Stello}, {Bedding}, {Jacob}, {Gilliland}, {Edmonds},
  {Michel}, \& {Matthiesen}}]{Frandsen07}
{Frandsen}, S., {Bruntt}, H., {Grundahl}, F., {et~al.} 2007, \aap, 475, 991

\bibitem[{{Frandsen} {et~al.}(2002){Frandsen}, {Carrier}, {Aerts}, {Stello},
  {Maas}, {Burnet}, {Bruntt}, {Teixeira}, {de Medeiros}, {Bouchy}, {Kjeldsen},
  {Pijpers}, \& {Christensen-Dalsgaard}}]{FrandsenHya}
{Frandsen}, S., {Carrier}, F., {Aerts}, C., {et~al.} 2002, \aap, 394, L5

\bibitem[{{Fr{\"o}hlich} {et~al.}(1997){Fr{\"o}hlich}, {Andersen},
  {Appourchaux}, {Berthomieu}, {Crommelynck}, {Domingo}, {Fichot}, {Finsterle},
  {Gomez}, {Gough}, {Jimenez}, {Leifsen}, {Lombaerts}, {Pap}, {Provost},
  {Cortes}, {Romero}, {Roth}, {Sekii}, {Telljohann}, {Toutain}, \&
  {Wehrli}}]{VIRGO2}
{Fr{\"o}hlich}, C., {Andersen}, B.~N., {Appourchaux}, T., {et~al.} 1997,
  \solphys, 170, 1

\bibitem[{{Fr{\"o}hlich} {et~al.}(1995){Fr{\"o}hlich}, {Romero}, {Roth},
  {Wehrli}, {Andersen}, {Appourchaux}, {Domingo}, {Telljohann}, {Berthomieu},
  {Delache}, {Provost}, {Toutain}, {Crommelynck}, {Chevalier}, {Fichot},
  {D{\"a}ppen}, {Gough}, {Hoeksema}, {Jim{\'e}nez}, {G{\'o}mez}, {Herreros},
  {Cort{\'e}s}, {Jones}, {Pap}, \& {Willson}}]{VIRGO}
{Fr{\"o}hlich}, C., {Romero}, J., {Roth}, H., {et~al.} 1995, \solphys, 162, 101

\bibitem[{{Gabriel} {et~al.}(2002){Gabriel}, {Baudin}, {Boumier},
  {Garc{\'{\i}}a}, {Turck-Chi{\`e}ze}, {Appourchaux}, {Bertello}, {Berthomieu},
  {Charra}, {Gough}, {Pall{\'e}}, {Provost}, {Renaud}, {Robillot}, {Roca
  Cort{\'e}s}, {Thiery}, \& {Ulrich}}]{Gabriel02}
{Gabriel}, A.~H., {Baudin}, F., {Boumier}, P., {et~al.} 2002, \aap, 390, 1119

\bibitem[{{Gabriel} {et~al.}(1995){Gabriel}, {Grec}, {Charra}, {Robillot},
  {Roca Cort{\'e}s}, {Turck-Chi{\`e}ze}, {Bocchia}, {Boumier}, {Cantin},
  {Cesp{\'e}des}, {Cougrand}, {Cr{\'e}tolle}, {Dam{\'e}}, {Decaudin},
  {Delache}, {Denis}, {Duc}, {Dzitko}, {Fossat}, {Fourmond}, {Garc{\'{\i}}a},
  {Gough}, {Grivel}, {Herreros}, {Lagard{\`e}re}, {Moalic}, {Pall{\'e}},
  {P{\'e}trou}, {Sanchez}, {Ulrich}, \& {van der Raay}}]{GOLF}
{Gabriel}, A.~H., {Grec}, G., {Charra}, J., {et~al.} 1995, \solphys, 162, 61

\bibitem[{{Gabriel}(1994)}]{Gabriel94}
{Gabriel}, M. 1994, \aap, 287, 685

\bibitem[{{Gai} {et~al.}(2011){Gai}, {Basu}, {Chaplin}, \&
  {Elsworth}}]{Gaigrid}
{Gai}, N., {Basu}, S., {Chaplin}, W.~J., \& {Elsworth}, Y. 2011, \apj, 730, 63

\bibitem[{{Garc{\'{\i}}a}(2011)}]{Garciareview}
{Garc{\'{\i}}a}, R.~A. 2011, in Proceedings of the IX Scientific Meeting of the
  Spanish Astronomical Society, Highlights of Spanish Astrophysics VI, in press
  [arXiv:1101.0236v1]

\bibitem[{{Garc{\'{\i}}a} \& {Ballot}(2008)}]{backwards}
{Garc{\'{\i}}a}, R.~A. \& {Ballot}, J. 2008, \aap, 477, 611

\bibitem[{{Garc{\'{\i}}a} {et~al.}(2011{\natexlab{a}}){Garc{\'{\i}}a},
  {Ceillier}, {Campante}, {Davies}, {Mathur}, {Suarez}, {Ballot}, {Benomar},
  {Bonanno}, {Brun}, {Chaplin}, {Christensen-Dalsgaard}, {Deheuvels},
  {Elsworth}, {Handberg}, {Hekker}, {Jimenez}, {Karoff}, {Kjeldsen}, {Mathis},
  {Mosser}, {Palle}, {Pinsonneault}, {Regulo}, {Salabert}, {Silva Aguirre},
  {Stello}, {Thompson}, {Verner}, \& {the PE11 team of Kepler
  WG1}}]{Garciafast}
{Garc{\'{\i}}a}, R.~A., {Ceillier}, T., {Campante}, T.~L., {et~al.}
  2011{\natexlab{a}}, Astronomical Society of the Pacific, in press
  [arXiv:1109.6488v1]

\bibitem[{{Garc{\'{\i}}a} {et~al.}(2011{\natexlab{b}}){Garc{\'{\i}}a},
  {Hekker}, {Stello}, {Guti{\'e}rrez-Soto}, {Handberg}, {Huber}, {Karoff},
  {Uytterhoeven}, {Appourchaux}, {Chaplin}, {Elsworth}, {Mathur}, {Ballot},
  {Christensen-Dalsgaard}, {Gilliland}, {Houdek}, {Jenkins}, {Kjeldsen},
  {McCauliff}, {Metcalfe}, {Middour}, {Molenda-Zakowicz}, {Monteiro}, {Smith},
  \& {Thompson}}]{Garciacorrections}
{Garc{\'{\i}}a}, R.~A., {Hekker}, S., {Stello}, D., {et~al.}
  2011{\natexlab{b}}, \mnras, 414, L6

\bibitem[{{Garc{\'{\i}}a} {et~al.}(2010){Garc{\'{\i}}a}, {Mathur}, {Salabert},
  {Ballot}, {R{\'e}gulo}, {Metcalfe}, \& {Baglin}}]{Garciaactivity}
{Garc{\'{\i}}a}, R.~A., {Mathur}, S., {Salabert}, D., {et~al.} 2010, Science,
  329, 1032

\bibitem[{{Garc{\'{\i}}a} {et~al.}(2009){Garc{\'{\i}}a}, {R{\'e}gulo},
  {Samadi}, {Ballot}, {Barban}, {Benomar}, {Chaplin}, {Gaulme}, {Appourchaux},
  {Mathur}, {Mosser}, {Toutain}, {Verner}, {Auvergne}, {Baglin}, {Baudin},
  {Boumier}, {Bruntt}, {Catala}, {Deheuvels}, {Elsworth}, {Jim{\'e}nez-Reyes},
  {Michel}, {P{\'e}rez Hern{\'a}ndez}, {Roxburgh}, \& {Salabert}}]{Garcia09}
{Garc{\'{\i}}a}, R.~A., {R{\'e}gulo}, C., {Samadi}, R., {et~al.} 2009, \aap,
  506, 41

\bibitem[{{Gaulme} {et~al.}(2009){Gaulme}, {Appourchaux}, \& {Boumier}}]{MAP}
{Gaulme}, P., {Appourchaux}, T., \& {Boumier}, P. 2009, \aap, 506, 7

\bibitem[{{Gaulme} {et~al.}(2010){Gaulme}, {Deheuvels}, {Weiss}, {Mosser},
  {Moutou}, {Bruntt}, {Donati}, {Vannier}, {Guillot}, {Appourchaux}, {Michel},
  {Auvergne}, {Samadi}, {Baudin}, {Catala}, \& {Baglin}}]{HD46375}
{Gaulme}, P., {Deheuvels}, S., {Weiss}, W.~W., {et~al.} 2010, \aap, 524, A47

\bibitem[{{Gilliland}(2008)}]{Gilliland08}
{Gilliland}, R.~L. 2008, \aj, 136, 566

\bibitem[{{Gilliland} {et~al.}(2010{\natexlab{a}}){Gilliland}, {Brown},
  {Christensen-Dalsgaard}, {Kjeldsen}, {Aerts}, {Appourchaux}, {Basu},
  {Bedding}, {Chaplin}, {Cunha}, {De Cat}, {De Ridder}, {Guzik}, {Handler},
  {Kawaler}, {Kiss}, {Kolenberg}, {Kurtz}, {Metcalfe}, {Monteiro}, {Szab{\'o}},
  {Arentoft}, {Balona}, {Debosscher}, {Elsworth}, {Quirion}, {Stello},
  {Su{\'a}rez}, {Borucki}, {Jenkins}, {Koch}, {Kondo}, {Latham}, {Rowe}, \&
  {Steffen}}]{KAI}
{Gilliland}, R.~L., {Brown}, T.~M., {Christensen-Dalsgaard}, J., {et~al.}
  2010{\natexlab{a}}, \pasp, 122, 131

\bibitem[{{Gilliland} {et~al.}(1993){Gilliland}, {Brown}, {Kjeldsen},
  {McCarthy}, {Peri}, {Belmonte}, {Vidal}, {Cram}, {Palmer}, {Frandsen},
  {Parthasarathy}, {Petro}, {Schneider}, {Stetson}, \& {Weiss}}]{Gilliland93}
{Gilliland}, R.~L., {Brown}, T.~M., {Kjeldsen}, H., {et~al.} 1993, \aj, 106,
  2441

\bibitem[{{Gilliland} {et~al.}(2010{\natexlab{b}}){Gilliland}, {Jenkins},
  {Borucki}, {Bryson}, {Caldwell}, {Clarke}, {Dotson}, {Haas}, {Hall}, {Klaus},
  {Koch}, {McCauliff}, {Quintana}, {Twicken}, \& {van Cleve}}]{GillilandSC}
{Gilliland}, R.~L., {Jenkins}, J.~M., {Borucki}, W.~J., {et~al.}
  2010{\natexlab{b}}, \apjl, 713, L160

\bibitem[{{Gilliland} {et~al.}(2011){Gilliland}, {McCullough}, {Nelan},
  {Brown}, {Charbonneau}, {Nutzman}, {Christensen-Dalsgaard}, \&
  {Kjeldsen}}]{GillilandHST}
{Gilliland}, R.~L., {McCullough}, P.~R., {Nelan}, E.~P., {et~al.} 2011, \apj,
  726, 2

\bibitem[{{Gizon} \& {Solanki}(2003)}]{GizonSolanki}
{Gizon}, L. \& {Solanki}, S.~K. 2003, \apj, 589, 1009

\bibitem[{{Gizon} \& {Solanki}(2004)}]{GizonSolanki04}
{Gizon}, L. \& {Solanki}, S.~K. 2004, \solphys, 220, 169

\bibitem[{{Goldreich} \& {Keeley}(1977)}]{GoldreichKeeley}
{Goldreich}, P. \& {Keeley}, D.~A. 1977, \apj, 212, 243

\bibitem[{{Gough}(1986)}]{Gough86}
{Gough}, D.~O. 1986, in Hydrodynamic and Magnetodynamic Problems in the Sun and
  Stars, ed. {Y.~Osaki}, 117

\bibitem[{{Gough}(1993)}]{Gough93}
{Gough}, D.~O. 1993, in Astrophysical Fluid Dynamics -- Les Houches 1987, ed.
  {J.-P.~Zahn \& J.~Zinn-Justin}, 399--560

\bibitem[{{Grec} {et~al.}(1983){Grec}, {Fossat}, \& {Pomerantz}}]{Grec83}
{Grec}, G., {Fossat}, E., \& {Pomerantz}, M.~A. 1983, \solphys, 82, 55

\bibitem[{{Gregory}(2005)}]{GregoryBook}
{Gregory}, P.~C. 2005, {Bayesian Logical Data Analysis for the Physical
  Sciences: A Comparative Approach with `Mathematica' Support} (Cambridge
  University Press)

\bibitem[{{Grigahc{\`e}ne} {et~al.}(2010){Grigahc{\`e}ne}, {Antoci}, {Balona},
  {Catanzaro}, {Daszy{\'n}ska-Daszkiewicz}, {Guzik}, {Handler}, {Houdek},
  {Kurtz}, {Marconi}, {Monteiro}, {Moya}, {Ripepi}, {Su{\'a}rez},
  {Uytterhoeven}, {Borucki}, {Brown}, {Christensen-Dalsgaard}, {Gilliland},
  {Jenkins}, {Kjeldsen}, {Koch}, {Bernabei}, {Bradley}, {Breger}, {Di
  Criscienzo}, {Dupret}, {Garc{\'{\i}}a}, {Garc{\'{\i}}a Hern{\'a}ndez},
  {Jackiewicz}, {Kaiser}, {Lehmann}, {Mart{\'{\i}}n-Ruiz}, {Mathias},
  {Molenda-{\.Z}akowicz}, {Nemec}, {Nuspl}, {Papar{\'o}}, {Roth}, {Szab{\'o}},
  {Suran}, \& {Ventura}}]{Ahmedhybrids}
{Grigahc{\`e}ne}, A., {Antoci}, V., {Balona}, L.~A., {et~al.} 2010, \apjl, 713,
  L192

\bibitem[{{Grigahc{\`e}ne} {et~al.}(2005){Grigahc{\`e}ne}, {Dupret}, {Gabriel},
  {Garrido}, \& {Scuflaire}}]{AhmedTDC}
{Grigahc{\`e}ne}, A., {Dupret}, M.-A., {Gabriel}, M., {Garrido}, R., \&
  {Scuflaire}, R. 2005, \aap, 434, 1055

\bibitem[{{Grigahc{\`e}ne} {et~al.}(2011){Grigahc{\`e}ne}, {Dupret}, {Sousa},
  {Monteiro}, {Garrido}, {Scuflaire}, \& {Gabriel}}]{Grignonad}
{Grigahc{\`e}ne}, A., {Dupret}, M.-A., {Sousa}, S.~G., {et~al.} 2011, \apss, in
  press [arXiv:1112.5961v1]

\bibitem[{{Gruberbauer} {et~al.}(2009){Gruberbauer}, {Kallinger}, {Weiss}, \&
  {Guenther}}]{Gruberbauer09}
{Gruberbauer}, M., {Kallinger}, T., {Weiss}, W.~W., \& {Guenther}, D.~B. 2009,
  \aap, 506, 1043

\bibitem[{{Grundahl} {et~al.}(2009{\natexlab{a}}){Grundahl},
  {Christensen-Dalsgaard}, {Arentoft}, {Frandsen}, {Kjeldsen}, {J{\o}rgensen},
  \& {Kj{\ae}rgaard}}]{SONG}
{Grundahl}, F., {Christensen-Dalsgaard}, J., {Arentoft}, T., {et~al.}
  2009{\natexlab{a}}, Communications in Asteroseismology, 158, 345

\bibitem[{{Grundahl} {et~al.}(2009{\natexlab{b}}){Grundahl},
  {Christensen-Dalsgaard}, {Kjeldsen}, {J{\o}rgensen}, {Arentoft}, {Frandsen},
  \& {Kj{\ae}rgaard}}]{SONG2}
{Grundahl}, F., {Christensen-Dalsgaard}, J., {Kjeldsen}, H., {et~al.}
  2009{\natexlab{b}}, in Astronomical Society of the Pacific Conference Series,
  ed. {M.~Dikpati, T.~Arentoft, I.~Gonz{\'a}lez Hern{\'a}ndez, C.~Lindsey, \&
  F.~Hill}, Vol. 416, 579

\bibitem[{{Grundahl} {et~al.}(2007){Grundahl}, {Kjeldsen},
  {Christensen-Dalsgaard}, {Arentoft}, \& {Frandsen}}]{Grundahl07}
{Grundahl}, F., {Kjeldsen}, H., {Christensen-Dalsgaard}, J., {Arentoft}, T., \&
  {Frandsen}, S. 2007, Communications in Asteroseismology, 150, 300

\bibitem[{{Guenther} {et~al.}(2005){Guenther}, {Kallinger}, {Reegen}, {Weiss},
  {Matthews}, {Kuschnig}, {Marchenko}, {Moffat}, {Rucinski}, {Sasselov}, \&
  {Walker}}]{Guenther05}
{Guenther}, D.~B., {Kallinger}, T., {Reegen}, P., {et~al.} 2005, \apj, 635, 547

\bibitem[{{Handberg} \& {Campante}(2011)}]{HandCamp}
{Handberg}, R. \& {Campante}, T.~L. 2011, \aap, 527, A56

\bibitem[{{Harvey}(1985)}]{Harvey85}
{Harvey}, J.~W. 1985, in ESA Special Publication, Vol. 235, Future Missions in
  Solar, Heliospheric \& Space Plasma Physics, ed. {E.~Rolfe \& B.~Battrick},
  199--208

\bibitem[{{Harvey}(1988)}]{Harvey88}
{Harvey}, J.~W. 1988, in IAU Symposium, Vol. 123, Advances in Helio- and
  Asteroseismology, ed. {J.~Christensen-Dalsgaard \& S.~Frandsen}, 497

\bibitem[{{Harvey} {et~al.}(1993){Harvey}, {Duvall}, {Jefferies}, \&
  {Pomerantz}}]{Harvey93}
{Harvey}, J.~W., {Duvall}, Jr., T.~L., {Jefferies}, S.~M., \& {Pomerantz},
  M.~A. 1993, in Astronomical Society of the Pacific Conference Series,
  Vol.~42, GONG 1992. Seismic Investigation of the Sun and Stars, ed.
  {T.~M.~Brown}, 111

\bibitem[{{Hastings}(1970)}]{Hastings}
{Hastings}, W.~K. 1970, Biometrika, 57, 97

\bibitem[{{Hekker} {et~al.}(2011{\natexlab{a}}){Hekker}, {Basu}, {Stello},
  {Kallinger}, {Grundahl}, {Mathur}, {Garc{\'{\i}}a}, {Mosser}, {Huber},
  {Bedding}, {Szab{\'o}}, {De Ridder}, {Chaplin}, {Elsworth}, {Hale},
  {Christensen-Dalsgaard}, {Gilliland}, {Still}, {McCauliff}, \&
  {Quintana}}]{HekkerRGclusters}
{Hekker}, S., {Basu}, S., {Stello}, D., {et~al.} 2011{\natexlab{a}}, \aap, 530,
  A100

\bibitem[{{Hekker} {et~al.}(2010){Hekker}, {Broomhall}, {Chaplin}, {Elsworth},
  {Fletcher}, {New}, {Arentoft}, {Quirion}, \& {Kjeldsen}}]{Hekkerpipeline}
{Hekker}, S., {Broomhall}, A.-M., {Chaplin}, W.~J., {et~al.} 2010, \mnras, 402,
  2049

\bibitem[{{Hekker} {et~al.}(2011{\natexlab{b}}){Hekker}, {Elsworth}, {De
  Ridder}, {Mosser}, {Garc{\'{\i}}a}, {Kallinger}, {Mathur}, {Huber}, {Buzasi},
  {Preston}, {Hale}, {Ballot}, {Chaplin}, {R{\'e}gulo}, {Bedding}, {Stello},
  {Borucki}, {Koch}, {Jenkins}, {Allen}, {Gilliland}, {Kjeldsen}, \&
  {Christensen-Dalsgaard}}]{HekkerRG11}
{Hekker}, S., {Elsworth}, Y., {De Ridder}, J., {et~al.} 2011{\natexlab{b}},
  \aap, 525, A131

\bibitem[{{Hekker} {et~al.}(2009){Hekker}, {Kallinger}, {Baudin}, {De Ridder},
  {Barban}, {Carrier}, {Hatzes}, {Weiss}, \& {Baglin}}]{HekkerRG09}
{Hekker}, S., {Kallinger}, T., {Baudin}, F., {et~al.} 2009, \aap, 506, 465

\bibitem[{{Hoffleit}(1997)}]{Mira}
{Hoffleit}, D. 1997, Journal of the American Association of Variable Star
  Observers (JAAVSO), 25, 115

\bibitem[{{H{\"o}gbom}(1974)}]{Hogbom74}
{H{\"o}gbom}, J.~A. 1974, \aaps, 15, 417

\bibitem[{{Horne} \& {Baliunas}(1986)}]{HorneBaliunas}
{Horne}, J.~H. \& {Baliunas}, S.~L. 1986, \apj, 302, 757

\bibitem[{{Houdek}(2006)}]{Houdek06}
{Houdek}, G. 2006, in ESA Special Publication, Vol. 624, Proceedings of
  \emph{SOHO} 18/GONG 2006/HELAS I, Beyond the spherical Sun

\bibitem[{{Houdek} {et~al.}(1999){Houdek}, {Balmforth},
  {Christensen-Dalsgaard}, \& {Gough}}]{Houdek99}
{Houdek}, G., {Balmforth}, N.~J., {Christensen-Dalsgaard}, J., \& {Gough},
  D.~O. 1999, \aap, 351, 582

\bibitem[{{Houdek} \& {Gough}(2007)}]{HoudekGough07}
{Houdek}, G. \& {Gough}, D.~O. 2007, \mnras, 375, 861

\bibitem[{{Howe}(2009)}]{Howe09}
{Howe}, R. 2009, Living Reviews in Solar Physics, 6, 1

\bibitem[{{Huber} {et~al.}(2011{\natexlab{a}}){Huber}, {Bedding}, {Arentoft},
  {Gruberbauer}, {Guenther}, {Houdek}, {Kallinger}, {Kjeldsen}, {Matthews},
  {Stello}, \& {Weiss}}]{HuberProcyon}
{Huber}, D., {Bedding}, T.~R., {Arentoft}, T., {et~al.} 2011{\natexlab{a}},
  \apj, 731, 94

\bibitem[{{Huber} {et~al.}(2011{\natexlab{b}}){Huber}, {Bedding}, {Stello},
  {Hekker}, {Mathur}, {Mosser}, {Verner}, {Bonanno}, {Buzasi}, {Campante},
  {Elsworth}, {Hale}, {Kallinger}, {Silva Aguirre}, {Chaplin}, {De Ridder},
  {Garcia}, {Appourchaux}, {Frandsen}, {Houdek}, {Molenda-Zakowicz},
  {Monteiro}, {Christensen-Dalsgaard}, {Gilliland}, {Kawaler}, {Kjeldsen},
  {Broomhall}, {Corsaro}, {Salabert}, {Sanderfer}, {Seader}, \&
  {Smith}}]{Huberscaling}
{Huber}, D., {Bedding}, T.~R., {Stello}, D., {et~al.} 2011{\natexlab{b}}, \apj,
  in press [arXiv:1109.3460v1]

\bibitem[{{Huber} {et~al.}(2010){Huber}, {Bedding}, {Stello}, {Mosser},
  {Mathur}, {Kallinger}, {Hekker}, {Elsworth}, {Buzasi}, {De Ridder},
  {Gilliland}, {Kjeldsen}, {Chaplin}, {Garc{\'{\i}}a}, {Hale}, {Preston},
  {White}, {Borucki}, {Christensen-Dalsgaard}, {Clarke}, {Jenkins}, \&
  {Koch}}]{HuberRG10}
{Huber}, D., {Bedding}, T.~R., {Stello}, D., {et~al.} 2010, \apj, 723, 1607

\bibitem[{{Huber} {et~al.}(2009){Huber}, {Stello}, {Bedding}, {Chaplin},
  {Arentoft}, {Quirion}, \& {Kjeldsen}}]{Huberpipeline}
{Huber}, D., {Stello}, D., {Bedding}, T.~R., {et~al.} 2009, Communications in
  Asteroseismology, 160, 74

\bibitem[{{Ireland} {et~al.}(2008){Ireland}, {M{\'e}rand}, {ten Brummelaar},
  {Tuthill}, {Schaefer}, {Turner}, {Sturmann}, {Sturmann}, \&
  {McAlister}}]{PAVO}
{Ireland}, M.~J., {M{\'e}rand}, A., {ten Brummelaar}, T.~A., {et~al.} 2008, in
  Society of Photo-Optical Instrumentation Engineers Conference Series, Vol.
  7013, Presented at the Society of Photo-Optical Instrumentation Engineers
  (SPIE) Conference

\bibitem[{{Jaynes}(1987)}]{Jaynes87}
{Jaynes}, E.~T. 1987, in {Maximum-Entropy and Bayesian Analysis and Estimation
  Problems. Proceedings of the Third Workshop on Maximum Entropy and Bayesian
  Methods in Applied Statistics}, ed. C.~R. {Smith} \& G.~J. {Erickson}
  (Reidel, Dordrecht, Netherlands), 1--37

\bibitem[{{Jeffreys}(1961)}]{Jeffreys61}
{Jeffreys}, H. 1961, Theory of Probability, 3rd edn. (Oxford University Press)

\bibitem[{{Jenkins} \& {Watts}(1968)}]{JenkinsWatts}
{Jenkins}, G.~M. \& {Watts}, D.~G. 1968, {Spectral Analysis and its
  Applications} (San Francisco: Holden Day)

\bibitem[{{Jenkins} {et~al.}(2010){Jenkins}, {Caldwell}, {Chandrasekaran},
  {Twicken}, {Bryson}, {Quintana}, {Clarke}, {Li}, {Allen}, {Tenenbaum}, {Wu},
  {Klaus}, {Van Cleve}, {Dotson}, {Haas}, {Gilliland}, {Koch}, \&
  {Borucki}}]{JenkinsLC}
{Jenkins}, J.~M., {Caldwell}, D.~A., {Chandrasekaran}, H., {et~al.} 2010,
  \apjl, 713, L120

\bibitem[{{Jim{\'e}nez-Reyes} {et~al.}(2008){Jim{\'e}nez-Reyes}, {Chaplin},
  {Garc{\'{\i}}a}, {Appourchaux}, {Baudin}, {Boumier}, {Elsworth}, {Fletcher},
  {Lazrek}, {Leibacher}, {Lochard}, {New}, {R{\'e}gulo}, {Salabert}, {Toutain},
  {Verner}, \& {Wachter}}]{localfit}
{Jim{\'e}nez-Reyes}, S.~J., {Chaplin}, W.~J., {Garc{\'{\i}}a}, R.~A., {et~al.}
  2008, \mnras, 389, 1780

\bibitem[{{Kallinger} {et~al.}(2010{\natexlab{a}}){Kallinger}, {Mosser},
  {Hekker}, {Huber}, {Stello}, {Mathur}, {Basu}, {Bedding}, {Chaplin}, {De
  Ridder}, {Elsworth}, {Frandsen}, {Garc{\'{\i}}a}, {Gruberbauer}, {Matthews},
  {Borucki}, {Bruntt}, {Christensen-Dalsgaard}, {Gilliland}, {Kjeldsen}, \&
  {Koch}}]{KallingerRG10}
{Kallinger}, T., {Mosser}, B., {Hekker}, S., {et~al.} 2010{\natexlab{a}}, \aap,
  522, A1

\bibitem[{{Kallinger} {et~al.}(2010{\natexlab{b}}){Kallinger}, {Weiss},
  {Barban}, {Baudin}, {Cameron}, {Carrier}, {De Ridder}, {Goupil},
  {Gruberbauer}, {Hatzes}, {Hekker}, {Samadi}, \& {Deleuil}}]{KallingerRGCorot}
{Kallinger}, T., {Weiss}, W.~W., {Barban}, C., {et~al.} 2010{\natexlab{b}},
  \aap, 509, A77

\bibitem[{{Karoff}(2012)}]{faculaeKaroff}
{Karoff}, C. 2012, \mnras, in press [arXiv:1201.2539v1]

\bibitem[{{Karoff} {et~al.}(2010){Karoff}, {Campante}, \&
  {Chaplin}}]{Karoffpipeline}
{Karoff}, C., {Campante}, T.~L., \& {Chaplin}, W.~J. 2010, Astronomische
  Nachrichten, in press [arXiv:1003.4167v1]

\bibitem[{{Karoff} {et~al.}(2009){Karoff}, {Metcalfe}, {Chaplin}, {Elsworth},
  {Kjeldsen}, {Arentoft}, \& {Buzasi}}]{Karoffcycle}
{Karoff}, C., {Metcalfe}, T.~S., {Chaplin}, W.~J., {et~al.} 2009, \mnras, 399,
  914

\bibitem[{{Kendall} \& {Stuart}(1979)}]{KendallStuart}
{Kendall}, M. \& {Stuart}, A. 1979, {The Advanced Theory of Statistics. Vol.~2:
  Inference and Relationship}, 4th edn. (Macmillan)

\bibitem[{{Khintchine}(1934)}]{Khintchine}
{Khintchine}, A. 1934, Mathematische Annalen, 109, 604

\bibitem[{{Kiss} \& {Bedding}(2003)}]{KissBedding}
{Kiss}, L.~L. \& {Bedding}, T.~R. 2003, \mnras, 343, L79

\bibitem[{{Kjeldsen} \& {Bedding}(1995)}]{KjeldsenBedding95}
{Kjeldsen}, H. \& {Bedding}, T.~R. 1995, \aap, 293, 87

\bibitem[{{Kjeldsen} \& {Bedding}(2011)}]{KB11}
{Kjeldsen}, H. \& {Bedding}, T.~R. 2011, \aap, 529, L8

\bibitem[{{Kjeldsen} {et~al.}(2008{\natexlab{a}}){Kjeldsen}, {Bedding},
  {Arentoft}, {Butler}, {Dall}, {Karoff}, {Kiss}, {Tinney}, \&
  {Chaplin}}]{Kjeldsen08}
{Kjeldsen}, H., {Bedding}, T.~R., {Arentoft}, T., {et~al.} 2008{\natexlab{a}},
  \apj, 682, 1370

\bibitem[{{Kjeldsen} {et~al.}(2003){Kjeldsen}, {Bedding}, {Baldry}, {Bruntt},
  {Butler}, {Fischer}, {Frandsen}, {Gates}, {Grundahl}, {Lang}, {Marcy},
  {Misch}, \& {Vogt}}]{Kjeldsen03etaBoo}
{Kjeldsen}, H., {Bedding}, T.~R., {Baldry}, I.~K., {et~al.} 2003, \aj, 126,
  1483

\bibitem[{{Kjeldsen} {et~al.}(2005){Kjeldsen}, {Bedding}, {Butler},
  {Christensen-Dalsgaard}, {Kiss}, {McCarthy}, {Marcy}, {Tinney}, \&
  {Wright}}]{Kjeldsen05}
{Kjeldsen}, H., {Bedding}, T.~R., {Butler}, R.~P., {et~al.} 2005, \apj, 635,
  1281

\bibitem[{{Kjeldsen} {et~al.}(2008{\natexlab{b}}){Kjeldsen}, {Bedding}, \&
  {Christensen-Dalsgaard}}]{Kjeldsen08corrections}
{Kjeldsen}, H., {Bedding}, T.~R., \& {Christensen-Dalsgaard}, J.
  2008{\natexlab{b}}, \apjl, 683, L175

\bibitem[{{Kjeldsen} {et~al.}(1995){Kjeldsen}, {Bedding}, {Viskum}, \&
  {Frandsen}}]{Kjeldsen95etaBoo}
{Kjeldsen}, H., {Bedding}, T.~R., {Viskum}, M., \& {Frandsen}, S. 1995, \aj,
  109, 1313

\bibitem[{{Koch} {et~al.}(2010){Koch}, {Borucki}, {Basri}, {Batalha}, {Brown},
  {Caldwell}, {Christensen-Dalsgaard}, {Cochran}, {DeVore}, {Dunham},
  {Gautier}, {Geary}, {Gilliland}, {Gould}, {Jenkins}, {Kondo}, {Latham},
  {Lissauer}, {Marcy}, {Monet}, {Sasselov}, {Boss}, {Brownlee}, {Caldwell},
  {Dupree}, {Howell}, {Kjeldsen}, {Meibom}, {Morrison}, {Owen}, {Reitsema},
  {Tarter}, {Bryson}, {Dotson}, {Gazis}, {Haas}, {Kolodziejczak}, {Rowe}, {Van
  Cleve}, {Allen}, {Chandrasekaran}, {Clarke}, {Li}, {Quintana}, {Tenenbaum},
  {Twicken}, \& {Wu}}]{Kepler2}
{Koch}, D.~G., {Borucki}, W.~J., {Basri}, G., {et~al.} 2010, \apjl, 713, L79

\bibitem[{{Koen}(1999)}]{Koen99}
{Koen}, C. 1999, \mnras, 309, 769

\bibitem[{{Kumar} {et~al.}(1988){Kumar}, {Franklin}, \& {Goldreich}}]{Kumar88}
{Kumar}, P., {Franklin}, J., \& {Goldreich}, P. 1988, \apj, 328, 879

\bibitem[{{Lamb}(1932)}]{Lamb32}
{Lamb}, H. 1932, {Hydrodynamics}, 6th edn. (Dover Publications)

\bibitem[{{Leavitt}(1908)}]{Leavitt1908}
{Leavitt}, H.~S. 1908, Annals of Harvard College Observatory, 60, 87

\bibitem[{{Leavitt} \& {Pickering}(1912)}]{LeavittPickering}
{Leavitt}, H.~S. \& {Pickering}, E.~C. 1912, Harvard College Observatory
  Circular, 173, 1

\bibitem[{{Ledoux}(1951)}]{Ledoux51}
{Ledoux}, P. 1951, \apj, 114, 373

\bibitem[{{Leighton} {et~al.}(1962){Leighton}, {Noyes}, \&
  {Simon}}]{Leighton62}
{Leighton}, R.~B., {Noyes}, R.~W., \& {Simon}, G.~W. 1962, \apj, 135, 474

\bibitem[{{Libbrecht}(1992)}]{Libbrecht92}
{Libbrecht}, K.~G. 1992, \apj, 387, 712

\bibitem[{{Liddle}(2009)}]{Liddle09}
{Liddle}, A.~R. 2009, Annual Review of Nuclear and Particle Science, 59, 95

\bibitem[{{Lomb}(1976)}]{Lomb76}
{Lomb}, N.~R. 1976, \apss, 39, 447

\bibitem[{{Marti{\'c}} {et~al.}(1999){Marti{\'c}}, {Schmitt}, {Lebrun},
  {Barban}, {Connes}, {Bouchy}, {Michel}, {Baglin}, {Appourchaux}, \&
  {Bertaux}}]{Martic99}
{Marti{\'c}}, M., {Schmitt}, J., {Lebrun}, J., {et~al.} 1999, \aap, 351, 993

\bibitem[{{Mathur} {et~al.}(2010{\natexlab{a}}){Mathur}, {Garc{\'{\i}}a},
  {Catala}, {Bruntt}, {Mosser}, {Appourchaux}, {Ballot}, {Creevey}, {Gaulme},
  {Hekker}, {Huber}, {Karoff}, {Piau}, {R{\'e}gulo}, {Roxburgh}, {Salabert},
  {Verner}, {Auvergne}, {Baglin}, {Chaplin}, {Elsworth}, {Michel}, {Samadi},
  {Sato}, \& {Stello}}]{Mathur10}
{Mathur}, S., {Garc{\'{\i}}a}, R.~A., {Catala}, C., {et~al.}
  2010{\natexlab{a}}, \aap, 518, A53

\bibitem[{{Mathur} {et~al.}(2010{\natexlab{b}}){Mathur}, {Garc{\'{\i}}a},
  {R{\'e}gulo}, {Creevey}, {Ballot}, {Salabert}, {Arentoft}, {Quirion},
  {Chaplin}, \& {Kjeldsen}}]{Mathurpipeline}
{Mathur}, S., {Garc{\'{\i}}a}, R.~A., {R{\'e}gulo}, C., {et~al.}
  2010{\natexlab{b}}, \aap, 511, A46

\bibitem[{{Mathur} {et~al.}(2011){Mathur}, {Handberg}, {Campante},
  {Garc{\'{\i}}a}, {Appourchaux}, {Bedding}, {Mosser}, {Chaplin}, {Ballot},
  {Benomar}, {Bonanno}, {Corsaro}, {Gaulme}, {Hekker}, {R{\'e}gulo},
  {Salabert}, {Verner}, {White}, {Brand{\~a}o}, {Creevey}, {Do{\u g}an},
  {Elsworth}, {Huber}, {Hale}, {Houdek}, {Karoff}, {Metcalfe},
  {Molenda-{\.Z}akowicz}, {Monteiro}, {Thompson}, {Christensen-Dalsgaard},
  {Gilliland}, {Kawaler}, {Kjeldsen}, {Quintana}, {Sanderfer}, \&
  {Seader}}]{BoogieTigger}
{Mathur}, S., {Handberg}, R., {Campante}, T.~L., {et~al.} 2011, \apj, 733, 95

\bibitem[{{Matthews} {et~al.}(2004){Matthews}, {Kuschnig}, {Guenther},
  {Walker}, {Moffat}, {Rucinski}, {Sasselov}, \& {Weiss}}]{MatthewsProcyon}
{Matthews}, J.~M., {Kuschnig}, R., {Guenther}, D.~B., {et~al.} 2004, \nat, 430,
  51

\bibitem[{{Mayor} {et~al.}(2003){Mayor}, {Pepe}, {Queloz}, {Bouchy},
  {Rupprecht}, {Lo Curto}, {Avila}, {Benz}, {Bertaux}, {Bonfils}, {Dall},
  {Dekker}, {Delabre}, {Eckert}, {Fleury}, {Gilliotte}, {Gojak}, {Guzman},
  {Kohler}, {Lizon}, {Longinotti}, {Lovis}, {Megevand}, {Pasquini}, {Reyes},
  {Sivan}, {Sosnowska}, {Soto}, {Udry}, {van Kesteren}, {Weber}, \&
  {Weilenmann}}]{HARPS}
{Mayor}, M., {Pepe}, F., {Queloz}, D., {et~al.} 2003, The Messenger, 114, 20

\bibitem[{{Mazumdar} {et~al.}(2006){Mazumdar}, {Basu}, {Collier}, \&
  {Demarque}}]{Mazumdar06}
{Mazumdar}, A., {Basu}, S., {Collier}, B.~L., \& {Demarque}, P. 2006, \mnras,
  372, 949

\bibitem[{{Mazumdar} \& {Michel}(2010)}]{Mazumdar10}
{Mazumdar}, A. \& {Michel}, E. 2010, Astronomische Nachrichten, in press
  [arXiv:1004.2739v1]

\bibitem[{{Metcalfe} {et~al.}(2007){Metcalfe}, {Dziembowski}, {Judge}, \&
  {Snow}}]{Metcalfecycle}
{Metcalfe}, T.~S., {Dziembowski}, W.~A., {Judge}, P.~G., \& {Snow}, M. 2007,
  \mnras, 379, L16

\bibitem[{{Metcalfe} {et~al.}(2010){Metcalfe}, {Monteiro}, {Thompson},
  {Molenda-{\.Z}akowicz}, {Appourchaux}, {Chaplin}, {Do{\u g}an},
  {Eggenberger}, {Bedding}, {Bruntt}, {Creevey}, {Quirion}, {Stello},
  {Bonanno}, {Silva Aguirre}, {Basu}, {Esch}, {Gai}, {Di Mauro}, {Kosovichev},
  {Kitiashvili}, {Su{\'a}rez}, {Moya}, {Piau}, {Garc{\'{\i}}a}, {Marques},
  {Frasca}, {Biazzo}, {Sousa}, {Dreizler}, {Bazot}, {Karoff}, {Frandsen},
  {Wilson}, {Brown}, {Christensen-Dalsgaard}, {Gilliland}, {Kjeldsen},
  {Campante}, {Fletcher}, {Handberg}, {R{\'e}gulo}, {Salabert}, {Schou},
  {Verner}, {Ballot}, {Broomhall}, {Elsworth}, {Hekker}, {Huber}, {Mathur},
  {New}, {Roxburgh}, {Sato}, {White}, {Borucki}, {Koch}, \&
  {Jenkins}}]{MetcalfeGemma}
{Metcalfe}, T.~S., {Monteiro}, M.~J.~P.~F.~G., {Thompson}, M.~J., {et~al.}
  2010, \apj, 723, 1583

\bibitem[{{Metropolis} {et~al.}(1953){Metropolis}, {Rosenbluth}, {Rosenbluth},
  {Teller}, \& {Teller}}]{Metropolis}
{Metropolis}, N., {Rosenbluth}, A.~W., {Rosenbluth}, M.~N., {Teller}, A.~H., \&
  {Teller}, E. 1953, \jcp, 21, 1087

\bibitem[{{Michel} {et~al.}(2008){Michel}, {Baglin}, {Auvergne}, {Catala},
  {Samadi}, {Baudin}, {Appourchaux}, {Barban}, {Weiss}, {Berthomieu},
  {Boumier}, {Dupret}, {Garcia}, {Fridlund}, {Garrido}, {Goupil}, {Kjeldsen},
  {Lebreton}, {Mosser}, {Grotsch-Noels}, {Janot-Pacheco}, {Provost},
  {Roxburgh}, {Thoul}, {Toutain}, {Tiph{\`e}ne}, {Turck-Chieze}, {Vauclair},
  {Vauclair}, {Aerts}, {Alecian}, {Ballot}, {Charpinet}, {Hubert},
  {Ligni{\`e}res}, {Mathias}, {Monteiro}, {Neiner}, {Poretti}, {Renan de
  Medeiros}, {Ribas}, {Rieutord}, {Cort{\'e}s}, \& {Zwintz}}]{Michel08}
{Michel}, E., {Baglin}, A., {Auvergne}, M., {et~al.} 2008, Science, 322, 558

\bibitem[{{Miglio} \& {Montalb{\'a}n}(2005)}]{MiglioMont}
{Miglio}, A. \& {Montalb{\'a}n}, J. 2005, \aap, 441, 615

\bibitem[{{Miglio} {et~al.}(2009){Miglio}, {Montalb{\'a}n}, {Baudin},
  {Eggenberger}, {Noels}, {Hekker}, {De Ridder}, {Weiss}, \&
  {Baglin}}]{MiglioRG09}
{Miglio}, A., {Montalb{\'a}n}, J., {Baudin}, F., {et~al.} 2009, \aap, 503, L21

\bibitem[{{Miglio} {et~al.}(2010){Miglio}, {Montalb{\'a}n}, {Carrier}, {De
  Ridder}, {Mosser}, {Eggenberger}, {Scuflaire}, {Ventura}, {D'Antona},
  {Noels}, \& {Baglin}}]{Miglio10}
{Miglio}, A., {Montalb{\'a}n}, J., {Carrier}, F., {et~al.} 2010, \aap, 520, L6

\bibitem[{{Monteiro} {et~al.}(2000){Monteiro}, {Christensen-Dalsgaard}, \&
  {Thompson}}]{Monteiro00}
{Monteiro}, M.~J.~P.~F.~G., {Christensen-Dalsgaard}, J., \& {Thompson}, M.~J.
  2000, \mnras, 316, 165

\bibitem[{{Monteiro} {et~al.}(2002){Monteiro}, {Christensen-Dalsgaard}, \&
  {Thompson}}]{Monteiro02}
{Monteiro}, M.~J.~P.~F.~G., {Christensen-Dalsgaard}, J., \& {Thompson}, M.~J.
  2002, in ESA Special Publication, Vol. 485, Stellar Structure and Habitable
  Planet Finding, ed. {B.~Battrick, F.~Favata, I.~W.~Roxburgh, \& D.~Galadi},
  291--298

\bibitem[{{Montgomery} \& {O'Donoghue}(1999)}]{MontDon}
{Montgomery}, M.~H. \& {O'Donoghue}, D. 1999, Delta Scuti Star Newsletter, 13,
  28

\bibitem[{{Moreira} {et~al.}(2005){Moreira}, {Appourchaux}, {Berthomieu}, \&
  {Toutain}}]{Moreira2005}
{Moreira}, O., {Appourchaux}, T., {Berthomieu}, G., \& {Toutain}, T. 2005,
  \mnras, 357, 191

\bibitem[{{Mosser} \& {Appourchaux}(2009)}]{Mosserpipeline}
{Mosser}, B. \& {Appourchaux}, T. 2009, \aap, 508, 877

\bibitem[{{Mosser} {et~al.}(2008){Mosser}, {Appourchaux}, {Catala}, {Buey}, \&
  {the SIAMOIS Team}}]{SIAMOIS}
{Mosser}, B., {Appourchaux}, T., {Catala}, C., {Buey}, J., \& {the SIAMOIS
  Team}. 2008, Journal of Physics Conference Series, 118, 012042

\bibitem[{{Mosser} {et~al.}(2011{\natexlab{a}}){Mosser}, {Barban},
  {Montalb{\'a}n}, {Beck}, {Miglio}, {Belkacem}, {Goupil}, {Hekker}, {De
  Ridder}, {Dupret}, {Elsworth}, {Noels}, {Baudin}, {Michel}, {Samadi},
  {Auvergne}, {Baglin}, \& {Catala}}]{MosserRGmixed}
{Mosser}, B., {Barban}, C., {Montalb{\'a}n}, J., {et~al.} 2011{\natexlab{a}},
  \aap, 532, A86

\bibitem[{{Mosser} {et~al.}(2011{\natexlab{b}}){Mosser}, {Belkacem}, {Goupil},
  {Michel}, {Elsworth}, {Barban}, {Kallinger}, {Hekker}, {De Ridder}, {Samadi},
  {Baudin}, {Pinheiro}, {Auvergne}, {Baglin}, \& {Catala}}]{MosserRGuni}
{Mosser}, B., {Belkacem}, K., {Goupil}, M.-J., {et~al.} 2011{\natexlab{b}},
  \aap, 525, L9

\bibitem[{{Mosser} {et~al.}(2010){Mosser}, {Belkacem}, {Goupil}, {Miglio},
  {Morel}, {Barban}, {Baudin}, {Hekker}, {Samadi}, {De Ridder}, {Weiss},
  {Auvergne}, \& {Baglin}}]{MosserRG}
{Mosser}, B., {Belkacem}, K., {Goupil}, M.-J., {et~al.} 2010, \aap, 517, A22

\bibitem[{{Mosser} {et~al.}(2005){Mosser}, {Bouchy}, {Catala}, {Michel},
  {Samadi}, {Th{\'e}venin}, {Eggenberger}, {Sosnowska}, {Moutou}, \&
  {Baglin}}]{Mosser05}
{Mosser}, B., {Bouchy}, F., {Catala}, C., {et~al.} 2005, \aap, 431, L13

\bibitem[{{Mosser} {et~al.}(2009){Mosser}, {Michel}, {Appourchaux}, {Barban},
  {Baudin}, {Boumier}, {Bruntt}, {Catala}, {Deheuvels}, {Garc{\'{\i}}a},
  {Gaulme}, {Regulo}, {Roxburgh}, {Samadi}, {Verner}, {Auvergne}, {Baglin},
  {Ballot}, {Benomar}, \& {Mathur}}]{Mosser09}
{Mosser}, B., {Michel}, E., {Appourchaux}, T., {et~al.} 2009, \aap, 506, 33

\bibitem[{{Naef} {et~al.}(2001){Naef}, {Mayor}, {Pepe}, {Queloz}, {Santos},
  {Udry}, \& {Burnet}}]{Naefplanet}
{Naef}, D., {Mayor}, M., {Pepe}, F., {et~al.} 2001, \aap, 375, 205

\bibitem[{{Neyman} \& {Pearson}(1933)}]{NeymanPearson}
{Neyman}, J. \& {Pearson}, E.~S. 1933, Royal Society of London Philosophical
  Transactions Series A, 231, 289

\bibitem[{{Nigam} \& {Kosovichev}(1998)}]{NigKos98}
{Nigam}, R. \& {Kosovichev}, A.~G. 1998, \apjl, 505, L51

\bibitem[{{Nigam} {et~al.}(1998){Nigam}, {Kosovichev}, {Scherrer}, \&
  {Schou}}]{Nigam98}
{Nigam}, R., {Kosovichev}, A.~G., {Scherrer}, P.~H., \& {Schou}, J. 1998,
  \apjl, 495, L115

\bibitem[{{North} {et~al.}(2007){North}, {Davis}, {Bedding}, {Ireland},
  {Jacob}, {O'Byrne}, {Owens}, {Robertson}, {Tango}, \& {Tuthill}}]{North}
{North}, J.~R., {Davis}, J., {Bedding}, T.~R., {et~al.} 2007, \mnras, 380, L80

\bibitem[{{Nyquist}(1928)}]{Nyquist}
{Nyquist}, H. 1928, Transactions of the AIEE, 47, 617

\bibitem[{{Osaki}(1975)}]{Osaki75}
{Osaki}, J. 1975, \pasj, 27, 237

\bibitem[{{Parseval des Ch\^enes}(1806)}]{Parseval}
{Parseval des Ch\^enes}, M.-A. 1806, M\'emoires pr\'esent\'es \`a l'Institut
  des Sciences, Lettres et Arts, pars divers savans, et lus dans ses
  assembl\'ees. Sciences, math\'ematiques et physiques. (Savans \'etrangers.),
  1, 638

\bibitem[{{Peligrad} \& {Wu}(2010)}]{PeligradWu}
{Peligrad}, M. \& {Wu}, W.~B. 2010, Annals of Probability, 38, 2009

\bibitem[{{Pereira} \& {Lopes}(2005)}]{PereiraLopes05}
{Pereira}, T.~M.~D. \& {Lopes}, I.~P. 2005, \apj, 622, 1068

\bibitem[{{Pereira} {et~al.}(2007){Pereira}, {Su{\'a}rez}, {Lopes},
  {Mart{\'{\i}}n-Ruiz}, {Amado}, {Garrido}, {Rodr{\'{\i}}guez}, {Costa},
  {Rolland}, {Arellano Ferro}, \& {Sareyan}}]{PereiraLopes07}
{Pereira}, T.~M.~D., {Su{\'a}rez}, J.~C., {Lopes}, I.~P., {et~al.} 2007, \aap,
  464, 659

\bibitem[{{Pourbaix} {et~al.}(2002){Pourbaix}, {Nidever}, {McCarthy}, {Butler},
  {Tinney}, {Marcy}, {Jones}, {Penny}, {Carter}, {Bouchy}, {Pepe}, {Hearnshaw},
  {Skuljan}, {Ramm}, \& {Kent}}]{Pourbaix}
{Pourbaix}, D., {Nidever}, D., {McCarthy}, C., {et~al.} 2002, \aap, 386, 280

\bibitem[{{Powell}(1964)}]{Powell}
{Powell}, M.~J.~D. 1964, Computer Journal, 7, 155

\bibitem[{{Press} \& {Rybicki}(1989)}]{PressRybicki}
{Press}, W.~H. \& {Rybicki}, G.~B. 1989, \apj, 338, 277

\bibitem[{{Quirion} {et~al.}(2010){Quirion}, {Christensen-Dalsgaard}, \&
  {Arentoft}}]{Quiriongrid}
{Quirion}, P.-O., {Christensen-Dalsgaard}, J., \& {Arentoft}, T. 2010, \apj,
  725, 2176

\bibitem[{{Rao}(1945)}]{Rao}
{Rao}, C.~R. 1945, Bulletin of the Calcutta Mathematical Society, 37, 81

\bibitem[{{Reegen}(2004)}]{Reegen04}
{Reegen}, P. 2004, in IAU Symposium, Vol. 224, The A-Star Puzzle, ed.
  {J.~Zverko, J.~Ziznovsky, S.~J.~Adelman, \& W.~W.~Weiss}, 791--798

\bibitem[{{Roberts} {et~al.}(1987){Roberts}, {Lehar}, \& {Dreher}}]{Roberts87}
{Roberts}, D.~H., {Lehar}, J., \& {Dreher}, J.~W. 1987, \aj, 93, 968

\bibitem[{{Roxburgh}(2009{\natexlab{a}})}]{Roxburghpipeline}
{Roxburgh}, I.~W. 2009{\natexlab{a}}, \aap, 506, 435

\bibitem[{{Roxburgh}(2009{\natexlab{b}})}]{Roxburgh09}
{Roxburgh}, I.~W. 2009{\natexlab{b}}, \aap, 493, 185

\bibitem[{{Roxburgh} \& {Vorontsov}(2003)}]{RoxVor03}
{Roxburgh}, I.~W. \& {Vorontsov}, S.~V. 2003, \aap, 411, 215

\bibitem[{{Saio}(1981)}]{Saio81}
{Saio}, H. 1981, \apj, 244, 299

\bibitem[{{Salabert} {et~al.}(2011){Salabert}, {Ballot}, \&
  {Garc{\'{\i}}a}}]{Salabert2011}
{Salabert}, D., {Ballot}, J., \& {Garc{\'{\i}}a}, R.~A. 2011, \aap, 528, A25

\bibitem[{{Salabert} {et~al.}(2010){Salabert}, {Garc{\'{\i}}a}, \&
  {Mathur}}]{Salabertlikeratio}
{Salabert}, D., {Garc{\'{\i}}a}, R.~A., \& {Mathur}, S. 2010, Astronomische
  Nachrichten, in press [arXiv:1003.6076v3]

\bibitem[{{Samadi} {et~al.}(2007){Samadi}, {Georgobiani}, {Trampedach},
  {Goupil}, {Stein}, \& {Nordlund}}]{Samadi07}
{Samadi}, R., {Georgobiani}, D., {Trampedach}, R., {et~al.} 2007, \aap, 463,
  297

\bibitem[{{Samadi} {et~al.}(2002){Samadi}, {Goupil}, \& {Houdek}}]{Samadi02}
{Samadi}, R., {Goupil}, M.-J., \& {Houdek}, G. 2002, \aap, 395, 563

\bibitem[{{Scargle}(1982)}]{Scargle82}
{Scargle}, J.~D. 1982, \apj, 263, 835

\bibitem[{{Scargle}(1989)}]{Scargle89}
{Scargle}, J.~D. 1989, \apj, 343, 874

\bibitem[{{Schou}(1992)}]{Schou92}
{Schou}, J. 1992, PhD thesis, Aarhus University, Aarhus, Denmark

\bibitem[{{Schou} \& {Brown}(1994)}]{SchouBrown94}
{Schou}, J. \& {Brown}, T.~M. 1994, \aaps, 107, 541

\bibitem[{{Schou} \& {Buzasi}(2001)}]{SchouBuzasi01}
{Schou}, J. \& {Buzasi}, D.~L. 2001, in ESA Special Publication, Vol. 464,
  \emph{SOHO} 10/GONG 2000 Workshop: Helio- and Asteroseismology at the Dawn of
  the Millennium, ed. {A.~Wilson \& P.~L.~Pall{\'e}}, 391--394

\bibitem[{{Schuster}(1905)}]{Schuster}
{Schuster}, A. 1905, Proceedings of the Royal Society of London, 77, 136

\bibitem[{{Scott} \& {Berger}(2010)}]{ScottBerger}
{Scott}, J.~G. \& {Berger}, J.~O. 2010, Annals of Statistics, 38, 2587

\bibitem[{{Sellke} {et~al.}(2001){Sellke}, {Bayarri}, \& {Berger}}]{Sellke2001}
{Sellke}, T., {Bayarri}, M.~J., \& {Berger}, J.~O. 2001, American Statistician,
  55, 62

\bibitem[{{Severino} {et~al.}(2001){Severino}, {Magr{\`\i}}, {Oliviero},
  {Straus}, \& {Jefferies}}]{Severino}
{Severino}, G., {Magr{\`\i}}, M., {Oliviero}, M., {Straus}, T., \& {Jefferies},
  S.~M. 2001, \apj, 561, 444

\bibitem[{{Shannon}(1949)}]{Shannon}
{Shannon}, C.~E. 1949, Proceedings of the Institute of Radio Engineers, 37, 10

\bibitem[{{Shapley}(1914)}]{Shapley1914}
{Shapley}, H. 1914, \apj, 40, 448

\bibitem[{{Silva Aguirre} {et~al.}(2011{\natexlab{a}}){Silva Aguirre},
  {Ballot}, {Serenelli}, \& {Weiss}}]{Victorconvcore}
{Silva Aguirre}, V., {Ballot}, J., {Serenelli}, A.~M., \& {Weiss}, A.
  2011{\natexlab{a}}, \aap, 529, A63

\bibitem[{{Silva Aguirre} {et~al.}(2011{\natexlab{b}}){Silva Aguirre},
  {Chaplin}, {Ballot}, {Basu}, {Bedding}, {Serenelli}, {Verner}, {Miglio},
  {Monteiro}, {Weiss}, {Appourchaux}, {Bonanno}, {Broomhall}, {Bruntt},
  {Campante}, {Casagrande}, {Corsaro}, {Elsworth}, {Garc{\'{\i}}a}, {Gaulme},
  {Handberg}, {Hekker}, {Huber}, {Karoff}, {Mathur}, {Mosser}, {Salabert},
  {Sch{\"o}nrich}, {Sousa}, {Stello}, {White}, {Christensen-Dalsgaard},
  {Gilliland}, {Kawaler}, {Kjeldsen}, {Houdek}, {Metcalfe},
  {Molenda-{\.Z}akowicz}, {Thompson}, {Caldwell}, {Christiansen}, \&
  {Wohler}}]{Victordifferential}
{Silva Aguirre}, V., {Chaplin}, W.~J., {Ballot}, J., {et~al.}
  2011{\natexlab{b}}, \apjl, 740, L2

\bibitem[{{Smeyers}(1968)}]{Smeyers68}
{Smeyers}, P. 1968, Annales d'Astrophysique, 31, 159

\bibitem[{{Soriano} \& {Vauclair}(2008)}]{SorianoVauclair}
{Soriano}, M. \& {Vauclair}, S. 2008, \aap, 488, 975

\bibitem[{{Stahn} \& {Gizon}(2008)}]{StahnGizon}
{Stahn}, T. \& {Gizon}, L. 2008, \solphys, 251, 31

\bibitem[{{Stello} {et~al.}(2010){Stello}, {Basu}, {Bruntt}, {Mosser},
  {Stevens}, {Brown}, {Christensen-Dalsgaard}, {Gilliland}, {Kjeldsen},
  {Arentoft}, {Ballot}, {Barban}, {Bedding}, {Chaplin}, {Elsworth},
  {Garc{\'{\i}}a}, {Goupil}, {Hekker}, {Huber}, {Mathur}, {Meibom},
  {Sangaralingam}, {Baldner}, {Belkacem}, {Biazzo}, {Brogaard}, {Su{\'a}rez},
  {D'Antona}, {Demarque}, {Esch}, {Gai}, {Grundahl}, {Lebreton}, {Jiang},
  {Jevtic}, {Karoff}, {Miglio}, {Molenda-{\.Z}akowicz}, {Montalb{\'a}n},
  {Noels}, {Roca Cort{\'e}s}, {Roxburgh}, {Serenelli}, {Silva Aguirre},
  {Sterken}, {Stine}, {Szab{\'o}}, {Weiss}, {Borucki}, {Koch}, \&
  {Jenkins}}]{StelloNGC6819}
{Stello}, D., {Basu}, S., {Bruntt}, H., {et~al.} 2010, \apjl, 713, L182

\bibitem[{{Stello} {et~al.}(2007){Stello}, {Bruntt}, {Kjeldsen}, {Bedding},
  {Arentoft}, {Gilliland}, {Nuspl}, {Kim}, {Kang}, {Koo}, {Lee}, {Sterken},
  {Lee}, {Jensen}, {Jacob}, {Szab{\'o}}, {Frandsen}, {Csubry}, {Dind},
  {Bouzid}, {Dall}, \& {Kiss}}]{Stello07}
{Stello}, D., {Bruntt}, H., {Kjeldsen}, H., {et~al.} 2007, \mnras, 377, 584

\bibitem[{{Stello} {et~al.}(2009{\natexlab{a}}){Stello}, {Chaplin}, {Basu},
  {Elsworth}, \& {Bedding}}]{Stello09}
{Stello}, D., {Chaplin}, W.~J., {Basu}, S., {Elsworth}, Y., \& {Bedding}, T.~R.
  2009{\natexlab{a}}, \mnras, 400, L80

\bibitem[{{Stello} {et~al.}(2009{\natexlab{b}}){Stello}, {Chaplin}, {Bruntt},
  {Creevey}, {Garc{\'{\i}}a-Hern{\'a}ndez}, {Monteiro}, {Moya}, {Quirion},
  {Sousa}, {Su{\'a}rez}, {Appourchaux}, {Arentoft}, {Ballot}, {Bedding},
  {Christensen-Dalsgaard}, {Elsworth}, {Fletcher}, {Garc{\'{\i}}a}, {Houdek},
  {Jim{\'e}nez-Reyes}, {Kjeldsen}, {New}, {R{\'e}gulo}, {Salabert}, \&
  {Toutain}}]{Stellogrid}
{Stello}, D., {Chaplin}, W.~J., {Bruntt}, H., {et~al.} 2009{\natexlab{b}},
  \apj, 700, 1589

\bibitem[{{Tarrant} {et~al.}(2007){Tarrant}, {Chaplin}, {Elsworth},
  {Spreckley}, \& {Stevens}}]{Tarrant07}
{Tarrant}, N.~J., {Chaplin}, W.~J., {Elsworth}, Y., {Spreckley}, S.~A., \&
  {Stevens}, I.~R. 2007, \mnras, 382, L48

\bibitem[{{Tarrant} {et~al.}(2008){Tarrant}, {Chaplin}, {Elsworth},
  {Spreckley}, \& {Stevens}}]{Tarrant08}
{Tarrant}, N.~J., {Chaplin}, W.~J., {Elsworth}, Y., {Spreckley}, S.~A., \&
  {Stevens}, I.~R. 2008, \aap, 483, L43

\bibitem[{{Tassoul}(1980)}]{Tassoul80}
{Tassoul}, M. 1980, \apjs, 43, 469

\bibitem[{{ten Brummelaar} {et~al.}(2005){ten Brummelaar}, {McAlister},
  {Ridgway}, {Bagnuolo}, {Turner}, {Sturmann}, {Sturmann}, {Berger}, {Ogden},
  {Cadman}, {Hartkopf}, {Hopper}, \& {Shure}}]{CHARA}
{ten Brummelaar}, T.~A., {McAlister}, H.~A., {Ridgway}, S.~T., {et~al.} 2005,
  \apj, 628, 453

\bibitem[{{Torrence} \& {Compo}(1998)}]{TorrenceCompo}
{Torrence}, C. \& {Compo}, G.~P. 1998, Bulletin of the American Meteorological
  Society, 79, 61

\bibitem[{{Toutain} \& {Appourchaux}(1994)}]{ToutainAppour94}
{Toutain}, T. \& {Appourchaux}, T. 1994, \aap, 289, 649

\bibitem[{{Uytterhoeven} {et~al.}(2008){Uytterhoeven}, {Mathias}, {Poretti},
  {Rainer}, {Mart{\'{\i}}n-Ruiz}, {Rodr{\'{\i}}guez}, {Amado}, {Le Contel},
  {Jankov}, {Niemczura}, {Pollard}, {Brunsden}, {Papar{\'o}}, {Costa},
  {Valtier}, {Garrido}, {Su{\'a}rez}, {Kilmartin}, {Chapellier},
  {Rodr{\'{\i}}guez-L{\'o}pez}, {Marin}, {Aceituno}, {Casanova}, {Rolland}, \&
  {Olivares}}]{Uytter08}
{Uytterhoeven}, K., {Mathias}, P., {Poretti}, E., {et~al.} 2008, \aap, 489,
  1213

\bibitem[{{Vandakurov}(1967)}]{Vandakurov67}
{Vandakurov}, Y.~V. 1967, \azh, 44, 786

\bibitem[{{Vauclair} {et~al.}(2008){Vauclair}, {Laymand}, {Bouchy}, {Vauclair},
  {Hui Bon Hoa}, {Charpinet}, \& {Bazot}}]{VauclairHor}
{Vauclair}, S., {Laymand}, M., {Bouchy}, F., {et~al.} 2008, \aap, 482, L5

\bibitem[{{Verner} {et~al.}(2011{\natexlab{a}}){Verner}, {Chaplin}, {Basu},
  {Brown}, {Hekker}, {Huber}, {Karoff}, {Mathur}, {Metcalfe}, {Mosser},
  {Quirion}, {Appourchaux}, {Bedding}, {Bruntt}, {Campante}, {Elsworth},
  {Garc{\'{\i}}a}, {Handberg}, {R{\'e}gulo}, {Roxburgh}, {Stello},
  {Christensen-Dalsgaard}, {Gilliland}, {Kawaler}, {Kjeldsen}, {Allen},
  {Clarke}, \& {Girouard}}]{KIC2}
{Verner}, G.~A., {Chaplin}, W.~J., {Basu}, S., {et~al.} 2011{\natexlab{a}},
  \apjl, 738, L28

\bibitem[{{Verner} {et~al.}(2006){Verner}, {Chaplin}, \&
  {Elsworth}}]{Vernerglitches}
{Verner}, G.~A., {Chaplin}, W.~J., \& {Elsworth}, Y. 2006, \apj, 638, 440

\bibitem[{{Verner} {et~al.}(2011{\natexlab{b}}){Verner}, {Elsworth}, {Chaplin},
  {Campante}, {Corsaro}, {Gaulme}, {Hekker}, {Huber}, {Karoff}, {Mathur},
  {Mosser}, {Appourchaux}, {Ballot}, {Bedding}, {Bonanno}, {Broomhall},
  {Garc{\'{\i}}a}, {Handberg}, {New}, {Stello}, {R{\'e}gulo}, {Roxburgh},
  {Salabert}, {White}, {Caldwell}, {Christiansen}, \&
  {Fanelli}}]{Vernercomparison}
{Verner}, G.~A., {Elsworth}, Y., {Chaplin}, W.~J., {et~al.} 2011{\natexlab{b}},
  \mnras, 415, 3539

\bibitem[{{Verner} \& {Roxburgh}(2011)}]{Vernerpipeline}
{Verner}, G.~A. \& {Roxburgh}, I.~W. 2011, Astronomische Nachrichten, in press
  [arXiv:1104.0631v1]

\bibitem[{{Walker}(2008)}]{Walker08}
{Walker}, G.~A.~H. 2008, Journal of Physics Conference Series, 118, 012013

\bibitem[{{Walker} {et~al.}(2003){Walker}, {Matthews}, {Kuschnig}, {Johnson},
  {Rucinski}, {Pazder}, {Burley}, {Walker}, {Skaret}, {Zee}, {Grocott},
  {Carroll}, {Sinclair}, {Sturgeon}, \& {Harron}}]{WalkerMOST}
{Walker}, G.~A.~H., {Matthews}, J.~M., {Kuschnig}, R., {et~al.} 2003, \pasp,
  115, 1023

\bibitem[{{White} {et~al.}(2011{\natexlab{a}}){White}, {Bedding}, {Stello},
  {Appourchaux}, {Ballot}, {Benomar}, {Bonanno}, {Broomhall}, {Campante},
  {Chaplin}, {Christensen-Dalsgaard}, {Corsaro}, {Do{\u g}an}, {Elsworth},
  {Fletcher}, {Garc{\'{\i}}a}, {Gaulme}, {Handberg}, {Hekker}, {Huber},
  {Karoff}, {Kjeldsen}, {Mathur}, {Mosser}, {Monteiro}, {R{\'e}gulo},
  {Salabert}, {Silva Aguirre}, {Thompson}, {Verner}, {Morris}, {Sanderfer}, \&
  {Seader}}]{obsdiagramsWhite}
{White}, T.~R., {Bedding}, T.~R., {Stello}, D., {et~al.} 2011{\natexlab{a}},
  \apjl, 742, L3

\bibitem[{{White} {et~al.}(2011{\natexlab{b}}){White}, {Bedding}, {Stello},
  {Christensen-Dalsgaard}, {Huber}, \& {Kjeldsen}}]{theodiagramsWhite}
{White}, T.~R., {Bedding}, T.~R., {Stello}, D., {et~al.} 2011{\natexlab{b}},
  \apj, in press [arXiv:1109.3455v1]

\bibitem[{{White} {et~al.}(2010){White}, {Brewer}, {Bedding}, {Stello}, \&
  {Kjeldsen}}]{White10}
{White}, T.~R., {Brewer}, B.~J., {Bedding}, T.~R., {Stello}, D., \& {Kjeldsen},
  H. 2010, Communications in Asteroseismology, 161, 39

\bibitem[{{Wiener}(1930)}]{Wiener}
{Wiener}, N. 1930, Acta Mathematica, 55, 117

\bibitem[{{Wilks}(1938)}]{Wilks38}
{Wilks}, S.~S. 1938, Annals of Mathematical Statistics, 9, 60

\bibitem[{{Woodard}(1984)}]{Woodard84}
{Woodard}, M.~F. 1984, PhD thesis, University of California, San Diego, USA

\bibitem[{{Zechmeister} \& {K{\"u}rster}(2009)}]{Zechmeister}
{Zechmeister}, M. \& {K{\"u}rster}, M. 2009, \aap, 496, 577

\bibitem[{{Zhevakin}(1963)}]{Zhevakin1963}
{Zhevakin}, S.~A. 1963, \araa, 1, 367

\bibitem[{{Zwintz} {et~al.}(1999){Zwintz}, {Kuschnig}, {Weiss}, {Gray}, \&
  {Jenkner}}]{Zwintz99}
{Zwintz}, K., {Kuschnig}, R., {Weiss}, W.~W., {Gray}, R.~O., \& {Jenkner}, H.
  1999, \aap, 343, 899

\end{thebibliography}

\end{footnotesize}
\end{multicols}








\begin{appendices}

\chapter{}\label{vichinature}
\includepdf[pages=-]{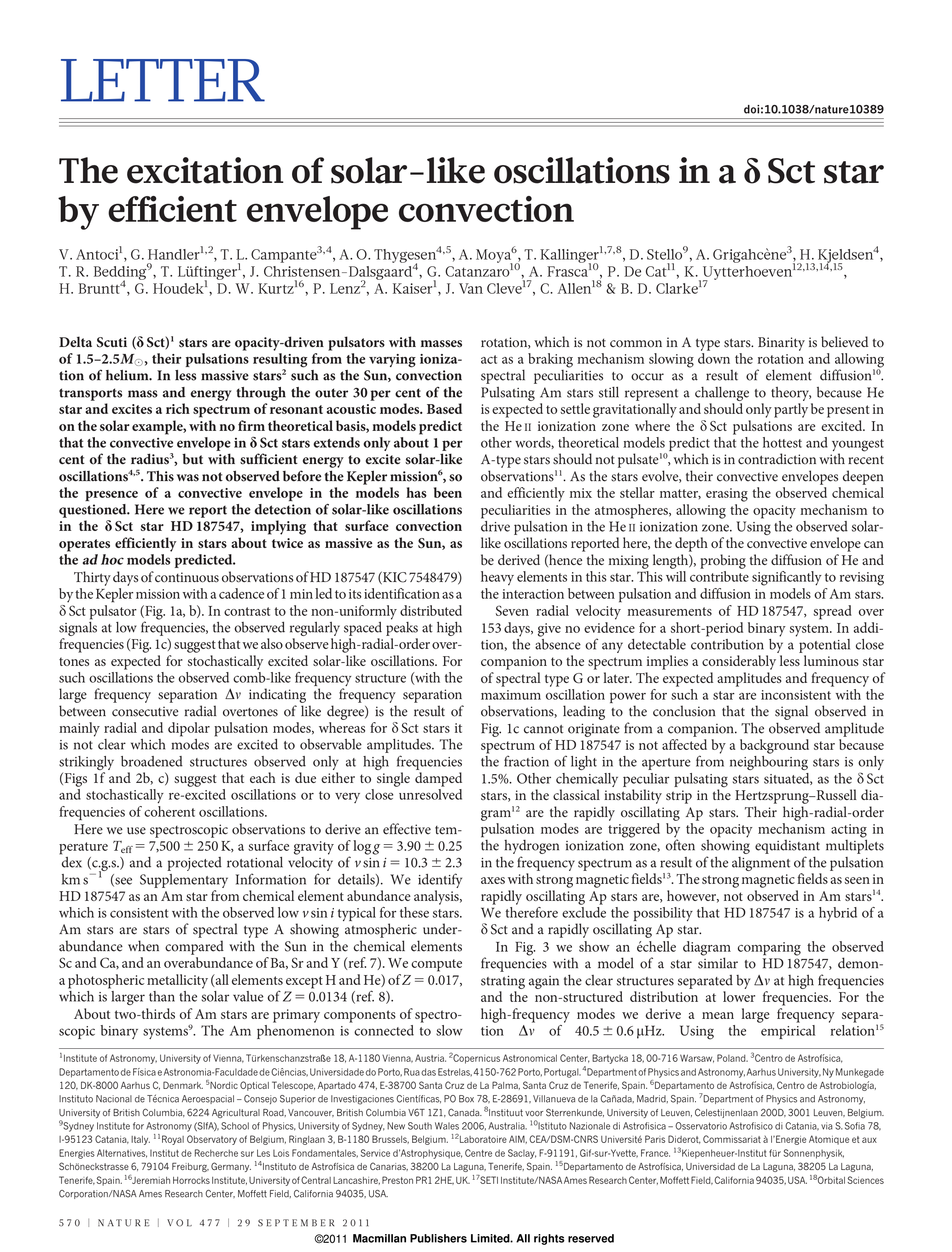}
\thispagestyle{plain}

\chapter{}\label{beddingprocyon}
\includepdf[pages=-]{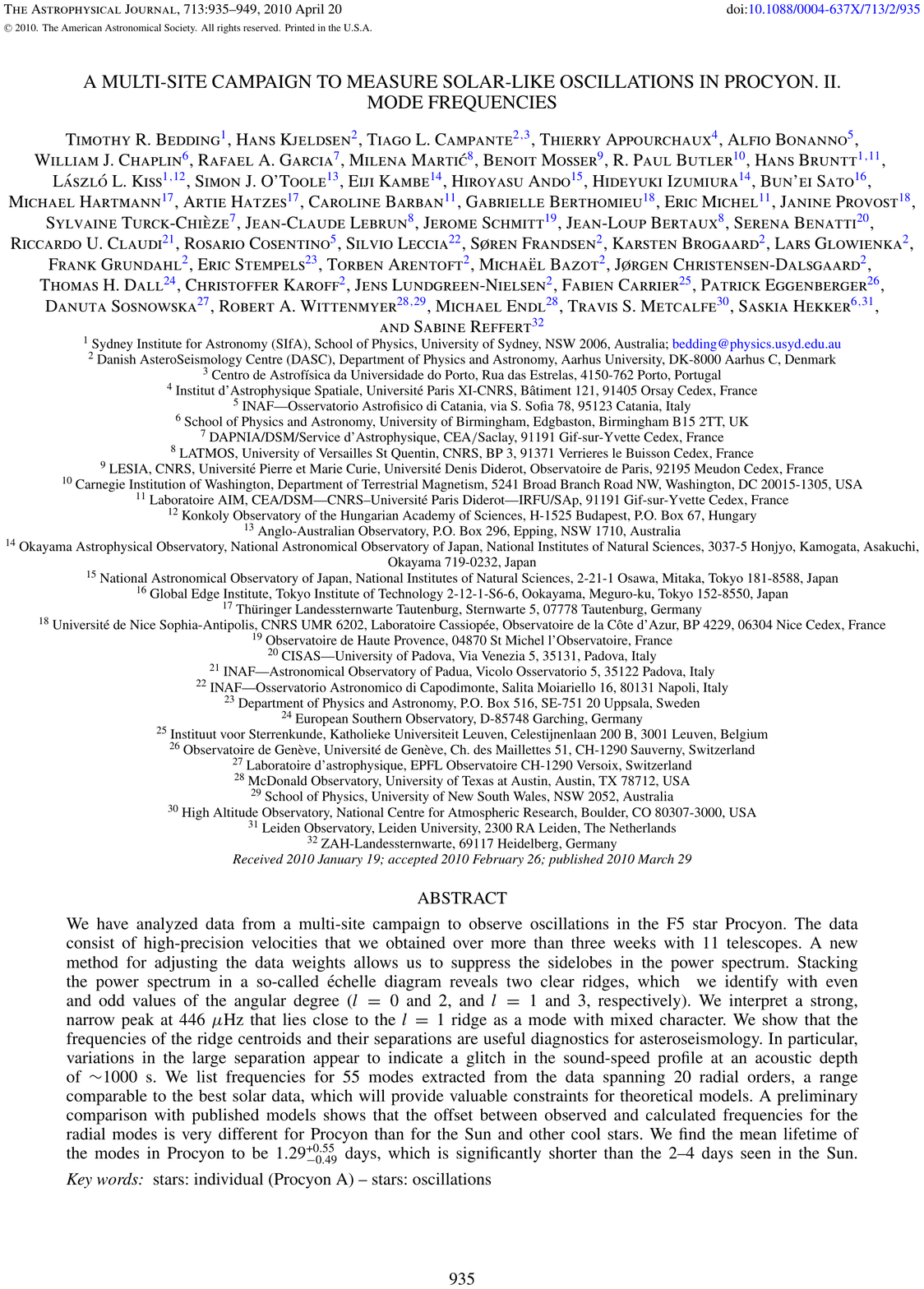}

\chapter{}\label{campantekepler}
\includepdf[pages=-]{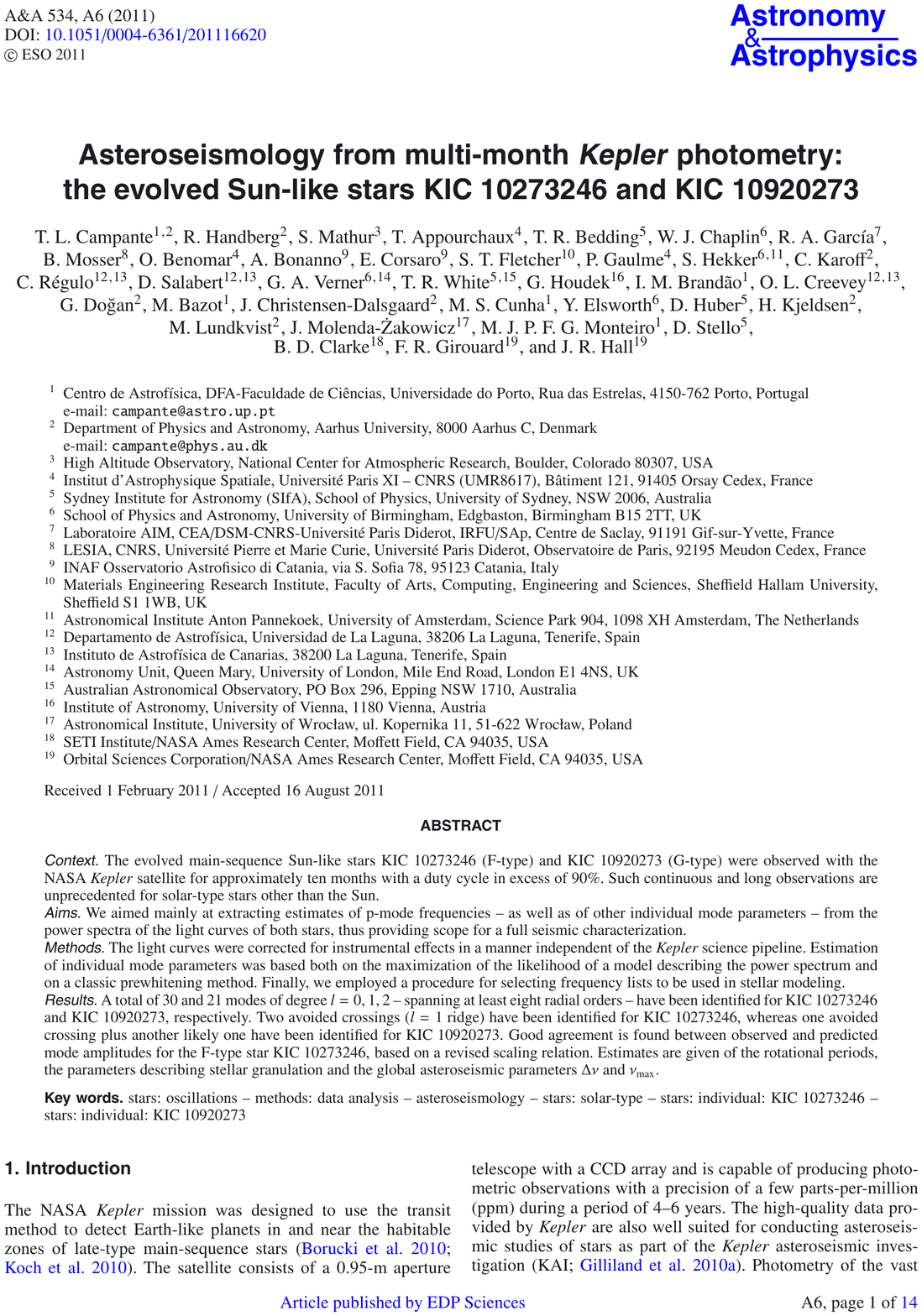}
\thispagestyle{plain}

\chapter{}\label{handcamp}
\includepdf[pages=-]{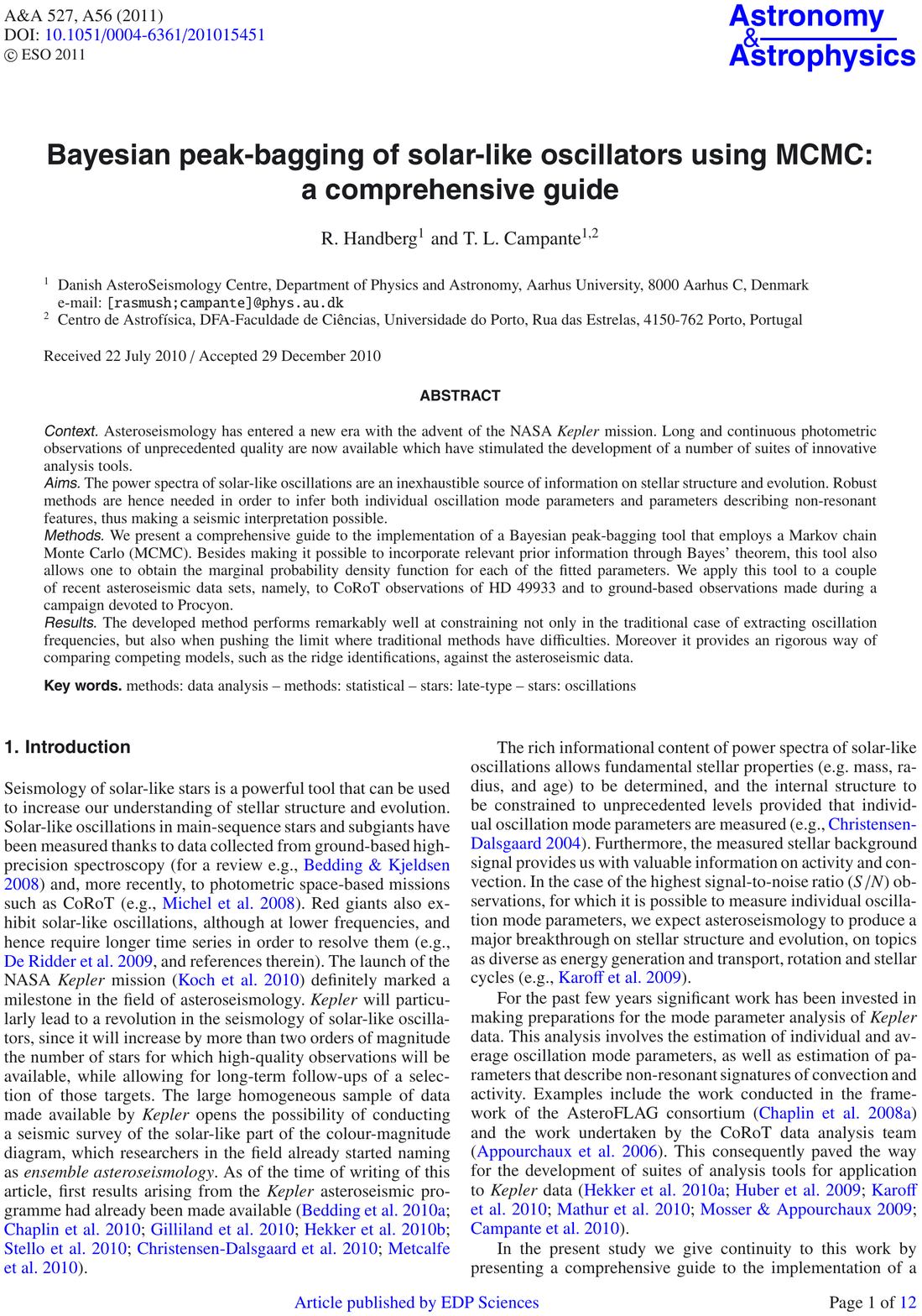}
\thispagestyle{plain}

\chapter{}\label{campanteautocov}
\includepdf[pages=-]{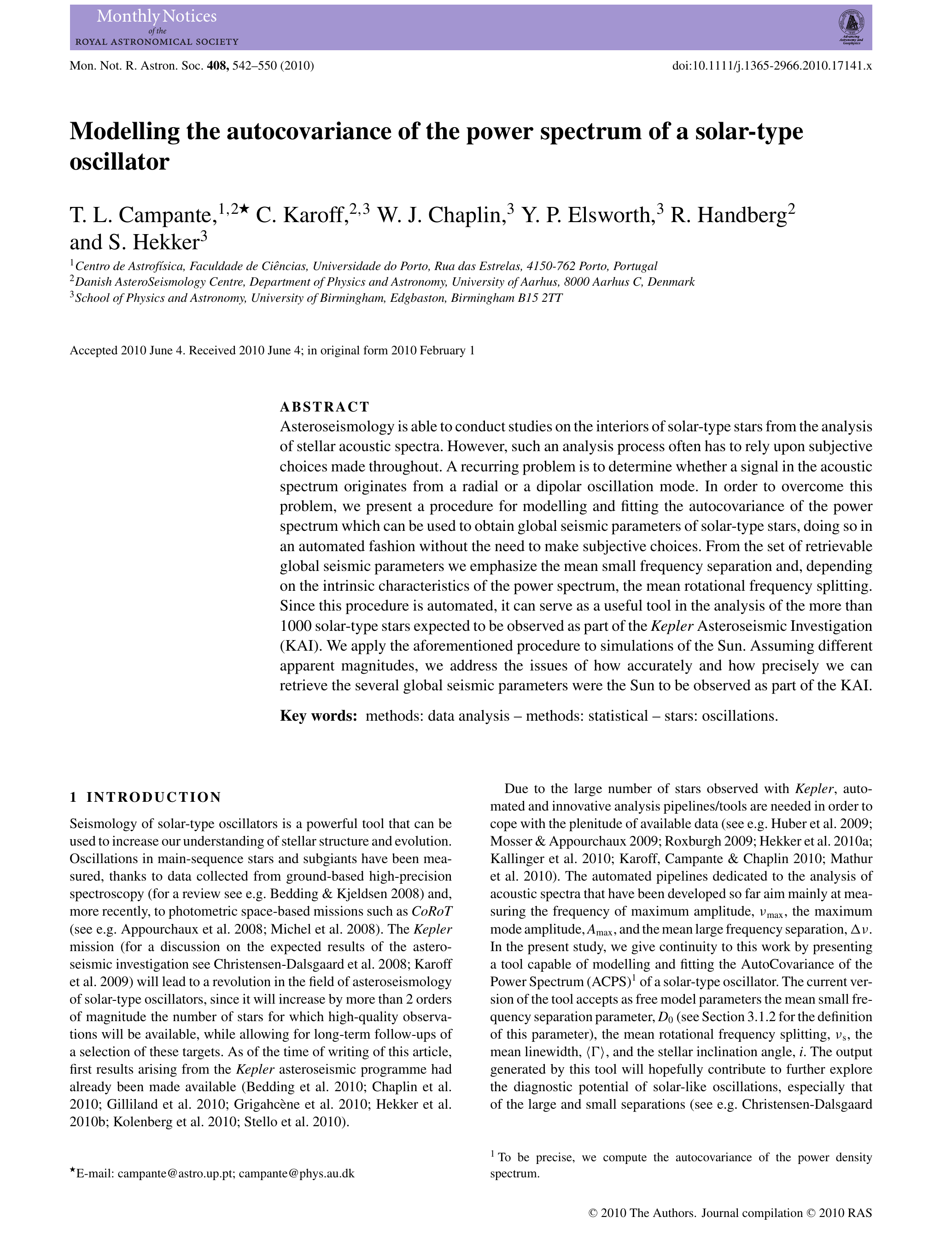}

\end{appendices}

\end{document}